\documentclass{interact}
\usepackage{natbib}
\usepackage[T1]{fontenc}
\usepackage[ngerman, english]{babel} %For \glq and \grq <http://texnical-designs.com/quotation-marks-and-the-latex-dirtytalk-package/>
\PassOptionsToPackage{hyphens}{url}\usepackage{hyperref}
\expandafter\def\csname ver@amsmath.sty\endcsname{2019/01/01} %For mathtools package
\expandafter\def\csname aligned@a\endcsname{} %For mathtools package
\usepackage{mathtools} %For \overbracket and \underbracket
\usepackage{orcidlink}
\usepackage{enumerate}
\usepackage{subcaption} % For \begin{subfigure} + for \ContinuedFloat - subfigures of a figure on multiple pages
\usepackage{tikz}
\graphicspath{{./Graphics/}}
\usepackage{bm}
\usepackage{multirow}
\usepackage{multicol}
\usepackage{array}
\newcolumntype{L}{>{$}l<{$}} % math-mode version of "l" column type
\newcolumntype{C}{>{$}c<{$}} % math-mode version of "c" column type
\newcolumntype{N}{>{\bfseries}p{0.04\textwidth}}
\newcolumntype{M}[1]{>{\centering\arraybackslash}p{#1}}
\usepackage{makecell} %For \thead
\usepackage{footnote} %For \footnote inside tabular
\makesavenoteenv{tabular}
\usepackage[flushleft]{threeparttable} %For note after tabular
\usepackage{longtable}
\usepackage[multiple]{footmisc} %To separate footnote numbers
\usepackage{transparent}
\usepackage{textgreek} % For \textomega

\usepackage{collectbox}
\newcommand{\tightbox}{%
    \collectbox{%
        \setlength{\fboxsep}{1pt}%
        \fcolorbox{red}{white}{\BOXCONTENT}%
    }%
}

\usepackage{chngcntr}
\theoremstyle{definition}
\newtheorem{syl}{Syllogism}
\counterwithin*{syl}{section}

\DeclareMathOperator{\termlogicassertion}{\text{?}}
\DeclareMathOperator{\termlogicA}{A}
\DeclareMathOperator{\termlogicE}{E}
\DeclareMathOperator{\termlogicI}{I}
\DeclareMathOperator{\termlogicO}{O}
\DeclareMathOperator{\termlogicAumlaut}{\text{Ä}}
\DeclareMathOperator{\termlogicEumlaut}{\text{Ë}}
\DeclareMathOperator{\termlogicIumlaut}{\text{Ï}}
\DeclareMathOperator{\termlogicOumlaut}{\text{Ö}}

\newcommand\metaeq{\mathrel{~|\!\!=\!\!|~}}
\newcommand\metale{\mathrel{~|\!\!=}}
\newcommand\metanleq{~\bm{\not}\mathrel{|\!\!=}}

\newcommand\variabletextvisiblespace[1][.3em]{%
  \mbox{\kern.06em\vrule height.3ex}%
  \vbox{\hrule width#1}%
  \hbox{\vrule height.3ex}}

\begin{document}

\title{Proofs of valid categorical syllogisms in one diagrammatic and two symbolic axiomatic systems}

\author{
\name{Antonielly Garcia Rodrigues\textsuperscript{a} and Eduardo Mario Dias\textsuperscript{b}}
\affil{\textsuperscript{a}Brazilian Health Regulatory Agency -- ANVISA, Gerência-Geral de Tecnologia da Informação (GGTIN), SIA Trecho 05 - Guará, Brasília-DF, Brazil. ZIP 71205-050.\\\orcidlink{0000-0001-5911-975X}ORC{\color{teal}ID} 0000-0001-5911-975X\\antonielly@gmail.com\vspace{0.25cm}\\\textsuperscript{b}University of São Paulo, Avenida Professor Luciano Gualberto, Travessa 3, 158, Sala A2-18, Cidade Universitária, São Paulo-SP, Brazil. ZIP 05508-900.\\\orcidlink{0000-0002-2104-1747}ORC{\color{teal}ID} 0000-0002-2104-1747\\emdias@pea.usp.br}
}

\maketitle

\begin{abstract}
Gottfried Leibniz embarked on a research program to prove all the Aristotelic categorical syllogisms by diagrammatic and algebraic methods. He succeeded in proving them by means of Euler diagrams, but didn't produce a manuscript with their algebraic proofs. We demonstrate how key excerpts scattered across various Leibniz's drafts on logic contained sufficient ingredients to prove them by an algebraic method --which we call the Leibniz-Cayley (LC) system-- without having to make use of the more expressive and complex machinery of first-order quantificational logic. In addition, we prove the classic categorical syllogisms again by a relational method --which we call the McColl-Ladd (ML) system-- employing categorical relations studied by Hugh McColl and Christine Ladd. Finally, we show the connection of ML and LC with Boolean algebra, proving that ML is a consequence of LC, and that LC is a consequence of the Boolean lattice axioms, thus establishing Leibniz's historical priority over George Boole in characterizing and applying (a sufficient fragment of) Boolean algebra to effectively tackle categorical syllogistic.%This paper fits squarely into the Leibnizian/Boolean research program of algebraically formulating and proving all classic categorical syllogisms which are the focus of the tradition of logic founded by Aristotle and nowadays known as ``term logic''. We mathematically prove all classic categorical syllogisms by means of easy-to-understand techniques developed in the Sturm-Venn diagrammatic tradition and Leibniz-Boole algebraic tradition, without having to make use of the more expressive and complex machinery of first-order quantificational logic pioneered by Gottlob Frege and, independently, by Charles Sanders Peirce. We consolidate the required formulations, axioms and proofs of the 24 classic categorical syllogisms into one diagrammatic system --the Euler system-- and two algebraic axiomatic systems --the Leibniz-Cayley system (LC) and the McColl-Ladd system (ML)-- and prove that ML is a strict consequence of LC, and that LC is a strict consequence of the Boolean lattice axioms.
\end{abstract}


\section{Introduction}

This paper is about some interesting ways, aligned with the ``algebra of logic'' tradition, one can prove the 24 classic categorical syllogistic moods. It offers the following contributions:

\begin{enumerate}
    \item A reproduction of short \emph{diagrammatic} proofs of the categorical syllogisms, pioneered by \citet{leibniz:1686e} and didactically expressed in modern form by \citet{piesk:2017}. (This is not a novel contribution\footnote{We provide a minor contribution here by suggesting that interpreting the boundaries (and not only the shaded areas) of classes in Euler and Venn diagrams as the empty class is convenient, as it perfectly fits some elementary laws of the algebra of sets.} of this paper, but is useful to introduce key concepts and motivate the following sections, which contain novel contributions.)

    \item The compilation in two Tables (\ref{tab:relations_with_identity} and \ref{tab:relations_with_single_symbol}) of the various ways one can algebraically represent the fundamental Aristotelic relations (and \citeauthor{de_morgan:1846}'s extensions) by means of elementary algebraic operations and relations: union, intersection, complementation; the empty class, the universe; equality, subset and superset, disjointness and exhaustion, and the negation of those relations.

    \item The characterization of two \emph{symbolic} axiomatic systems, LC (for Leibniz-Cayley) and ML (for McColl-Ladd), and the systematic proof of all the 24 classic categorical syllogistic moods in them. LC, an algebraic system, is built upon intersection, complementation, identity, non-identity, and the empty class. ML, a relational system, is built upon complementation and the subset and conjointness relations.

    \item Tables pointing out which syllogistic moods require which axioms of LC and ML (Sections \ref{subsec:LC_matrix} and \ref{subsec:ML_matrix}).

    \item A discussion of key excerpts of Leibniz's drafts which demonstrates that, in the late 17th century, he presented nearly all building blocks for algebraically proving the 24 classic categorical from axioms -- doing away with the misconception that the solution of categorical syllogistic in the algebra of logic tradition had to wait the invention of Boolean algebra in the 19th century.

    \item A proof that ML is strictly less expressive than LC, which itself is strictly less expressive than Boolean algebra -- thus forming a hierarchy of logical systems based on expressiveness.

    \item Historical notes throughout the text, which can guide History of Logic researchers to key primary literature for their investigations.
\end{enumerate}

The categorical syllogism proofs expressed by means of Euler and Venn diagrams by \citet{piesk:2017} (reproduced in Section~\ref{sec:euler}) sparkled our motivation for the research that led to this paper. Faced with those elegant diagrammatic proofs, we were irresistibly attracted to the intellectual exercise of proving the same 24 categorical syllogisms in the (Boolean) algebra of (term\footnote{In this paper, we use ``term'' to designate an extensional class, which secondarily happens to be associated to a linguistic entity such as subject or predicative. A (classificatory) term is a reference --a label-- to a class in extension. We follow \citeauthor{de_morgan:1846}'s (\citeyear{de_morgan:1846}) extensional approach: ``\emph{A term, or name, is merely the word which it is lawful to apply to any one of a collection of objects of thought [...].}''. (For a dissenting view on ``terms'' in logic, see \citet[pp.~124,125-126]{waragai:2007}, \citet{kulicki:2012}, and \citet[p.~130]{lukaziewicz:1957}.)}) logic. As we delved deep into the literature of the field of algebraic term logic, we learned that this is one of the historical goals of the Leibnizian-Boolean research program --a research goal which dates back to at least April 1679 \citep[pp.~43-44]{leibniz:1679a}--, which is, surprisingly, only partially completed to this day -- by neglect, not because (with tools available nowadays) it is a tough problem. We stand on the shoulders of giants to comprehensively document solutions to this goal by employing modern notation and concepts refined and matured over centuries of hard work by symbolic logicians.

Most elements required to establish the axioms for all the diagrammatic and algebraic systems we describe here were anticipated in unpublished drafts by Gottfried Wilhelm Leibniz in the 17th century -- long before George \citet{boole:1847} and his intellectual successors came to the scene. A few of these elements had already been published and were in principle accessible to 19th century mathematical logic researchers \mbox{\citep{erdmann:1840, gerhardt:1890, peckhaus:2018}}, but others came to the public eye thanks to Louis Couturat's examination and publication, in the 20th century, of selected drafts on logic which were on Leibniz's \emph{Nachlass} in Hanover \citep{couturat:1903}, which have fortunately been preserved to the present day. Because many relevant drafts of Leibniz's were unknown in the 19th century, there were a lot of independent reinventions of his ideas by various pioneers in the Boolean research program\footnote{Leibniz explored both extensional and intensional interpretations of the same abstract, deductive logical system. The intensional interpretation, actually favored by the rationalist Leibniz across many of his manuscripts \citep[pp.~13-18,32,35-37,73-74,186-187,213-215,231,322-323,327-330,377,382-385]{lewis:1918}\citep[points~11-12,7,17]{leibniz:1679b}, is very interesting in itself. In this paper, however, we are concerned with characterizing noteworthy properties of (fragments of) purely deductive, formal, \emph{extensional} term logics, supportive of arbitrarily assembled classes, whose constituent elements are left implicit in the systems.

The brand of algebra of extensional categorical syllogistic we discuss at length in this paper assumes the existence of logical classes/``categories''. Some philosophers find them problematic and sometimes don't accept (or at least attempt to work around) them \citep{boolos:1985, ongley:2013, klement:2010}.
}.%Leibniz favored an intensional interpretation: \citep[pp.~264-267]{ferrari:1989}

The tradition of term logic was initiated circa 350 BCE by the founding master \citet{owen:1853a}(\citeyear{owen:1853b}), who introduced four fundamental categorical relation forms and enumerated (and proved by means of logical methods) most of the classic categorical syllogisms we algebraically prove in this paper. Aristotle had key insights such as earliest recorded use of literal \emph{placeholders} (``variables'') in what would millennia later be recognized as a branch of algebra\footnote{The use of literal placeholders in numerical algebra would have to wait the independent reinvention by Jordanus de Nemore circa 1,225 CE, that is, approximately 1,600 years after Aristotle \citep{de_nemore:1225, turner:1983}.} \citep[pp.~7-8]{lukaziewicz:1957}\citep[p.~12]{patzig:1968}\citep[pp.~23,34,39]{bar-am:2008}\citep{braem:1475}.%https://books.google.com/books?id=dWp2HUQl-0UC&pg=PA39&dq=%22Aristotle+is+the+first+that+we+know+of+to+introduce+variables+explicitly%22&redir_esc=y
%https://en.wikipedia.org/wiki/Logical_form#History

The 24 classic categorical syllogisms employ four possibilities of fundamental categorical relations originally investigated by Aristotle in [gr] ``\emph{Peri Hermeneias}'' / [la] ``\emph{De Interpretatione}'' / [en] ``On Interpretation'' (\citeyear[Chapter~7]{owen:1853a}) and in [gr] ``\emph{Analytica Protera}'' / [la] ``\emph{Analytica Priora}'' / [en] ``Prior Analytics'' \citep{owen:1853b}, and popularized by Boethius \citep{parsons:2021} through linguistic expressions loosely similar to the following:
\begin{itemize}
    \item $\mathbf{b}\termlogicA \mathbf{c}$: Every \textbf{b} is \textbf{c}.
    \item $\mathbf{b}\termlogicE \mathbf{c}$: No \textbf{b} is \textbf{c}. (Every \textbf{b} is not \textbf{c}.)
    \item $\mathbf{b}\termlogicI \mathbf{c}$: At least one\footnote{We follow \citet[pp.~6-11,18]{beziau:2012} in adopting ``At least one'' rather than ``Some''. It represents the set-theoretically precise notion of ``inhabitation'' or ``presence'' (often miscalled ``existence'', even --by habit-- in this paper.). Moreover, unlike ``Some'', ``At least one'' invites the generalization toward numerical syllogisms \citep[pp.~5-6,225-228,249]{pratt-hartmann:2023}, an exciting and active research topic pioneered by \citet[pp.~384,406]{de_morgan:1846}.} \textbf{b} is \textbf{c}.
    \item $\mathbf{b}\termlogicO \mathbf{c}$: At least one \textbf{b} is not \textbf{c}.
\end{itemize}

Aristotle's syllogistic was the earliest simultaneous treatment of \emph{generality} (``Every \textbf{b} is \textbf{c}.'') and \emph{existence} (``At least one \textbf{b} is \textbf{c}.'') in logic, and he had already recognized how those notions are intimately related in that generality means the lack (or non-existence) of a counterexample \citep{von_plato:2021}\footnote{This ancient observation by Aristotle is precisely formalized (with some abuse of notation) in modern first-order logic as

$\underset{x\in \mathbf{I}}{\forall}\!:~x\in \mathbf{b} \Rightarrow x\in \mathbf{c} \quad\metaeq\quad \underset{x\in \mathbf{I}}{\overline\exists}\!:~x\in \mathbf{b} \land x\notin \mathbf{c}$

In term logic, it is expressed much more simply as

$\mathbf{b}\termlogicA \mathbf{c} \metaeq \mathbf{b}\overline\termlogicO \mathbf{c}$

that is, $\termlogicA$ and $\termlogicO$ are contradictory.}\footnote{In this paper, we decided to adopt the following notation:

\vspace{-\topsep}
\begin{itemize}%[topsep=-\parskip]
    \item  ``\!\!$\metaeq$\!\!'' is the metalogical relation of equivalence, which indicates that the syntactic derivation is \emph{bi}directionally valid. It was intentionally chosen to be an ``equals'' symbol sandwiched between two vertical bars. It is most often expressed in sequent calculus and proof theory by the ``$\dashv\vdash$'' symbol.%https://math.stackexchange.com/questions/1164127/meaning-of-dashv-vdash
    %dyadic turnstile, monadic turnstile and two turnstiles: [pp.~185-186] J. Michael Dunn; Gary M. Hardegree. Algebraic Methods in Philosophical Logic. Oxford University Press. Oxford:UK. 2001. https://books.google.com/books?id=-AokWhbILUIC&pg=PA186
    %two turnstiles - pp. 28,17; monadic turnstile - p. 35: Colin Allen; Michael Hand. Logic Primer, 2nd edition. MIT Press. Cambridge:USA. 2001. https://books.google.com/books?id=RSTYAgAAQBAJ&pg=PA28&dq=sequent+turnstile
    %from dyadic to monadic turnstile - a specialization of the deductive theorem: https://books.google.com/books?id=5Funj1Zaau0C&pg=PA70&dq=%22the+following+equivalence%22

    \item ``$\!\!\metale$'' is the metalogical relation of \emph{uni}directional syntactic derivation, most often expressed in sequent calculus and proof theory by the turnstile symbol (``$\vdash$''). The previous symbol (``\!\!$\metaeq$\!\!'') was a ``sandwich''; this one (``$\!\!\metale$'') is ``half a sandwich''. We find it an unfortunate historical accident that the double turnstile symbol (``$\vDash$''), that visually looks like our ``\!\!$\metale$'', is commonly used in the model theory literature to indicate semantic, not syntactic entailment. It would have been nice if the denotation of ``$\vdash$'' and ``$\vDash$'' were swapped.%https://math.stackexchange.com/questions/239363/notation-for-a-does-not-imply-b#239371

    \item ``,'' is the metalogical ``and'' operation.

    \item ``${\scriptstyle|\!\lor\!|}$'' is the metalogical ``or'' operation, following our convention of going to the meta, metameta, metametameta level and so on by progressively adding vertical bars around a symbol.
\end{itemize}%\vspace{-0.5cm}

%hilbert:1922: pp. 18-19 of the PDF file
\citet[pp.~174-175]{hilbert:1922} employed the term ``metamathematics'', back then in the narrow sense of (finitary) proof theory (``\emph{Beweistheorie}''). In the Polish school of logic, \citet{lukaziewicz:1930} Łukasiewicz and Tarski (1930) employed the words [de]``\emph{Metalogik}'' ([en]``metalogic'') and [de]``\emph{metalogischen}'' ([en]``metalogical'') as synonyms to ``metamathematics'' and ``metamathematical'', in the wider sense of ``(pertaining to the) theory of deduction''. The ``object language vs. metalanguage'' distinction was made explicit by \citet[pp.~167-168,154\{fn.~1\}]{tarski:1933}, who, in a later article (\citeyear[p.~402]{tarski:1936}), credited his doctoral advisor Stanisław Leśniewski with pioneering it. Moreover, in the latter article, the words ``metalogical'' and ``metalanguage'' appear in the same page: 407.}.

More than two millennia after Aristotle, Leibniz, knowledgeable in both numerical algebra and term logic, decided to research the possibility of turning term logic into an algebra, making use of the fact that both exact sciences already employed literal placeholders back in his days. These were pioneering feats in what we nowadays call ``Boolean algebra''.

Key ideas beyond Leibniz's for achieving a symbolic algebra of term logic powerful enough to offer various insightful ways to prove classic categorical syllogisms were advanced by \citet{cayley:1871}, \citet{mccoll:1877}, \citet{ladd:1883}, and \citet{mitchell:1883}, and are summarized in this paper.

Our terminology divides the categorical syllogism proof methods into:

\begin{enumerate}
    \item \emph{Diagrammatic}, where representations and proofs are topologically visual rather than symbolic;

    \item \emph{Algebraic}, where transformations involving ``$=$'', ``$\neq$'' and dyadic operations (\emph{functions}) on classes are employed in a way somewhat familiar to numerical algebra students at middle school;
    %operational/equational

    \item \emph{Relational}, where axioms involving dyadic \emph{relations} among classes other than ``$=$'' and ``$\neq$'' are used, and proofs employ either free-form logical entailment (from logic) or strict-form \emph{composition} of relations (from relation algebra);

    \item \emph{Refutatory}, where concepts and tools typical in propositional logic or involving the Theorem K from relation algebra predominate;
    %conclusion denial / backtracking

    \item \emph{Quantificational}, where concepts and tools from first-order monadic quantificational logic are employed.

\end{enumerate}

Aristotle was the earliest to prove many classic categorical syllogisms, using logical \emph{refutatory} methods involving consequence denial such as ``\emph{reductio ad absurdum}'' and what was later called ``proof by regression''\footnote{Useful in various contexts, for instance to show that \emph{modus ponens} and \emph{modus tollens} can be derived from each other, and to establish alternative definitions for antisymmetric relation in order theory:

$b\preceq c,~b\overbracket[0.6pt][0.7ex]{\neq c\quad\!\metale\quad c\npreceq} b$

$b\preceq c,~c\underbracket[0.6pt][0.7ex]{\preceq b\quad\!\metale\quad b=}c$}. Leibniz proved some categorical syllogisms with diagrammatic methods \citep{leibniz:1686e} and beautifully explained Aristotle's method of proof by regression \citep{leibniz:1682}, where a premise and the conclusion of a valid assertion are transposed, generating a new valid assertion. \citet[pp.~87-89]{de_morgan:1847} repeated this application of regression to categorical syllogistic. An evolved variant of regression is the method of ``inconsistent triads'' or ``antilogisms'' by \citet{ladd:1883}\citep[pp.~108-110]{lewis:1918}\citep[pp.~2-3]{green:1991}. In another evolutionary direction from proof by regression, De Morgan proved the Theorem K in relation algebra \citep[p.~344]{de_morgan:1860}\citep[pp.~434-435]{maddux:1991}\citep[pp.~242-243,416-417]{schroeder:1895}. But all these methods intentionally use tools of propositional logic instead of algebraic transformations.
%Proof by regression
%https://books.google.com/books?id=HscAAAAAMAAJ&pg=PA87&dq=%22every+universal+syllogism+has+two+particular%22
%De Syllogismo Categorico ex Inclusione Exclusioneve Terminorum, point~5
%https://archive.org/details/diephilosophisc01leibgoog/page/208/mode/1up https://www.uni-muenster.de/Leibniz/DatenVI5/op1120.pdf https://www.uni-muenster.de/Leibniz/bd_6_5_einzel.html
%pp. 2-4: Charles Sanders Peirce. Memoranda a propósito del silogismo aristotélico. http://www.unav.es/gep/PeirceMemoranda1866.pdf
%Theorem K
%https://books.google.com/books?id=2vIIAAAAIAAJ&pg=PA344&dq=%22I+shall+call+this+result+theorem+K%22+%22opponent+syllogisms%22

Leibniz made across private drafts various insightful attempts to devise an \emph{algebraic} method for proving categorical syllogisms\footnote{Leibniz not only algebraized Aristotelic logic; in fact, Leibniz's logic goes beyond categorical syllogistic, as \citet[p.~10-14,18-19,34,36,39]{malink:2019} show.
%[Related to bc<=b and to a<=b, a<=c |= a<=bc]
%A est B et B est C, ergo A est C, nam A est BY.
%[https://www-jstor-org.ez67.periodicos.capes.gov.br/stable/40694037?seq=25#metadata_info_tab_contents]
}, and even correctly proved the categorical syllogism Barbara-1 in an algebraic fashion \citep[pp.~229-230, Axioma 1]{leibniz:1690a}. With a masterful ability, he correctly identified most concepts and notions needed for this task, and thus almost completed the goal on his own. Unfortunately, he missed a small piece of the puzzle -- one of the suitable algebraic representations of particular categorical relations which was provided almost two centuries later by \citet{cayley:1871}\citep[p.~473]{valencia:2004}. Moreover, he identified individual axioms needed to complete this task, but they are scattered across some of his manuscripts -- we organize them in a centralized fashion in this paper.

The earliest \emph{relational} system we could find whose explicit goal (successfully achieved in the same paper) was to prove the set of classic categorical syllogisms was devised by Hugh \citet{mccoll:1877}\footnote{Almost one century later, \citet{tamaki:1974} employed \citeauthor{mccoll:1877}'s relations \{$\subseteq,~\nsubseteq$\} and proved the classic categorical syllogisms again, without adopting Boethius' connexive thesis for ``$\subseteq$'' and assuming existential import for the categorical relations \{$\termlogicA,~\termlogicI$\} though not for \{$\termlogicE,~\termlogicO$\}.}. Thus, McColl deserves the credit of being the earliest to satisfactorily achieve this historical goal of the Leibnizian-Boolean research program (though with a method different from the algebraic one elected by Leibniz and Boole). Much remains to be said in this regard -- McColl's is not the only relational method possible, as we will show. We can obtain further insights at the problem by looking at other methods.
%Ladd-Franklin (1889, pp.~562-563<https://www.jstor.org/stable/1411857?seq=20>) recognized McColl's priority in the ``logic of the non-symmetrical affirmative copula `all a is b' ''.

Proofs of all the 24 classic categorical syllogisms in monadic first-order quantificational logic are known; this exercise has been done countless times\footnote{See, for example, \citet{tennant:2014} and \citet{metamath:2021}. We can even find an implementation of the proof search of the first-order logic representation of categorical syllogistic in a programming language \citep{argyraki:2019}.}. But we claim that first-order logic is a too heavy machinery for tackling those simple 3-sentence argument forms. Although insightful and very welcome to our portfolio of knowledge, we should not feel satisfied by that solution; it feels like using a bazooka to kill a fly. Instead of invoking all the power and complexity of quantificational-functional reasoning, we offer alternative methods of algebraic proofs instead. According to Anellis, ``[...] early efforts'' at algebraizing Aristotelic syllogistic after Gottfried Leibniz and before George Boole ``proved to be incomplete and abortive. Contemporary efforts to arithmeticize or algebraize Aristotelic syllogistic still persist'' to this day \citep{anellis:2007}. Ours is such a solution, or rather a catalog of various alternative solutions.

Unlike the original Aristotelic tradition of term logic and Leibniz's advanced attempts at devising an algebra of term logic, we will adopt in this paper term logic \emph{without} existential import, rather than assuming that a ``term'' class is necessarily inhabited by default. As a consequence, whenever a class is inhabited, we will have to explicitly declare so through a premise.

Throughout the paper we provide copious citations and footnotes for historically relevant materials as early as we could find to the origins of key insights and fundamental building blocks to construct the symbolic approaches to classic categorical syllogistic which we consider in this paper. Our remarks are not intended to present the history of the concepts, tools and ideas envisioned by the pioneers \emph{on their own terms} for achieving their own goals, but to record instead the \emph{origins} of the ingredients we use and \emph{repurpose} for the categorical syllogistic theorems we have the goal of proving in this paper. Here, historical remarks put into context the ingredients of our modernized presentation of proofs in algebraic categorical syllogistic.

\section{Preliminaries -- diagrams involving one or two terms}
%[Assigning a set-theoretic semantics to syntactical terms, operations and relations in term logic]
%Semantics assigns computational meaning to valid strings in a programming language syntax.
%[https://en.wikipedia.org/wiki/Semantics_(computer_science)]
%[...] Concepts (or [General, Classificatory] Terms, as it will in [... many] cases be more convenient to call them) [...]
%[p. 415: Alfred Sidgwick. Reviewed Work: Logik. Eine Untersuchung der Principien der Erkenntniss und der Methoden wissenschaftlicher Forschung. by Wilhelm Wundt. In: Mind, Vol. 5, No. 19 (July 1880), pp. 409--424. https://www-jstor-org.ez67.periodicos.capes.gov.br/stable/2246401?seq=7]
%[...] all figures of whatever shape must necessarily represent *extent*. [...]
%[p. 418: Alfred Sidgwick. Reviewed Work: Logik. Eine Untersuchung der Principien der Erkenntniss und der Methoden wissenschaftlicher Forschung. by Wilhelm Wundt. In: Mind, Vol. 5, No. 19 (July 1880), pp. 409--424. https://www-jstor-org.ez67.periodicos.capes.gov.br/stable/2246401?seq=10]
%On logics, see also -- pp. 619-620: https://www.jstor.org/stable/27897187?seq=6

\subsection{A subclass from a universe}

\begin{figure}[ht]
    \centering
    \begin{subfigure}[t]{0.22\textwidth}
        \includegraphics[scale=0.2]{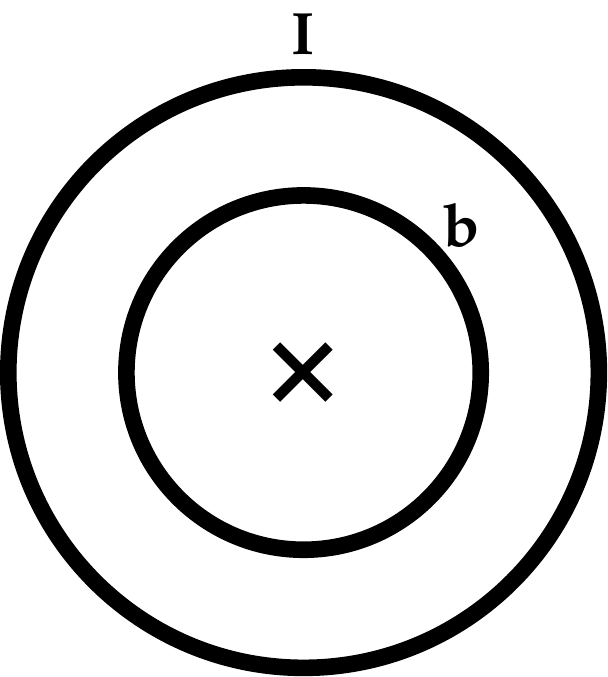}
        \caption{$\mathbf{b}\termlogicI \mathbf{b}$: $b$ is inhabited. (Venn diagram.)}
    \end{subfigure}
    \begin{subfigure}[t]{0.22\textwidth}
        \includegraphics[scale=0.2]{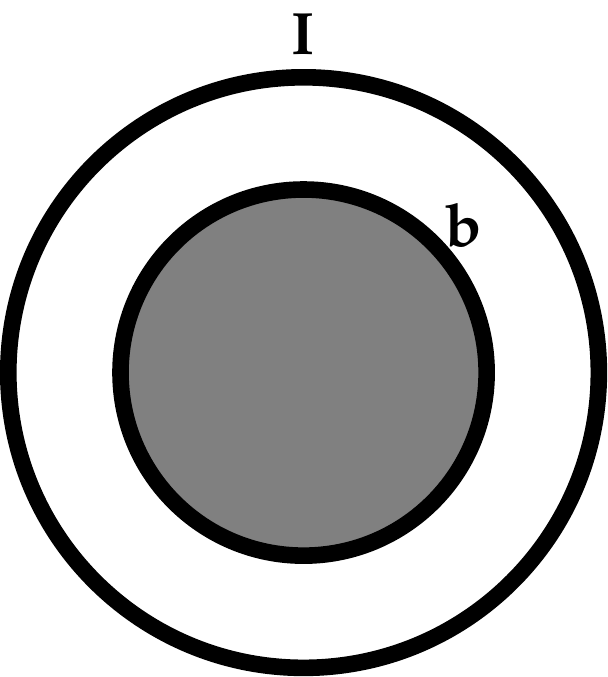}
        \caption{$\mathbf{b}\termlogicE \mathbf{b}$: $b$ is empty. (Venn diagram.)}
    \end{subfigure}
    \begin{subfigure}[t]{0.22\textwidth}
        \includegraphics[scale=0.2]{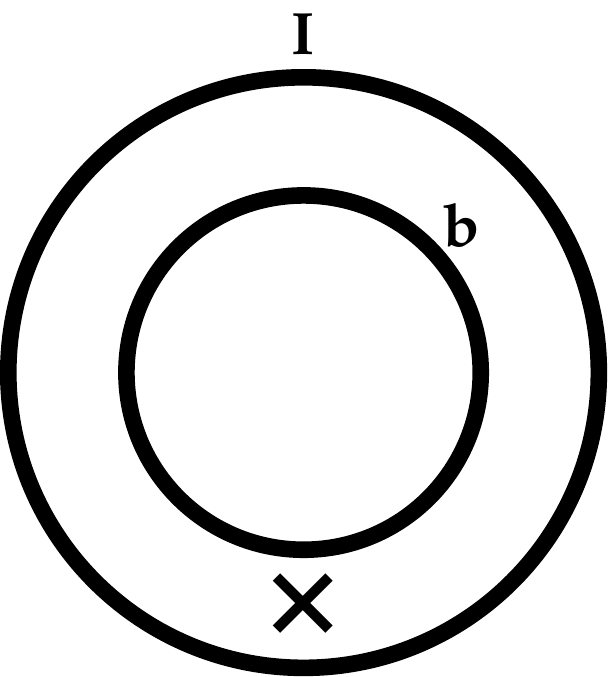}
        \caption{$\mathbf{b'}\termlogicI \mathbf{b'}$: $b'$ is inhabited. (Venn diagram.)}
    \end{subfigure}
    \begin{subfigure}[t]{0.22\textwidth}
        \includegraphics[scale=0.2]{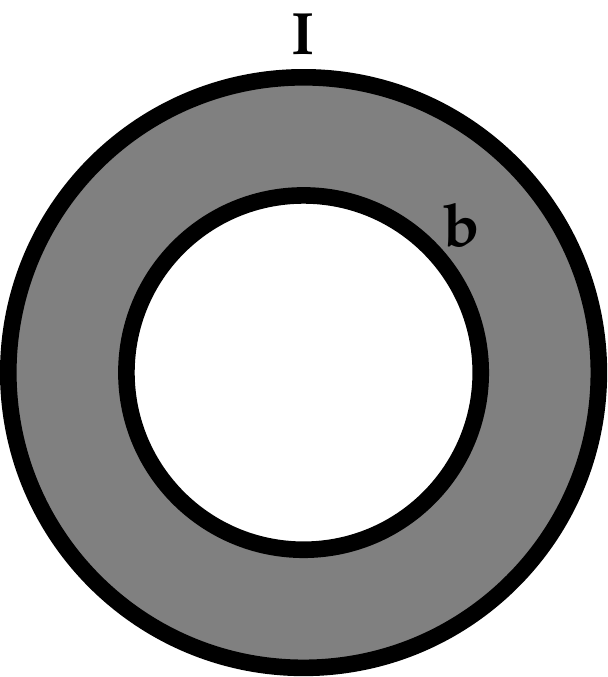}
        \caption{$\mathbf{b'}\termlogicE \mathbf{b'}$: $b'$ is empty. (Venn diagram.)}
    \end{subfigure}
    \caption{Four possibilities of fundamental relations involving a single term of a universe class -- Venn and Euler diagrams.}
    \label{fig:empty_inhabited_one_term}
\end{figure}

The ``smallest'' independent regions in a \citeauthor{venn:1880} diagram (\citeyear{venn:1880}) are called \emph{minterms}. Each minterm is either \emph{inhabited} or \emph{empty}, that is, it either has or does not have at least one element. Figure~\ref{fig:empty_inhabited_one_term} shows the Venn diagram possibilities involving either an inhabitation or an emptiness mark for each minterm inside a universe of discourse\footnote{In this paper, ``\textbf{I}'' was intentionally chosen to represent the un\textbf{I}verse of discourse class to avoid using the initial ``U'', which might be confused with the union operator ``$\cup$'', and also because ``\textbf{I}'' resembles the digit ``1'', just like the empty set symbol ``$\bm{\varnothing}$'' resembles the digit ``0''. Both digits play a major role in Boolean algebra and in anything nowadays referred to as ``digital''.} \textbf{I} with one specially designated subclass/term \textbf{b}. Here the minterms happen to be \textbf{b} itself and its complement, \textbf{b}$'$. When we do not know whether a minterm is inhabited or empty, we leave it blank, in order to indicate lack of information on our part.
%De Morgan (1847) stated that the complement of the universe is the empty class: https://books.google.com/books?id=HscAAAAAMAAJ&pg=PA120&dq=%22It+will+however+be%22+%22By+u%22

\subsection{Representing two terms from a universe}

\begin{figure}[ht]
    \centering
%    \begin{venndiagram2sets}[tikzoptions={thick}, labelA=\textbf{b},labelB=\textbf{c}, labelNotAB=$b^\prime c^\prime$, labelOnlyA= $bc^\prime$, labelOnlyB = $b^\prime c$, labelAB=$bc$]
%    \end{venndiagram2sets}
    \includegraphics[scale=0.2]{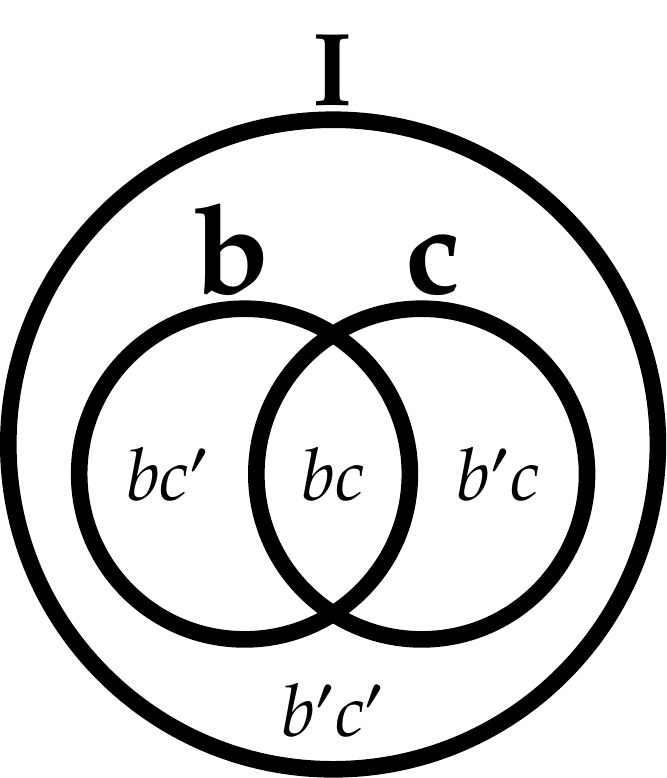}
    \caption{Four minterms of a universe class that has two generating terms as subclasses.}
    \label{fig:four_minterms}
\end{figure}

Let's draw a Venn (and also Euler) diagram representing a universe class that has two terms \textbf{b} and \textbf{c} as subclasses (Figure~\ref{fig:four_minterms}). In a translation from set algebra to term logic, we will call the sets/classes \textbf{b} and \textbf{c} the ``subject'' and the ``predicative'' terms, respectively. The Venn diagram shows that this configuration gives rise to four minterms\footnote{Throughout this paper, we assume for notational convenience that the juxtaposition of two terms, $\mathbf{b}\mathbf{c}$, represents the intersection of the classes they refer to, $\mathbf{b}\!\cap\!\mathbf{c}$.}: $bc, bc', b'c, b'c'$. Here the notion of complement is again demonstrated to be fundamental. And again a blank minterm indicates that we do not have the knowledge whether that minterm is inhabited or empty.

\begin{figure}
    \centering
    \begin{subfigure}[t]{0.22\textwidth}
%        \begin{venndiagram2sets}[tikzoptions={scale=0.5,thick}, labelA=\textbf{b},labelB=\textbf{c}, labelAB=\Large\texttimes]
%        \end{venndiagram2sets}
        \includegraphics[scale=0.2]{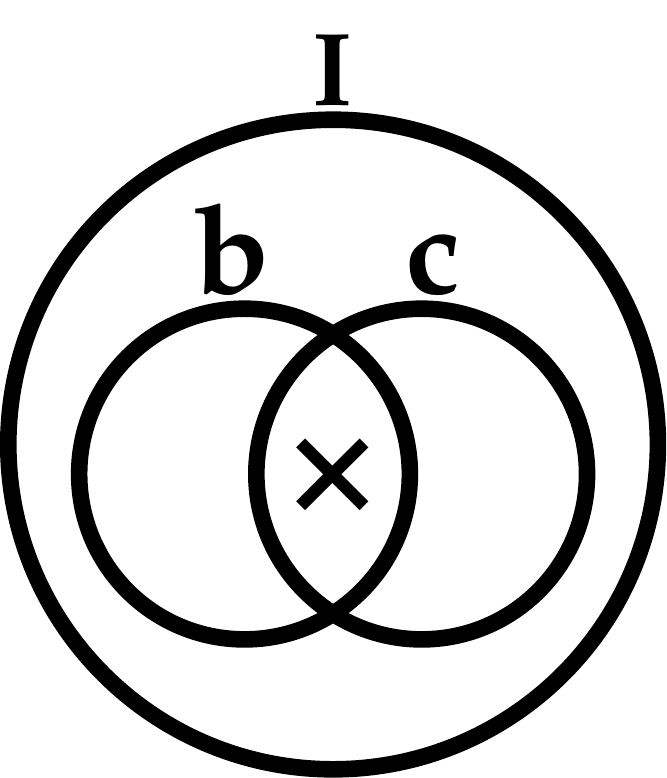}
        \caption{$\mathbf{b}\termlogicI \mathbf{c}$: $bc$ is inhabited. (Venn diagram.)}
    \end{subfigure}
    \begin{subfigure}[t]{0.22\textwidth}
%        \begin{venndiagram2sets}[tikzoptions={scale=0.5,thick}, labelA=\textbf{b},labelB=\textbf{c}]
%            \fillACapB
%        \end{venndiagram2sets}
        \includegraphics[scale=0.2]{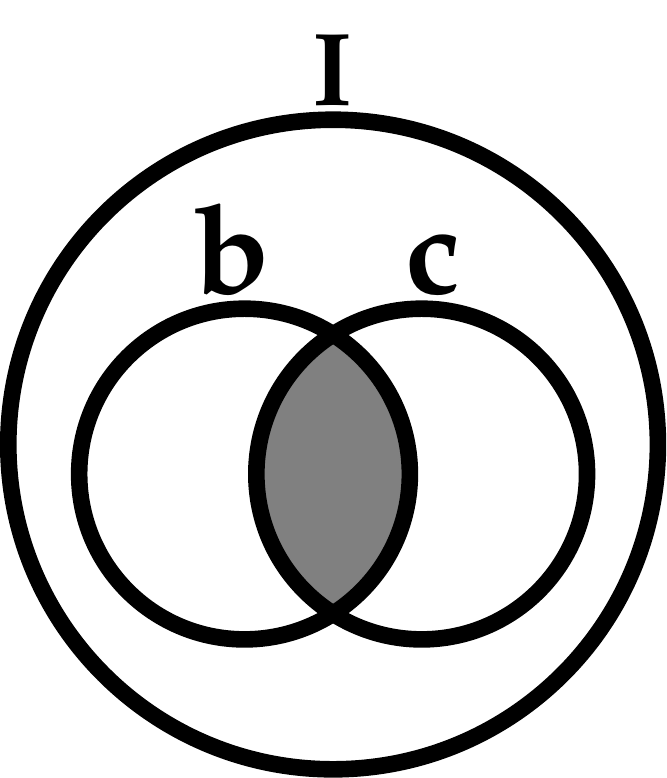}
        \caption{$\mathbf{b}\termlogicE \mathbf{c}$: $bc$ is empty. (Venn diagram.)}
    \end{subfigure}
    \begin{subfigure}[t]{0.22\textwidth}
%        \begin{venndiagram2sets}[tikzoptions={scale=0.5,thick}, labelA=\textbf{b},labelB=\textbf{c}, labelOnlyA=\Large\texttimes]
%        \end{venndiagram2sets}
        \includegraphics[scale=0.2]{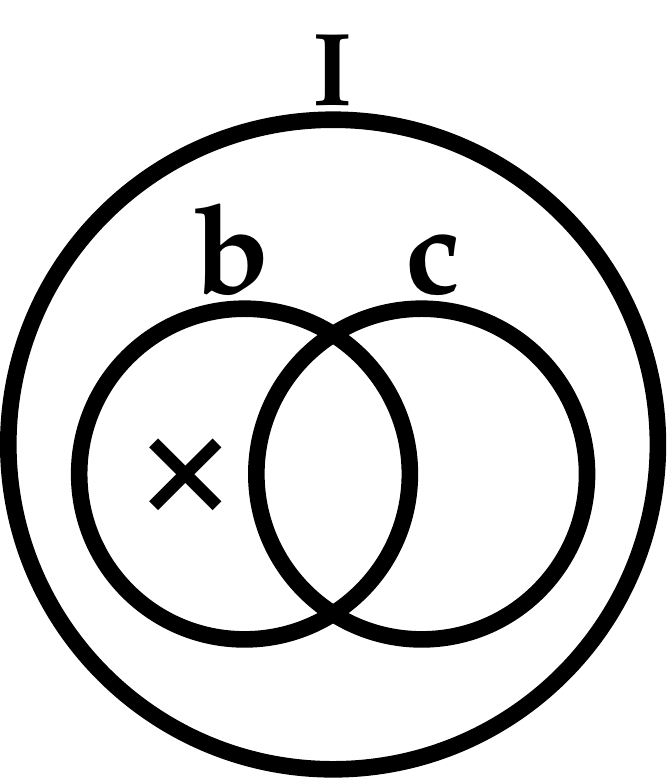}
        \caption{$\mathbf{b}\termlogicO \mathbf{c}$: $bc'$ is inhabited. (Venn diagram.)}
    \end{subfigure}
    \begin{subfigure}[t]{0.22\textwidth}
%        \begin{venndiagram2sets}[tikzoptions={scale=0.5,thick}, labelA=\textbf{b},labelB=\textbf{c}]
%            \fillOnlyA
%        \end{venndiagram2sets}
        \includegraphics[scale=0.2]{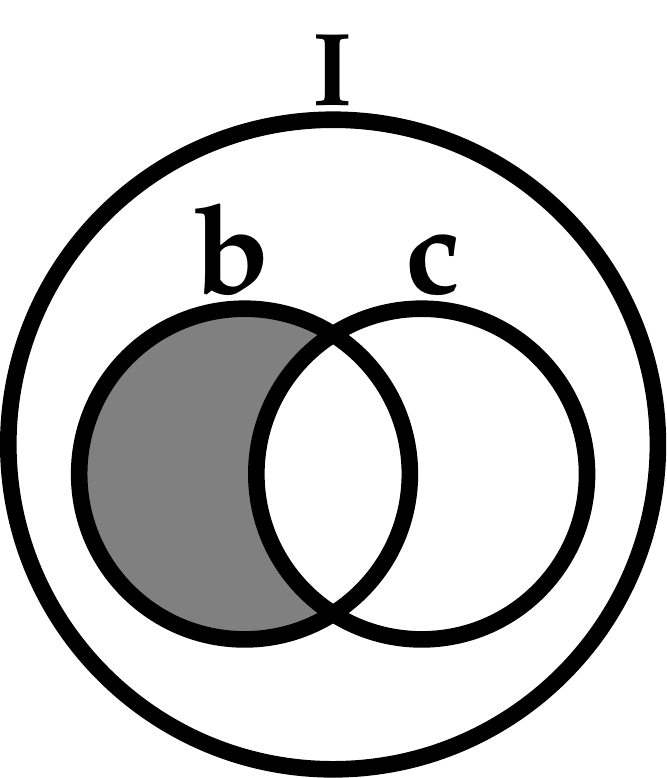}
        \caption{$\mathbf{b}\termlogicA \mathbf{c}$: $bc'$ is empty. (Venn diagram.)}
    \end{subfigure}

    \begin{subfigure}[t]{0.22\textwidth}
        \includegraphics[scale=0.2]{Graphics/at_least_one_b_is_c-venn+euler.pdf}
        \caption{$\mathbf{b}\termlogicI \mathbf{c}$: $bc$ is inhabited. (Euler diagram.)}
    \end{subfigure}
    \begin{subfigure}[t]{0.22\textwidth}
        \includegraphics[scale=0.2]{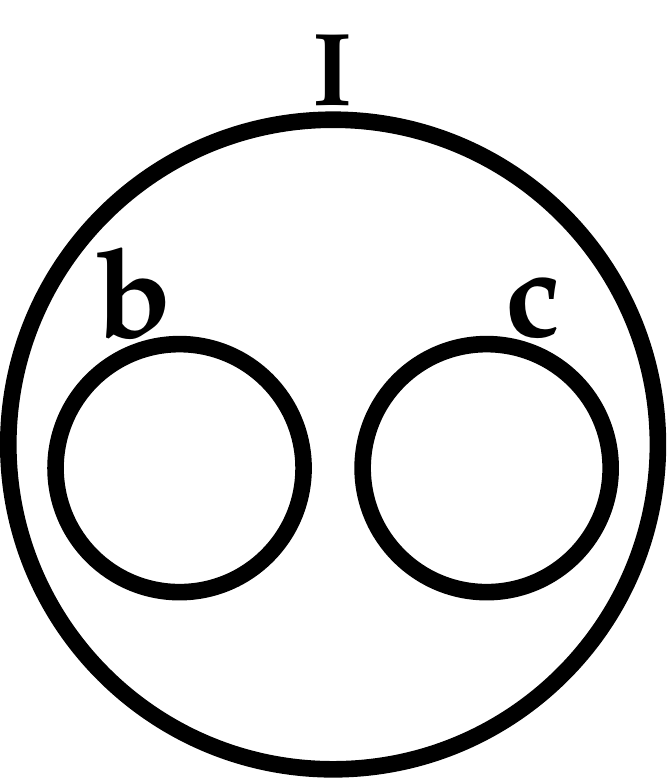}
        \caption{$\mathbf{b}\termlogicE \mathbf{c}$: $bc$ is empty. (Euler diagram.)}
    \end{subfigure}
    \begin{subfigure}[t]{0.22\textwidth}
        \includegraphics[scale=0.2]{Graphics/at_least_one_b_is_not_c-venn+euler.pdf}
        \caption{$\mathbf{b}\termlogicO \mathbf{c}$: $bc'$ is inhabited. (Euler diagram.)}
    \end{subfigure}
    \begin{subfigure}[t]{0.22\textwidth}
        \includegraphics[scale=0.2]{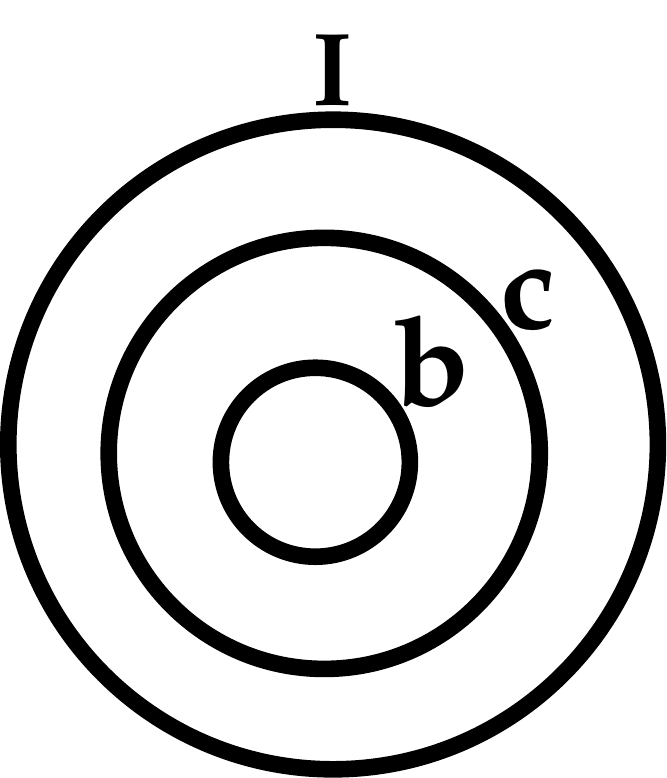}
        \caption{$\mathbf{b}\termlogicA \mathbf{c}$: $bc'$ is empty. (Euler diagram.)}
    \end{subfigure}

\hrule

    \begin{subfigure}[t]{0.22\textwidth}
        \includegraphics[scale=0.2]{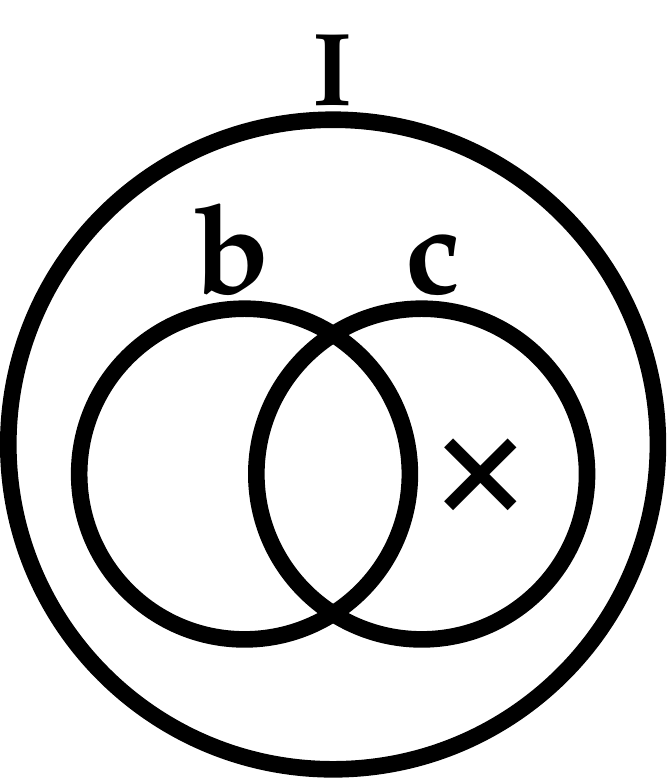}
        \caption{$\mathbf{b}\termlogicOumlaut \mathbf{c}$: $b'c$ is inhabited. (Venn diagram.)}
    \end{subfigure}
    \begin{subfigure}[t]{0.22\textwidth}
        \includegraphics[scale=0.2]{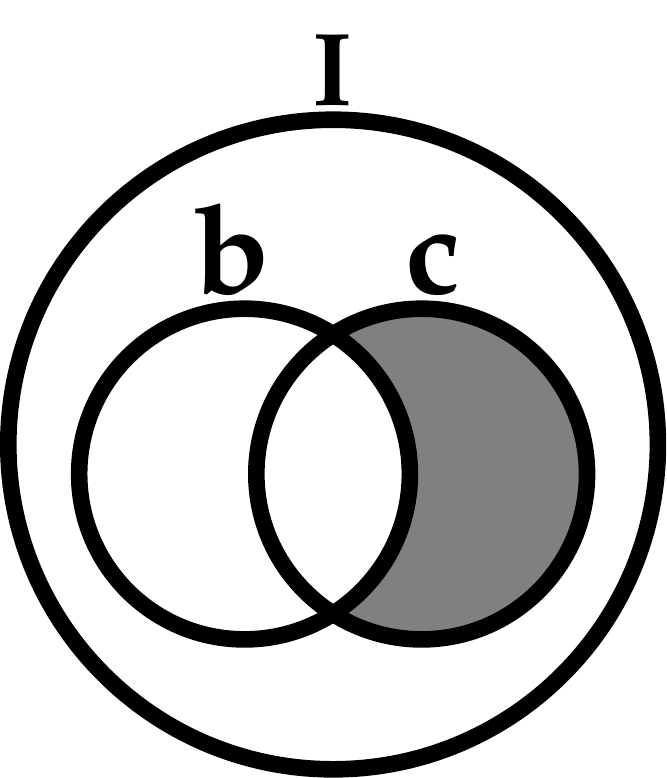}
        \caption{$\mathbf{b}\termlogicAumlaut \mathbf{c}$: $b'c$ is empty. (Venn diagram.)}
    \end{subfigure}
    \begin{subfigure}[t]{0.22\textwidth}
        \includegraphics[scale=0.2]{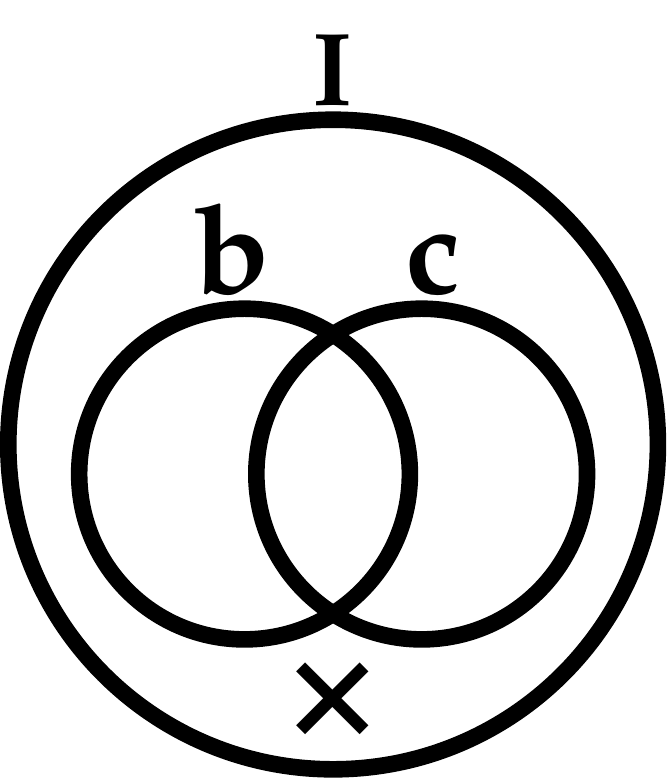}
        \caption{$\mathbf{b}\termlogicIumlaut \mathbf{c}$: $b'c'$ is inhabited. (Venn diagram.)}
    \end{subfigure}
    \begin{subfigure}[t]{0.22\textwidth}
        \includegraphics[scale=0.2]{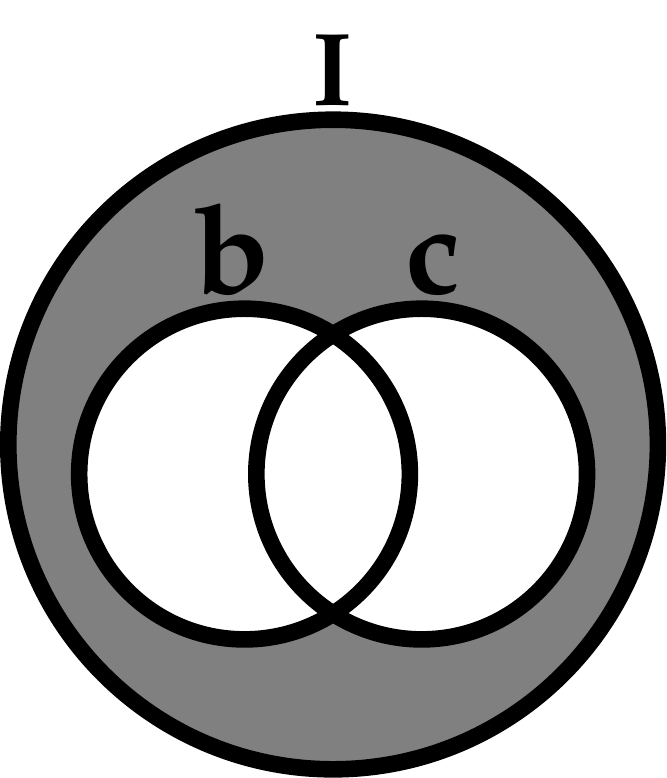}
        \caption{$\mathbf{b}\termlogicEumlaut \mathbf{c}$: $b'c'$ is empty. (Venn diagram.)}
    \end{subfigure}

    \begin{subfigure}[t]{0.22\textwidth}
        \includegraphics[scale=0.2]{Graphics/at_least_one_non-b_is_c-venn+euler.pdf}
        \caption{$\mathbf{b}\termlogicOumlaut \mathbf{c}$: $b'c$ is inhabited. (Euler diagram.)}
    \end{subfigure}
    \begin{subfigure}[t]{0.22\textwidth}
        \includegraphics[scale=0.2]{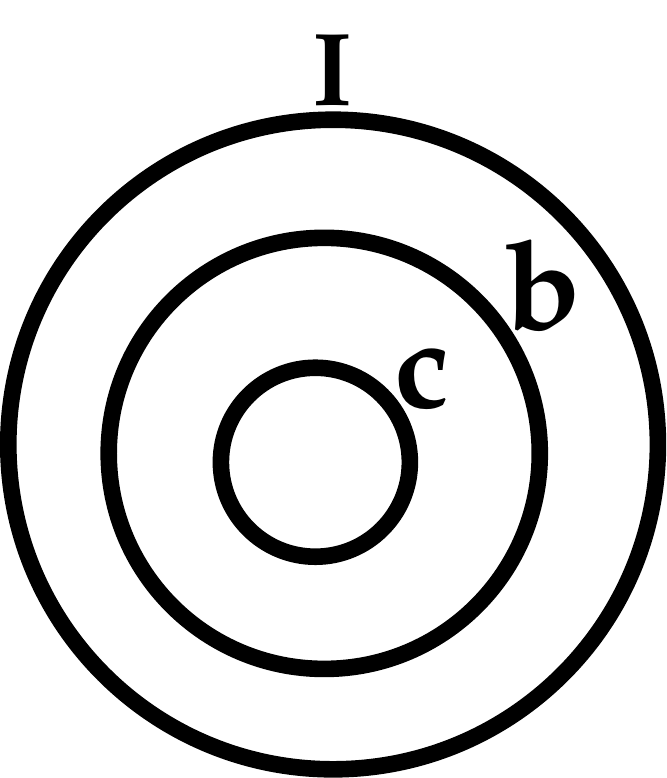}
        \caption{$\mathbf{b}\termlogicAumlaut \mathbf{c}$: $b'c$ is empty. (Euler diagram.)}
    \end{subfigure}
    \begin{subfigure}[t]{0.22\textwidth}
        \includegraphics[scale=0.2]{Graphics/at_least_one_non-b_is_not_c-venn+euler.pdf}
        \caption{$\mathbf{b}\termlogicIumlaut \mathbf{c}$: $b'c'$ is inhabited. (Euler diagram.)}
    \end{subfigure}
    \begin{subfigure}[t]{0.22\textwidth}
        \includegraphics[scale=0.2]{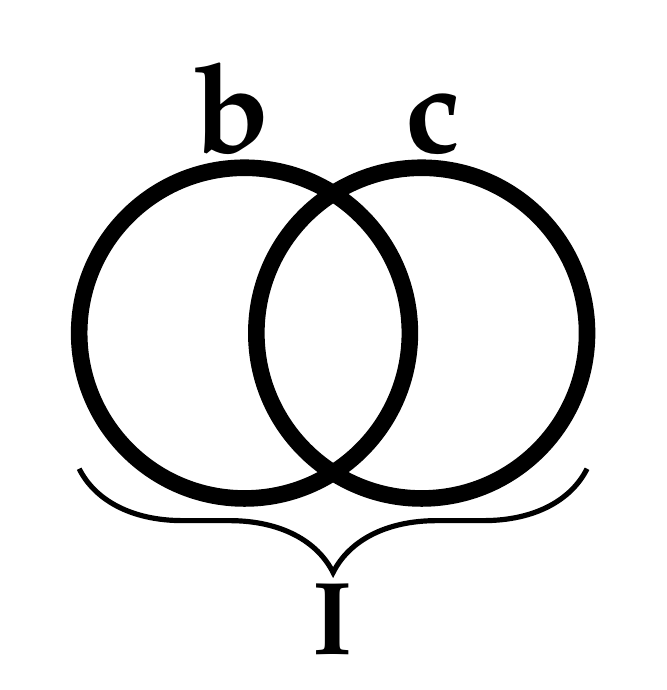}
        \caption{$\mathbf{b}\termlogicEumlaut \mathbf{c}$: $b'c'$ is empty. (Euler diagram.)}
    \end{subfigure}
    \caption{Eight possibilities of fundamental relations between two terms of a universe class, one minterm at a time -- Venn and Euler diagrams.}
    \label{fig:empty_inhabited_two_terms}
\end{figure}

Figure~\ref{fig:empty_inhabited_two_terms} enumerates the two possibilities (emptiness or inhabitation mark) for each of the four minterms independently considered for both Venn\footnote{Some introductory textbooks include these Venn diagrams and some corresponding algebraic symbology. See for instance \citet[pp.~106-109]{copi:2016}.} and Euler diagrams, showing how they are distinct by direct contrast\footnote{Those possibilities could be portrayed by \citeauthor{carroll:1886} diagrams (\citeyear[pp.~44,28]{carroll:1886}) as well, which we decided to leave out of the scope of this paper.}. It shows eight dyadic relations in the form $\mathbf{b}\termlogicassertion \mathbf{c}$, where ``$\termlogicassertion$'' is the relation symbol, which are the fundamental components of what is called De Morgan's syllogistic\footnote{In order to preserve the order adopted for our Venn and Euler diagrams, we will enumerate in our Tables the fundamental categorical relations in De Morgan's syllogistic in the order $\termlogicI,\termlogicE,\termlogicO,\termlogicA,\termlogicOumlaut,\termlogicAumlaut,\termlogicIumlaut,\termlogicEumlaut$ rather than in the more conventional order $\termlogicA,\termlogicE,\termlogicI,\termlogicO,\termlogicEumlaut,\termlogicAumlaut,\termlogicOumlaut,\termlogicIumlaut$. The convention of adopting the umlaut to represent the inverse relation (e.g.: $\mathbf{b}\termlogicIumlaut\mathbf{c} \metaeq \mathbf{b}'\termlogicI\mathbf{c'}$) is from \citet{menne:1957}(\citeyear{menne:1962}).} -- an extension of Aristotle's syllogistic\footnote{Aristotle's syllogistic, which include the 24 classic categorical syllogisms we prove in this paper in Sections~\ref{sec:euler}, \ref{sec:LC} and \ref{sec:ML}, involve the initial four possibilities only, which deal with the two minterms that are subclasses of the subject (\textbf{b}): $bc$ and $bc'$.

Historically, however, Aristotle also alluded to the possibility of categorical relations with a negated subject, such as ``Every non-man is just'' \citep[Chapter~10]{owen:1853a} -- millennia later they would become the interest of systematic study.} \citep[p.~381]{de_morgan:1846}\citep[pp.~60-61]{de_morgan:1847}.%dyadic relations in the form $\mathbf{b}\termlogicassertion \mathbf{c}$ involving two sets (or ``terms'', in the jargon of term logic). The initial component of the ordered pair $(b,c)$ is called the ``subject'', and the final component is called the ``predicative''. ``$\termlogicassertion$'' is the relation symbol.
%See also:
%https://books.google.com/books?id=UD-eAgAAQBAJ&pg=PA36&dq=%22De+Morgan's+First+Notions%22
%De Morgan (1839): Close to De Morgan's syllogistic + Hamilton-like quantifiers + complementation
%pp. 10,16-17
%https://books.google.com/books?id=vcEwAQAAMAAJ&pg=PA10
%Read 2 pages:
%https://books.google.com/books?id=Shc8boMnrBoC&pg=RA2-PA130&dq=Hamilton+Aristotle

Venn and Euler diagrams make it visible that every class being depicted is a subclass of the universe class ($\mathbf{b}\!\subseteq\!\mathbf{I}$). However, a challenge in drawing both Euler and Venn diagrams is how to represent the facts that the empty class is a subclass of every class being depicted ($\bm{\varnothing}\subseteq\!\mathbf{b}$) and that the intersection of any class with the empty one is the latter ($\mathbf{b}\!\cap\!\bm{\varnothing}=\bm{\varnothing}$)\footnote{The empty class is the only class simultaneously ``included in'' and ``excluded from'' every conceivable class, that is, $\mathbf{b}\!\subseteq\!\mathbf{c}$ and $\mathbf{b}\!\bm{\not}\!\Cap\:\mathbf{c}$ if and only if $\mathbf{b}=\bm{\varnothing}$. Proof:
%[...] Instead of writing
%``$A\subseteq B$'' (A is-wholly B) and ``$A\nsubseteq B$'' (A is-not-wholly B or A is-partly-not B)
%we might write
%``$A\not\!\Cap B$'' (A is-wholly-not B) and ``$A\Cap B$'' (A is-partly B).
%[...]
%The sign ``$\not\!\Cap$'' is a wedge, sign of exclusion. ``$A\not\!\Cap B$'' is to be read [...] A is [wholly] excluded from B. The sign ``$\Cap$'' is an incompleted wedge, sign of incomplete exclusion. ``$A\Cap B$'' is to be read "A is in [at least in] part B", "A is not-wholly excluded from B." [...]
%[...] ``$\Cap$'' connects terms that exist, while ``$\not\!\Cap$'' connects terms which may be non-existent. [...]
%Propositions expressed with the copula ``$\subseteq$'' are called inclusions; propositions expressed with the copula ``$\not\!\Cap$'' may be called exclusions.
%[...]
%The copulas ``$\not\!\Cap$'' and ``$\Cap$'' with the symbol ``$\mathbf{I}$'' give means for expressing the total non-existence and the partial existence of expressions of any degree of complexity. [...]
%A universal proposition does not imply the existence of its subject; therefore ``$x\not\!\Cap\bm{\varnothing}$'' = ``x (if there is any x) is not non-existent,'' -- a proposition which is true whatever x may be.
%A particular proposition does imply the existence of its subject ; therefore ``$x\Cap\bm{\varnothing}$'' = ``x exists, and at the same time does not exist,'' -- a proposition which is false whatever X may be.
%[pp. 27-29: https://archive.org/details/bub_gb_V7oIAAAAQAAJ/page/n33/mode/2up]

\begin{multicols}{2}

\begin{tabular}{c}
$\,{{\color{blue}\mathbf{b}}\!\subseteq\!\mathbf{c} \metaeq \color{blue}\mathbf{b}}\mathbf{c}={\color{blue}\mathbf{b}}$ \qquad ${\color{blue}\mathbf{b}}\!\bm{\not}\!\Cap\:\mathbf{c} \metaeq {\color{blue}\mathbf{b}}\mathbf{c}={\color{red}\bm{\varnothing}}$\\\hline
\{If $\mathbf{b}\!\subseteq\!\mathbf{c}$ and $\mathbf{b}\!\bm{\not}\!\Cap\:\mathbf{c}$, then ...\}\qquad\qquad\qquad\!\!\quad\\
\qquad${\color{blue}\mathbf{b}}\!\subseteq\!\mathbf{c},\ {\color{blue}\mathbf{b}}\!\bm{\not}\!\Cap\:\mathbf{c} \metaeq {\color{blue}\mathbf{b}}\mathbf{c}={\color{blue}\mathbf{b}},\ {\color{blue}\mathbf{b}}\mathbf{c}={\color{red}\bm{\varnothing}}$\\
${\color{blue}\mathbf{b}}\!\subseteq\!\mathbf{c},\ {\color{blue}\mathbf{b}}\!\bm{\not}\!\Cap\:\mathbf{c} \metaeq {\color{blue}\mathbf{b}}={\color{blue}\mathbf{b}}\mathbf{c}={\color{red}\bm{\varnothing}}$\\
${\color{blue}\mathbf{b}}\!\subseteq\!\mathbf{c},\ {\color{blue}\mathbf{b}}\!\bm{\not}\!\Cap\:\mathbf{c} \metaeq\, {\!\color{blue}\mathbf{b}}={\color{red}\bm{\varnothing}}$.\qquad\,\qquad\\
{\color{white}.}\;\;\;\{... $\mathbf{b}=\bm{\varnothing}$.\}\qquad\qquad\qquad\qquad\qquad\qquad\qquad\qquad
\end{tabular}\pagebreak

\begin{tabular}{c}
$\,{{\color{blue}\mathbf{b}}\!\subseteq\!\mathbf{c} \metaeq \color{blue}\mathbf{b}}\mathbf{c}={\color{blue}\mathbf{b}}$ \qquad ${\color{blue}\mathbf{b}}\!\bm{\not}\!\Cap\:\mathbf{c} \metaeq {\color{blue}\mathbf{b}}\mathbf{c}={\color{red}\bm{\varnothing}}$\\\hline
\{If $\mathbf{b}=\bm{\varnothing}$, then ...\}\qquad\qquad\qquad\qquad\qquad\!\!\quad\\
$\,{{\color{blue}\bm{\varnothing}}\!\subseteq\!\mathbf{c} \metaeq \!\color{blue}\bm{\varnothing}}\mathbf{c}={\color{blue}\bm{\varnothing}}$ \qquad $\!\!{\color{blue}\bm{\varnothing}}\!\bm{\not}\!\Cap\:\mathbf{c} \metaeq \!{\color{blue}\bm{\varnothing}}\mathbf{c}={\color{red}\bm{\varnothing}}$\\
${\color{blue}\bm{\varnothing}}\!\subseteq\!\mathbf{c},\,{\color{blue}\bm{\varnothing}}\!\bm{\not}\!\Cap\:\mathbf{c} \!\metaeq\! {\color{blue}\bm{\varnothing}}\mathbf{c}=\bm{\varnothing}{\color{white}.}$\qquad\!\!\!\quad\\
\qquad$\;{\color{blue}\bm{\varnothing}}\mathbf{c}=\bm{\varnothing} \!\metaeq\! {\color{blue}\bm{\varnothing}}\!\subseteq\!\mathbf{c},\,{\color{blue}\bm{\varnothing}}\!\bm{\not}\!\Cap\:\mathbf{c}$.\\
{\color{white}.}\;\;\;\{... $\mathbf{b}\!\subseteq\!\mathbf{c}$ and $\mathbf{b}\!\bm{\not}\!\Cap\:\mathbf{c}$.\}\qquad\qquad\qquad\qquad\qquad\qquad
\end{tabular}

\end{multicols}

Therefore, ``every \textbf{b} is \textbf{c}'' and ``no \textbf{b} is \textbf{c}'' simultaneously if and only if $\mathbf{b}=\bm{\varnothing}$.

Notice that, as a consequence of the definition of ``$\!\bm{\not}\!\Cap$'', $\bm{\varnothing}\!\bm{\not}\!\Cap\:\mathbf{c} \,\metaeq\, \bm{\varnothing}\!\cap\!\mathbf{c}=\bm{\varnothing}$.}. We propose a solution which does not require modifying diagrams (for instance, by adding a new marker), but merely requires us to change how we look at them.
%[...] John Venn['s ...] interpretation, another treatment of syllogistic reasoning [(other than symbolic algebra)], takes b, c, d, ... [(terms)] to be regions in space. In this interpretation, the product of b and c is interpreted as the region common to regions b and c, and the sum of b and c is the set of all points that belong to either b or c. The formula $b \subseteq c$ represents the proposition that region b is contained in region c. The null region is represented by $\bm{\varnothing}$ and the universal space is represented by $\mathbf{I}$. This interpretation leads to a diagrammatic method for testing the validity of syllogisms that remains well-known to logic students and is included in many contemporary texts on symbolic logic.
%[p. 460: I. Susan Russinoff. The Syllogism's Final Solution. In: The Bulletin of Symbolic Logic, Vol. 5, No. 4 (Dec., 1999), pp. 451--469. https://people.math.ethz.ch/~halorenz/4students/Literatur/RussinoffSyllo.pdf#page=11 https://www-jstor-org.ez67.periodicos.capes.gov.br/stable/421118?seq=10]

It is convenient to consider the boundary\footnote{In another mathematical context, topos theory, Lawvere deals with a strongly related concept named ``intrinsic boundary'' or ``co-Heyting boundary'' \citep{ncatlab:2016}\citep{lawvere:1991}\citep[p.~127]{pagliani:1998}.} of the representation of a given class (say, $\mathbf{b}$) as the intersection between what is ``inside'' ($\mathbf{b}$) and what is ``outside'' ($\mathbf{b}'$) that class. As $\mathbf{b}\mathbf{b}'=\bm{\varnothing}$, we feel we are justified in adopting the convention that the boundary of a class stands for the empty class. The boundary of a class is reasonably considered as an integral part of its visual representation; this is a convenient representation for the fact that $\bm{\varnothing}\!\subseteq\!\mathbf{b}$. Moreover, in a Venn or Euler diagram, the intersection between a class and its border --which, as we have said, stands for the empty class-- is visually realized as the border itself. Thus, the fact that $\mathbf{b}\!\cap\!\bm{\varnothing}=\bm{\varnothing}$ is also neatly represented.
%co-Heyting boundary (topology)
%https://proofwiki.org/wiki/Definition:Boundary_(Topology)
%https://proofwiki.org/wiki/Boundary_is_Intersection_of_Closure_with_Closure_of_Complement
%https://proofwiki.org/wiki/Definition:Closure_(Topology)
%https://proofwiki.org/wiki/Definition:Limit_Point/Topology/Set

As borders are present in any Venn or Euler diagram, the empty class is always visibly depicted, and usually more than once, for boundaries are non-contiguous among classes --most notably the boundary of the universe \textbf{I} and the boundary of any of its subclasses (\textbf{b}, \textbf{c} and so on). This would make the representation of the empty class ``fragmented'' in a diagram. This may at first sight look inelegant, but it actually portrays in an elegant fashion an interesting property of the empty class: assuming idempotence and associativity of union ($\cup$), this ``discontiguous border line interpretation''\footnote{In a mereotopological analysis of Euler and Venn diagrams, as boundaries represent the empty class (whether they touch each other or not), RCC-5 \citep{cohn:1994} becomes a valid degeneration of RCC-8 \citep{randell:1992, bennett:1994} in this context.} is algebraically supported by the identity $\bm{\varnothing}=\bm{\varnothing}\cup\bm{\varnothing}=\bm{\varnothing}\cup\bm{\varnothing}\cup...~\bm{\varnothing}=bb'\cup cc'\cup...~zz'$. We can see also see that the empty class is made of pure boundaries when it is represented by the notation ``$\{\!\}$'', and, in the representation of an inhabited class such as ``$\{x\}$'', the borders (braces) are discontiguous.%Discontiguous boundary
%https://proofwiki.org/wiki/Boundary_of_Union_of_Separated_Sets_equals_Union_of_Boundaries
%https://www.emathzone.com/tutorials/general-topology/boundary-point-of-a-set.html
%https://en.wikipedia.org/wiki/Boundary_(topology)
%RCC
%https://www.researchgate.net/figure/Relations-identified-by-RCC5-RCC8-9-intersection-model-and-their-correspondence_fig1_235996110 https://www.researchgate.net/figure/Comparing-different-granularities-in-the-9-intersection-model-and-RCC-5-and-RCC-8_fig1_220649981 slide42:https://slideplayer.com/slide/697792/ https://www.scielo.org.mx/scielo.php?script=sci_arttext&pid=S1870-90442016000100031#:~:text=RCC5 https://www.researchgate.net/figure/Region-connection-calculus-8-RCC8-It-handles-qualitative-spatial-representation-and_fig1_271840886 https://www.researchgate.net/figure/The-RCC8-relations-and-their-conceptual-neighbourhood-diagram_fig1_257518293 pp.110-111:https://perso.telecom-paristech.fr/bloch/AIC/articles/Chen-Cohn2015.pdf https://www.semanticscholar.org/paper/An-ontological-analysis-of-vague-motion-verbs%2C-with-D'Odorico/d3d0b2c18eda0728d0a7558cacf88996e4ce3c87/figure/2 https://www.researchgate.net/figure/RCC5-relations-and-RCC5-lattice_fig1_225391728 https://www.researchgate.net/figure/RCC8-relations-and-lattice_fig2_225391728
%Another connection between RCC and BL
%s. 4.2 RCC models are Boolean algebras - pp. 72-75 (10-13 of the PDF file): https://doi.org/10.1016/S0304-3975(99)00156-5

In addition, in this interpretation a shaded region in a Venn diagram can be seen as a thicker expansion of a boundary, engulfing an entire region. (A diagram designer who wants to stress this would select the same color for class boundaries and for shaded regions -- which usually isn't done for aesthetic reasons only.)

\begin{figure}[ht]
    \centering
    \includegraphics[scale=0.6]{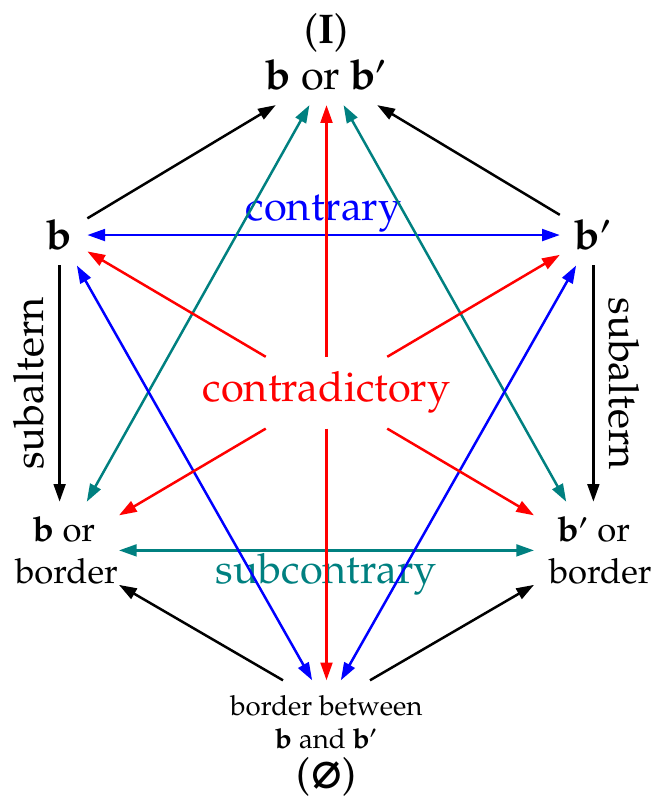}
    \caption{A logical hexagon of opposition modelling the interaction between complementary classes and their shared border.}
    \label{fig:logical_hexagon_of_opposition-border_between_complementary_classes}
\end{figure}

The interaction between complementary classes and their shared border can be modelled by a logical hexagon of opposition -- Figure~\ref{fig:logical_hexagon_of_opposition-border_between_complementary_classes}, loosely inspired by \citet[pp.~38-39, picture~53]{beziau:2012}.

\begin{figure}[ht]
    \centering
    \begin{subfigure}[t]{0.84\textwidth}
        \centering
        \includegraphics[scale=0.2]{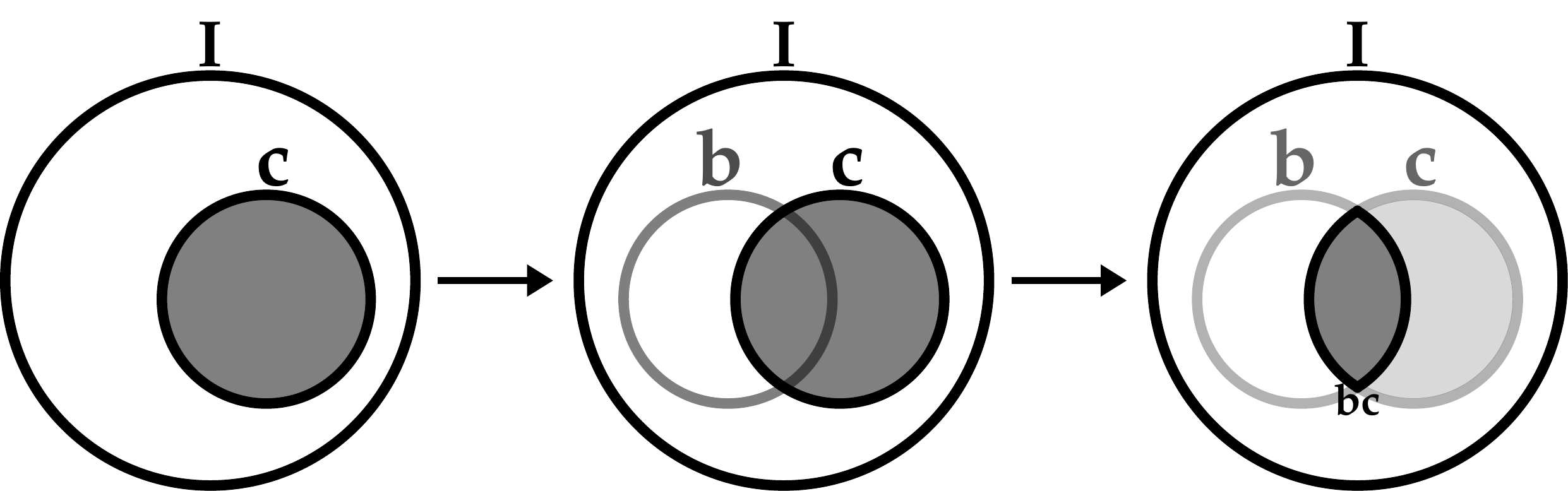}
        \caption{$c$ is empty, therefore $bc$ is empty. (Venn diagram.)}
        \label{fig:empty_b}
    \end{subfigure}
    \begin{subfigure}[t]{0.84\textwidth}
        \centering
        \includegraphics[scale=0.2]{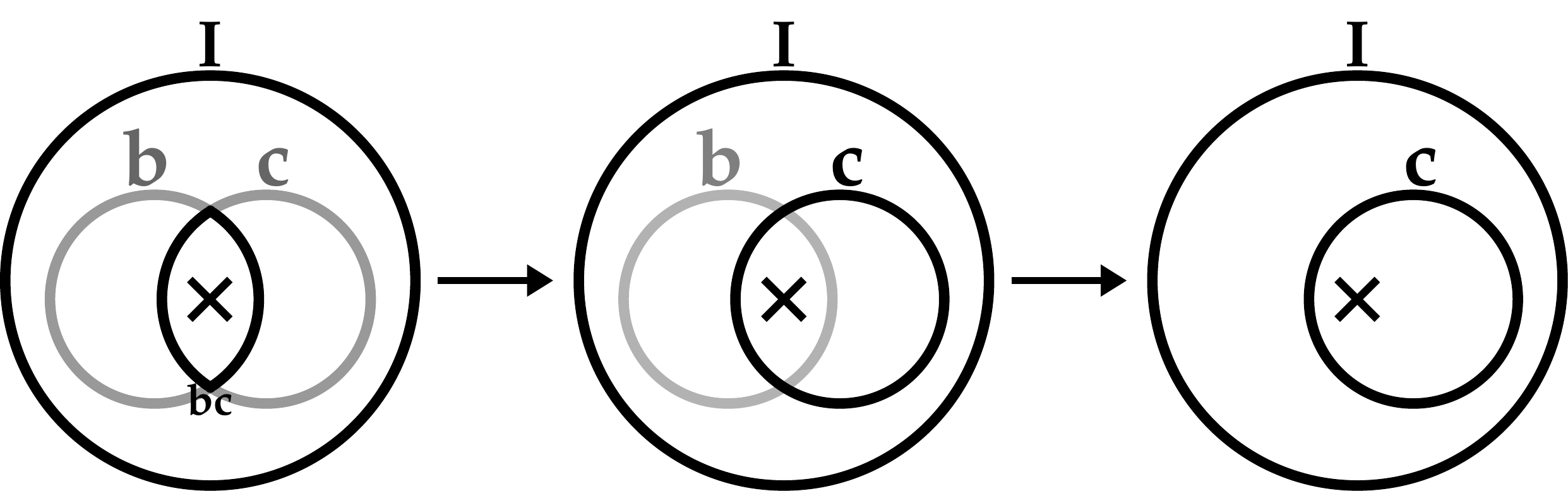}
        \caption{$bc$ is inhabited, therefore $c$ is inhabited. (Venn diagram.)}
        \label{fig:inhabited_bc}
    \end{subfigure}
    \caption{Emptiness and inhabitation markers for a minterm in a universe class that has two generating terms.}
    \label{fig:empty_b_or_inhabited_bc}
\end{figure}

As shown in Figures~\ref{fig:empty_b} and~\ref{fig:inhabited_bc}, the notational choices for the Venn diagram were deliberately made in order to visually represent the following desired properties (which are true propositions in set algebra, in its generalization, Boolean algebra, and even in multiset algebra):

\begin{enumerate}[i.]
	\item $b\bm{\varnothing}=\bm{\varnothing}$. Therefore,\newline~$c=\bm{\varnothing} \metale bc=\bm{\varnothing}$.
	\item The contrapositive of the previous logical assertion:\newline~$bc\neq\bm{\varnothing} \metale c\neq\bm{\varnothing}$.
\end{enumerate}
%0=b0, 0=c |= 0=bc %((b/\c)->d)<=>(b->(c->d))
%
%c=0 |= bc=0
%Alternative proof, which does not apply substitution of similars, but interchangeability of identicals instead:
%c=0
%bc=b0 {p=q |= f(p)=f(q)}, p<-c, q<-0, f(x)<-bx
%bc=0 {b0=0}

%In the modern square of opposition:
%bAc, bEc |=| bAc, bAc' |= b=0
%b=/=0 |= (bAc, bAc')' |=| bA'c |v| bA'c' |=| bA'c |v| bE'c |=| bOc |v| bIc
%
%bAc, cEc |= bEb {empty class entails empty subclass}
%bIb, bAc |= cIc {inhabited class entails inhabited superclass}

In order to achieve this, in Venn diagrams the emptiness marker --the shade-- was intentionally chosen because it occupies a region as ``wide'' as possible (a whole class) and spreads all the way inside, but not outside that region (that is, the marker fills all of its subclasses down to the empty class, but not superclasses). In contrast, the inhabitation marker --{\Large\textbf\texttimes}-- was purposely chosen due to having the opposite properties: it occupies a ``minimum'' area inside a minterm and affects a region as large as possible --all the regions that contain it (that is, all the superclasses of that class, up to the universe class). These smart design decisions\footnote{The earliest explicit textual description we could find of the \emph{rationale} for the desired features of emptiness \emph{and} inhabitation marks (including alternative inhabitation marks) for term logic diagrams is by \citet[pp.~599-600]{venn:1883}. %(6-7 of the PDF file.)
%Alternative inhabitation, also known as existential exhaustion

\citet[p.~126, ``Lettre CV'' from February 24th, 1761]{euler:1770} employed a special marker, ``{\huge $\ast$}'', to indicate that a classificatory term is inhabited. He therefore deserves credit for the earliest use we could find of the inhabitation marker in logical diagrams.
%Redundancy - credit to Euler
%521 \citet[p.~126, ``Lettre CV'' from February 24th, 1761]{euler:1770} employed a special marker
%550 In his diagrammatic notation, a missing
%576 \citet[p.~126, ``Lettre CV'' from February 24th, 1761]{euler:1770} began using a notational

In a draft circa 1903 --CP~4.359-4.363 \citep[pp.~307--312]{cp:1960}, %p. 1395 of the PDF file
referred to as MS.~479 in the ``Robin catalog'' at \guillemotleft\url{https://peirce.sitehost.iu.edu/robin/robin_fm/logic.htm}\guillemotright)--, Peirce drew Venn diagrams for O, I and Ï relations using a cross as inhabitation marker, and on CP~4.349 %pp. 1383-1384 of the PDF file
he describes the procedure of representing existence in a Venn diagram (using a dot rather than a cross, however). Peirce's draft CP~4.359-4.363 is also the earliest source we could find for the graphic display of the \emph{alternative inhabitation} representation, {\Large\textbf\texttimes}\!\!\!\!{\large\textbf{-----}}\!\!\!\!{\Large\textbf\texttimes}, used in this paper only later, in Figure~\ref{fig:logical_hexagon-categorical_relations}. (See also \citet{hammer:1995}, \citet{pietarinen:2016}(\citeyear[pp.~84-100]{pietarinen:2021}) and \citet{shin:2018}.)%Figs. 37-43 (p. 1401 of the PDF file) prove Darii-1. Figs. 44-48 (p. 1401 of the PDF file) prove "Frisesomorum"<https://www.treccani.it/enciclopedia/frisesomorum_%28Dizionario-di-filosofia%29/><https://books.google.com/books?id=rt0MBQAAQBAJ&pg=PA349><https://books.google.com/books?id=D8vANpjnKHEC&pg=PA129&dq=duabus+%22in+Frisesomorum%22><https://encyclo-philo.fr/syllogisme-a#:~:text=Frisesomorum><http://www.eventoj.hu/steb/vortaroj/filozofia-vortaro/FORMO-1.html> ("IEÖ-1"): mIp, sEm |= sÖp.

Decades later, the inhabitation mark and the alternative inhabitation representation are employed for proving categorical syllogisms by means of Euler and Venn diagrams by \citet[pp.~176,183-184]{lewis:1918}, %(182,189-190 of the PDF file).
and the inhabitation mark is employed for Venn diagrams by \citet[p.~70]{quine:1950}(\citeyear[pp.~98,102-110]{quine:1982}).} make them semiotically appropriate notations to graphically represent the two properties we want.

Notice that the arrows in Figure~\ref{fig:empty_b_or_inhabited_bc} are unidirectional. The converse is not necessarily true.

\section{Euler system}
\label{sec:euler}

In this paper, we are concerned with categorical syllogisms that comply to a rigid form that follows these rules:

\begin{itemize}
    \item Three terms are involved: \textbf{s}, \textbf{m} and \textbf{p}\footnote{Standing for \textbf{s}ubject of the conclusion, \textbf{m}ediator (or \textbf{m}iddle term), which does not appear in the conclusion, and \textbf{p}redicative of the conclusion.}.
    \item There are two premises, where one involves \textbf{s} and \textbf{m}, and the other involves \textbf{m} and \textbf{p}.
    \item There may be an additional premise asserting that a given term among \textbf{s}, \textbf{m} and \textbf{p} is necessarily inhabited.
    \item There are one or two conclusions involving \textbf{s} and \textbf{p}.
\end{itemize}

In the limited logic we are concerned with, Euler diagrams are expressive enough for our needs. Informal diagrammatic proofs\footnote{These proofs are informal not because they are diagrammatic, but because we have not explicitly enumerated here the axioms and inference rules required by this logic system.} of the 24 classic categorical syllogisms by means of Euler diagrams, taken from \citet{piesk:2017}\footnote{A variant of this Euler diagram representation of categorical syllogisms is offered by \citet{flage:2002}.}\footnote{For two of the moods, we adopt the names ``Baroko-2'' and ``Bokardo-3'' with `k' rather than `c' to preserve compatibility with the rationale for the name of \citeauthor{de_morgan:1860}'s ``Theorem K'' (\citeyear[p.~344]{de_morgan:1860}).}, are shown in Figure~\ref{fig:euler_proofs} (split into three parts).

All diagrammatic categorical syllogism proofs include a \emph{logical elimination} step -- the dropping of information irrelevant to the conclusion \citep[p.~602]{venn:1883}.

Leonhard Euler, arguably the greatest mathematician ever\footnote{To see just a single example of Euler's impressive achievements, an easy-to-understand problem he devised \emph{and} solved, popularly called the ``Seven Bridges of Königsberg'', is the founding point of \emph{two} major branches of mathematics at once: Graph Theory and Topology.}, has the merit of popularizing this kind of diagram, after a series of didactic tutorials he wrote in French for educating a young princess was published in 1770 in the form a textbook, titled ``\emph{Lettres a une princesse d'Allemagne sur divers sujets de physique \& de philosophie}'' \citep[pp.~99--131, ``Lettre CII'' from February 14th, 1761 -- ``Lettre CV'' from February 24th, 1761]{euler:1770}, which became a best-seller at the time. However, he was definitely \textbf{not} the inventor of the kind of diagram which nowadays bears his name. The earliest occurrence we could find of this kind of diagram is a book from 1661 --a century earlier than Euler's letters-- by \citet[p.~86]{sturm:1661}. Also, around 1686 --decades before Euler was born--, Leibniz applied this kind of diagram to deduce categorical syllogisms, in a draft written in Latin and nowadays known as ``\emph{De Formae Logicae comprobatione per linearum ductus}'' \citep{leibniz:1686e}\footnote{\citet{vacca:1899} rediscovered in the library of Hannover the draft where Leibniz anticipated Euler in the use of ``Euler diagrams'' for logic reasoning. Later, on Vacca's advice \citep[Préface, p.~I]{couturat:1903}\citep{luciano:2012}, Couturat went to the library of Hannover to research Leibniz's manuscripts on logic and then published that insightful draft by \citet{leibniz:1686e}.}, with a presentation that loosely resembles the modern one by \citet{piesk:2017}\footnote{Most of Leibniz's proofs are correct. (His diagrammatic Ferio-1 configuration, for instance, is almost correct, although he made up for it later in Ferison-3; and his ``Fessapmo'', or Fesapo-4 configuration, is not fully correct, although Felapton-3 is).

In his diagrammatic notation, a missing refinement adopted by Euler decades later would have made Leibniz's configurations clearer and less ambiguous: the employment of a symbol analogous to ``\!\!{\Large\textbf\texttimes}\!'' to mark a classificatory term as inhabited \citep[p.~126, ``Lettre CV'' from February 24th, 1761]{euler:1770} -- which perhaps the great Leibniz would never have thought of because, in alignment with Aristotle, his logic assumed that all classificatory terms of interest were necessarily inhabited \citep{leibniz:1686e}:

``\emph{Undes patet omnes imperfectos alterutro modo ex perfectae figurae modis derivari vel addendo praemissae superfluam quantitatem, vel demendo conclusioni utilem.}'' (Hence it is clear that all imperfect moods can be derived from the moods of a perfect figure, either by adding the superfluous quantity to the premises, or by weakening the useful conclusion.)

(The inhabitation assumption, or existential import, appears in various other drafts, for instance in \citet[p.~233]{leibniz:1690b}.)

Nevertheless, Leibniz's diagrammatic configurations are very good for the rigor standards from that age.}. He also realized that the final diagrammatic configuration is the same for similar figures from different moods, which vary from each other only by conversion of premise relations. At the end of his draft essay, Leibniz cites \citeauthor{sturm:1661} and mentions he had read his book when he was young. (For further historical remarks on ``Euler'' diagrams, see \citet{lemanski:2017}(\citeyear{lemanski:2018}) and \citet{bennett:2015}.)

\citet[p.~126, ``Lettre CV'' from February 24th, 1761]{euler:1770} began using a notational device to explicitly mark inhabited classificatory terms in his diagrams, and in the next page (127) he adopted the interpretation that, for a blank minterm/region inside a term circle, it is uncertain whether it is inhabited or not. He had the same agnostic position for the blank minterms/regions inside a term circle in the next example (ibid., pp.~128-130). Those pieces of evidence combined suggest that Euler didn't assume existential import for classificatory terms, and didn't consider that universal assertions declared terms to be inhabited -- only particular assertions did.

However, there are other pieces of evidence contradicting this conclusion. In the next letter, \citet[pp.~136--139, ``Lettre CVI'' from February 28th, 1761]{euler:1770} enumerated 19 classic categorical syllogisms in the following order: Barbara-1, Darii-1, Celarent-1, Ferio-1, Camestres-2, Baroko-2, Cesare-2, Festino-2, Darapti-3, Disamis-3, Datisi-3, Felapton-3, Ferison-3, Bokardo-3, Bamalip-4, Dimatis-4, Calemes-4, Fesapo-4, and Fresison-4. He did not provide their proofs, though; it seems he decided to leave the proofs as an exercise to the student, since he provided examples of proofs in his previous letter \citep[pp.~124--139, ``Lettre CV'' from February 24th, 1761]{euler:1770}. Of these syllogisms, Darapti-3, Felapton-3, Bamalip-4, and Fesapo-4 require an additional existential premise in a logical system lacking existential import.

Missing from \citeauthor{euler:1770}'s enumeration are all and only the classic categorical syllogisms having a ``weakened''/subaltern (from universal to particular) conclusion obtained from other syllogisms: Barbari-1, Celaront-1, Cesaro-2, Camestros-2, and Calemos-4. All these categorical syllogisms require an additional existential premise in a logical system lacking existential import.

We suspect that \citeauthor{euler:1770} simply copied the enumeration of 19 classic categorical syllogisms from another source and trusted the enumeration to be correct, rather than trying to prove them all using his diagrammatic notation. Had Euler tried to prove them all, he would have discovered that some categorical syllogisms in the two-premise form are invalid when existential import is not implicitly assumed for universal categorical assertions. We don't know what would have been his reaction to this information: would he have embraced the lack of existential import as an improvement on Aristotelic logic --like \citet[pp.~283-286]{brentano:1874}\citep{land:1876} did more than a century later--, or would he have tried to ``fix'' his system to accommodate tradition?

The Euler diagrammatic system has been shown here as a motivation for introducing the algebraic and relational systems that follow. The purpose is to show that, while diagrammatic proofs for all 24 classic categorical syllogisms have long been known, we also need algebraic proofs of the same theorems for new insights. As our focus is on justifying and providing the algebraic proofs, we will explicitly describe in this paper neither the axioms nor the inference rules for the Euler diagrammatic system. (See the list of open problems in Section~\ref{sec:future_work}.)

\begin{figure}[tb]
    %\hspace*{-0.53in}
    \centering
    %\raggedright
    \begin{subfigure}[t]{0.55\textwidth}
        \scalebox{0.10}{
            \includegraphics{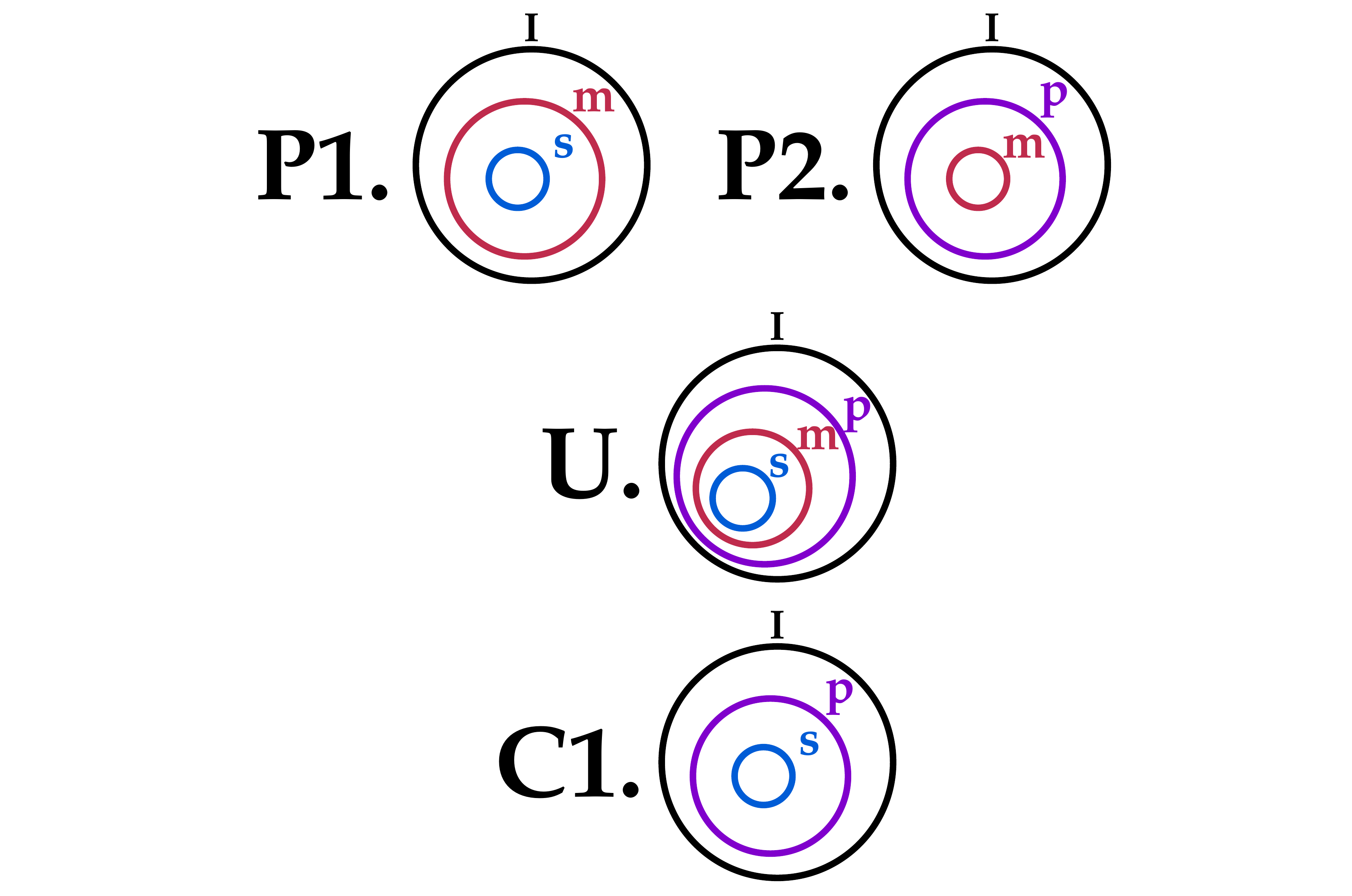}
        }
        \caption{Barbara-1.}
    \end{subfigure}%
    \begin{subfigure}[t]{0.55\textwidth}
        \scalebox{0.10}{
            \includegraphics{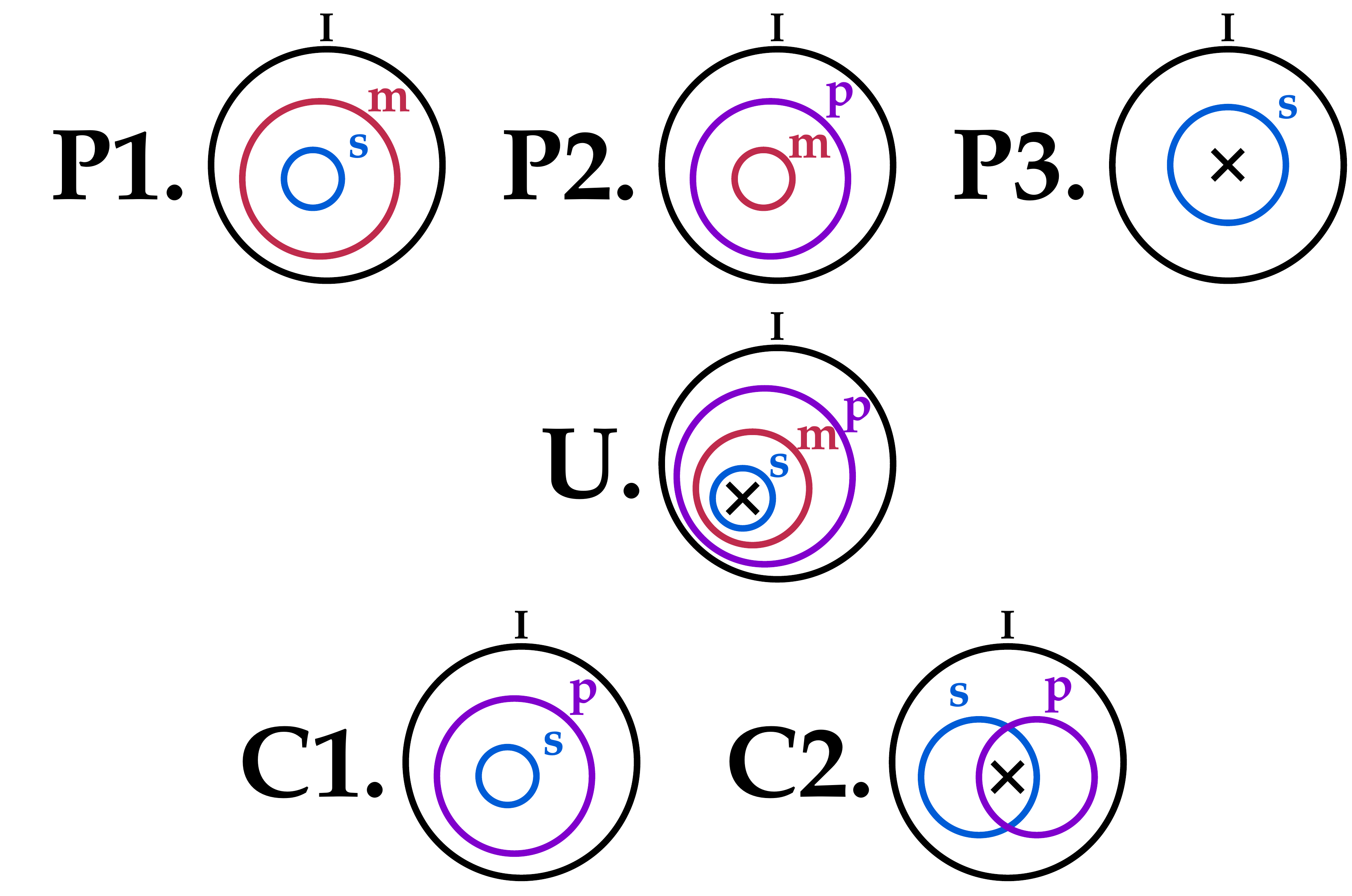}
        }
        \caption{Barbari-1.}
    \end{subfigure}
    \begin{subfigure}[t]{0.55\textwidth}
        \scalebox{0.10}{
            \includegraphics{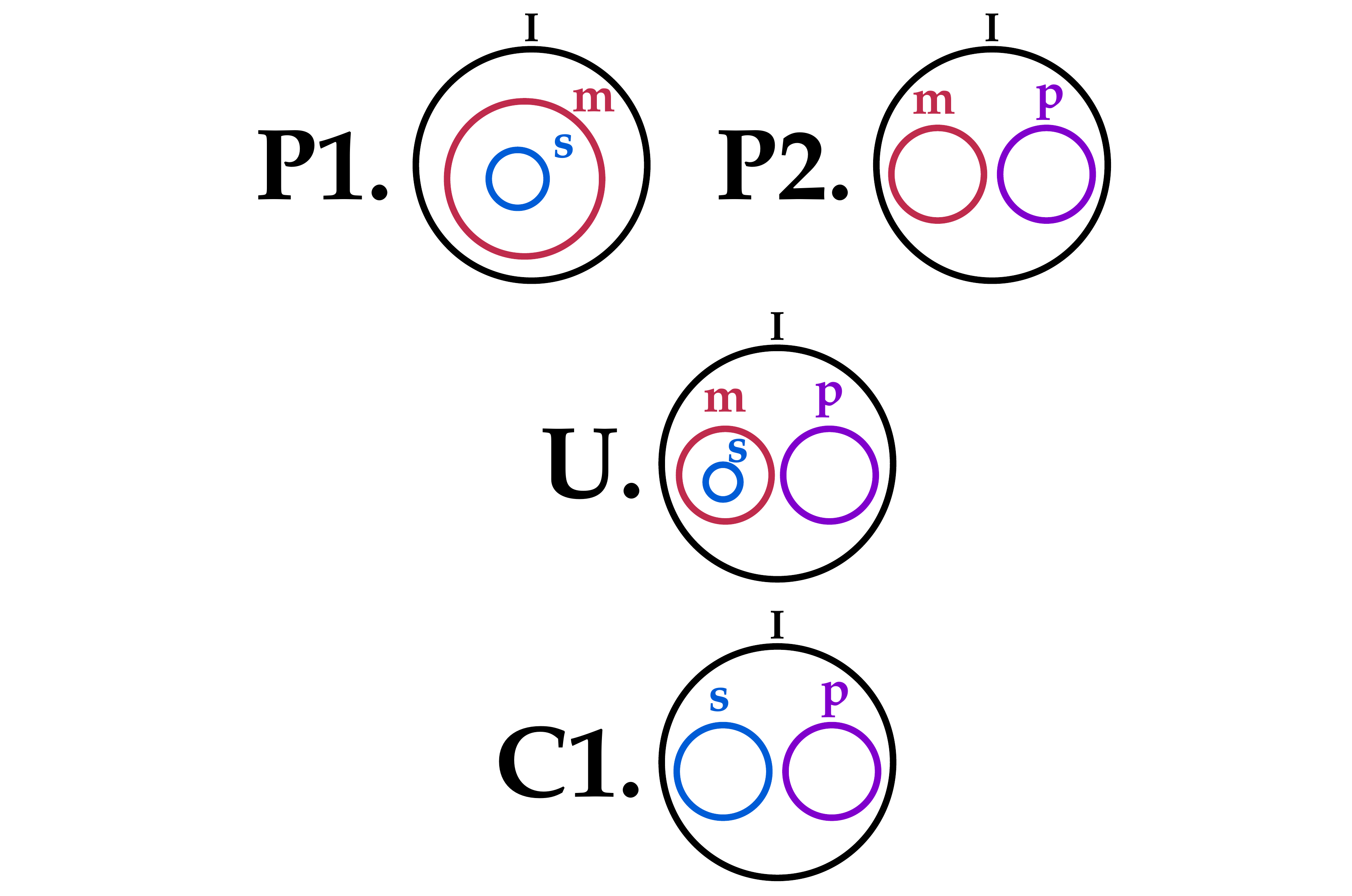}
        }
        \caption{Celarent-1 and Cesare-2.}
    \end{subfigure}%
    \begin{subfigure}[t]{0.55\textwidth}
        \scalebox{0.10}{
            \includegraphics{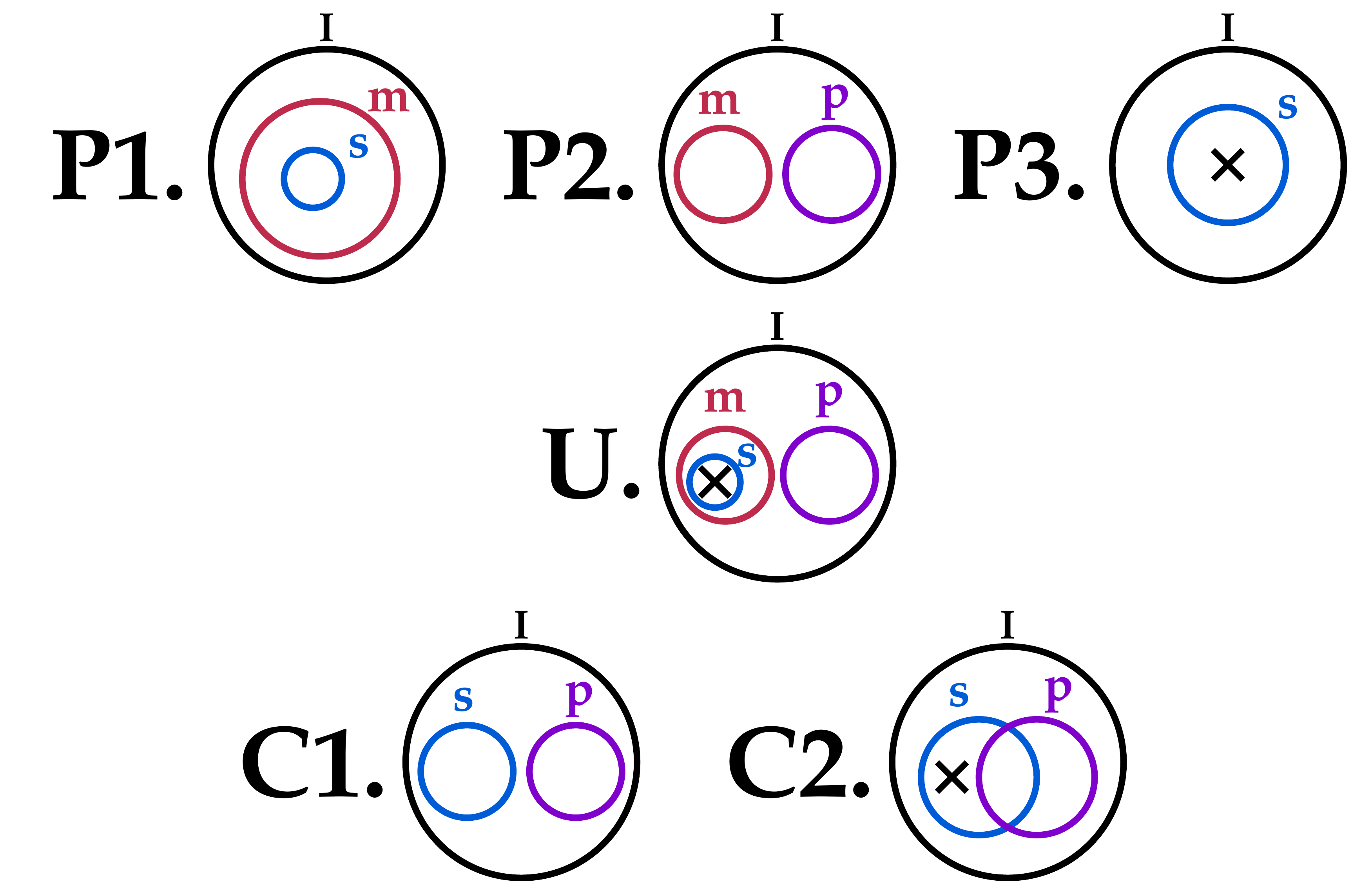}
        }
        \caption{Celaront-1 and Cesaro-2.}
    \end{subfigure}
    \begin{subfigure}[t]{0.55\textwidth}
        \scalebox{0.10}{
            \includegraphics{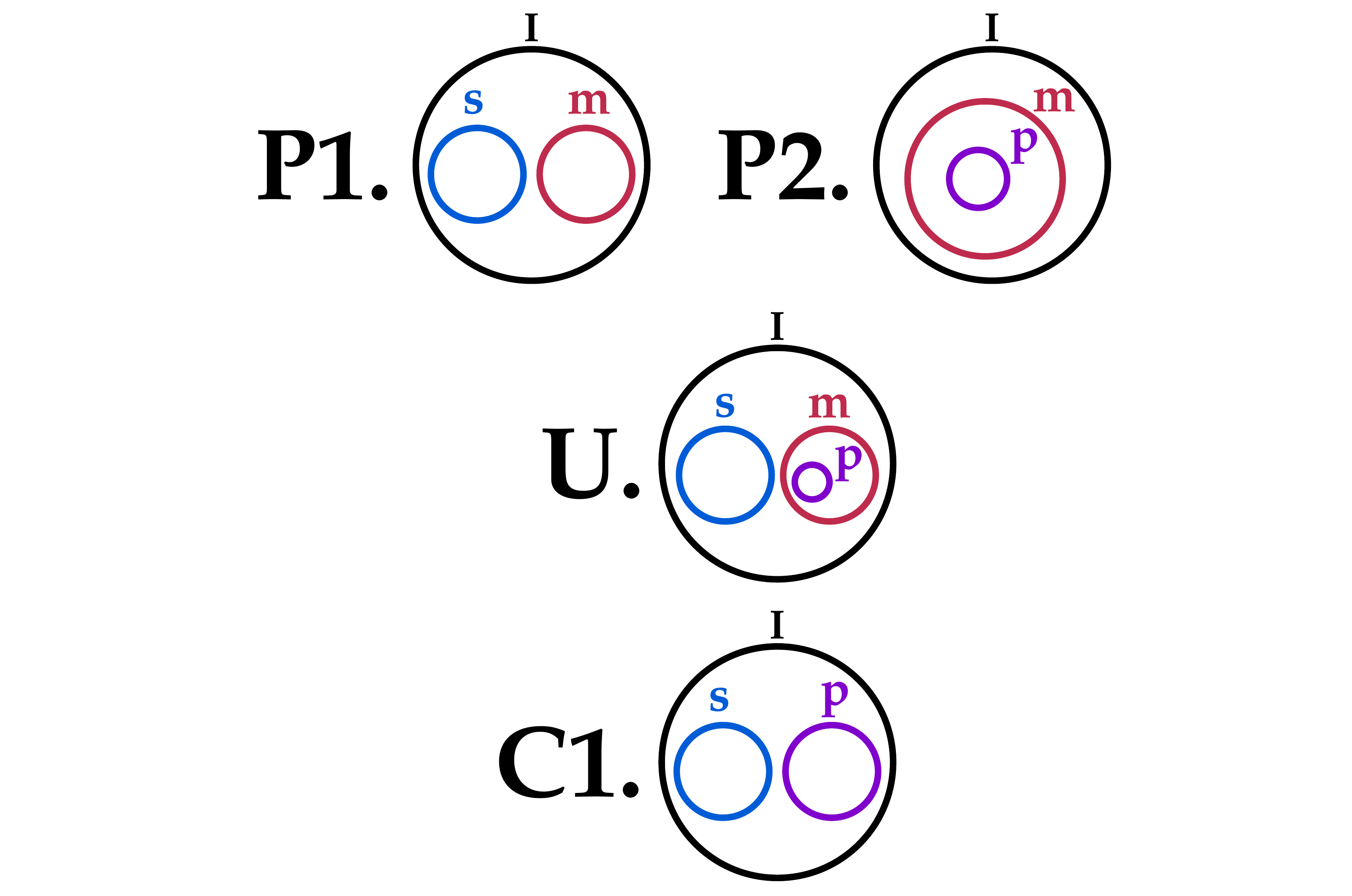}
        }
        \caption{Camestres-2 and Calemes-4.}
    \end{subfigure}%
    \begin{subfigure}[t]{0.55\textwidth}
        \scalebox{0.10}{
            \includegraphics{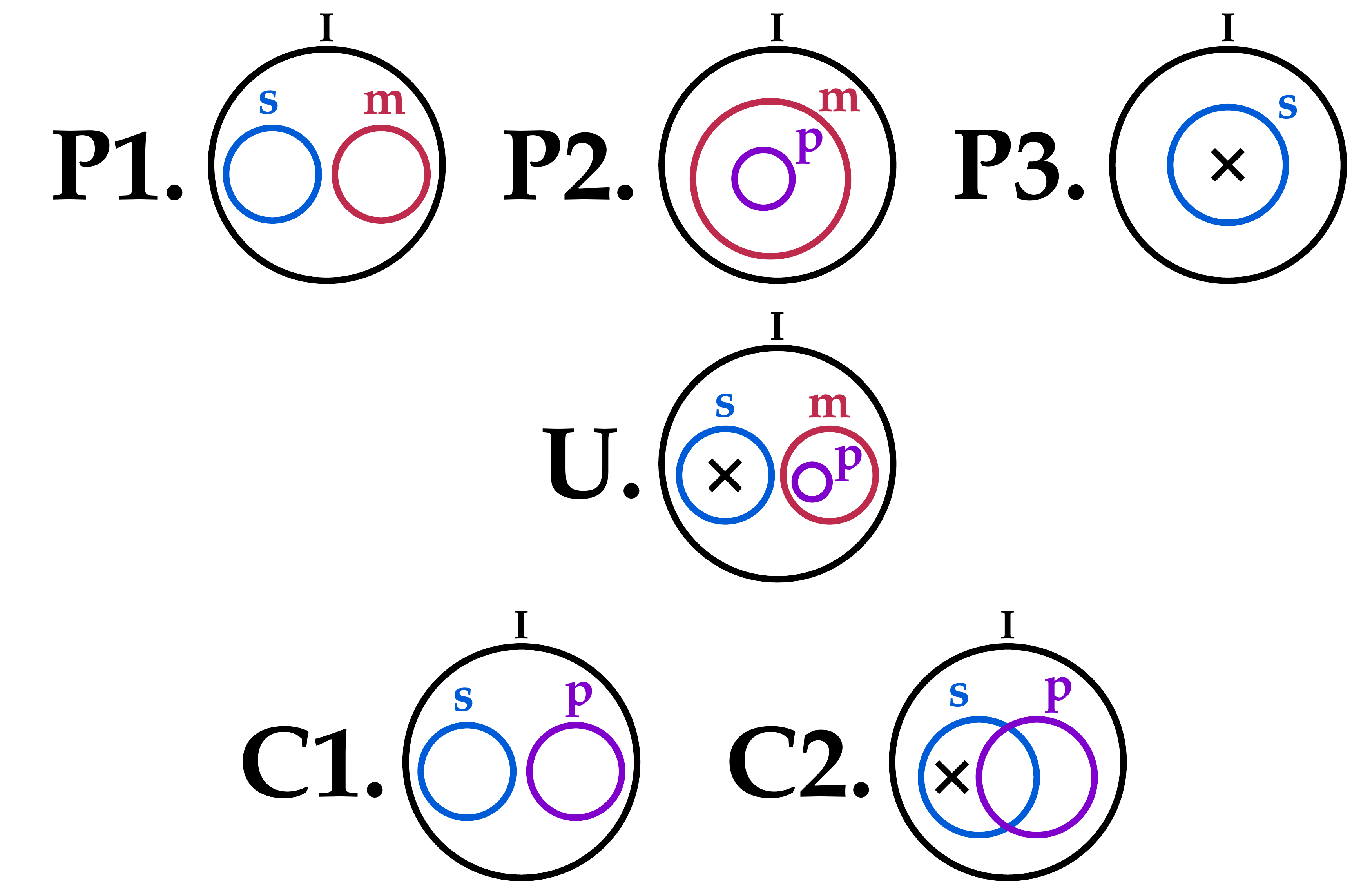}
        }
        \caption{Camestros-2 and Calemos-4.}
    \end{subfigure}
    \caption{Proofs from \citet{piesk:2017} of syllogisms by means of Euler diagrams (part 1 of 3).}
    \label{fig:euler_proofs}
\end{figure}

\begin{figure}[tb]
\ContinuedFloat
    \centering
    \begin{subfigure}[t]{0.55\textwidth}
        \scalebox{0.10}{
            \includegraphics{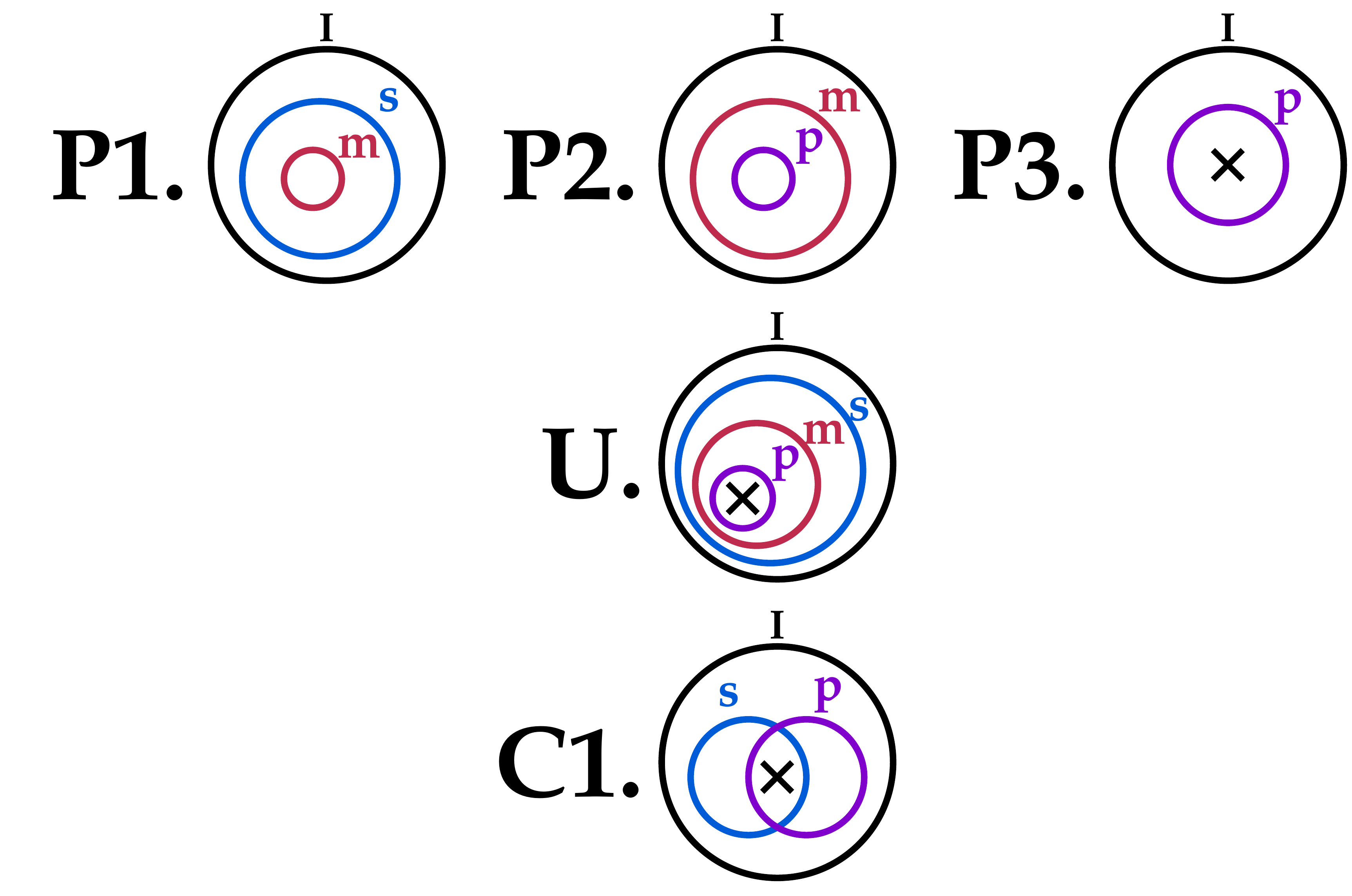}
        }
        \caption{Bamalip-4.}
    \end{subfigure}%
    \begin{subfigure}[t]{0.55\textwidth}
        \scalebox{0.10}{
            \includegraphics{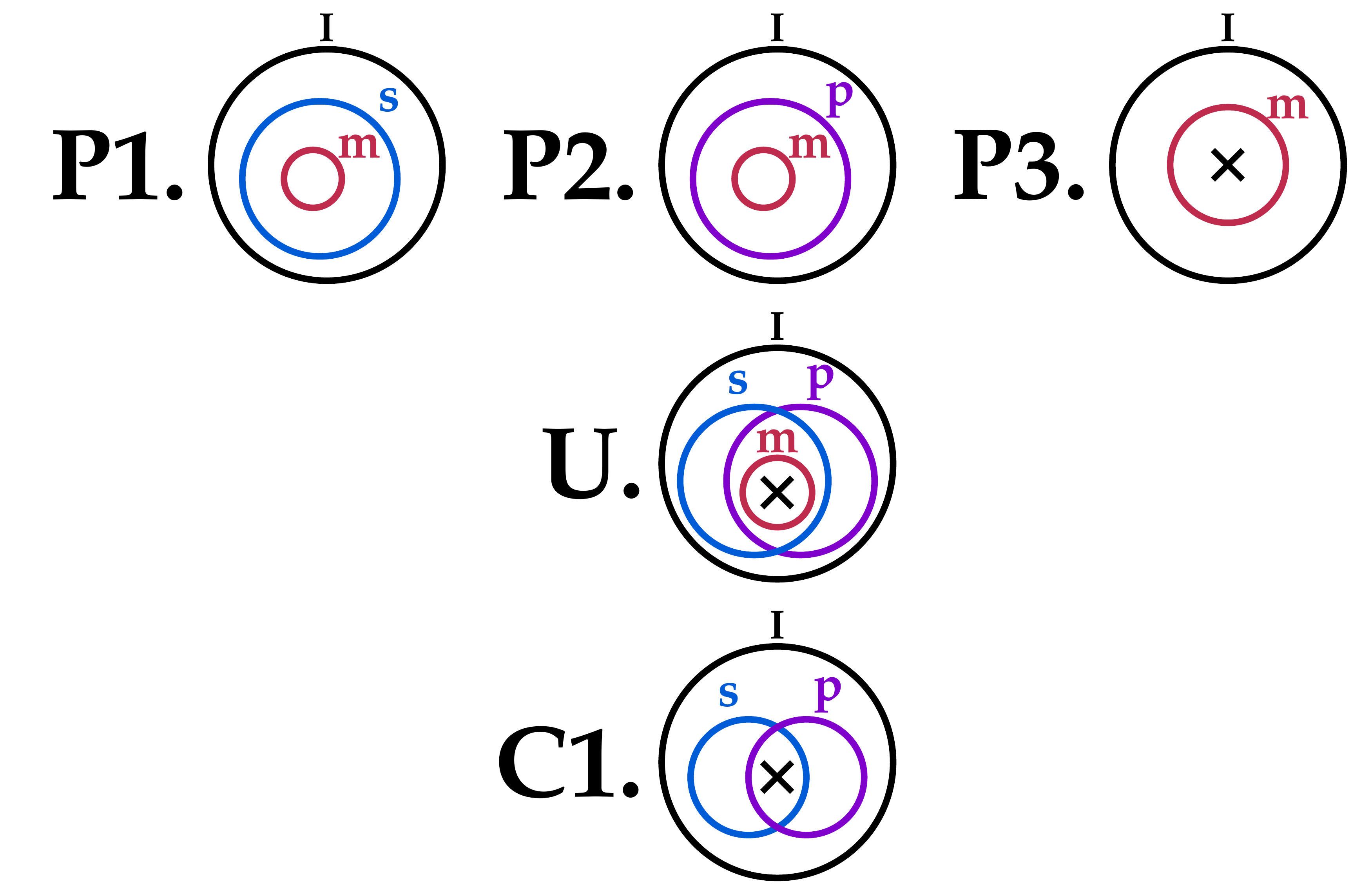}
        }
        \caption{Darapti-3.}
    \end{subfigure}
    \begin{subfigure}[t]{0.55\textwidth}
        \scalebox{0.10}{
            \includegraphics{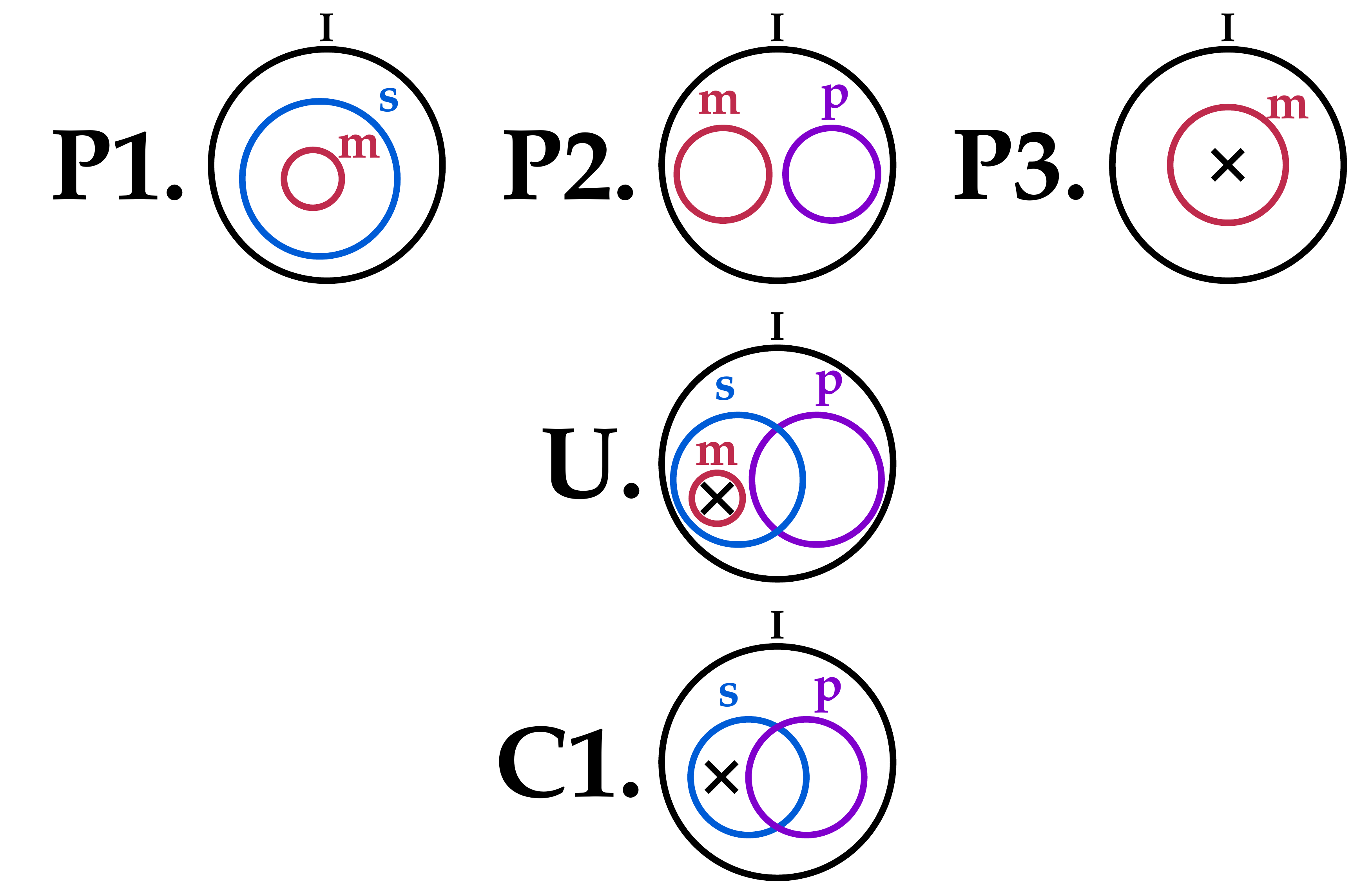}
        }
        \caption{Felapton-3 and Fesapo-4.}
    \end{subfigure}%
    \begin{subfigure}[t]{0.55\textwidth}
        \scalebox{0.10}{
            \includegraphics{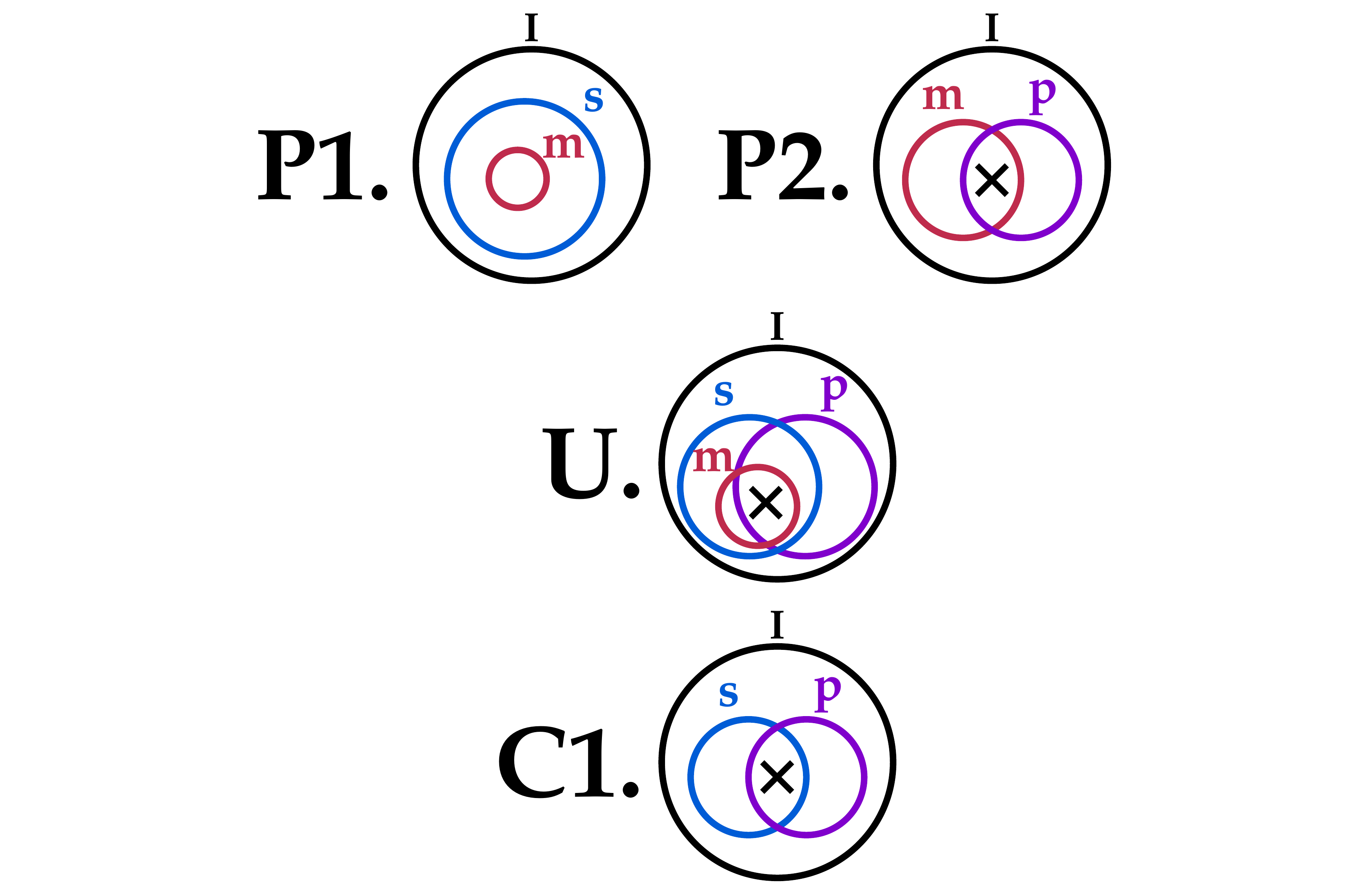}
        }
        \caption{Disamis-3 and Dimatis-4.}
    \end{subfigure}
    \caption{Proofs from \citet{piesk:2017} of syllogisms by means of Euler diagrams (part 2 of 3).}
\end{figure}

\begin{figure}[tb]
\ContinuedFloat
    \centering
    \begin{subfigure}[t]{0.55\textwidth}
        \scalebox{0.10}{
            \includegraphics{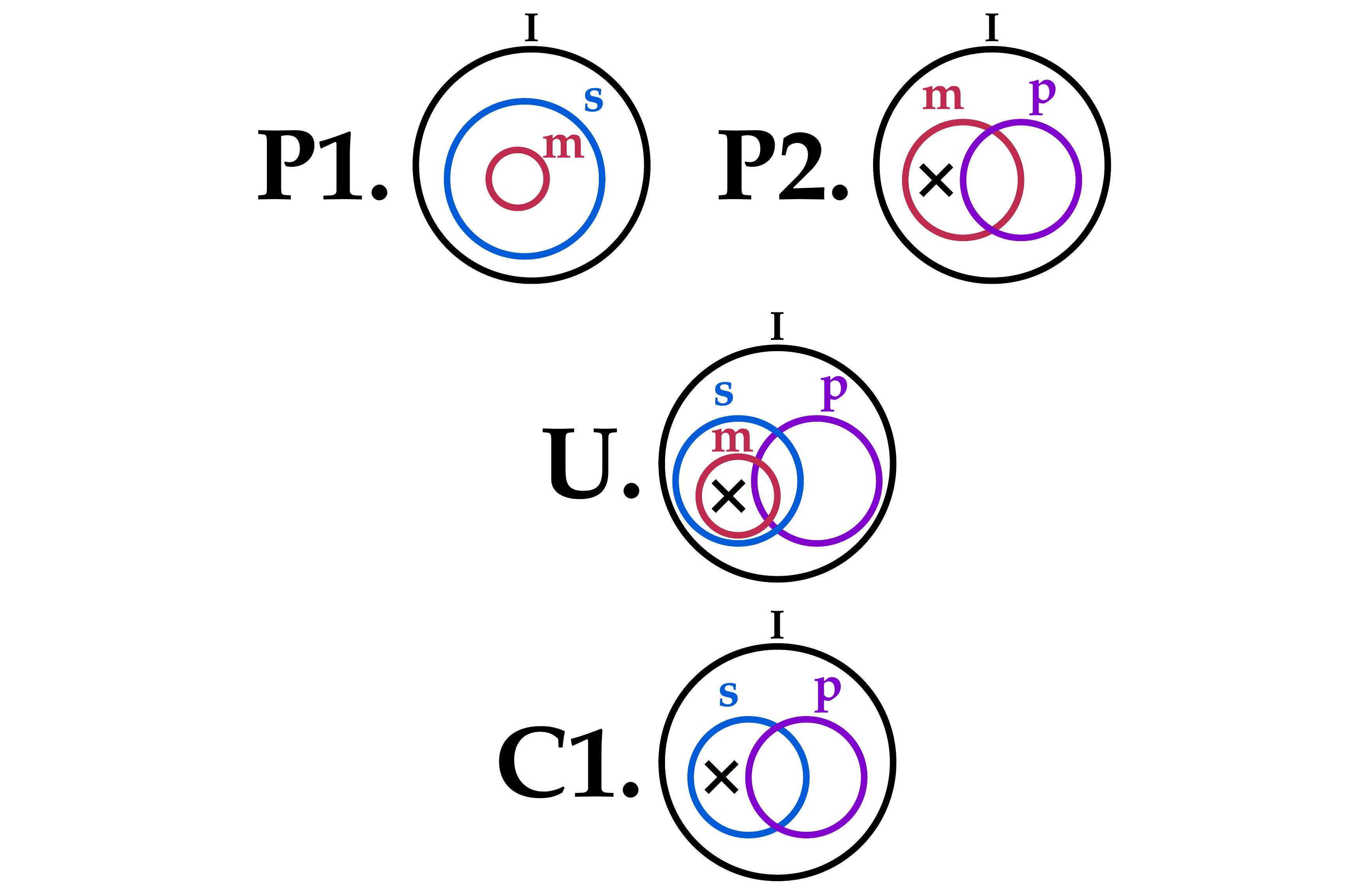}
        }
        \caption{Bokardo-3.}
    \end{subfigure}%
    \begin{subfigure}[t]{0.55\textwidth}
        \scalebox{0.10}{
            \includegraphics{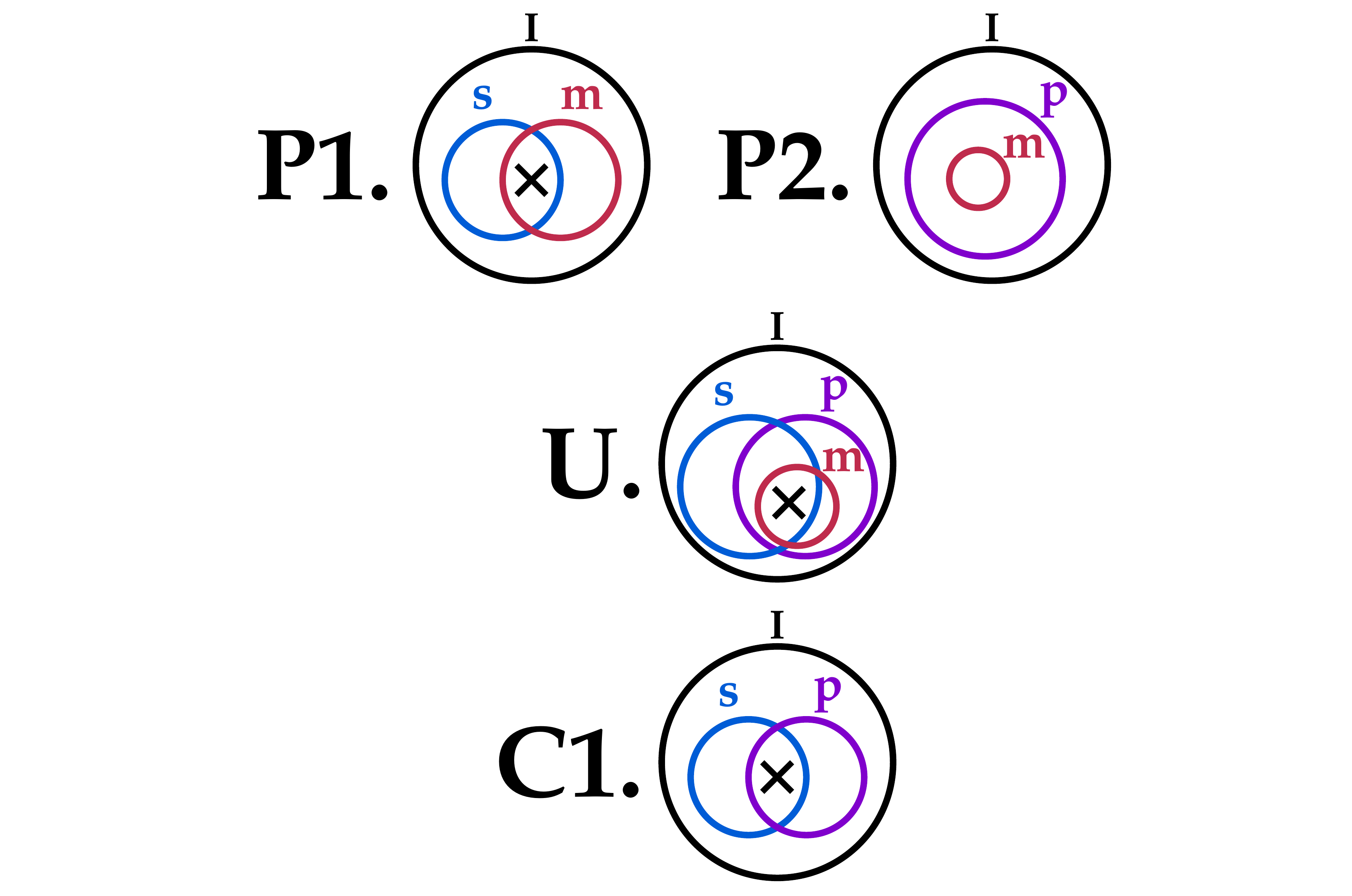}
        }
        \caption{Darii-1 and Datisi-3.}
    \end{subfigure}
    \begin{subfigure}[t]{0.55\textwidth}
        \scalebox{0.10}{
            \includegraphics{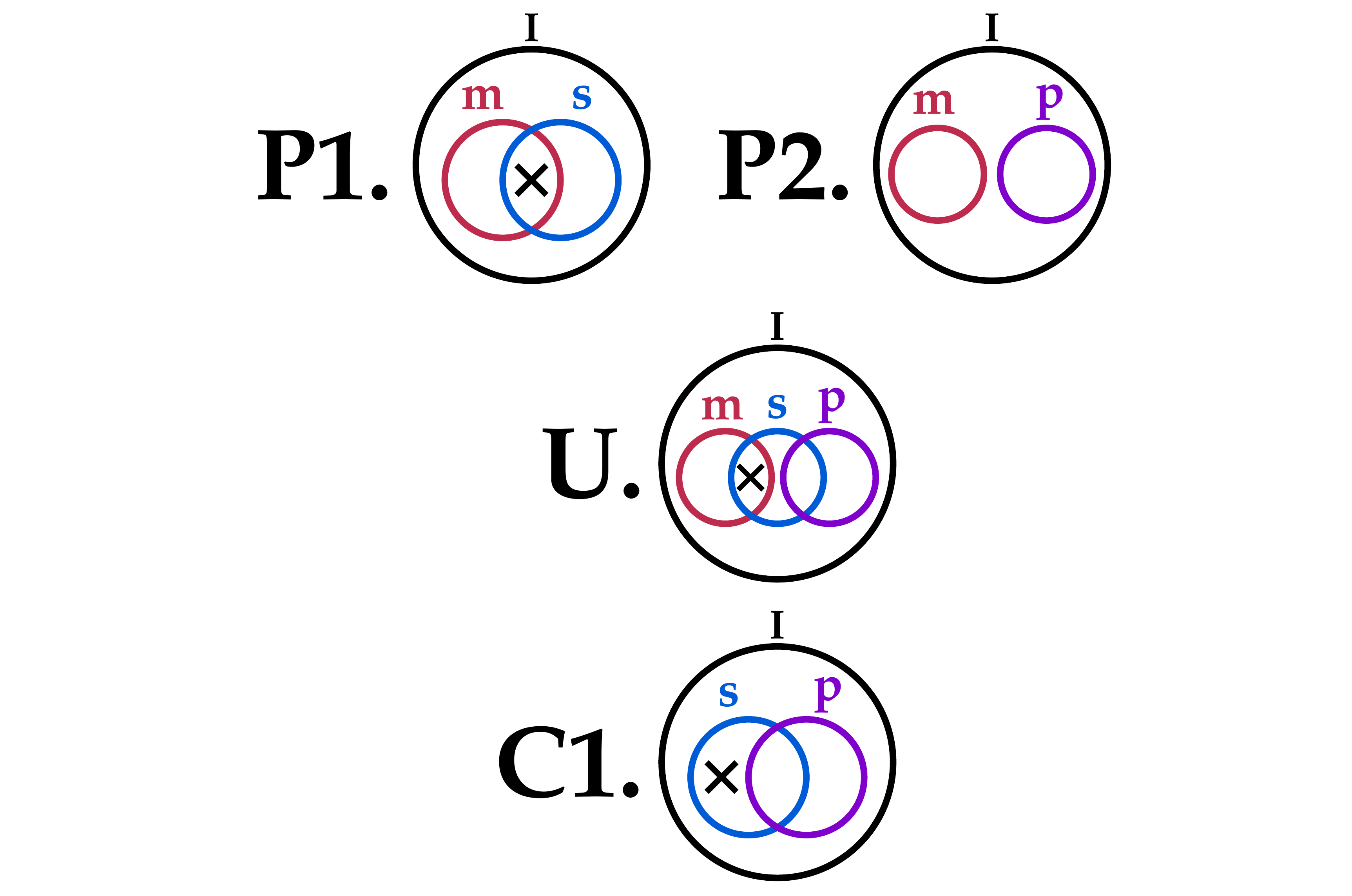}
        }
        \caption{Ferio-1, Festino-2, Ferison-3 and Fresison-4.}
    \end{subfigure}%
    \begin{subfigure}[t]{0.55\textwidth}
        \scalebox{0.10}{
            \includegraphics{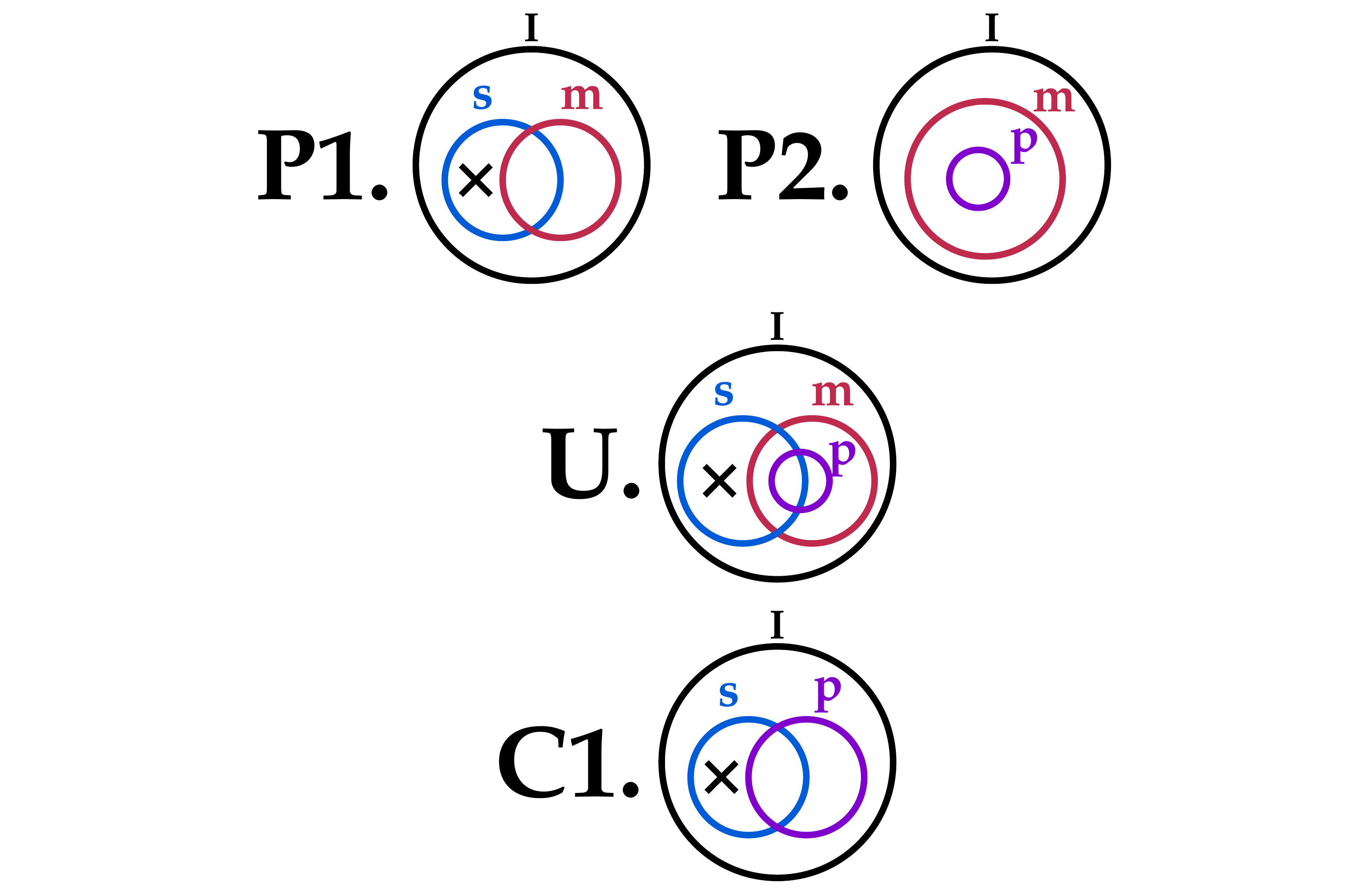}
        }
        \caption{Baroko-2.}
    \end{subfigure}
    \caption{Proofs from \citet{piesk:2017} of syllogisms by means of Euler diagrams (part 3 of 3).}
\end{figure}

\section{Algebraic and relational representations}
\label{sec:algebraic_and_relational_representations}

Tables~\ref{tab:relations_with_identity} and~\ref{tab:relations_with_single_symbol} show various alternative algebraic or relational representations for each fundamental categorical relation\footnote{The identities for $\mathbf{b}\termlogicA\mathbf{c}$ and $\mathbf{b}\termlogicE\mathbf{c}$ in Table~\ref{tab:relations_with_identity} were enumerated by Robert \citet[p.~20, points~40-41]{grassmann:1872}.

The coining of some of the column names for Table~\ref{tab:relations_with_single_symbol} was loosely inspired by \citet[p.~79]{ladd:1890}. Reverse is the inverse of the obverse, like in numismatics.%?seq=5

Missing from Table~\ref{tab:relations_with_single_symbol} (for space reasons) are the representations ``$\mathbf{c}'\termlogicassertion\mathbf{b}$'' and ``$\mathbf{c}\termlogicassertion\mathbf{b}'$'', which are the ``converse of the obverse'' and the ``converse of the reverse'', respectively.}. The symbols have their usual meanings in set algebra. Juxtaposition of terms means intersection of classes.
%64 propositions with [b'?|c'?]R[b'?|c'?] (4*8*2), but only 8 equivalence classes
%https://archive.org/details/studiesandexerci029427mbp/page/141/mode/1up
%See also:
%p. 51. Ivo Thomas. CS(n): An extension of CS. In: Albert Menne. Logico-Philosophical Studies. https://link-springer-com.ez67.periodicos.capes.gov.br/content/pdf/10.1007%2F978-94-010-3649-8.pdf
%R - 8 possibilities: <= <=' ^ ^' U U' >= >='
%bIc: b^c b<='c' b'>='c b'U'c' c^b c'>='b c<='b' c'U'b'
%(8 representations with [b'?|c'?]R[b'?|c'?] for bIc)
%Therefore, 8*8=64 -- perfectly fitting a chess board.
%Every fundamental relation (e.g. bIc) employs half of the 8 possibilites for R, but each of them twice.

The symbols ``$\Cap$''\textbf{/}``$\!\bm{\not}\!\Cap$'' and ``$\Cup$''\textbf{/}``$\!\bm{\not}\!\,\!\Cup$'' mean that their terms are conjoint/disjoint and exhaustive/exclusionary, respectively. They are so defined:

$\mathbf{b}\Cap\mathbf{c} \metaeq \mathbf{b}\cap\mathbf{c}\neq\bm{\varnothing}$

$\mathbf{b}\!\bm{\not}\!\Cap\;\mathbf{c} \metaeq \mathbf{b}\cap\mathbf{c}=\bm{\varnothing}$

$\mathbf{b}\Cup\!\;\mathbf{c} \metaeq \mathbf{b}\cup\mathbf{c}=\textbf{I}$

$\mathbf{b}\!\bm{\not}\!\,\!\Cup\ \mathbf{c} \metaeq \mathbf{b}\cup\mathbf{c}\neq\textbf{I}$.

The corresponding categorical relations are mutually connected by the \citeauthor{de_morgan:1847}'s laws:

$\mathbf{b}\Cap\mathbf{c} = \mathbf{b}'\!\bm{\not}\!\Cup\;\mathbf{c}'$

$\mathbf{b}\!\bm{\not}\!\Cap\;\mathbf{c} = \mathbf{b}'\Cup \mathbf{c}'$
%https://iep.utm.edu/sheffers/#:~:text=metalanguage)%3A

$\mathbf{b}\Cup\mathbf{c} = \mathbf{b}'\!\bm{\not}\!\Cap\;\mathbf{c}'$

$\mathbf{b}\!\bm{\not}\!\Cup\;\mathbf{c} = \mathbf{b}'\Cap\mathbf{c}'$.

The analogous relation to ``$\!\bm{\not}\!\Cap$'' in propositional logic is an assertion involving Sheffer's stroke (``nand'') operation \citep[pp.~9,10]{janssen-lauret:2023}. The analogous relation to ``$\!\bm{\not}\!\Cup$'' in propositional logic is an assertion involving Peirce's arrow (``nor'') operation.
%https://books.google.com/books?id=pRugEAAAQBAJ&pg=PT30

The ``$\Cap$'' and ``$\Cup$'' relations are not as often used as ``$\subseteq$'' and ``$\supseteq$''\footnote{Leibniz developed a logic of containment which employed the ``$\subseteq$'' relation \citep[p.~1]{malink:2019}, writing it as ``\emph{est}''. \citet[pp.~71-72]{segner:1740} adopted symbols that meant ``$\subset$'', ``$\supset$'' and ``$=$'' -- though not symbols corresponding to ``$\subseteq$'' and ``$\supseteq$''. (He also adopted a symbol corresponding to the modern ``$\Cap$'', and a symbol standing for the monadic operation of class complementation.) As we can see, \citeauthor{segner:1740} was more fond of the symbolic tradition than Leibniz, despite offering a superficial treatment of logic which doesn't come close to Leibniz's deep conceptual analyses. \citeauthor{segner:1740}'s novel contributions were simply symbolic notations for some categorical relations.} in the literature about set algebra, but are just as important. \citeauthor{ladd:1883} developed in her Doctoral thesis (\citeyear{ladd:1883}) the earliest in-depth study we could find about the ``$\Cap$'' and ``$\!\bm{\not}\!\Cap$'' relations\footnote{\citeauthor{ladd:1883} actually adopted the symbols ``{\large $\vee$}'' and ``{\large $\overline{\vee}$}'', perhaps influenced by \citeauthor{boole:1847}'s (\citeyear[pp.~21--22]{boole:1847}) use of ``v'' --which was also cited by \citet[p.~229]{wundt:1880}-- to represent ``some'' (``at least one'') in his unsuccessful attempt to algebraically treat particular categorical relations; there is at least a curious resemblance among both forms \citep{halsted:1883}\citep[p.~97]{mitchell:1883}. This might be confusing for an uninitiated reader of \citeauthor{ladd:1883}'s thesis since, in modern notation, ``$\vee$'' is often used in Logic with the meaning ``\emph{or}''.

Like \citeauthor{ladd:1883}'s original notation, the modern one has the semiotic advantage of suggesting symmetrical relations:
%Segner is the earliest logician who emphasized symbolism, according to Venn: https://archive.org/details/symboliclogic01venngoog/page/n453/mode/1up
%Complementation notation, by Segner: https://books.google.com/books?id=FdE6AAAAMAAJ&pg=PA184&dq=%22one+of+the%22+earliest https://archive.org/details/symboliclogic01venngoog/page/n225/mode/1up
%Ladd, Venn, ``v'', and particular relations represented as equalities and inequalities:
%https://www.jstor.org/stable/2011661?seq=3#metadata_info_tab_contents
%In pp. 29,31, Schröder (1873), "Lehrbuch der Arithmetik und Algebra für Lehrer und Studirende", adopts the notation ``(=)'' to represent conjointness. https://books.google.com/books?id=OvMGAAAAYAAJ&pg=PA29 https://archive.org/details/bub_gb_8JYAAAAAMAAJ/page/n44/mode/1up https://www.digitale-sammlungen.de/de/view/bsb11160379?page=47

\begin{minipage}{\textwidth}
$\textbf{b}\Cap\textbf{c} \metaeq \textbf{c}\Cap\textbf{b}$

$\textbf{b}\!\bm{\not}\!\Cap\:\textbf{c} \metaeq \textbf{c}\!\bm{\not}\!\Cap\:\textbf{b}$.
\end{minipage}
%[Schröder commenting on the antisymmetry of McColl's relations "<" and ">" and on the symmetry of Ladd's relations "<>" and "<|>":]
%Es ist demnach auch "()" and "(|)" ein Paar von Beziehungszeichen, welches für die Logik des Umfanges ausreichen würde, während mit einem derselben allein (im frühern Sinne) nicht auszukommen wäre. Ebendieser, die sie nur äusserlich anders gestaltet, bedient sich Miss Ladd in [1].
%[pp. 120,122-123: https://books.google.com/books?id=g9lLAAAAYAAJ&pg=PA123&dq=%22Es+ist+demnach%22]

She was not the earliest adopter of symbols for the categorical relations ``$\bm{\not}\!\Cap$'' and ``$\Cap$'', though. In \citeyear{mounyer:1646} --more than two centuries before \citeauthor{ladd:1883}'s Doctoral thesis--, \citet[pp.~254-263]{mounyer:1646} had already adopted a dedicated symbol for the ``$\!\bm{\not}\!\Cap$'' relation -- they employed ``X''. It is used, for instance, in ``\emph{Theorema 52}'' (ibid., p.~257), where they state Celarent-1, albeit with the converse conclusion. (In the same book [ibid., p.~254], by the way, there appears the earliest occurrence of a truth table we could find.) \citet[pp.~71-72,83]{segner:1740} adopted the same symbol ``X'' for the relation which we represent by the modern notation ``$\Cap$''. \citet[pp.~244-248]{wundt:1880} chose iconic (semiotic-considerate) symbols for disjointness and conjointness: ``$)($'', ``$\between$''. The same conjointness symbol was adopted much later by \citet[p.~59]{menne:1962}\citep[p.~238]{novak:1980}. In 1881, Ladd was already aware of \citeauthor{wundt:1880}'s writings on logic \citep[p.~10]{pietarinen:2015} and developed her logic upon the two mentioned relations which Wundt had assigned dedicated symbols to \citep[p.~17, fn.~1]{ladd:1883}.
%Ladd and Wundt - p. 22, fn. 3: https://www.researchgate.net/publication/272264511_Christine_Ladd-Franklin's_and_Victoria_Welby's_correspondence_with_Charles_Peirce https://doi.org/10.1515/sem-2013-0052
%For ")(" - Left half circle: https://www.fileformat.info/info/unicode/char/1f907/index.htm https://en.wikipedia.org/wiki/Supplemental_Symbols_and_Pictographs#Chart
%Right half circle: https://graphicdesign.stackexchange.com/questions/159549/what-is-the-codepoint-of-the-right-half-circle-in-unicode

\citeauthor{ladd:1883}'s notation did not distinguish the ``$\!\bm{\not}\!\!\Cap$'' relation between terms from the metalogical ``\emph{nand}'' relation between formulae; the object level vs. metalevel distinction was not typical in that era. The metalogical ``\emph{nand}'' relation is a noteworthy alternative to the ``$\!\!\metale$'' assertion, having many interesting properties, many of which have been discovered by \citet{ladd:1883}, such as symmetry and free transposition; we feel that, after the publication of \citeauthor{ladd:1883}'s thesis, the community of logicians has not explored that relation as deeply as they should have done.}. In the same book where \citeauthor{ladd:1883}'s Doctoral thesis was published \citep{peirce:1883}\footnote{\citeauthor{ladd:1883} and \citeauthor{mitchell:1883} were both supervised by Charles Sanders \citeauthor{peirce:1883}, the most important and influential American logician of the 19th century, and the editor of the book which contains their Doctoral theses, among others.}, \citeauthor{mitchell:1883}'s Doctoral thesis (\citeyear{mitchell:1883})\citep[p.~5]{green:1991}\citep[p.~601]{venn:1883} was published, defining the ``$\Cup$'' relation and its complement, ``$\!\bm{\not}\!\!\Cup$''\footnote{For ``$\Cup$'', \citet[p.~75]{mitchell:1883} originally adopted the syntax ``$(\textbf{b}+\textbf{c})_1$''; for ``$\!\bm{\not}\!\Cup$'', he (p.~97) adopted the syntax ``$(\textbf{b}+\textbf{c})_q$''. \citeauthor{mitchell:1883} also made use of \citeauthor{ladd:1883}'s relations ``$\Cap$'' and ``$\!\bm{\not}\!\!\Cap$'', though he adopted the syntax ``$(\textbf{b}\textbf{c})_u$'' and ``$(\textbf{b}\textbf{c})_0$'', respectively (pp.~75,97), where ``$u$'' means ``at least one in the \textbf{u}niverse of discourse ($\mathbf{I}$)''. Decades earlier than Mitchell, \citet[p.~381]{de_morgan:1846}(\citeyear[pp.~60-61]{de_morgan:1847}) had presented all the 8 relations of the extended Aristotelic syllogistic. \citet[p.~184]{mccoll:1877} also had mentioned the categorical relations which later became the focus of \citeauthor{mitchell:1883}'s investigation. More than a century after Mitchell, \citet{dekker:2015} explored \citeauthor{de_morgan:1846}'s syllogistic and adopted ``$x$'' and ``$y$'' for ``$\Cup$'' and ``$\!\bm{\not}\!\Cup$'', respectively.
}. Some researchers have used the modern notation which we adopt here for the relation symbols, which is beneficial to humans as corroborated by empirical cognitive psychology research \citep{wege:2020}. For instance, \citet[pp.~11,5-7]{icard:2014} uses ``$\Cap$'' and a not too different symbol for ``$\Cup$''.

\citeauthor{ladd:1883}'s relations ``$\Cap$'' and ``$\!\bm{\not}\!\!\Cap$'' were decades later also employed by \citet[pp.~11-12]{rescher:1954}, although with their meanings exchanged. As he noticed, Leibniz in a sense anticipated later researchers (such as \citeauthor{ladd:1883}) in understanding the importance of the relations ``$\Cap$'' and ``$\!\bm{\not}\!\Cap$'' for logic, through his notions of ``communicating'' and ``incommunicating'' terms and a few theorems he enunciated which employ these notions \citep{leibniz:1686b}\citep[pp.~268--269]{leibniz:1686d}\citep{leibniz:1686f}\citep{lenzen:2014a}\citep[pp.~17-18]{lewis:1918}. However, unlike \citeauthor{ladd:1883}'s, Leibniz's approach is pre-symbolic: he did not dedicate any specific logical symbol to represent these complementary relations.

In the preface to the book he edited, \citet[p.~v]{peirce:1883} remarks ({\color{blue}annotations} are ours):%https://archive.org/details/bub_gb_V7oIAAAAQAAJ/page/n6/mode/1up

\begin{quote}\guillemotleft Miss Ladd and Mr. Mitchell also use two signs expressive of simple relations involving existence and non-existence; but in their choice of these relations they diverge both from McColl and me, and from one another. In fact, of the eight simple relations of terms signalized by De Morgan, Mr. McColl and I have chosen two {\color{blue}(``$\subseteq$'' vs. ``$\nsubseteq$'')}, Miss Ladd two others {\color{blue}(``$\bm{\not}\!\Cap$'' vs. ``$\Cap$'')}, Mr. Mitchell a fifth and sixth {\color{blue}(``$\Cup$'' vs. ``$\bm{\not}\!\Cup$'')}. {\color{blue}(Missing: ``$\supseteq$'' vs. ``$\nsupseteq$'', the converse relations to ``$\subseteq$'' and ``$\nsubseteq$'', respectively.)} The logical world is thus in a situation to weigh the advantages and disadvantages of the different systems.\guillemotright\end{quote}

\begin{table}[tb]

\begin{subtable}[t]{\textwidth}
\footnotesize
\makebox[\textwidth]{\begin{tabular}{|@{\hspace{0.2em}}M{1.0cm}@{\hspace{0.2em}}|@{\hspace{0.4em}}M{1.6cm}|@{\hspace{0.4em}}M{1.6cm}|@{\hspace{0.4em}}M{1.6cm}|@{\hspace{0.4em}}M{1.6cm}|@{\hspace{0.4em}}M{1.86cm}|@{\hspace{0.0em}}|@{\hspace{0.4em}}M{1.6cm}|}
\hline

Catego-rical relation & In terms of intersection and $\bm{\varnothing}$ & In terms\ \ \ \ of union and \textbf{I} & In terms of intersection and subject & In terms of union and subject & In terms of in-\ tersection and predicative & In terms of union and predicative \\\hline\hline

\multirow{2}{*}{$\mathbf{b}\termlogicI \mathbf{c}$} & \tightbox{$\color{blue}{\mathbf{b}\mathbf{c}\neq\bm{\varnothing}}$} & $\mathbf{b}'\!\cup\! \mathbf{c}'\neq\textbf{I}$ & $\mathbf{b}\mathbf{c}'\neq \mathbf{b}$ & $\mathbf{b}'\!\cup\! \mathbf{c}\neq \mathbf{b}'$ & $\mathbf{b}'\mathbf{c}\neq \mathbf{c}$ & $\mathbf{b}\!\cup\! \mathbf{c}'\neq \mathbf{c}'$ \\

& $\color{blue}{\mathbf{b}\mathbf{c}\supset\bm{\varnothing}}$ & $\mathbf{b}'\!\cup\! \mathbf{c}'\subset\textbf{I}$ & $\mathbf{b}\mathbf{c}'\subset \mathbf{b}$ & $\mathbf{b}'\!\cup\! \mathbf{c}\supset \mathbf{b}'$ & $\mathbf{b}'\mathbf{c}\subset \mathbf{c}$ & $\mathbf{b}\!\cup\! \mathbf{c}'\supset \mathbf{c}'$ \\\hline

\multirow{2}{*}{$\mathbf{b}\termlogicE \mathbf{c}$} & \tightbox{$\color{blue}{\mathbf{b}\mathbf{c}=\bm{\varnothing}}$} & $\mathbf{b}'\!\cup\! \mathbf{c}'=\textbf{I}$ & $\mathbf{b}\mathbf{c}'=\mathbf{b}$ & $\mathbf{b}'\!\cup\! \mathbf{c}=\mathbf{b}'$ & $\mathbf{b}'\mathbf{c}=\mathbf{c}$ & $\mathbf{b}\!\cup\! \mathbf{c}'=\mathbf{c}'$ \\

& $\color{blue}{\mathbf{b}\mathbf{c}\subseteq\bm{\varnothing}}$ & $\mathbf{b}'\!\cup\! \mathbf{c}'\supseteq\textbf{I}$ & $\mathbf{b}\mathbf{c}'\supseteq \mathbf{b}$ & $\mathbf{b}'\!\cup\! \mathbf{c}\subseteq \mathbf{b}'$ & $\mathbf{b}'\mathbf{c}\supseteq \mathbf{c}$ & $\mathbf{b}\!\cup\! \mathbf{c}'\subseteq \mathbf{c}'$ \\\hline\hline

\multirow{2}{*}{$\mathbf{b}\termlogicO \mathbf{c}$} & $\color{blue}{\mathbf{b}\mathbf{c}'\neq\bm{\varnothing}}$ & $\mathbf{b}'\!\cup\! \mathbf{c}\neq\textbf{I}$ & $\mathbf{b}\mathbf{c}\neq \mathbf{b}$ & $\mathbf{b}'\!\cup\! \mathbf{c}'\neq \mathbf{b}'$ & $\mathbf{b}'\mathbf{c}'\neq \mathbf{c}'$ & $\mathbf{b}\!\cup\! \mathbf{c}\neq \mathbf{c}$ \\

& $\color{blue}{\mathbf{b}\mathbf{c}'\supset\bm{\varnothing}}$ & $\mathbf{b}'\!\cup\! \mathbf{c}\subset\textbf{I}$ & $\mathbf{b}\mathbf{c}\subset \mathbf{b}$ & $\mathbf{b}'\!\cup\! \mathbf{c}'\supset \mathbf{b}'$ & $\mathbf{b}'\mathbf{c}'\subset \mathbf{c}'$ & \tightbox{$\mathbf{b}\!\cup\! \mathbf{c}\supset \mathbf{c}$} \\\hline

\multirow{2}{*}{$\mathbf{b}\termlogicA \mathbf{c}$} & $\color{blue}{\mathbf{b}\mathbf{c}'=\bm{\varnothing}}$ & $\mathbf{b}'\!\cup\! \mathbf{c}=\textbf{I}$ & $\mathbf{b}\mathbf{c}=\mathbf{b}$ & $\mathbf{b}'\!\cup\! \mathbf{c}'=\mathbf{b}'$ & $\mathbf{b}'\mathbf{c}'=\mathbf{c}'$ & $\mathbf{b}\!\cup\! \mathbf{c}=\mathbf{c}$ \\

& $\color{blue}{\mathbf{b}\mathbf{c}'\subseteq\bm{\varnothing}}$ & $\mathbf{b}'\!\cup\! \mathbf{c}\supseteq\textbf{I}$ & $\mathbf{b}\mathbf{c}\supseteq \mathbf{b}$ & $\mathbf{b}'\!\cup\! \mathbf{c}'\subseteq \mathbf{b}'$ & $\mathbf{b}'\mathbf{c}'\supseteq \mathbf{c}'$ & \tightbox{$\mathbf{b}\!\cup\! \mathbf{c}\subseteq \mathbf{c}$} \\\hline\hline

\multirow{2}{*}{$\mathbf{b}\termlogicOumlaut \mathbf{c}$} & $\color{blue}{\mathbf{b}'\mathbf{c}\neq\bm{\varnothing}}$ & $\mathbf{b}\!\cup\! \mathbf{c}'\neq\textbf{I}$ & $\mathbf{b}'\mathbf{c}'\neq \mathbf{b}'$ & $\mathbf{b}\!\cup\! \mathbf{c}\neq \mathbf{b}$ & $\mathbf{b}\mathbf{c}\neq \mathbf{c}$ & $\mathbf{b}'\!\cup\! \mathbf{c}'\neq \mathbf{c}'$ \\

& $\color{blue}{\mathbf{b}'\mathbf{c}\supset\bm{\varnothing}}$ & $\mathbf{b}\!\cup\! \mathbf{c}'\subset\textbf{I}$ & $\mathbf{b}'\mathbf{c}'\subset \mathbf{b}'$ & $\mathbf{b}\!\cup\! \mathbf{c}\supset \mathbf{b}$ & \tightbox{$\mathbf{b}\mathbf{c}\subset \mathbf{c}$} & $\mathbf{b}'\!\cup\! \mathbf{c}'\supset \mathbf{c}'$ \\\hline

\multirow{2}{*}{$\mathbf{b}\termlogicAumlaut \mathbf{c}$} & $\color{blue}{\mathbf{b}'\mathbf{c}=\bm{\varnothing}}$ & $\mathbf{b}\!\cup\! \mathbf{c}'=\textbf{I}$ & $\mathbf{b}'\mathbf{c}'=\mathbf{b}'$ & $\mathbf{b}\!\cup\! \mathbf{c}=\mathbf{b}$ & $\mathbf{b}\mathbf{c}=\mathbf{c}$ & $\mathbf{b}'\!\cup\! \mathbf{c}'=\mathbf{c}'$ \\

& $\color{blue}{\mathbf{b}'\mathbf{c}\subseteq\bm{\varnothing}}$ & $\mathbf{b}\!\cup\! \mathbf{c}'\supseteq\textbf{I}$ & $\mathbf{b}'\mathbf{c}'\supseteq \mathbf{b}'$ & $\mathbf{b}\!\cup\! \mathbf{c}\subseteq \mathbf{b}$ & \tightbox{$\mathbf{b}\mathbf{c}\supseteq \mathbf{c}$} & $\mathbf{b}'\!\cup\! \mathbf{c}'\subseteq \mathbf{c}'$ \\\hline\hline

\multirow{2}{*}{$\mathbf{b}\termlogicIumlaut \mathbf{c}$} & $\color{blue}{\mathbf{b}'\mathbf{c}'\neq\bm{\varnothing}}$ & \tightbox{$\mathbf{b}\!\cup\! \mathbf{c}\neq\textbf{I}$} & $\mathbf{b}'\mathbf{c}\neq \mathbf{b}'$ & $\mathbf{b}\!\cup\! \mathbf{c}'\neq \mathbf{b}$ & $\mathbf{b}\mathbf{c}'\neq \mathbf{c}'$ & $\mathbf{b}'\!\cup\! \mathbf{c}\neq \mathbf{c}$ \\

& $\color{blue}{\mathbf{b}'\mathbf{c}'\supset\bm{\varnothing}}$ & $\mathbf{b}\!\cup\! \mathbf{c}\subset\textbf{I}$ & $\mathbf{b}'\mathbf{c}\subset \mathbf{b}'$ & $\mathbf{b}\!\cup\! \mathbf{c}'\supset \mathbf{b}$ & $\mathbf{b}\mathbf{c}'\subset \mathbf{c}'$ & $\mathbf{b}'\!\cup\! \mathbf{c}\supset \mathbf{c}$ \\\hline

\multirow{2}{*}{$\mathbf{b}\termlogicEumlaut \mathbf{c}$} & $\color{blue}{\mathbf{b}'\mathbf{c}'=\bm{\varnothing}}$ & \tightbox{$\mathbf{b}\!\cup\! \mathbf{c}=\textbf{I}$} & $\mathbf{b}'\mathbf{c}=\mathbf{b}'$ & $\mathbf{b}\!\cup\! \mathbf{c}'=\mathbf{b}$ & $\mathbf{b}\mathbf{c}'=\mathbf{c}'$ & $\mathbf{b}'\!\cup\! \mathbf{c}=\mathbf{c}$ \\

& $\color{blue}{\mathbf{b}'\mathbf{c}'\subseteq\bm{\varnothing}}$ & $\mathbf{b}\!\cup\! \mathbf{c}\supseteq\textbf{I}$ & $\mathbf{b}'\mathbf{c}\supseteq \mathbf{b}'$ & $\mathbf{b}\!\cup\! \mathbf{c}'\subseteq \mathbf{b}$ & $\mathbf{b}\mathbf{c}'\supseteq \mathbf{c}'$ & $\mathbf{b}'\!\cup\! \mathbf{c}\subseteq \mathbf{c}$ \\\hline

\end{tabular}}
\caption{Algebraic representations of categorical relations with identity and non-identity.}
\label{tab:relations_with_identity}
%\end{table}
\end{subtable}
\begin{subtable}[t]{\textwidth}
\footnotesize
\makebox[\textwidth]{\begin{tabular}{|@{\hspace{0.2em}}M{1.0cm}@{\hspace{0.2em}}|@{\hspace{0.4em}}M{1.3cm}!{\vrule width 1pt}@{\hspace{0.4em}}M{1.3cm}!{\vrule width 1pt}@{\hspace{0.4em}}M{1.3cm}|@{\hspace{0.4em}}M{1.3cm}|@{\hspace{0.4em}}M{1.3cm}!{\vrule width 1pt}@{\hspace{0.4em}}M{1.3cm}|@{\hspace{0.4em}}M{1.3cm}|}
\hline

%Catego-rical relation & Default symbology ($\mathbf{b}\termlogicassertion \mathbf{c}$) & Comple-mentary ($\sim\!\!(\mathbf{b}\termlogicassertion \mathbf{c})$) to & Reverse ($\mathbf{b}'\termlogicassertion \mathbf{c}$) of & Obverse ($\mathbf{b}\termlogicassertion \mathbf{c}'$) of & Inverse ($\mathbf{b}'\termlogicassertion \mathbf{c}'$) of & Converse ($\mathbf{c}\termlogicassertion \mathbf{b}$) of & Contra-positive ($\mathbf{c}'\termlogicassertion \mathbf{b}'$) of \\\hline\hline
Catego-rical relation & Plain symbology:\ \ \ $\mathbf{b}\termlogicassertion \mathbf{c}$ & Comple-mented representation: $\sim\!\!(\mathbf{b}\termlogicassertion \mathbf{c}$) & Obverse representation: $\mathbf{b}\termlogicassertion \mathbf{c}'$ & Reverse representation: $\mathbf{b}'\termlogicassertion \mathbf{c}$ & Inverse representation: $\mathbf{b}'\termlogicassertion \mathbf{c}'$ & Converse representation: $\mathbf{c}\termlogicassertion \mathbf{b}$ & Contra-positive representation: $\mathbf{c}'\termlogicassertion \mathbf{b}'$ \\\hline\hline

\begin{tabular}{c}\\[-8pt]$\mathbf{b}\termlogicI \mathbf{c}$\end{tabular} & 
%\collectbox{\setlength{\fboxsep}{1pt}\fcolorbox{red}{white}{$\color{blue}{\mathbf{b}\Cap \mathbf{c}}$}} &
\tightbox{$\color{blue}{\mathbf{b}\Cap \mathbf{c}}$} & $\sim\!\!(\mathbf{b}\!\bm{\not}\!\Cap~ \mathbf{c})$ & $\mathbf{b}\nsubseteq \mathbf{c}'$ & $\mathbf{b}'\nsupseteq \mathbf{c}$ & $\mathbf{b}'\!\bm{\not}\!\Cup~ \mathbf{c}'$ & $\mathbf{c}\Cap \mathbf{b}$ & $\mathbf{c}'\!\bm{\not}\!\Cup~ \mathbf{b}'$ \\\hline

\begin{tabular}{c}\\[-8pt]$\mathbf{b}\termlogicE \mathbf{c}$\end{tabular} & \tightbox{$\color{blue}{\mathbf{b}\!\bm{\not}\!\Cap~ \mathbf{c}}$} & $\sim\!\!(\mathbf{b}\Cap \mathbf{c})$ & $\mathbf{b}\subseteq \mathbf{c}'$ & $\mathbf{b}'\supseteq \mathbf{c}$ & $\mathbf{b}'\Cup \mathbf{c}'$ & $\mathbf{c}\!\bm{\not}\!\Cap~ \mathbf{b}$ & $\mathbf{c}'\Cup \mathbf{b}'$ \\\hline\hline

\begin{tabular}{c}\\[-8pt]$\mathbf{b}\termlogicO \mathbf{c}$\end{tabular} & \tightbox{$\mathbf{b}\nsubseteq \mathbf{c}$} & $\sim\!\!(\mathbf{b}\subseteq \mathbf{c})$ & $\color{blue}{\mathbf{b}\Cap \mathbf{c}'}$ & $\mathbf{b}'\!\bm{\not}\!\Cup~ \mathbf{c}$ & $\mathbf{b}'\nsupseteq \mathbf{c}'$ & $\mathbf{c}\nsupseteq \mathbf{b}$ & $\mathbf{c}'\nsubseteq \mathbf{b}'$ \\\hline

\begin{tabular}{c}\\[-8pt]$\mathbf{b}\termlogicA \mathbf{c}$\end{tabular} & \tightbox{$\mathbf{b}\subseteq \mathbf{c}$} & $\sim\!\!(\mathbf{b}\nsubseteq \mathbf{c})$ & $\color{blue}{\mathbf{b}\!\bm{\not}\!\Cap~ \mathbf{c}'}$ & $\mathbf{b}'\Cup \mathbf{c}$ & $\mathbf{b}'\supseteq \mathbf{c}'$ & $\mathbf{c}\supseteq \mathbf{b}$ & $\mathbf{c}'\subseteq \mathbf{b}'$ \\\hline\hline

\begin{tabular}{c}\\[-8pt]$\mathbf{b}\termlogicOumlaut \mathbf{c}$\end{tabular} & \tightbox{$\mathbf{b}\nsupseteq \mathbf{c}$} & $\sim\!\!(\mathbf{b}\supseteq \mathbf{c})$ & $\mathbf{b}\!\bm{\not}\!\Cup~ \mathbf{c}'$ & $\color{blue}{\mathbf{b}'\Cap \mathbf{c}}$ & $\mathbf{b}'\nsubseteq \mathbf{c}'$ & $\mathbf{c}\nsubseteq \mathbf{b}$ & $\mathbf{c}'\nsupseteq \mathbf{b}'$ \\\hline

\begin{tabular}{c}\\[-8pt]$\mathbf{b}\termlogicAumlaut \mathbf{c}$\end{tabular} & \tightbox{$\mathbf{b}\supseteq \mathbf{c}$} & $\sim\!\!(\mathbf{b}\nsupseteq \mathbf{c})$ & $\mathbf{b}\Cup \mathbf{c}'$ & $\color{blue}{\mathbf{b}'\!\bm{\not}\!\Cap~ \mathbf{c}}$ & $\mathbf{b}'\subseteq \mathbf{c}'$ & $\mathbf{c}\subseteq \mathbf{b}$ & $\mathbf{c}'\supseteq \mathbf{b}'$ \\\hline\hline

\begin{tabular}{c}\\[-8pt]$\mathbf{b}\termlogicIumlaut \mathbf{c}$\end{tabular} & \tightbox{$\mathbf{b}\!\bm{\not}\!\Cup~ \mathbf{c}$} & $\sim\!\!(\mathbf{b}\Cup \mathbf{c})$ & $\mathbf{b}\nsupseteq \mathbf{c}'$ & $\mathbf{b}'\nsubseteq \mathbf{c}$ & $\color{blue}{\mathbf{b}'\Cap \mathbf{c}'}$ & $\mathbf{c}\!\bm{\not}\!\Cup~ \mathbf{b}$ & $\mathbf{c}'\Cap \mathbf{b}'$ \\\hline

\begin{tabular}{c}\\[-8pt]$\mathbf{b}\termlogicEumlaut \mathbf{c}$\end{tabular} & \tightbox{$\mathbf{b}\Cup \mathbf{c}$} & $\sim\!\!(\mathbf{b}\!\bm{\not}\!\Cup~ \mathbf{c})$ & $\mathbf{b}\supseteq \mathbf{c}'$ & $\mathbf{b}'\subseteq \mathbf{c}$ & $\color{blue}{\mathbf{b}'\!\bm{\not}\!\Cap~ \mathbf{c}'}$ & $\mathbf{c}\Cup \mathbf{b}$ & $\mathbf{c}'\!\bm{\not}\!\Cap~ \mathbf{b}'$ \\\hline

\end{tabular}}
\caption{Equivalent relational representations of each categorical relation with a single term on each side of a single relation symbol.}
\label{tab:relations_with_single_symbol}
\end{subtable}

\caption{Algebraic and relational representations of categorical relations.}
\label{tab:algebraic_and_relational_representations}
\end{table}

For each row, the representations in Tables~\ref{tab:relations_with_single_symbol} and~\ref{tab:relations_with_identity} correspond as follows:

\begin{itemize}
    \item Representation with ``$\termlogicE$''\textbf{/}``$\!\bm{\not}\!\!\Cap$'' or ``$\termlogicI$''\textbf{/}``$\Cap$'' ({\color{blue}highlighted} in Table~\ref{tab:relations_with_single_symbol}): relation in terms of intersection and ``$\bm{\varnothing}$'' ({\color{blue}highlighted} in Table~\ref{tab:relations_with_identity}). (These represent the inhabitation/emptiness mark in the appropriate minterm of the respective Venn diagram in Figure~\ref{fig:empty_inhabited_two_terms}.)
    
    \item Representation with ``$\termlogicEumlaut$''\textbf{/}``$\Cup$'' or ``$\termlogicIumlaut$''\textbf{/}``$\!\bm{\not}\!\Cup$'': relation in terms of union and ``\textbf{I}''.
    
    \item Representation with ``$\termlogicAumlaut$''\textbf{/}``$\supseteq$'' or ``$\termlogicOumlaut$''\textbf{/}``$\nsupseteq$'': relation in terms of intersection and predicative (in both sides), or union and subject (in both sides).
    
    \item Representation with ``$\termlogicA$''\textbf{/}``$\subseteq$'' or ``$\termlogicO$''\textbf{/}``$\nsubseteq$'': relation in terms of union and predicative (in both sides), or intersection and subject (in both sides).
    
    \item The representations which mnemonically correspond to plain symbology are\\\tightbox{highlighted} in the respective rows of Tables~\ref{tab:relations_with_single_symbol} and~\ref{tab:relations_with_identity}.
\end{itemize}

Thus the 17th and 19th centuries gifted us with distinct systems of logic in their genesis which, once integrated and harmonized, are actually complementary points of view of the same algebra of term logic \citep{moktefi:2019}\footnote{Jevons argued that the relation ``$=$'' is the most fundamental one. \citet[p.~2]{peirce:1870} argued instead that ``$\subseteq$'' is more fundamental than ``='' -- \citet[p.~177]{mccoll:1877} would likely agree. And Ladd claimed the primacy of ``$\Cap$'' (and ``$\bm{\not}\!\Cap$''). For Leibniz's views, see \citet[pp.~28-33]{malink:2019}. We are free to be agnostic and consider them as complementary viewpoints that shed light on different aspects of the same term logic.}:%See also \citet[p.~21]{peirce:1880}: https://books.google.com/books?id=Hq4EAAAAYAAJ&pg=PA21
%Leibniz ("Notes de Calcul Logique", ca. 1678)<https://books.google.com/books?id=onpADwAAQBAJ&pg=PT199&dq=%22in+one+fragment+of+around%22 https://archive.org/details/opusculesetfrag00coutgoog/page/325/mode/1up> explicitly adopted on an occasion symbols for ``$\subseteq$'', ``$\supseteq$'', ``$\subset$'' and ``$=$''; their negations are represented by the symbols preceded by ``\emph{non} ''.

\begin{itemize}
    \item Diagrammatic (Leibniz, Venn, Carroll): Euler, Venn, and Carroll diagrams;

    \item Equational (Leibniz, Boole, Jevons, Cayley): $=$, $\neq$;
    %\item Equational (Leibniz, Boole, Jevons, Cayley): {\large $\bm{=}$} ({\tiny $\subseteq$ and $\supseteq$}), {\large $\bm{\neq}$} ({\tiny $\subset$ or $\supset$});

    \item Subsumptive (McColl, Peirce): $\subseteq$, $\nsubseteq$, $\nsupseteq$, $\supseteq$;

    \item Semicomplementary (Ladd, Mitchell): $\Cap$, $\bm{\not}\!\Cap$, $\bm{\not}\!\Cup$, $\Cup$.
\end{itemize}

After translating to relation algebra notation, the ``converse representation'' column from Table~\ref{tab:relations_with_single_symbol} becomes:

\vspace{-\topsep}
\begin{multicols}{2}\setlength{\columnseprule}{0pt}
\begin{itemize}\setlength\itemsep{0.25em}\small
    \item $\,\breve{\termlogicI}\,=\,\termlogicI$

    \item $\breve{\termlogicE}=\termlogicE$

    \item $\breve{\termlogicO}=\termlogicOumlaut$

    \item $\breve{\termlogicA}=\termlogicAumlaut$

    \item $\breve{\termlogicOumlaut}=\termlogicO$

    \item $\breve{\termlogicAumlaut}=\termlogicA$

    \item $\,\breve{\termlogicIumlaut}\,=\,\termlogicIumlaut$

    \item $\breve{\termlogicEumlaut}=\termlogicEumlaut$.
\end{itemize}
\end{multicols}

These converse representations can be employed to transform categorical syllogisms in \citeauthor{de_morgan:1846}'s syllogistic into ``perfect'', composition-friendly syllogisms in the ``first'' figure.

Table~\ref{tab:relations_with_identity} shows that at least one representation is available for each of the 8 categorical relations which avoids dealing with the complementation operation; we just have focus on this sublattice of the Boolean lattice generated by the atoms $\{b'c', b'c, bc', bc\}$ and structured by the $\{\cap, \cup\}$ operations:

$\bm{\varnothing} \subseteq \mathbf{b}\mathbf{c} \subseteq \{\mathbf{b},\mathbf{c}\} \subseteq \mathbf{b}\cup \mathbf{c} \subseteq \mathbf{I}$

\begin{figure}[ht]
    \centering
    \includegraphics[scale=1]{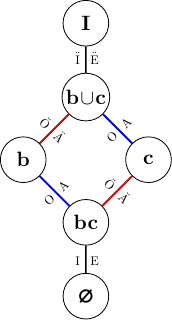}
    \caption{A sublattice of the Boolean lattice generated by the atoms $\{b'c', b'c, bc', bc\}$ where complementation is not employed.}
    \label{fig:sublattice}
\end{figure}

This partial order is also displayed by means of a Hasse diagram in Figure~\ref{fig:sublattice}. The representations using only the elements of that Hasse diagram and the ``$\subseteq$'' relation (or its converse, ``$\supseteq$'', to put the isolated term on the right for uniformity) are in Table~\ref{tab:no_class_complementation}.

\begin{table}[hb]
\begin{tabular}{|p{1.2cm}|c|c|p{2.8cm}|}
\hline
\multirow{2}{*}{Relation} & Extracted from the & Sublattice relation(s), & Respective re-\\
~ & sublattice relation(s) & rewritten ($\subseteq$/$\supseteq$) & presentation(s)\\\hline
$\mathbf{b}\termlogicI \mathbf{c}$ & \multirow{2}{*}{$\bm{\varnothing} \subseteq \mathbf{b}\mathbf{c}$} & \multirow{2}{*}{$\mathbf{b}\mathbf{c} \supseteq \bm{\varnothing}$} & $\mathbf{b}\mathbf{c} \supset \bm{\varnothing}$\\
$\mathbf{b}\termlogicE \mathbf{c}$ & ~ & ~ & $\mathbf{b}\mathbf{c} = \bm{\varnothing}$\\\hline
$\mathbf{b}\termlogicO \mathbf{c}$ & \multirow{2}{*}{$\mathbf{b}\mathbf{c}\ {\color{blue}\bm{\subseteq}}\ \mathbf{b}$, $\mathbf{c}\ {\color{blue}\bm{\subseteq}}\ \mathbf{b}\!\cup\!\mathbf{c}$} & \multirow{2}{*}{$\mathbf{b}\mathbf{c}\ {\color{blue}\bm{\subseteq}}\ \mathbf{b}$, $\mathbf{b}\!\cup\!\mathbf{c}\ {\color{blue}\bm{\supseteq}}\ \mathbf{c}$} & $\mathbf{b}\mathbf{c} \subset \mathbf{b}$, $\mathbf{b}\!\cup\!\mathbf{c} \supset \mathbf{c}$\\
$\mathbf{b}\termlogicA \mathbf{c}$ & ~ & ~ & $\mathbf{b}\mathbf{c} = \mathbf{b}$, $\mathbf{b}\!\cup\!\mathbf{c} = \mathbf{c}$\\\hline
$\mathbf{b}\termlogicOumlaut \mathbf{c}$ & \multirow{2}{*}{$\mathbf{b}\mathbf{c}\ {\color{red}\bm{\subseteq}}\ \mathbf{c}$, $\mathbf{b}\ {\color{red}\bm{\subseteq}}\ \mathbf{b}\!\cup\!\mathbf{c}$} & \multirow{2}{*}{$\mathbf{b}\mathbf{c}\ {\color{red}\bm{\subseteq}}\ \mathbf{c}$, $\mathbf{b}\!\cup\!\mathbf{c}\ {\color{red}\bm{\supseteq}}\ \mathbf{b}$} & $\mathbf{b}\mathbf{c} \subset \mathbf{c}$, $\mathbf{b}\!\cup\!\mathbf{c} \supset \mathbf{b}$\\
$\mathbf{b}\termlogicAumlaut \mathbf{c}$ & ~ & ~ & $\mathbf{b}\mathbf{c} = \mathbf{c}$, $\mathbf{b}\!\cup\!\mathbf{c} = \mathbf{b}$\\\hline
$\mathbf{b}\termlogicIumlaut \mathbf{c}$ & \multirow{2}{*}{$\mathbf{b}\!\cup\!\mathbf{c} \subseteq \mathbf{I}$} & \multirow{2}{*}{$\mathbf{b}\!\cup\!\mathbf{c} \subseteq \mathbf{I}$} & $\mathbf{b}\!\cup\!\mathbf{c} \subset \mathbf{I}$\\
$\mathbf{b}\termlogicEumlaut \mathbf{c}$ & ~ & ~ & $\mathbf{b}\!\cup\!\mathbf{c} = \mathbf{I}$\\\hline
\end{tabular}
\caption{Representations avoiding class complementation.}
\label{tab:no_class_complementation}
\end{table}

It is interesting to notice that what are often (and controversially) called for some reason the three (or four) ``laws of thought''\footnote{Various other fundamental laws of classical logic are enumerated in Section~\ref{subsubsec:Boolean_lattice}.} \citep[pp.~86-87,77]{ladd:1890}\citep{peirce:1901}\citep{leibniz:1679c}\citep[p.~31]{ladd:1883}\citep[p.~48]{richeri:1761}\citep[p.~259,~point~8; p.~261,~point~3]{leibniz:1686d} are just different representations of ``$\mathbf{b}\termlogicA\mathbf{b}$ and $\mathbf{b}\termlogicAumlaut\mathbf{b}$'', as shown in Table~\ref{tab:laws_of_thought}.
%See also (read 2 pages):
%https://books.google.com/books?id=zwC9AAAAIAAJ&pg=PA3&dq=%22letter+to+Clarke%22+%22Leibniz+regards%22+%22Addenda+ad+Specimen+calculi+universalis%22
%Ladd's two laws of identity - See also:
%pp. 17-18
%https://archive.org/details/studiesinlogic00gilmgoog/page/n34/mode/1up
%pp. 533,538
%https://projecteuclid.org/journals/notre-dame-journal-of-formal-logic/volume-62/issue-3/What-Problem-Did-Ladd-Franklin-Think-She-Solved/10.1215/00294527-2021-0026.full
%
%b<=b
%bb'=0
%b'Ub=I
%bb=b
%b'Ub'=b'
%b'b'=b'
%bUb=b
%
%b>=b
%b'b=0
%bUb'=I
%b'b'=b'
%bUb=b
%bb=b
%b'Ub'=b'
%
%b^2=b |=| b(1-b)=0 <p. 112: Boole (1854), https://books.google.com/books?id=SWgLVT0otY8C&pg=PA112&dq=%22obeyed,+the+equation%22>

From Table~\ref{tab:laws_of_thought}, and by looking up in Table~\ref{tab:relations_with_single_symbol} which plain relation corresponds to the contrapositive or converse representation, we can also see that

\begin{enumerate}
    \item $\mathbf{b}$ and $\mathbf{c}$ are \emph{identical}\footnote{\label{fn:positive_and_negative_identity}By employing Boolean algebra laws, \citet[pp.~42-43, points 112-113]{jevons:1864} offers an interesting proof that positive identity entails negative identity and vice-versa ($\mathbf{b}=\mathbf{c} \metaeq \mathbf{b}'=\mathbf{c}'$). %https://books.google.com/books?id=WVMOAAAAYAAJ&pg=PA42&dq=%22Let+A+=+B%22
    %b = b -- The need for stability (and for consistency in substitution)
    %https://books.google.com/books?id=WVMOAAAAYAAJ&pg=PA6&dq=sameness
    Leibniz offered an elegant (and simpler) proof by using only interchangeability of identicals and involution of complementation \citep[point~11]{leibniz:1690c}\citep[p.~266]{lenzen:2018b}.
    %Also Leibniz considered this metaequivalence to be axiomatically true in the end of the page: https://archive.org/details/opusculesetfrag00coutgoog/page/234/mode/1up
    % In point 8: https://archive.org/details/opusculesetfrag00coutgoog/page/235/mode/1up
    %And in point 11: https://archive.org/details/opusculesetfrag00coutgoog/page/421/mode/1up
    %Proof by contraposition - Read 2 pages
    %https://www-jstor-org.ez67.periodicos.capes.gov.br/stable/40694037?seq=26#metadata_info_tab_contents
} ($\mathbf{b}=\mathbf{c}$) if and only if $\mathbf{b}\termlogicA\mathbf{c}$ ($\mathbf{b}\subseteq\mathbf{c}$) and $\mathbf{b}\termlogicAumlaut\mathbf{c}$ ($\mathbf{b}\supseteq\mathbf{c}$).%b<=c, c'>=b'|=|b>=c

    \item $\mathbf{b}$ and $\mathbf{c}$ are \emph{complementary} ($\mathbf{c}=\mathbf{b}'$) if and only if $\mathbf{b}\termlogicE\mathbf{c}$ ($\mathbf{b}\!\!\bm{\not}\!\Cap~\!\mathbf{c}$) and $\mathbf{b}\termlogicEumlaut\mathbf{c}$ ($\mathbf{b}\Cup\mathbf{c}$).%b^'c, cUb|=|bUc
    
    \item $\mathbf{b}$ and $\mathbf{c}$ are simultaneously \emph{identical and complementary} ($\mathbf{b}=\mathbf{c}=\mathbf{b}'$) if and only if the universe is empty ($\mathbf{I}=\bm{\varnothing}$)\footnote{To prove this we also assume idempotence of juxtaposition/$\cap$ and $\cup$.}, and thus in such a special case everything degenerates into emptiness (monovalent algebra).%In the complementarity laws: b'<-b. In this scenario, all four laws are simultaneously satisfied.
    %Monovalent algebra - correspondence between the one-point lattice/ring/magma and the empty universe of discourse - pp. 53,90 (17,54 of the PDF file), theorem 13: Stone (1935), https://www.ams.org/journals/tran/1936-040-01/S0002-9947-1936-1501865-8/S0002-9947-1936-1501865-8.pdf
    %Logic is not committed to the existence of any element in the universe, being ontologically neutral: no existence statement is a theorem of any of these logical systems<https://plato.stanford.edu/entries/lesniewski/#:~:text=logic%20not%20being>. See also: <https://books.google.com/books?id=IMgg0Uc00I4C&pg=PA318&dq=%22logic+should+be+ontologically+neutral%22+%22no+individuals%22>.
\end{enumerate}

\begin{table}[tb]
\begin{tabular}{p{3.23cm}p{3.82cm}p{3.73cm}}
%\hline
Identity laws: & \begin{tabular}{@{}l@{}}Plain symbology\\$\textbf{b}\subseteq\textbf{b},~\textbf{b}\supseteq\textbf{b}$\\Positive identity\\(``Identity'')\end{tabular} & \begin{tabular}{@{}l@{}}Inverse representation\\$\textbf{b}'\supseteq\textbf{b}',~\textbf{b}'\subseteq\textbf{b}'$\\Negative identity\\(\emph{Usually omitted})\end{tabular}\\%\hline
 & & \\%\hline
Complementarity/\ Contradiction laws: & \begin{tabular}{@{}l@{}}Obverse representation\\$\textbf{b}\bm{\not}\!\Cap~\textbf{b}',~\textbf{b}'\bm{\not}\!\Cap~\textbf{b}$\\Disjointness\\(``[Non-]Contradiction'')\end{tabular} & \begin{tabular}{@{}l@{}}Reverse representation\\$\textbf{b}'\Cup\textbf{b},~\textbf{b}\Cup\textbf{b}'$\\Exhaustion\\(``Excluded third'')\end{tabular}\\\hline
\end{tabular}%interdisjointness, mutual disjointness, mutual exclusion; coexhaustion, joint exhaustion
\caption{``Laws of thought'': variant forms of {\tiny\guillemotleft}$\mathbf{b}\termlogicA\mathbf{b},~ \mathbf{b}\termlogicAumlaut\mathbf{b}${\tiny\guillemotright} according to the $\mathbf{b}\termlogicA\mathbf{c}$ and $\mathbf{b}\termlogicAumlaut\mathbf{c}$ rows (where $\mathbf{c}\leftarrow\mathbf{b}$) from Table~\ref{tab:relations_with_single_symbol} \citep[pp.~86-87,77]{ladd:1890}.}
\label{tab:laws_of_thought}
\end{table}

\begin{figure}[ht]
    \centering
    \includegraphics[scale=0.75]{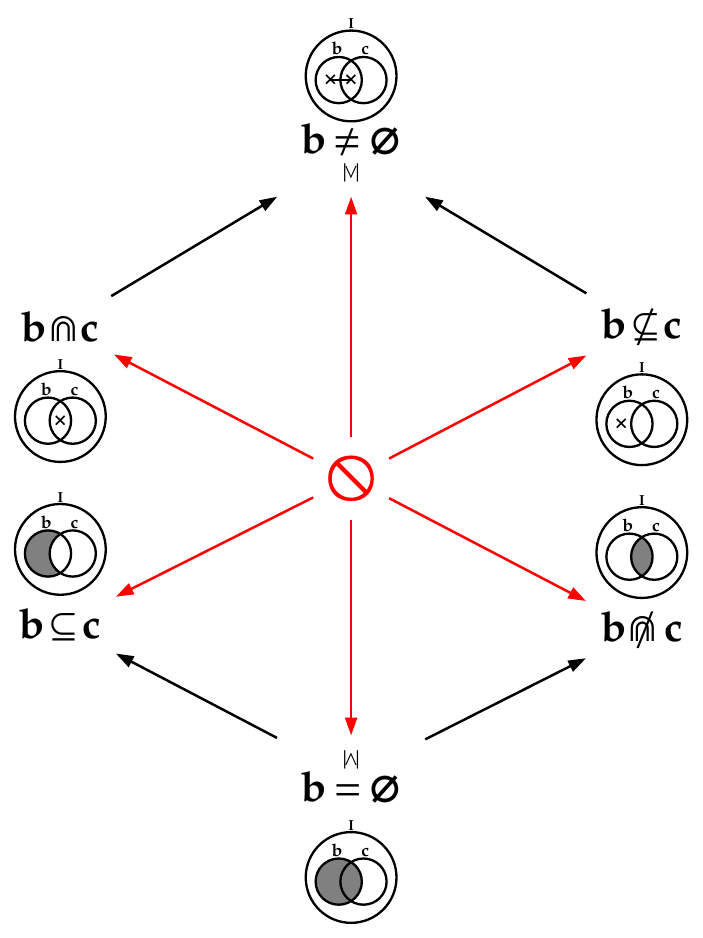}
    \caption{A logical ``hexagon'' (which almost looks like a ``cube'' due to optical illusion) showing the categorical relations $\mathbf{b}\termlogicI\mathbf{c}$, $\mathbf{b}\termlogicO\mathbf{c}$, $\mathbf{b}\termlogicA\mathbf{c}$, and $\mathbf{b}\termlogicE\mathbf{c}$, in the absence of existential import for universal ones.}
    \label{fig:logical_hexagon-categorical_relations}
\end{figure}

Finally, we can combine the symbolic and diagrammatic notations (Venn diagrams) to draw a logical ``hexagon'' that highlights mutually contradictory relations (Figure~\ref{fig:logical_hexagon-categorical_relations}). {\color{red}\textbf{Red}} bidirectional arrows in a straight line indicate mutual contradiction, whereas {\color{black}\textbf{black}} unidirectional arrows indicate implication. Moreover, ``$\mathbf{b}\!=\!\bm{\varnothing}$'' is equivalent to ``$\mathbf{b}\!\subseteq\!\mathbf{c}~{\scriptstyle|\!\land\!|}~\mathbf{b}\!\bm{\not}\!\Cap\,\mathbf{c}$'', and ``$\mathbf{b}\!\neq\!\bm{\varnothing}$'' is equivalent to ``$\mathbf{b}\!\Cap\!\mathbf{c}~{\scriptstyle|\!\lor\!|}~\mathbf{b}\!\nsubseteq\!\mathbf{c}$''.

\section{Replacement and inference rules}
\label{sec:inference_rules}
%https://en.wikipedia.org/wiki/Equational_logic#Syllogism
%https://web.archive.org/web/19970618005333/http://www.cs.cornell.edu/home/gries/Logic/Equational.html#Inference

In the description of each algebraic system, we make the axioms explicit. As a pedantic remark in the interest of precision, we describe here the replacement and inference rules adopted by the algebraic systems discussed in this paper.

An inference rule they assume is \emph{substitution of equivalent expressions}\footnote{An informal, example-driven treatment is given by \citet[pp.~9\{point~23\},18-19,27]{jevons:1864}.}. If we know that $expr_1 = expr_2$, then we can replace $expr_1$ for $expr_2$ (or vice-versa) anywhere the other expression appears:

$expr_1 = expr_2,~rel~\qquad~ \metale \qquad~rel[expr_2\leftarrow expr_1]$

For instance, suppose that $(\mathbf{b}\mathbf{c})\mathbf{d}=\mathbf{f}\mathbf{g}$ and $(\mathbf{f}\mathbf{g})\mathbf{h}\neq \mathbf{j}$; then $((\mathbf{b}\mathbf{c})\mathbf{d})\mathbf{h}\neq \mathbf{j}$.

In particular, by making $rel = expr_2~\termlogicassertion~expr_3$, where ``$\termlogicassertion$'' is a dyadic relation\footnote{Jevons --the modern founder of what we nowadays call ``Boolean algebra'', which differs in some key aspects from \citeauthor{boole:1847}'s original algebra \citep[pp.~61,82-83,119-121,139-140]{hailperin:1986}\citep[pp.~74,78]{lewis:1918}-- called this specialized rule the ``law of sameness'', ``substitution of equals'' or ``\textbf{\emph{substitution of similars}}'' \citep[pp.~V,16-27]{jevons:1869}\citep[pp.~7-13,29-30,41; pp.~53-54, point~138; pp.~73-74, point~172]{jevons:1864}\citep[p.~75]{lewis:1918}\citep[pp.~28-33]{malink:2019}.
%Condillac (1780) said something vaguely along the lines of equational reasoning (identifying true propositions with equations), but not as forcibly - and he does not mention the word "logic" in this sequence of pages of his book:
%pp. 112-125: https://books.google.com/books?id=7YpQAAAAYAAJ&pg=PA112&dq=%22It+is+analysis+which%22

Years before Jevon's inference adopted a generic, arbitrary dyadic relation which we have designated by ``$\termlogicassertion$'', \citet[p.~18]{boole:1847}, likely repeating Whately \citep[p.~74]{jevons:1869}, discussed more specialized versions of this 3-expression inference rule, in which the concrete relations ``$=$'' and ``$\neq$'' instantiate the generic relation.
}:%https://archive.org/details/diephilosophisc01leibgoog/page/237/mode/1up, Prop.~4,5,6,9,10,11
%https://books.google.com/books?id=KcryCQAAQBAJ&pg=PA816, points~5-6
%https://books.google.com/books?id=KcryCQAAQBAJ&pg=PA877
%https://www-jstor-org.ez67.periodicos.capes.gov.br/stable/2251094?seq=3#metadata_info_tab_contents
%Caramuel: https://books.google.com/books?id=fURPAAAAcAAJ&pg=PA466

$expr_1 = expr_2,~expr_2~\termlogicassertion~expr_3~\quad~ \metale ~\quad~expr_1~\termlogicassertion~expr_3$

(From which the transitivity of ``='' is straightforwardly derived\footnote{Two other notable properties of ``$=$'' --as it is an equivalence relation-- are symmetry (which we use in many of our proofs) and reflexivity (which we didn't need to employ in any occasion). Symmetry, reflexivity and transitivity of the ``$=$'' relation were explicitly stated in logic by \citet[pp.~9-11, points~22,24,27]{jevons:1864}. More than two millennia earlier, \citet{euclid:300bce} explicitly stated reflexivity and transitivity (though not symmetry) of ``$=$'' through his ``common notions'' 4 and 1, respectively, and also stated some corollaries of the principle of interchangeability of identicals through his ``common notions'' 2 and 3.}.)
%Symmetry of ``='': https://stackoverflow.com/questions/2300093/how-do-you-flip-an-equation-exchange-lhs-with-rhs-in-maple

Specializing it even more, we obtain Leibniz's principle of \emph{interchangeability of identicals} \citep[p.~94, \emph{Definitio}~1]{leibniz:1686f}\citep{forrest:2018} \citep[p.~18]{boole:1847}\citep[pp.~16-18]{jevons:1864}:%Also, an example by Jevons that unfortunately is racist, as it assumes that black-skinned humans are an alien race, ``the other'': \citep[pp.~34-35]{jevons:1869}
%Another source containing the formula: https://books.google.com/books?id=JfaqBgAAQBAJ&pg=PA149&dq=%22indiscernibility+of+identicals%22+second-order+logic
%https://books.google.com/books?id=GK8SEAAAQBAJ&pg=PA124&dq=%22Second-order+logic+also%22
%https://www.encyclopedia.com/philosophy-and-religion/philosophy/philosophy-terms-and-concepts/identity#:~:text=identity%20is%20reflexive
%Siva Somayyajula; Chirag Bharadwaj. Implementing Constructive Logics with OCaml. 2016.
%p. 4 of the PDF file
%https://ssomayyajula.github.io/files/impl_constr_logic_final_paper.pdf
%[Substitutivity of identicals]
%María Manzano; Manuel Crescencio Moreno. IDENTITY, EQUALITY, NAMEABILITY AND COMPLETENESS. PART II. In: Bulletin of the Section of Logic, Volume 47/3 (2018), pp. 141-158. http://dx.doi.org/10.18778/0138-0680.47.3.01
%p. 143 (p. 3 of the PDF file)
%http://pldml.icm.edu.pl/pldml/element/bwmeta1.element.ojs-doi-10_18778_0138-0680_47_3_01/c/4712-4176.pdf
%[Indiscernibility of identicals - an objection from intensional reasoning: possible worlds (p. 1)]
%https://www.jstor.org/stable/27744932
%[...] Leibniz used the principle of the identity of indiscernibles in arguments for a variety of metaphysical doctrines, including the impossibility of Newtonian absolute space. [Note by Antonielly: Relations of distance between objects.]
%[https://www.britannica.com/science/identity-of-indiscernibles]

$x_1 = x_2,~f(x_2)=y~\qquad~ \metale \qquad~f(x_1)=y$

or, more succinctly,

$x_1 = x_2~\qquad~ \metale \qquad~f(x_1)=f(x_2)$
%https://math.stackexchange.com/questions/2932631/algebra-can-you-think-of-each-side-of-the-equation-as-a-term#2932658

Another applicable inference rule is \emph{substitution of placeholder terms}: if we take any axiom or theorem having the term $\mathbf{x}$ as a ``generic'' placeholder (where it is explicitly or implicitly understood that it is universally quantified) and replace it by the term or expression $\mathbf{w}$ --in effect, a relabelling, which we are always free to do\footnote{This is analogous to \textalpha-conversion in \textlambda-calculus \citep[p.~355, postulate~I]{church:1932}.}, as long as we consistently do it for all occurrences of $\mathbf{x}$ --, the result will also be true:

$x\leftarrow w~\qquad~ \metale \qquad~rel[x\leftarrow w]$

For instance, as we will see later, the Leibniz-Cayley system has the axiom $\mathbf{b}\mathbf{c}=\mathbf{c}\mathbf{b}$; a consequence is that $\mathbf{b}\mathbf{m}=\mathbf{m}\mathbf{b}$, by consistently relabelling $\mathbf{c}$ as $\mathbf{m}$.

More generally, this can be done with various placeholder terms at once\footnote{On the other hand, in general we cannot replace a term for a generic placeholder \emph{expression}, even if we do it consistently. For instance:

$(\mathbf{b}\mathbf{c})\mathbf{b} = \mathbf{b}\mathbf{c}$ \qquad\qquad\quad(Universally true.)

$\quad\mathbf{d}~\mathbf{b} = ~\mathbf{d}$~ ~\{$\mathbf{d}\leftarrow\mathbf{b}\mathbf{c}$\} (Not true in general!)}:

$x\leftarrow w, z\leftarrow y, ..., q\leftarrow p~\qquad~ \metale \qquad~rel[x\leftarrow w, z\leftarrow y, ..., q\leftarrow p]$

Another inference rule we adopt is \emph{modus ponens}, \emph{implication elimination} or \emph{detachment}\footnote{Or its generalization, the \emph{cut rule}.}:

$(r1 \metale r2) ~|,\!|~ r1 \quad |\!\!\!\!\metale \quad r2$.

For instance, the Leibniz-Cayley system has the axiom $\mathbf{b}\mathbf{c}\neq\bm{\varnothing} \metale \mathbf{b}\neq\bm{\varnothing}$. Thus, in the cases where we know that $\mathbf{b}\mathbf{c}\neq\bm{\varnothing}$, we are justified in deducing $\mathbf{b}\neq\bm{\varnothing}$.

In summary, the only inference rules we need are are substitution of equivalent expressions, substitution of placeholder terms, and \emph{modus ponens}.

In the proofs that follow, all required proof steps are made explicit; no shortcuts are taken.

\section{Leibniz-Cayley (LC) system}
\label{sec:LC}

\subsection{Leibniz-Cayley system representations}
\label{subsec:LC_representations}

The representations of categorical assertions in the Leibniz-Cayley system make use only of one dyadic operation (which we interpret as intersection), one monadic operation (complement), two dyadic relations (equality, difference), and one constant object (the empty class). The following representations of fundamental categorical assertions are favored in the Leibniz-Cayley system as we define it:

\begin{tabular}{LlC}
\mathbf{A} & \makecell[l]{Every \textbf{b} is \textbf{c}.\\} & bc=b\\\hline
\mathbf{E} & \makecell[l]{No \textbf{b} is \textbf{c}.\\(Every \textbf{b} is not-\textbf{c}.)} & bc'\!=b\\\hline
\mathbf{I} & \makecell[l]{At least one \textbf{b} is \textbf{c}.\\} & bc\neq\bm{\varnothing}\\\hline
\mathbf{O} & \makecell[l]{At least one \textbf{b} is not \textbf{c}.\\(At-least-one \textbf{b} is not-\textbf{c}.)} & bc'\!\neq\bm{\varnothing}\\\hline
\mathbf{*} & \makecell[l]{At least one \textbf{b} exists.\\(\textbf{b} is-not empty.)} & b\neq\bm{\varnothing}
\end{tabular}

By means of the ``substitution of similars'' inference rule (Section~\ref{sec:inference_rules}) --and with no need of any axiom--, these representations straightforwardly lead to the ``subalternation laws'' from traditional Aristotelic logic when the subject term is inhabited:

%$\mathbf{b}\mathbf{c}=\mathbf{b},\ \mathbf{b}\neq\bm{\varnothing} \quad\metale\quad\ \mathbf{b}\mathbf{c}\neq\bm{\varnothing}$
$bc=b,\ b\neq\bm{\varnothing} \quad\metale\quad\ bc\neq\bm{\varnothing}$

$bc'\!=b,\ b\neq\bm{\varnothing} \quad\metale\quad\ \!bc'\!\neq\bm{\varnothing}$.

Leibniz has the merit of being the earliest logician we could find to algebraically represent the universal assertions as we do here, and came very, very close to representing particular assertions in the LC fashion, as we will show in the next paragraphs. His various strategies for representing particular assertions are very insightful: a short summary is made by \citet[p.~165]{brown:2012}, and further comments are made by \citet[p.~15]{lewis:1918} and \citet[pp.~691-692]{malink:2016}. In \citeyear{cayley:1871}, \citeauthor{cayley:1871} devised a proper way of representing particular assertions by means of non-identity relations with respect to ``$\bm{\varnothing}$'' \citep{cayley:1871}\citep[p.~473]{valencia:2004}, in a break with his predecessor \citet{boole:1847}.\footnote{In \citeyear{peirce:1870} --one year before \citeauthor{cayley:1871}--, \citet[pp.~57-58]{peirce:1870}, in a minor remark in the context of his paper, proposed representing particular categorical assertions such as ``At least one \textbf{b} is \textbf{c}.'' as ``$bc\supset\bm{\varnothing}$''. Cayley represented it instead as ``$bc\neq\bm{\varnothing}$''. So, Peirce deserves the credit of providing before Cayley a proper way of representing particular assertions by means of comparison relations involving ``$\bm{\varnothing}$''. We found no evidence that Cayley was aware of Peirce's discovery when he published his paper.

We prefer to adopt \citeauthor{cayley:1871}'s representation here because for this section we want a system with only ``$=$'' and ``$\neq$'' (coincidence and non-coincidence), not ``$\subseteq$'', ``$\supset$'', or other super-/sub-classhood relations. In addition, we feel it is fair to pay special homage to Cayley in the Leibnizian research program on term logic because his paper, unlike \citeauthor{peirce:1870}'s, adopts as its central concern the algebraic representation of classic categorical assertions.}.

Leibniz offered algebraic representations of categorical assertions in various drafts\footnote{In his research over decades, Leibniz toyed with various attempts of logical systems. As a faithful Aristotelic logic traditionalist, he gave preference to constructing an \emph{intensional} logic of \emph{concepts} rather than an extensional logic of classes (a preference later shared by Frege, who invented a brand of quantificational logic that modelled concepts as Boolean-valued functions -- which he employed to enunciate his Basic Law V that inspired the discovery of Russell's paradox and led to intense research towards axiomatic set theory, type theory, lambda calculus and combinatory logic) and often assumed \emph{existential import}. (These are interesting features on their own, however they are not within the scope of this paper.) He was a 17th-century rationalist Germanic/continental logician concerned with \emph{organized} concepts or ideas, not a 19th-century British logician committed to working with \emph{arbitrarily formed} extensional classes \citep[p.~14,35-37]{lewis:1918}. He also drew symbolic treatment parallels of term logic and propositional logic --anticipating \citet{boole:1847}--, and dealt with some notions of modal logic.

There is no single manuscript where Leibniz does \emph{everything} the way \emph{we} want. A typical situation is that, in a manuscript, Leibniz often has an insight that represents a progress towards our desired end stage, and then, after not having completed the entire puzzle, backtracks to try another direction, undoing the progress towards what we want. Then in another manuscript he documents another important insight, but does not combine it with a good insight he previously abandoned. It was state-of-the-art research at that age. He invents virtually all the required pieces of our puzzle, but the pieces are scattered across different boxes and in each box they are mixed with pieces that are incompatible to our puzzle. In fairness, he wasn't trying to achieve exactly our goals. But this means that, in order to understand how Leibniz contributed so much to the \emph{extensional} algebra of categorical syllogisms involving \emph{classes} which \emph{lack} existential import by default, we are forced to cherry-pick particularly noteworthy passages from his drafts, ignoring much of the original context surrounding those snippets, and assemble excerpts from different drafts to bring a Frankenstein's monster alive, adopting our 21st-century prejudices as a guide to picking and choosing and combining.

A quote mining exercise does not make a sound historical research, however. The reader is warned that Leibniz's drafts as a whole are far \emph{more} nuanced than the details we focus on: the real Leibniz the logician is far richer in insights and thoughts than our ``extensional Leibniz'' \citep[pp.~13-14]{lewis:1918}. (The same can be said about our convenient quotations from other early symbolic logicians, such as Jevons, Ladd and Mitchell.) Fortunately, professional historians of logic have plenty of rich material to explore and comment on all the nuances of the real Leibniz for many decades to come.

Nevertheless, even our impoverished, extensional Leibniz is enough for us to appreciate how prolific Leibniz was as a source of great insights. We provide strong evidence that Leibniz is a tremendously skilled founding master of the \emph{algebra} of logic. Indeed, every time we revisit Leibniz's drafts, the master teaches us something new about logic which we passed over in previous readings.
%We have performed a quote mining exercise, strategically including convenient excerpts (and conveniently excluding parts of excerpts) to gather, in a didactically ordered fashion, the concepts and tools we need to present a modernized presentation of algebraic categorical syllogistic. We don't falsify anything we present as facts, and we never credit Leibniz with anything he has not done; on the other hand, our selective use of information does not let distracting facts (pointing to other directions) get in the way of a good story. Thus, our presentation does not abide to the high standards which professional historians are required to. The reader is warned that Leibniz's drafts as a whole are far \emph{more} nuanced than the details we focus on: the real Leibniz the logician is far richer in insights and thoughts than our ``extensional Leibniz''. Fortunately, professional historians of logic have plenty of rich material to explore and comment on all the nuances of the real Leibniz for many decades to come.
%Proverb:
%https://twitter.com/ClareWoodhead3/status/1472150715682344962
%https://books.google.com/books?id=LT4qBgAAQBAJ&pg=PA142&lpg=PA142&dq=%22never+lets+the+facts+get+in+the+way+of+a+good+story%22
}. In a representative draft, \citet[pp.~235-236]{leibniz:1690b} wrote (the symbolic decoding \textbf{{\color{blue}highlighted}} inside square brackets are ours):

\begin{quote}\guillemotleft (5)A $\infty$ non non A. {\color{blue}[$\mathbf{a}=(\mathbf{a}')'$]}
%Other restatements of the involution law:
%https://books.google.com/books?id=KcryCQAAQBAJ&pg=PA877
%https://books.google.com/books?id=KcryCQAAQBAJ&pg=PA879

[...]

(7) AB $\infty$ BA {\color{blue}[$\mathbf{a}\mathbf{b}=\mathbf{b}\mathbf{a}$]}.

[...]

(12) Coincidunt A $\infty$ AB et non B $\infty$ non B non A.

{\color{blue}[$\mathbf{a}\!=\!\mathbf{a}\mathbf{b} \metaeq \mathbf{b}'\!=\!\mathbf{b'}\mathbf{a}'$]}

(13) [...] Universalis affirmativa {\color{blue}[$\mathbf{a}\termlogicA\mathbf{b}$:]} sic exprimi potest:

A $\infty$ AB {\color{blue}[$\mathbf{a}=\mathbf{a}\mathbf{b}$]} [...]

Particularis affirmativa {\color{blue}[$\mathbf{a}\termlogicI\mathbf{b}$:]} sic:

[...] AB est Ens {\color{blue}[$\mathfrak{I}\langle\mathbf{a}\mathbf{b}\rangle$]} [...] vel {\color{blue}[or]}

A non$\infty$ A non B {\color{blue}[$\mathbf{a}\neq\mathbf{a}\mathbf{b}'$]}.

Universalis negativa: Nullum A est B {\color{blue}[$\mathbf{a}\termlogicE\mathbf{b}$:]}, sic:

[...] A $\infty$ A non B {\color{blue}[$\mathbf{a}=\mathbf{a}\mathbf{b}'$]} seu {\color{blue}[or]} AB est non Ens {\color{blue}[$\overline{\mathfrak{I}}\langle\mathbf{a}\mathbf{b}\rangle$]}.

Particularis negativa: Quoddam A est non B {\color{blue}[$\mathbf{a}\termlogicO\mathbf{b}$:]},

A non$\infty$ AB {\color{blue}[$\mathbf{a}\neq\mathbf{a}\mathbf{b}$]}, vel {\color{blue}[or]} A non B est Ens {\color{blue}[$\mathfrak{I}\langle\mathbf{a}\mathbf{b}'\rangle$]}. [...]\guillemotright\end{quote}

In Leibniz's era, the symbols ``$\infty$'' and ``='' were adopted by different authors to represent the equality relation; Leibniz adopted the former convention. For our modern ``$\neq$'' he adopted ``\emph{non}$\infty$''\footnote{The employment of symbols which are nowadays nonstandard was not unusual up to the 19th century. For instance, in Robert \citeauthor{grassmann:1872}'s treatise on Logic (\citeyear[p.~8]{grassmann:1872}), the symbol ``$\gtrless$'' --a combination of ``$>$'' and ``$<$'' \citep[p.~158]{grattan-guinness:2000}-- was employed instead of ``$\neq$'' to represent non-equality.}. In both sides of each (in)equation, literal symbols, accompanied or not by a negation particle (``\emph{non}''), are employed. In contrast, in other important points of this excerpt, Leibniz consciously adopted the expressions ``\emph{est Ens}'' and ``\emph{est non Ens}'' after literal symbols. This is strong evidence that by ``\emph{[non]}$\infty$'' and ``\emph{est [non] Ens}'' Leibniz meant distinct relations in this and some other excerpts of his drafts. In our interpretation, Leibniz adopted ``\emph{est Ens}'' (``is an entity''\footnote{The literal translation of [la]~``\emph{ens}'' is [pt,es,it]~``\emph{ente}'', [en]~``entity''.

In some manuscripts, e.g. \citep[pp.~391-395, points~144-146,148-155,165,167-169,171]{leibniz:1686a}, Leibniz adopts the word ``\emph{res}'' (``thing'') rather than or in alternation with ``\emph{Ens}'' (``entity''). Sometimes, e.g. \citep[pp.~398-399, points~199-200]{leibniz:1686a}, Leibniz simply adopts ``\emph{est}'' (``is'') rather than ``\emph{est Ens}'', and ``\emph{non est}'' (``is not'') rather than ``\emph{est non Ens}''.
%146,148-150,199
%https://books.google.com/books?id=onpADwAAQBAJ&pg=PT202&dq=%22199+of+the%22
%The title page of that paper:
%Leibniz's Transformation of the Theory of the Syllogism into an Algebra of Concepts
%https://books.google.com/books?id=onpADwAAQBAJ&pg=PT169
%https://br1lib.org/book/5889515/d89013
%[Correct, but there is equivocation due to polysemy]
%si AB non est, A est non B [ab=0 |=| a<=b']
%[https://books.google.com/books?id=hT_nBQAAQBAJ&pg=PA507&dq=%22si+AB+non+est%22+%22A+est+non+B%22 https://www-jstor-org.ez67.periodicos.capes.gov.br/stable/40694037?seq=20#metadata_info_tab_contents]
%[Universal assertions turned singular]
%Qui non est a est non a.
%Qui non est non a est a.
%[https://www-jstor-org.ez67.periodicos.capes.gov.br/stable/40694037?seq=23#metadata_info_tab_contents]
%The use of natural language imposes additional obstacles for clear reasoning
%https://books.google.com/books?id=bx7cgMCLyjsC&q=logical
}) to indicate that its subject was an inhabited term, and ``\emph{est non Ens}'' (``is a non-entity'') to indicate that its subject was a non-inhabited (that is, extensionally empty) term. These are monadic relations, which we respectively represent in symbolic notation by ``$\mathfrak{I}\langle s\rangle$'' and ``$\overline{\mathfrak{I}}\langle s\rangle$''\footnote{We chose the character ``$\mathfrak{I}$'' for this monadic relation because it is the initial character of ``$\mathfrak{Inhabit\bar{a}tus}$''/``$\mathfrak{Inhabited}$'' and also the vowel corresponding to the particular affirmative categorical dyadic relation ``$\termlogicI$'', ensuring that ``$\mathbf{b}\termlogicI\mathbf{c} \metaeq \mathfrak{I}\langle \mathbf{b}\mathbf{c}\rangle$''. The complement of ``$\mathfrak{I}$'' is the monadic relation ``$\overline{\mathfrak{I}}$''. One may think of ``$\mathfrak{I}$'' and ``$\overline{\mathfrak{I}}$'' by the mnemonics ``is'' and ``is not'' in English, respectively.

Almost two centuries after Leibniz, \citet[pp.~29-30]{ladd:1883}\citep[p.~598]{venn:1883} reinvented these monadic relations by employing the representations ``$s\,\Cap$'' and ``$s\!\!\not\!\Cap$'' to respectively stand for ``$s\Cap\mathbf{I}$'' and ``$s\!\not\!\Cap\;\mathbf{I}$''. If one asserts some thing exists (does not exist), then one asserts it exists (does not exist) within the universe.}.

In this excerpt, Leibniz correctly proposed the following representations\footnote{\label{fn:leibniz_system_representations}The \textbf{{\color{blue}highlighted}} symbolic encodings are the ones we would select for a pure Leibniz's system, since they would fit the axioms enumerated in Section~\ref{subsec:LC_axioms}.}:
 \begin{itemize}
     \item $\mathbf{a}\termlogicA\mathbf{b}$: {\color{blue}$\mathbf{a}=\mathbf{a}\mathbf{b}$}

     \item $\mathbf{a}\termlogicE\mathbf{b}$: {\color{blue}$\mathbf{a}=\mathbf{a}\mathbf{b}'$}; $\overline{\mathfrak{I}}\langle\mathbf{a}\mathbf{b}\rangle$

     \item $\mathbf{a}\termlogicI\mathbf{b}$: \ $\mathbf{a}\neq\mathbf{a}\mathbf{b}'$; {\color{blue}$\mathfrak{I}\langle\mathbf{a}\mathbf{b}\rangle$}

     \item $\mathbf{a}\termlogicO\mathbf{b}$: $\mathbf{a}\neq\mathbf{a}\mathbf{b}$; \ {\color{blue}$\!\mathfrak{I}\langle\mathbf{a}\mathbf{b}'\rangle$}.
 \end{itemize}
%aEB in both forms again: p. 262

Leibniz didn't mention in this particular excerpt ``$\overline{\mathfrak{I}}\langle\mathbf{a}\mathbf{b}'\rangle$'' as an alternative representation for ``$\mathbf{a}\termlogicA\mathbf{b}$'', although he fixed this omission in other drafts (\citeyear{leibniz:1691a, leibniz:1691b}). Also noteworthy is that he didn't shy away of using the ``$\neq$'' relation in the representation of particular categorical assertions, unlike \citet[pp.~21--22]{boole:1847} more than 150 years later, who only attempted to represent all categorical assertions with equations (using ``='').
%Another manuscript where Leibniz mentions ``$\overline{\mathfrak{I}}\langle\mathbf{a}\mathbf{b}'\rangle$'': https://archive.org/details/diephilosophisc00gerhgoog/page/n227/mode/1up https://books.google.com/books?id=br6EuI4j5mEC&pg=PA84&dq=%22id+est+aequivalent+AB+et+A%22

As we can see, in this short excerpt Leibniz identified the importance of the term combination/intersection and complementation operations, the equality/coincidence (``$\infty$'') and difference/non-coincidence (``\emph{non}$\infty$'') relations, the involution of complementation ($a\!\!=\!\!(a')'$), the commutativity of intersection ($ab\!=\!ba$), inhabitation (``\emph{Ens}'') and non-inhabitation/emptiness (``\emph{non-Ens}'', nonexistent, in a loose, non-literal translation\footnote{Ramon \citet[p.~162]{llull:1993}, a major intellectual influence on the young Leibniz, wrote the following on the contrast between ``\emph{Ens}'' and ``\emph{non-Ens}'' and the power of imagination:

``\emph{Si extra intellectum nullum non ens est ens, solus intellectus facit non ens.}''

``As there is no non-being outside of the intellect, then only intellect creates non-being.''
%A sentence according to the same template "As there is ..., then ... .": https://www.europarl.europa.eu/doceo/document/CRE-6-2008-09-25_EN.html#:~:text=As%20there%20is%20a

(Literally: ``If outside the intellect no non-entity is [an] entity, only intellect makes [a] non-entity.'')}).

Later, \citet{cayley:1871} proposed the following representations\footnote{The \textbf{{\color{blue}highlighted}} algebraic encodings of particular categorical assertions are the ones we have selected for the Leibniz-Cayley system instead of Leibniz's relational representations of particular categorical assertions.}:

 \begin{itemize}
     \item $\mathbf{a}\termlogicA\mathbf{b}$: $\mathbf{a}\mathbf{b}'\!=\!\bm{\varnothing}$

     \item $\mathbf{a}\termlogicE\mathbf{b}$: $\mathbf{a}\mathbf{b}=\bm{\varnothing}$

     \item $\mathbf{a}\termlogicI\mathbf{b}$: \ {\color{blue}$\mathbf{a}\mathbf{b}\neq\bm{\varnothing}$}

     \item $\mathbf{a}\termlogicO\mathbf{b}$: {\color{blue}$\mathbf{a}\mathbf{b}'\!\neq\!\bm{\varnothing}$}.
 \end{itemize}

Notice that Leibniz's monadic relations ``$\mathfrak{I}\langle s\rangle$'' and ``$\overline{\mathfrak{I}}\langle s\rangle$'' respectively correspond\footnote{Of course, here we are projecting our 21-century extensional goals on Leibniz's excerpts. Leibniz was actually dealing with modal logic concepts when he was talking about ``\emph{est Ens}'' and variants according to \citet[p.~5]{lenzen:1987}:

\begin{quote}\footnotesize{\guillemotleft [...] there is abundant textual evidence to show that at least as applied to terms, i.e. to concepts, Leibniz always uses `\emph{est Ens}' as synonymous with `\emph{est Possibile}' [...]. Accordingly `\emph{est non-Ens}' means the same as `\emph{non est Ens}' or `\emph{est impossibilis}' [...].

[...]

[...] on the whole, there is overwhelming evidence showing that Leibniz expresses the possibility-operator `\emph{A est possibile}' equally by means of `\emph{A est Res}', `\emph{A est Ens}' or even `\emph{A est}'. [...]\guillemotright}\end{quote}

\citet[p.~94]{lenzen:2004b} reinforces this assertion of the synonymy of ``\emph{est}'', ``\emph{est Ens}'', ``\emph{est res}'' and ``\emph{est possibile}'', and he complements the previous quote with a short commentary (\citeyear[p.~74]{lenzen:2004a}) on some metaphysical goals to which Leibniz applied his brand of modal reasoning.

Reinforcing this preference for a logic of possibility over a logic of actuality, Leibniz also justified the validity of ``conversion by limitation (\emph{per accidens})'' (an Aristotelic logic law akin to ``subalternation'') in terms of possibility \citep[p.~101]{leibniz:1691a}\citep[p.~211-212]{leibniz:1691b}.
%[Read until the beginning of the next page]
%https://books.google.com/books?id=4AsTDAAAQBAJ&pg=PA118&dq=%22Leibniz+himself+was+well%22 https://dspace.unitus.it/bitstream/2067/2591/1/MODERN%20LOGIC%20VERY%20LAST%20PRE%20PEER%20REVIEW.pdf

We bother providing our unorthodox interpretation because our goal is to extract from Leibniz's manuscripts concepts and tools that lead us to our modern algebra of categorical syllogisms rather than understanding Leibniz's logic on its own terms.} to \citeauthor{cayley:1871}'s dyadic relations ``$s\neq\bm{\varnothing}$'' and ``$s=\bm{\varnothing}$''. Moreover, notice that all of Cayley's representations are about either the emptiness or the inhabitation of some combination of terms.

%\footnote{\citet[p.~84]{mitchell:1883} hinted at what we can now understand as the relationship between the \emph{lack of existential import} and the law of \emph{substitution of similars} (Section~\ref{sec:inference_rules}): ``[...] a particular conclusion cannot be drawn from universal premises, since a particular proposition implies the existence of its subject, while a universal does not.'' See also the subject inhabitation law LC6 in Section~\ref{subsec:LC_axioms}.%https://archive.org/details/studiesinlogic00gilmgoog/page/n105/mode/1up
%}

We will show that a pure Leibniz's system --which adopts Leibniz's algebraic representations for universal categorical assertions and Leibniz's monadic relational representations for particular categorical assertions, together with taking as axioms some laws stated by Leibniz, which are enumerated in Section~\ref{subsec:LC_axioms}-- suffices to prove all 24 categorical syllogisms (see the proofs in Section~\ref{subsec:LC_proofs}). For this narrow purpose, \citeauthor{cayley:1871}'s (\citeyear{cayley:1871}) representations of the fundamental categorical assertions are superfluous.

Nevertheless, the Leibniz-Cayley system adopts \citeauthor{cayley:1871}'s algebraic representations (with ``$\neq\!\bm{\varnothing}$'') for particular categorical assertions rather than Leibniz's monadic relational representations (with ``$\mathfrak{I}$'') because: we prefer to construct a system with only (in)equations rather than a hybrid algebraic/relational system; it makes it easier to algebraically justify the axiom LC3 (Section~\ref{subsec:LC_axioms}); and, more importantly, because LC3 in the representation adopted by LC (rather than as represented a pure Leibniz's system) is straightforward to prove as a theorem in Boolean algebra (Section~\ref{subsubsec:LC_from_Boolean_lattice}).

Unfortunately, Leibniz chose to keep the note containing that key excerpt as a private draft, perhaps feeling his system hadn't yet achieved far enough results to his liking; it was published only in 1903 by the diligent editor \citet{couturat:1903}, decades after the results of \citet[p.~21]{boole:1847} and \citet{cayley:1871}, taken in combination, were published containing the algebraic representations of categorical assertions in the LC system\footnote{\label{fn:est_ens}We stress that Cayley explicitly called attention to the correspondence between ``$\mathbf{a}\termlogicI\mathbf{b}$'' and ``$\mathbf{a}\mathbf{b}\!\neq\!\bm{\varnothing}$'', and that of ``$\mathbf{a}\termlogicO\mathbf{b}$'' and ``$\mathbf{a}\mathbf{b}'\!\neq\!\bm{\varnothing}$''. Leibniz arguably pioneered such correspondences only for those --e.g. \citet[p.~358, point~17]{couturat:1901}, \citet[pp.~527,532]{marciszewski:1984} and \citet[p.~199]{sotirov:1999}-- who perform a reconstruction of his logic that treats ``\emph{est}'' as synonym to ``\emph{aequivalent}'' (``$=$''), ``\emph{non-Ens}'' as the empty \emph{class}, and forces the leap ``\emph{est Ens $\!\metaeq\!$ non est non-Ens}'' for particular categorical assertions, which Leibniz \textbf{didn't} perform in the drafts we consulted\textsuperscript{$\bm{\ast}$}.

Both Boole and Cayley algebraically represented ``$\mathbf{b}\termlogicE\mathbf{c}$'' as ``$\mathbf{b}\mathbf{c}\!=\!\bm{\varnothing}$''. In Boolean algebra (Section~\ref{subsubsec:Boolean_lattice}), we can derive the LC representation as follows:
%It can also be derived from Boole's representation b=vc' <https://archive.org/details/investigationofl00boolrich/page/228/mode/2up?view=theater>. Compare to <https://archive.org/details/mathematicalanal00booluoft/page/25/mode/1up> and <p. xxxii (36 of the PDF file): https://www.math.uwaterloo.ca/~snburris/htdocs/MAL_Nov_20_2022.pdf>, clearly mindful of the doctrine of the quantification of the predicate.
%By Boole's original algebra:
%bc=0 |= b-bc=b-0 |=| b(1-c)=b |=| bc'=b %Justifiable by: b-(bc) = b*(bc)' = b*(b'+c') = 0+bc' = bc'
%b(1-c)=b |=| b-bc=b |=| b-b=bc |=| 0=bc

$\mathbf{b}\mathbf{c}=\bm{\varnothing} \quad\metale\quad \mathbf{b}\mathbf{c}\cup\mathbf{b}\mathbf{c}'=\bm{\varnothing}\cup\mathbf{b}\mathbf{c}' \quad\metaeq\quad \mathbf{b}(\mathbf{c}\cup\mathbf{c}')=\mathbf{b}\mathbf{c}' \quad\metaeq\quad \mathbf{b}\mathbf{I}=\mathbf{b}\mathbf{c}' \quad\metaeq\quad \mathbf{b}=\mathbf{b}\mathbf{c}'$.

And the converse:

$\mathbf{b}=\mathbf{b}\mathbf{c}' \quad\metale\quad \mathbf{b}\mathbf{c}=(\mathbf{b}\mathbf{c}')\mathbf{c} \quad\metaeq\quad \mathbf{b}\mathbf{c}=\mathbf{b}(\mathbf{c}'\mathbf{c}) \quad\metaeq\quad \mathbf{b}\mathbf{c}=\mathbf{b}\bm{\varnothing} \quad\metaeq\quad \mathbf{b}\mathbf{c}=\bm{\varnothing}$.

Therefore,

$\mathbf{b}\mathbf{c}=\bm{\varnothing} \quad\metaeq\quad \mathbf{b}=\mathbf{b}\mathbf{c}'$.

($\bm{\ast}$) It is easy to be misled because the copular verb ``\emph{est}'' is polysemous. Leibniz himself took advantage of --and sometimes was confused \citep[pp.~67-68,74]{lenzen:2018a} by-- the reuse of ``\emph{est}'' with different meanings \citep[pp.~118-119]{levey:2011}\citep{lenzen:1986}\citep[pp.~4,9]{rescher:1954}.

Regarding uses of ``\emph{est Ens}'' with distinct meanings, Leibniz on some occasions explicitly employed ``\emph{Ens}'' and ``\textbf{\emph{Nihil}}'' (``nothing''/``empty'') as complementary, such as in the following excerpt on metaphysics, discussed in detail by \citet[pp.~14-16]{koszkalo:2017}:%CTRLf: to postulate

``\emph{Essentia ablata existentia aut est ens reale aut nihil. Si nihil, aut non fuit in creaturis, quod absurdum; aut non distincta ab existentia fuit, quod intendo.}''

(``Essence taken away from existence is \textbf{either a real entity or nothing}. If it is nothing, either it was not in creatures, which is absurd; or it was not distinct from existence, which I intend.'')
%Scotus:
%Conceptus communis univocus si contrahatur, oportet quod contrahatur per aliquod additum, illud additum aut est ens, aut non ens; quia illud quod contrahit aliud, oportet quod sit extra rationem eius. Sed nihil est extra rationem entis; si est non ens, non contrahit.
%[Lukáš Novák. CONCEPTUAL ATOMISM, "APORIA GENERIS" AND A WAY OUT FOR LEIBNIZ AND THE ARISTOTELIANS. http://www.skaut.org/ln/docs/aporiageneris.pdf]

We can also exercise our creativity and explore the polysemy of ``\emph{est}'' by performing the following loose (and historically inaccurate) interpretations:

%\begin{table}[hb]
%\resizebox{\textwidth}{!}{
\begin{tabular}{|l|ll|ll|}
\cline{2-5}
\multicolumn{1}{c|}{} & \multicolumn{2}{c|}{\textbf{Leibniz, Ladd}} & \multicolumn{2}{c|}{\textbf{McColl, Cayley}} \\\hline
\textbf{Universal} & bc {\tiny\guillemotleft}\textbf{{\color{blue}\emph{non} est} {\transparent{0.3}\variabletextvisiblespace[1.5em] }{\color{blue}{\transparent{0.4}[}Ens{\transparent{0.4}]}}}{\tiny\guillemotright}. & ${\color{blue}\overbracket[0.14ex][0.6ex]{\overline{\mathfrak{I}}}^{\text{\guillemotleft...\guillemotright}}}\!\!\!\langle\mathbf{b}\mathbf{c}\rangle$, $\mathbf{b}\mathbf{c}\!\!{\color{blue}\overbracket[0.14ex][0.6ex]{\bm{\not}\!\Cap\text{I}}^{\text{\guillemotleft...\guillemotright}}}$ & bc {\tiny\guillemotleft}\textbf{{\transparent{0.3}\variabletextvisiblespace[1.5em]} {\color{red}est}}{\tiny\guillemotright} ~non-Ens. & \!\!$\mathbf{b}\mathbf{c}{\!\!\color{red}\overbracket[0.14ex][0.6ex]{\subseteq}^{\text{\guillemotleft...\guillemotright}}}\!\!\bm{\varnothing}$, $\mathbf{b}\mathbf{c}{\!\!\color{red}\overbracket[0.14ex][0.6ex]{=\!\!\!\!\!\!{\transparent{0}\subseteq}}^{\text{\guillemotleft...\guillemotright}}}\!\!\bm{\varnothing}$\\\hline
%& \!\!$\mathbf{b}\mathbf{c}{\color{red}\overbracket[0.14ex][0.6ex]{\subseteq\!\!\bm{\varnothing}}^{\text{\guillemotleft...\guillemotright}}}$, $\mathbf{b}\mathbf{c}{\color{red}\overbracket[0.14ex][0.6ex]{=\!\!\bm{\varnothing}\!\!\!\!\!{\transparent{0}\subseteq}}^{\text{\guillemotleft...\guillemotright}}}$\\\hline

\textbf{Particular} & bc {\tiny\guillemotleft}\textbf{{\transparent{0.3}\variabletextvisiblespace[1.5em]} {\color{blue}est} {\transparent{0.3}\variabletextvisiblespace[1.5em] }{\color{blue}{\transparent{0.4}[}Ens{\transparent{0.4}]}}}{\tiny\guillemotright}. & ${\color{blue}\overbracket[0.14ex][0.6ex]{\mathfrak{I}}^{\text{\guillemotleft...\guillemotright}}}\!\!\!\langle\mathbf{b}\mathbf{c}\rangle$, $\mathbf{b}\mathbf{c}\!\!{\color{blue}\overbracket[0.14ex][0.6ex]{\;\!\Cap\text{I}}^{\text{\guillemotleft...\guillemotright}}}$ & bc {\tiny\guillemotleft}\textbf{{\color{red}\emph{non} est}}{\tiny\guillemotright} ~non-Ens. & \!\!$\mathbf{b}\mathbf{c}{\!\!\color{red}\overbracket[0.14ex][0.6ex]{\nsubseteq}^{\text{\guillemotleft...\guillemotright}}}\!\!\bm{\varnothing}$, $\mathbf{b}\mathbf{c}{\!\!\color{red}\overbracket[0.14ex][0.6ex]{\neq}^{\text{\guillemotleft...\guillemotright}}}\!\!\bm{\varnothing}$\\\hline
%& bc {\tiny\guillemotleft}\textbf{{\color{red}\emph{non} est}}{\tiny\guillemotright} ~non-Ens. & \!\!$\mathbf{b}\mathbf{c}{\color{red}\overbracket[0.14ex][0.6ex]{\nsubseteq\!\!\bm{\varnothing}}^{\text{\guillemotleft...\guillemotright}}}$, $\mathbf{b}\mathbf{c}{\color{red}\overbracket[0.14ex][0.6ex]{\neq\!\!\bm{\varnothing}}^{\text{\guillemotleft...\guillemotright}}}$\\\hline
%x>=y, x<='y |= x>y
%x is a superset of y, at least one x is not y |= x is a proper superset of y
\end{tabular}%}
\begin{center}Different choices of primary notion: \emph{\guillemotleft\textbf{est Ens}}\guillemotright~(``exists'') vs. \emph{\guillemotleft\textbf{est}\guillemotright~non-Ens} (``is empty'').\end{center}
%\caption{Different choices of primary notion: \emph{\guillemotleft\textbf{est Ens}}\guillemotright~vs. \emph{\guillemotleft\textbf{est}\guillemotright~non-Ens}.}
%\label{tab:est_ens}
%\end{table}
}.

This historical accident might lead to the impression that the tradition of the algebra of logic was pioneered by Boole in the chronology of publications, if not in the chronology of ideas. However, some other drafts on logic by Leibniz were published in a book \citep{erdmann:1840} years before \citeauthor{boole:1847}'s \citeyear{boole:1847} pioneering treatise on logic. In particular, a draft by Leibniz (\citeyear{leibniz:1691a, leibniz:1691b}) contained the following excerpt (the symbolic decoding \textbf{{\color{blue}highlighted}} inside square brackets is ours):

\begin{quote}\guillemotleft Reductio mea vetus talis fuit:

Universalis Affirmativa: Omne A est B {\color{blue}[$\mathbf{a}\termlogicA\mathbf{b}$:]}, id est aequivalent AB et A {\color{blue}[$\mathbf{a}\mathbf{b}=\mathbf{a}$]} seu {\color{blue}[or]} A non B est non-Ens {\color{blue}[$\overline{\mathfrak{I}}\langle\mathbf{a}\mathbf{b}'\rangle$]}.

Particularis Negativa: Quoddam A non est B {\color{blue}[$\mathbf{a}\termlogicO\mathbf{b}$:]} seu non aequivalent AB et A {\color{blue}[$\mathbf{a}\mathbf{b}\neq\mathbf{a}$]} seu {\color{blue}[or]} A non B est Ens {\color{blue}[$\mathfrak{I}\langle\mathbf{a}\mathbf{b}'\rangle$]}.

At Universalis Negativa: Nullum A est B {\color{blue}[$\mathbf{a}\termlogicE\mathbf{b}$:]}, erit AB est non-Ens {\color{blue}[$\overline{\mathfrak{I}}\langle\mathbf{a}\mathbf{b}\rangle$]}.

Et Particularis Affirmativa: Quoddam A est B {\color{blue}[$\mathbf{a}\termlogicI\mathbf{b}$:]}, erit AB est Ens {\color{blue}[$\mathfrak{I}\langle\mathbf{a}\mathbf{b}\rangle$]}.

[...] aequivalent AB et BA {\color{blue}[$\mathbf{a}\mathbf{b}=\mathbf{b}\mathbf{a}$]}. [...]\guillemotright\end{quote}

%\makeatletter
%\newcommand{\oset}[3][0ex]{%
%  \mathrel{\mathop{#3}\limits^{
%    \vbox to#1{\kern-2\ex@
%    \hbox{$\scriptstyle#2$}\vss}}}}
%\makeatother

%``$\oset[0ex]{>}{\scriptstyle <}$''

Unlike the previous excerpt, here Leibniz didn't adopt the symbols ``$\infty$'' and ``\emph{non}$\infty$'', preferring instead to write ``\emph{aequivalent}'' and ``\emph{non aequivalent}'' in full. Notice that, again, before ``\emph{Ens}'' and ``\emph{non-Ens}'' he adopted neither ``$\infty$'' nor ``\emph{aequivalent}'', but the verb ``\emph{est}'', certainly to stress the distinction between the [in]equality relation and the relation indicated by ``\emph{est [non-]Ens}''.

In this excerpt, Leibniz correctly proposed the following representations\footref{fn:leibniz_system_representations}:

 \begin{itemize}
     \item $\mathbf{a}\termlogicA\mathbf{b}$: {\color{blue}$\mathbf{a}\mathbf{b}=\mathbf{a}$}; $\overline{\mathfrak{I}}\langle\mathbf{a}\mathbf{b}'\rangle$

     \item $\mathbf{a}\termlogicE\mathbf{b}$: $\overline{\mathfrak{I}}\langle\mathbf{a}\mathbf{b}\rangle$

     \item $\mathbf{a}\termlogicI\mathbf{b}$: {\color{blue}$\mathfrak{I}\langle\mathbf{a}\mathbf{b}\rangle$}

     \item $\mathbf{a}\termlogicO\mathbf{b}$: $\mathbf{a}\mathbf{b}\neq\mathbf{a}$; {\color{blue}$\mathfrak{I}\langle\mathbf{a}\mathbf{b}'\rangle$}.
 \end{itemize}

For our purposes, in this excerpt Leibniz only missed the representation of ``$\mathbf{a}\termlogicE\mathbf{b}$'' that would be required for a pure Leibniz's system compatible with the axioms in Section~\ref{subsec:LC_axioms}: ``{\color{blue}$\mathbf{a}\mathbf{b}'\!=\!\mathbf{a}$}''. But the main point is that Leibniz's research program on categorical syllogistic, some algebraic representations and laws, and some attempts at a complete system for proving all the classic categorical syllogisms had already been published --by the editor \citet{erdmann:1840}-- years before \citeauthor{boole:1847}'s \citeyear{boole:1847} book. This excerpt alone would suffice, in our view, to establish Leibniz as the founding master of the \emph{algebra} of logic in the chronology of publications too -- and not only in the chronology of ideas.

It is also worth it to point out that, despite doing his research almost two centuries earlier, Leibniz went farther than \citet{boole:1847} towards LC in this excerpt. He correctly identified a proper use of ``$\neq$'' in term logic (for instance, $\mathbf{a}\termlogicO\mathbf{b}$: $\mathbf{a}\mathbf{b}\!\neq\!\mathbf{a}$, as the quote shows), whereas Boole only employed ``$=$'' in his logic.

In other excerpts where Leibniz gets it wrong (from the perspective of extensional term logic \emph{without} existential import) by employing ``$=$~an inhabited term'' instead of ``$\neq\!\bm{\varnothing}$'', Boole miserably fails in the same way: Leibniz sometimes employed the letter ``Y'', ``Z'' or ``W'' to stand for a not-yet-determined class --e.g. in \citep[p.~234; p.~236, point~13]{leibniz:1690b}--, whereas \citet[pp.~21--22]{boole:1847} usually represented an arbitrary inhabited class by the letter ``$v$''. This corroborates \citeauthor{couturat:1901}'s remark (\citeyear[p.~386]{couturat:1901}) that Leibniz possessed almost all principles of the Boole-Schröder logic, and in some points he was even more advanced than Boole himself\footnote{On the other hand, Leibniz was in hindsight too conservative, clinging too much to the traditional Aristotelic paradigm of grammatically inspired \emph{term} logic. This may seem ironic, given that the algebra of logic he pioneered would surely be considered by 17th-century logicians a radical innovation in his time, had he published his drafts -- and indeed it was. However, it wasn't as predominantly symbolic as the 19th-century tradition initiated by Boole. Even for the algebraic logician Leibniz, logic still was more philosophical than algebraic, given features such as the handling of many sentences in verbose prose (instead of adopting purely symbolic representations), his decades-long dedication to the study of the logical meanings of ``\emph{est}'' (``is''), the absence of the insight of dealing with inhabited terms as extensionally non-empty (``\emph{non est non Ens}''), his bias toward intensionality and even modality, and the stubborn conservation of existential import for universal categorical assertions.}.

One of the likely reasons why both Leibniz and Boole insisted on employing ``$=$'' to particular assertions was to preserve the validity of the ``subalternation'' laws\footnote{An opinion we share with \citet[p.~140]{marciszewski:1995}.}\\
$\mathbf{b}\termlogicA\mathbf{c} \!\metale \mathbf{b}\termlogicI\mathbf{c}$\\
$\mathbf{b}\termlogicE\mathbf{c} \!\metale \mathbf{b}\termlogicO\mathbf{c}$\\
from Aristotelic logic -- after all,\\
$\mathbf{b}\mathbf{c}{\color{white}'}\!\!=\!\!\mathbf{b} \metale \mathbf{b}\mathbf{c}{\color{white}'}\!\!=\!\!v$\\
$\mathbf{b}\mathbf{c}'\!\!=\!\!\mathbf{b} \metale \mathbf{b}\mathbf{c}'\!\!=\!\!v$\\
make \emph{some} sense in algebraic reasoning\footnote{In first-order quantificational logic, it corresponds to (an instantiation of) the \emph{existential introduction} axiom:

$\mathbf{b} \in \mathcal{P}(\mathbf{I}) \metale\; \exists \emph{v}.~\emph{v} \in \mathcal{P}(\mathbf{I})$

where $\mathcal{P}(\mathbf{I})$ is the powerclass of $\mathbf{I}$.}. (For some supporting evidence, see \citet[p.~234]{leibniz:1690b}, \citet[pp.~213-214]{leibniz:1691b}, \citet[p.~102]{leibniz:1691a}, \citet[pp.~358-361, point~18]{couturat:1901}\footnote{Despite \citeauthor{couturat:1901}'s strong stance against existential import of universal categorical assertions.}, \citet[pp.~21-25]{boole:1847} and \citet[pp.~55, points~140-141; 57, point~144]{jevons:1864}.)
%Human is some kind of primate / Every human is (some) primate: sp=h
%Every human is (human) primate: hp=h
%Human primate is something / Some human is primate: hp=v
%Every human is primate $\metale$ Some human is primate

\citeauthor{boole:1847}'s desire to preserve existential import for all categorical assertions\footnote{\citet[pp.~26-30]{boole:1847} embraced ``conversion by limitation or \emph{per accidens}'' of ``$\termlogicA$'' into ``$\termlogicI$'' and of ``$\termlogicE$'' into ``$\termlogicO$''. This gives rise to difficulties and to a clunky algebraic system for categorical syllogistic, as shown by \citet[pp.~168-169,171]{makinson:2022}.} may have encouraged him to think of inhabitation as primary and of emptiness (non-inhabitation) as a derived, subordinate notion in term logic. This seems to be a reasonable and pragmatic choice of default condition for terms at first sight when we consider, like Aristotle, that we most often care to reason about existent things in the world, not nonexistent ones. However, a mathematical fact is that there are infinitely many inhabited classes, but (extensionally) a unique empty class\footnote{\citet[points~15-22,28-30,39]{leibniz:1686d} knew this; he enunciated this and other important facts about the empty class.
%Boole wanted his algebra to apply to the traditional Aristotelic logic, including conversion by limitation (per accidens)<pp. 24-26,27-28 (99-101,102-103 of the PDF file): https://www.math.uwaterloo.ca/~snburris/htdocs/MAL_Nov_20_2022.pdf>.
%Makinson refers to an oversight by Boole's observed by Burris: <p. 35 (110 of the PDF file): https://www.math.uwaterloo.ca/~snburris/htdocs/MAL_Nov_20_2022.pdf>.

Since the origination of axiomatic set theory (if not earlier), the fact that the empty class is unique is stated as a theorem (with various proofs), not as an axiom.

By the Boolean lattice axioms shown in Section~\ref{subsubsec:Boolean_lattice}, the empty class is the identity element with respect to the union operation. One can easily prove that the identity element associated to a dyadic operation is always unique. See \guillemotleft\url{https://proofwiki.org/wiki/Identity_is_Unique}\guillemotright.

It is also easy to prove that the empty class is unique by employing the properties of ``$\subseteq$'', or alternatively by employing the extensionality axiom from set theory. See \guillemotleft\url{https://proofwiki.org/wiki/Empty_Set_is_Unique}\guillemotright.}: his ``$v$'' doesn't have a well-determined referent, whereas ``$\bm{\varnothing}$'' does. Thus, equality (``$=$'') and difference (``$\neq$'') assertions are well-defined with ``$\bm{\varnothing}$'', but not with ``$v$''. Algebraic manipulations in LC benefit from this important property of ``$\bm{\varnothing}$'' by considering emptiness as the primary notion, and inhabitation as a synonym for non-emptiness, thus \emph{subverting} our initial disposition for considering inhabitation as a primitive notion\footnote{Notice that, whereas a purely algebraic treatment with ``$=\!\bm{\varnothing}$'' and ``$\neq\!\bm{\varnothing}$'' requires subverting Aristotle's reasonable choice of primary notion, Leibniz's relation ``$\mathfrak{I}$'' preserves it, as the table in footnote\footref{fn:est_ens} shows.}. In addition, \citeauthor{boole:1847}'s obsession with equational reasoning may have led him to overlook that, just like ``$\mathbf{b}\termlogicE\mathbf{c}$'' is contradictory to ``$\mathbf{b}\termlogicI\mathbf{c}$'', ``$\mathbf{b}\mathbf{c}\!=\!\bm{\varnothing}$'' is contradictory to ``$\mathbf{b}\mathbf{c}\!\neq\!\bm{\varnothing}$'', not to ``$\mathbf{b}\mathbf{c}\!=\!v$''. \citet{cayley:1871} realized what \citeauthor{boole:1847} (\citeyear{boole:1847, boole:1854}) overlooked\footnote{\citet[p.~94]{leibniz:1686f}, and most emphatically \citet[pp.~2-4,8-13,29-30]{jevons:1864}, in the pamphlet that founded modern Boolean algebra, anticipated \citet{cayley:1871} in identifying the importance of the negation of ``$=$'' --namely, the ``$\neq$'' relation-- for general deductive reasoning (though not for the representation of particular categorical relations with ``$\bm{\varnothing}$''). \citeauthor{jevons:1864} divided relations into:
%[Leibniz:]
%
%Definitio 1. Eadem sunt quorum unum polest substitui alteri salva veritate. Si sint A et B et A ingrediatur aliquam propositionem veram, et ibi in aliquo loco ipsius A pro ipso substituendo B fiat nova propositio eaque itidem vera, idque semper succedat in quacunque tali propositione, A et B dicuntur esse Eadem; et contra si Eadem sint A et B, procedet substitutio quam dixi. Eadem etiam vocantur coincidentia; aliquando tamen A quidem et A vocantur idem, A vero et B si sint eadem vocantur coincidentia.
%
%Defin. 2. Diversa sunt quae sunt non eadem, seu in quibus substitutio aliquando non succedit. Coroll. Unde etiam, quae non sunt diversa, sunt eadem.
%
%Charact. 1. A=B significat A et B esse eadem vel coincidentia.
%
%Charact. 2. A non = B vel B non = A significat A et B esse diversa.
%
%[https://archive.org/details/operaphilosophic00leibuoft/page/94/mode/1up https://archive.org/details/diephilosophisc01leibgoog/page/228/mode/1up]

\vspace{-\topsep}
\begin{itemize}
    \item \emph{affirmative}, having the ``$=$'' copula meaning ``equals'', ``extensionally coincides with'', or, in Jevons' parlance, ``is the same as''.

    \item \emph{negative}, having the ``$\neq$'' copula meaning ``does \emph{not} equal'', ``does \emph{not} extensionally coincide with'', or ``is \emph{not} the same as''.
\end{itemize}
\vspace{-\topsep}

One can intuitively understand Jevons' point of view, which generalizes two important relations from numerical algebra \citep[pp.~5,8,15-26,73]{jevons:1869}\citep[p.~7, point~15; p.~6, point~13; p.~86, point~203]{jevons:1864}. \citeauthor{cayley:1871}'s insight, however, is that, in the traditional jargon of categorical syllogistic, sentences with these copulae don't correspond to affirmative and negative ones, but respectively to \emph{universal} and \emph{particular} ones -- an insight later explained in prose by \citet[p.~596]{venn:1883}. In categorical syllogistic jargon, ``affirmative'' versus ``negative'' is a distinction in ``quality'' (in contrast to ``quantity'') revealed to be about (predicative) obversion --with some resemblance to the opposition of qualities in \citet[p.~83, point~193]{jevons:1864}--, and thus these adjectives as used in categorical syllogistic have different meanings from \citeauthor{jevons:1864}' usage.

Indeed, considering yet another distinct meaning, \citeauthor{jevons:1864}' ``affirmative'' sentences are in another sense always ``negative'', if by ``negative'' we now mean that they assert a \emph{\textbf{non}existence} stance ($b\!=\!\bm{\varnothing}$ / $\overline{\mathfrak{I}}\langle b\rangle$) \citep[p.~71, point~167]{jevons:1864}, as \citet[pp.~283-286]{brentano:1874}\citep{land:1876} later noticed. Likewise, ``negative'' sentences in \citeauthor{jevons:1864}' sense are in another sense always ``affirmative'': they assert something \emph{does exist} ($b\!\neq\!\bm{\varnothing}$ / $\mathfrak{I}\langle b\rangle$). (Which should be the ``primitive/affirmative/default'' notion: emptiness/nonexistence (``\emph{est~Nihil}'') or inhabitation/existence (``\emph{est~Ens}'')? Revisit the table in footnote\footref{fn:est_ens}.)

Given the three distinct meanings of ``affirmative''/``negative'', we should always make it clear which one we are referring to.}.
%There is only one empty class: Specimen Calculi Coincidentium, ca. 1686, point 20: https://books.google.com/books?id=KcryCQAAQBAJ&pg=PA817
%non Ens (BB'=0): \citep[p.~259, point~8; p.~261,~point~3]{leibniz:1686d}
%Nihil (M-M=0): \citep[Defin.~8]{leibniz:1686f}: Defin. 8. Compensatio est, cum idem ponitur et detrahitur in eodem, expressa cum expressa. Destructio est cum quid ob compensationem abjicitur, ut non amplius exprimatur et pro M-M ponendo Nihil.
%See also: De Calculo Irrepetibilium, ca. 1687. https://books.google.com/books?id=df_nBQAAQBAJ&pg=PA855 https://books.google.com/books?id=KcryCQAAQBAJ&pg=PA855, an unrelated fragment of which appears in https://archive.org/details/opusculesetfrag00coutgoog/page/256/mode/1up
%Nihil = Nullo

For comparison, if we want to preserve the validity of the subalternation laws in LC, they would have to be respectively expressed -- at the cost of a more complicated formula, roughly following \citet[pp.~167-168]{venn:1881} -- as\\
%In the 2nd edition, Venn (1894) repeated it, in a symbolism somewhat closer to ours <pp. 187-188: https://books.google.com/books?id=RjtbglWMcAoC&pg=PA187>.
$\mathbf{b}\mathbf{c}{\color{white}'}\!\!=\!\!\mathbf{b},~\mathbf{b}\neq\bm{\varnothing} \quad\metale\quad \mathbf{b}\mathbf{c}{\color{white}'}\neq\bm{\varnothing}$\\
$\mathbf{b}\mathbf{c}'\!\!=\!\!\mathbf{b},~\mathbf{b}\neq\bm{\varnothing} \quad\metale\quad \mathbf{b}\mathbf{c}'\neq\bm{\varnothing}$.\\
Alternatively, by transposing the latter premise:\\
$\mathbf{b}\mathbf{c}'\!\!=\!\!\mathbf{b} \quad\metale\quad\; \mathbf{b}\neq\bm{\varnothing}~{\scriptstyle|\!\lor\!|}~\mathbf{b}\mathbf{c}'\neq\bm{\varnothing}$ \{where ``${\scriptstyle|\!\lor\!|}$'' is the meta-level ``or''\}.

Thus, we would have to
\begin{itemize}
    \item adopt a definition of $\mathbf{b}\termlogicA\mathbf{c}$ which explicitly adds existential import of the subject as a constraint (so the definition of $\mathbf{b}\termlogicA\mathbf{c}$ would have to be composed of both left-hand-side premises); \emph{and} also

    \item \emph{either} add existential import of the subject to the definition of $\mathbf{b}\termlogicE\mathbf{c}$ --like \citet[p.~180, rule~18]{mccoll:1877} did\footnote{Storrs \label{fn:boethius_connexive_thesis}\citet[p.~349,347-348]{mccall:1967} explained that, to go from Barbara-1 to Barbari-1 without adding any further premise, we should accept the validity of $\termlogicA\!\!\rightarrow\!\!\termlogicI$ subalternation, which in a symbology from Table~\ref{tab:relations_with_single_symbol} would be\\
$\mathbf{s}\subseteq\mathbf{p} ~\metale\quad \!\!\mathbf{s}\nsubseteq\mathbf{p}'$\\
(\emph{``est~P'' est ``non~est non~P}'', Boethius' connexive thesis),\\
and he discussed the difficulties caused by this.

Hugh \citet[p.~180, rule~18]{mccoll:1877} also embraced this thesis, which entails existential import for universal categorical assertions.
%https://www-jstor-org.ez67.periodicos.capes.gov.br/stable/2251844?seq=4#metadata_info_tab_contents
%``\emph{s est p}'' \emph{est} ``\emph{s non est non p}'
%[non est non Ens = est Ens]
%p. 316: Juan Caramuel Lobkowitz (Ioannis Caramuelis). Caramuelis Praecursor Logicvs: Complectens Grammaticam Audacem, Cuius partes sunt tres, Methodica, Metrica, Critica. Francofurti:Sacrum Romanum Imperium. 1654. https://books.google.com/books?id=fURPAAAAcAAJ&pg=PA316
}-- \emph{or} abandon existential import of the subject for the definition of $\mathbf{b}\termlogicO\mathbf{c}$ --like \citet[pp.~191-192]{tamaki:1974} did. Pick your poison.
\end{itemize}

Alternatively, we could adopt algebraic definitions for ``$\mathbf{b}\termlogicA\mathbf{c}$'', ``$\mathbf{b}\termlogicE\mathbf{c}$'' and ``$\mathbf{b}\termlogicO\mathbf{c}$'' formed by a single premise only, with the consequence that the subalternation ``laws'' would no longer be universally applicable, but subject to an additional inhabitation constraint for the subject -- as we do in this paper.
%From a coincidence, you can't conclude an apartness, such as:
%b=c |=| b=/=c'
%https://www-jstor-org.ez67.periodicos.capes.gov.br/stable/40694037?seq=33#metadata_info_tab_contents
%Equality-only and equality-or-difference representations - Read 2 pages
%https://www-jstor-org.ez67.periodicos.capes.gov.br/stable/4543676?seq=12#metadata_info_tab_contents
%https://www.jstor.org/stable/2011661?seq=3#metadata_info_tab_contents

\subsection{Leibniz-Cayley system axioms}
\label{subsec:LC_axioms}

In order to prove all the classic categorical syllogisms, we have extracted from Leibniz's drafts on logic the following axioms to form LC:

\begin{tabular}{lLl}
\textbf{(LC1)} & bc=cb & \{commutativity {\tiny\citep[p.~235, point~7]{leibniz:1690b}}\}\\
\textbf{(LC2)} & b(cd)=(bc)d & \{associativity\footnotemark\newcounter{tablefootnote}\setcounter{tablefootnote}{\value{footnote}}\}\\
\textbf{(LC3)} & bc\neq\bm{\varnothing} \metale c\neq\bm{\varnothing} & \{predicative inhabitation\footnotemark\}\\%rightwise existential elimination
\textbf{(LC4)} & bc=b \metaeq c'b'=c' & \{subsumption contraposition {\tiny\citep[point~19]{leibniz:1690c}}\footnotemark\}\\
\textbf{(LC5)} & bc'=b \metaeq cb'=c & \{disjointness conversion\footnotemark\}
\end{tabular}
%Most LC axioms are about rearrangement. The only elimination axioms or rules in LC are LC3 and "expr1 = expr2, rel |= rel[expr2 <- expr1]" (used e.g. in Barbara-1).

\footnotetext[\value{tablefootnote}]{We haven't found an explicit statement of this law by Leibniz, although he implicitly made use of it in a proof \citep[pp.~229-230, Axioma 1]{leibniz:1690a}. The earliest explicit statement we could find of the associative law in an algebra is by William Rowan \citet[p.~430]{hamilton:1843}. In the context of the algebra of logic, it was explicitly stated by \citet[p.~251]{peirce:1867}.}

\addtocounter{tablefootnote}{1}\footnotetext[\value{tablefootnote}]{In Leibniz's manuscripts on logic, we failed to find a direct assertion like

\emph{A Nihil = Nihil} \{$a\bm{\varnothing}=\bm{\varnothing}$\},

although we cannot discard it is present in some form somewhere. Nevertheless, we can deduce it from other assertions scattered throughout his manuscripts:

$a\cup\bm{\varnothing} = a$ \citep[p.~267, point~28]{leibniz:1686d}

$\bm{\varnothing}\cup a = a$ \{$b\cup n = n\cup b$ \citep[p.~237, Axiom.~1]{leibniz:1687}\}

$\bm{\varnothing}\subseteq a$ \{$a\!\cup\! y\!=\! c \;\,\metaeq\; a\!\subseteq\! c$ \citep[p.~265, points~9-10]{leibniz:1686d}\}%Also in \citep[p.~311]{leibniz:1686e}

$\bm{\varnothing}a=\bm{\varnothing}$ \{$a\!\subseteq\! b \;\,\metaeq\; ab\!=\!a$ \citep[p.~236, point~13]{leibniz:1690b}\}

$a\bm{\varnothing}=\bm{\varnothing}$ \{$ab=ba$ \citep[p.~235, point~7]{leibniz:1690b}\}

From this, we can deduce the predicative inhabitation law:

\qquad\qquad\ \,$b\bm{\varnothing}\!=\!\bm{\varnothing}$ \{placeholder relabelling: $a\leftarrow b$\}

$c\!=\!\bm{\varnothing} \;\,\metale\; bc\!=\!\bm{\varnothing}$ \{substitution of similars\}%leftwise nonexistential introduction

$bc\!\neq\!\bm{\varnothing} \metale\;\,\; c\!\neq\!\bm{\varnothing}$ \{transposition (from propositional logic)\}.

In Leibniz's original system, which adopts for particular assertions the representations ``$\mathfrak{I}\langle s\rangle$'' and ``$\overline{\mathfrak{I}}\langle s\rangle$'' instead of ``$s\!\!\neq\!\!\bm{\varnothing}$'' and ``$s\!\!=\!\!\bm{\varnothing}$'' respectively, we can prove an equivalent law to LC3, though the proof is different. Unlike LC, \citet[point~5]{leibniz:1686c} assumed existential import of the generating terms: ``\emph{A est, id est A est Ens.}'' (``$\mathfrak{I}\langle a\rangle$''). Thus,

$\qquad\quad\metale\;\;\;\; \mathfrak{I}\langle c\rangle$ \{existential import\}

$\mathfrak{I}\langle bc\rangle \quad\!\!\!\metale\;\;\;\; \mathfrak{I}\langle c\rangle$ \{antecedent introduction before a true consequent\}.

In a system like Leibniz's original one but without existential import, one could simply declare the latter law as an alternative axiom to LC3 by fiat, diagrammatically justified by Figure~\ref{fig:inhabited_bc}, although if we didn't also assume ``$\mathfrak{I}\langle s\rangle \;\metaeq\; s\!\neq\!\bm{\varnothing}$'', the axiom would be wanting of a satisfactory \emph{algebraic} justification. \citet[p.~34]{ladd:1883} also states this law.}
%p. 32:
%b non-Ens, c non-Ens |=| bUc non-Ens
%b Ens |v| c Ens |=| bUc Ens
%p. 34:
%Therefore,
%bUc non-Ens |= b non-Ens
%b non-Ens |= bc non-Ens
%b Ens |= bUc Ens
%bc Ens |= b Ens
%p. 33:
%Therefore,
%(bUc non-Ens) is incompatible with (b Ens) {since bUc non-Ens |= b non-Ens}
%(bc Ens) is incompatible with (b non-Ens) {since b Ens |= bUc Ens}

\addtocounter{tablefootnote}{1}\footnotetext[\value{tablefootnote}]{In the representation of categorical assertions by means of single vowels, it corresponds to ``A-contraposition'' ($\mathbf{b}\termlogicA \mathbf{c} \metaeq \mathbf{c'}\termlogicA \mathbf{b'}$). It symbolically represents contravariance/antitonicity of subclasshood under complementation.}

\addtocounter{tablefootnote}{1}\footnotetext[\value{tablefootnote}]{In the representation of categorical assertions by means of single vowels, it corresponds to ``E-conversion'' ($\mathbf{b}\termlogicE \mathbf{c} \metaeq \mathbf{c}\termlogicE \mathbf{b}$). Leibniz stated that $(\mathbf{b}')'=\mathbf{b}$, which, together with the LC4 axiom, would suffice to deduce LC5 as a theorem.

By adopting both subsumption contraposition and disjointness conversion as axioms, we make the involution law, $(\mathbf{b}')'=\mathbf{b}$, superfluous for the strict purpose of proving classical categorical syllogisms.}

The following is a convenient lemma to shorten the proofs of some valid mood/figure pairs:

\begin{tabular}{lLl}
%\textbf{(LC6)} & \makecell[l]{b=c \metale bd=cd\\\scriptstyle{\text{Proof: }b=c {~|\!=~} db=dc {~|\!=\!|~} bd=dc {~|\!=\!|~} bd=cd}} & \{right restriction\}\\
\textbf{(LC6)} & \makecell[l]{bc\neq\bm{\varnothing} \metale b\neq\bm{\varnothing}\\\scriptstyle{\text{Proof: }bc\neq\bm{\varnothing} {~|\!=\!|~} cb\neq\bm{\varnothing} {~|\!=~} b\neq\bm{\varnothing}}} & \{subject inhabitation\}%leftwise existential elimination
\end{tabular}

Thus, it is fair to say that by the end of 1690 Leibniz had already figured out all the laws required to work as axioms of an algebraic system of categorical syllogistic -- a research he seriously undertook from at least 1679 \citep[pp.~43-44]{leibniz:1679a} on\footnote{LC1 and LC2 are two of the three algebraic axioms of semilattices (Section~\ref{subsubsec:Boolean_lattice}). Interestingly, idempotence ($bb=b$), which \citet[p.~235, point~6]{leibniz:1690b} also explicitly stated, is not required in LC. These three laws, together with LC3, are the bounded (meet-)semilattice axioms.}.

Regarding subject inhabitation (LC6) and predicative inhabitation (LC3), each one can be proved from the other by using commutativity (LC1). Thus, any of them might have been chosen as an axiom. The motivation for LC choosing LC3 rather than LC6 is explained in Section~\ref{subsec:LC_matrix}.

Notice that LC didn't adopt axioms of subalternation:

%$\textbf{b}\textbf{c}=\textbf{b} \quad\metaeq\quad \textbf{b}\textbf{c}\neq\bm{\varnothing}$,
$bc=b \quad\metaeq\quad bc\neq\bm{\varnothing}$

$bc'\!=b \quad\metaeq\quad \!bc'\!\neq\bm{\varnothing}$,

which would imply (by LC6) existential import for universal categorical assertions:

%$\textbf{b}\textbf{c}=\textbf{b} \quad\metaeq\quad \textbf{b}\neq\bm{\varnothing}$.
$bc=b \quad\metaeq\quad b\neq\bm{\varnothing}$

$bc'\!=b \quad\metaeq\quad b\neq\bm{\varnothing}$.

The lack of this axiom makes LC fully compatible to Boolean algebra, as we will prove in Section~\ref{subsubsec:LC_from_Boolean_lattice}. However, due to this intentional omission, 9 of the 24 classic categorical syllogisms require not just two but three premises for them to be valid in this system, just like in the Euler system (Section~\ref{sec:euler}).

These axioms stand on their own as relations between formal, abstract objects which can be manipulated according to the inference rules the system is subject to. In this purely formal sense, the operands can be seen as just ``terms'' (in mathematical expressions), ``names'' or individuating, arbitrary ``labels''. Given these axioms, a skilled middle-school algebra student could trace and maybe even derive on her own the proofs of our theorems without being informed of the context (what these operands and operations refer to). If we had no intended direction and just wanted to derive the theorems of any formal system to see where they lead us to, the chosen axioms would be as arbitrary as any other. For us, the legitimacy of the system as a \emph{logic} which we are interested in studying is justified by the fact that the Euler system, involving \emph{classes} represented by the Euler (and Venn) diagrams shown in Section~\ref{sec:euler} and by \citet{piesk:2017}, is a fully compliant model or interpretation of the axiomatic system we defined, since we want to explore different approaches to prove the same theorems, the classic categorical syllogisms without existential import on universal assertions.

\subsection{Syllogism proofs in the Leibniz-Cayley system}
\label{subsec:LC_proofs}

Here are the proofs of categorical syllogisms in LC. None of them has more than 5 steps.

\begin{multicols}{2}
\begin{syl}
\textbf{Barbara-1:}\\
\scriptsize
\begin{tabular}{N @{\hspace{1.2\tabcolsep}} L @{\hspace{1.2\tabcolsep}} l}
P1. & sm=s & \{Every s is m.\}\\
P2. & mp=m & \{Every m is p.\}\\
\cline{1-2}
C1. & sp=s & \{Every s is p.\}\\
\multicolumn{3}{l}{\textsc{~Proof.}}\\
\multicolumn{3}{l}{(by \citet[pp.~229-230, Axioma 1]{leibniz:1690a},}\\
\multicolumn{3}{l}{once we make associativity explicit:)}\\[4pt]
S3. & s(mp)=s & (P1), (P2)\\[4pt]
S4. & (sm)p=s & (S3),~\begin{tabular}{@{}l@{}}\{$b(cd)=(bc)d$\}\\~$b\leftarrow s, c\leftarrow m, d\leftarrow p$\end{tabular}\\[7pt]
S5. & \mathbf{sp=s} & (S4), (P1). Therefore: (C1).
\end{tabular}
%\citet[p.~264]{leibniz:1686d} offers a different proof.
%
%\color{white}\qedsymbol
\end{syl}
%\hrule
\begin{center}\textbf{*~*~*}\end{center}

\columnbreak\begin{syl}
\textbf{Barbari-1:}\\
\scriptsize
\begin{tabular}{N @{\hspace{1.2\tabcolsep}} L @{\hspace{1.2\tabcolsep}} l}
P1. & sm=s & \{Every s is m.\}\\
P2. & mp=m & \{Every m is p.\}\\
P3. & s\neq\bm{\varnothing} & \{At least one s exists.\}\\
\cline{1-2}
C1. & sp=s & \{Every s is p.\}\\
C2. & sp\neq\bm{\varnothing} & \{At least one s is p.\}\\
\multicolumn{3}{l}{\textsc{~Proof.}}\\[4pt]
S4. & s(mp)=s & (P1), (P2)\\[4pt]
S5. & (sm)p=s & (S4),~\begin{tabular}{@{}l@{}}\{$b(cd)=(bc)d$\}\\~$b\leftarrow s, c\leftarrow m, d\leftarrow p$\end{tabular}\\[7pt]
S6. & \mathbf{sp=s} & (S5), (P1). Therefore: (C1).\\[4pt]
S7. & \mathbf{sp\neq\bm{\varnothing}} & (S6), (P3). Therefore: (C2).
\end{tabular}
%
%\color{white}\qedsymbol
\end{syl}
%\hrule
\begin{center}\textbf{*~*~*}\end{center}

\columnbreak\begin{syl}
\textbf{Celarent-1:}\\
\scriptsize
\begin{tabular}{N @{\hspace{1.2\tabcolsep}} L @{\hspace{1.2\tabcolsep}} l}
P1. & sm=s & \{Every s is m.\}\\
P2. & mp'=m & \{No m is p.\}\\
\cline{1-2}
C1. & sp'=s & \{No s is p.\}\\
\multicolumn{3}{l}{\textsc{~Proof.}}\\[4pt]
\multicolumn{3}{l}{(by \citet[pp.~102]{whitehead:1898}, once we}\\
\multicolumn{3}{l}{make the use of associativity explicit:)}\\[4pt]
S3. & s(mp')=s & (P1), (P2)\\[4pt]
S4. & (sm)p'=s & (S3),~\begin{tabular}{@{}l@{}}\{$b(cd)=(bc)d$\}\\~$b\leftarrow s, c\leftarrow m, d\leftarrow p'$\end{tabular}\\[7pt]
S5. & \mathbf{sp'=s} & (S4), (P1). Therefore: (C1).
\end{tabular}
%
%\color{white}\qedsymbol
\end{syl}
%\hrule
\begin{center}\textbf{*~*~*}\end{center}

\begin{syl}
\textbf{Celaront-1:}\\
\scriptsize
\begin{tabular}{N @{\hspace{1.2\tabcolsep}} L @{\hspace{1.2\tabcolsep}} l}
P1. & sm=s & \{Every s is m.\}\\
P2. & mp'=m & \{No m is p.\}\\
P3. & s\neq\bm{\varnothing} & \{At least one s exists.\}\\
\cline{1-2}
C1. & sp'=s & \{No s is p.\}\\
C2. & sp'\neq\bm{\varnothing} & \{At least one s is not p.\}\\
\multicolumn{3}{l}{\textsc{~Proof.}}\\[4pt]
S4. & s(mp')=s & (P1), (P2)\\[4pt]
S5. & (sm)p'=s & (S4),~\begin{tabular}{@{}l@{}}\{$b(cd)=(bc)d$\}\\~$b\leftarrow s, c\leftarrow m, d\leftarrow p'$\end{tabular}\\[7pt]
S6. & \mathbf{sp'=s} & (S5), (P1). Therefore: (C1).\\[4pt]
S7. & \mathbf{sp'\neq\bm{\varnothing}} & (S6), (P3). Therefore: (C2).
\end{tabular}
%
%\color{white}\qedsymbol
\end{syl}
%\hrule
\begin{center}\textbf{*~*~*}\end{center}

\begin{syl}
\textbf{Camestres-2:}\\
\scriptsize
\begin{tabular}{N @{\hspace{1.2\tabcolsep}} L @{\hspace{1.2\tabcolsep}} l}
P1. & sm'=s & \{No s is m.\}\\
P2. & pm=p & \{Every p is m.\}\\
\cline{1-2}
C1. & sp'=s & \{No s is p.\}\\
\multicolumn{3}{l}{\textsc{~Proof.}}\\
S3. & m'p'=m' & (P2),~\begin{tabular}{@{}c@{}}\{$bc=b\metaeq c'b'=c'$\}\\$b\leftarrow p, c\leftarrow m$
\end{tabular}\\[7pt]
S4. & s(m'p')=s & (P1), (S3)\\[4pt]
S5. & (sm')p'=s & (S4),~\begin{tabular}{@{}l@{}}\{$b(cd)=(bc)d$\}\\~$b\leftarrow s, c\leftarrow m', d\leftarrow p'$\end{tabular}\\[7pt]
S6. & \mathbf{sp'=s} & (S5), (P1). Therefore: (C1).
\end{tabular}
%
%\color{white}\qedsymbol
\end{syl}
%\hrule
\begin{center}\textbf{*~*~*}\end{center}

\begin{syl}
\textbf{Camestros-2:}\\
\scriptsize
\begin{tabular}{N @{\hspace{1.2\tabcolsep}} L @{\hspace{1.2\tabcolsep}} l}
P1. & sm'=s & \{No s is m.\}\\
P2. & pm=p & \{Every p is m.\}\\
P3. & s\neq\bm{\varnothing} & \{At least one s exists.\}\\
\cline{1-2}
C1. & sp'=s & \{No s is p.\}\\
C2. & sp'\neq\bm{\varnothing} & \{At least one s is not p.\}\\
\multicolumn{3}{l}{\textsc{~Proof.}}\\
S4. & m'p'=m' & (P2),~\begin{tabular}{@{}c@{}}\{$bc=b\metaeq c'b'=c'$\}\\$b\leftarrow p, c\leftarrow m$
\end{tabular}\\[7pt]
S5. & s(m'p')=s & (P1), (S4)\\[4pt]
S6. & (sm')p'=s & (S5),~\begin{tabular}{@{}l@{}}\{$b(cd)=(bc)d$\}\\~$b\leftarrow s, c\leftarrow m', d\leftarrow p'$\end{tabular}\\[7pt]
S7. & \mathbf{sp'=s} & (S6), (P1). Therefore: (C1).\\[4pt]
S8. & \mathbf{sp'\neq\bm{\varnothing}} & (S7), (P3). Therefore: (C2).
\end{tabular}
%
%\color{white}\qedsymbol
\end{syl}
%\hrule
\begin{center}\textbf{*~*~*}\end{center}

%\end{multicols}

\columnbreak\begin{syl}
\textbf{Bamalip-4:}\\
\scriptsize
\begin{tabular}{N @{\hspace{1.2\tabcolsep}} L @{\hspace{1.2\tabcolsep}} l}
P1. & ms=m & \{Every m is s.\}\\
P2. & pm=p & \{Every p is m.\}\\
P3. & p\neq\bm{\varnothing} & \{At least one p exists.\}\\
\cline{1-2}
C1. & sp\neq\bm{\varnothing} & \{At least one s is p.\}\\
\multicolumn{3}{l}{\textsc{~Proof.}}\\[4pt]
S4. & p(ms)=p & (P2), (P1)\\[4pt]
S5. & (pm)s=p & (S4),~\begin{tabular}{@{}l@{}}\{$b(cd)=(bc)d$\}\\~$b\leftarrow p, c\leftarrow m, d\leftarrow s$\end{tabular}\\[7pt]
S6. & ps=p & (S5), (P2)\\[4pt]
S7. & sp=p & (S6),~\begin{tabular}{@{}l@{}}\{$bc=cb$\}\\~$b\leftarrow p, c\leftarrow s$\end{tabular}\\[7pt]
S8. & \mathbf{sp\neq\bm{\varnothing}} & (S7), (P3). Therefore: (C1).
\end{tabular}
%
%\color{white}\qedsymbol
\end{syl}
%\hrule
\makebox[0.5\textwidth][c]{\textbf{*~*~*}}

\begin{syl}
\textbf{Darapti-3:}\\
\scriptsize
\begin{tabular}{N @{\hspace{1.2\tabcolsep}} L @{\hspace{1.2\tabcolsep}} l}
P1. & ms=m & \{Every m is s.\}\\
P2. & mp=m & \{Every m is p.\}\\
P3. & m\neq\bm{\varnothing} & \{At least one m exists.\}\\
\cline{1-2}
C1. & sp\neq\bm{\varnothing} & \{At least one s is p.\}\\
\multicolumn{3}{l}{\textsc{~Proof.}}\\[4pt]
S4. & (ms)p=m & (P2), (P1)\\[4pt]
S5. & m(sp)=m & (S4),~\begin{tabular}{@{}l@{}}\{$b(cd)=(bc)d$\}\\~$b\leftarrow m, c\leftarrow s, d\leftarrow p$\end{tabular}\\[7pt]
S6. & m(sp)\neq\bm{\varnothing} & (S5), (P3)\\[4pt]
S7. & \mathbf{sp\neq\bm{\varnothing}} & (S6),~\begin{tabular}{@{}l@{}}\begin{tabular}{@{}c@{}}\{$bc\neq\bm{\varnothing} \metale c\neq\bm{\varnothing}$\}\\$b\leftarrow m, c\leftarrow sp$.\end{tabular}\\Therefore: (C1).\end{tabular}
\end{tabular}
%
%\color{white}\qedsymbol
\end{syl}
%\hrule
\makebox[0.5\textwidth][c]{\textbf{*~*~*}}

\begin{syl}
\textbf{Felapton-3:}\\
\scriptsize
\begin{tabular}{N @{\hspace{1.2\tabcolsep}} L @{\hspace{1.2\tabcolsep}} l}
P1. & ms=m & \{Every m is s.\}\\
P2. & mp'=m & \{No m is p.\}\\
P3. & m\neq\bm{\varnothing} & \{At least one m exists.\}\\
\cline{1-2}
C1. & sp'\neq\bm{\varnothing} & \{At least one s is not p.\}\\
\multicolumn{3}{l}{\textsc{~Proof.}}\\[4pt]
S4. & (ms)p'=m & (P2), (P1)\\[4pt]
S5. & m(sp')=m & (S4),~\begin{tabular}{@{}l@{}}\{$b(cd)=(bc)d$\}\\~$b\leftarrow m, c\leftarrow s, d\leftarrow p'$\end{tabular}\\[7pt]
S6. & m(sp')\neq\bm{\varnothing} & (S5), (P3)\\[4pt]
S7. & \mathbf{sp'\neq\bm{\varnothing}} & (S6),~\begin{tabular}{@{}l@{}}\begin{tabular}{@{}c@{}}\{$bc\neq\bm{\varnothing} \metale c\neq\bm{\varnothing}$\}\\$b\leftarrow m, c\leftarrow sp'$.\end{tabular}\\Therefore: (C1).\end{tabular}
\end{tabular}
%
%\color{white}\qedsymbol
\end{syl}
%\hrule
\makebox[0.5\textwidth][c]{\textbf{*~*~*}}

\begin{syl}
\textbf{Disamis-3:}\\
\scriptsize
\begin{tabular}{N @{\hspace{1.2\tabcolsep}} L @{\hspace{1.2\tabcolsep}} l}
P1. & ms=m & \{Every m is s.\}\\
P2. & mp\neq\bm{\varnothing} & \{At least one m is p.\}\\
\cline{1-2}
C1. & sp\neq\bm{\varnothing} & \{At least one s is p.\}\\
\multicolumn{3}{l}{\textsc{~Proof.}}\\[4pt]
S3. & (ms)p\neq\bm{\varnothing} & (P2), (P1)\\[4pt]
S4. & m(sp)\neq\bm{\varnothing} & (S3),~\begin{tabular}{@{}l@{}}\{$b(cd)=(bc)d$\}\\~$b\leftarrow m, c\leftarrow s, d\leftarrow p$\end{tabular}\\[7pt]
S5. & \mathbf{sp\neq\bm{\varnothing}} & (S4),~\begin{tabular}{@{}l@{}}\begin{tabular}{@{}c@{}}\{$bc\neq\bm{\varnothing} \metale c\neq\bm{\varnothing}$\}\\$b\leftarrow m, c\leftarrow sp$.\end{tabular}\\Therefore: (C1).\end{tabular}
\end{tabular}
%
%\color{white}\qedsymbol
\end{syl}
%\hrule
\makebox[0.5\textwidth][c]{\textbf{*~*~*}}

\begin{syl}
\textbf{Bokardo-3:}\\
\scriptsize
\begin{tabular}{N @{\hspace{1.2\tabcolsep}} L @{\hspace{1.2\tabcolsep}} l}
P1. & ms=m & \{Every m is s.\}\\
P2. & mp'\neq\bm{\varnothing} & \{At least one m is not p.\}\\
\cline{1-2}
C1. & sp'\neq\bm{\varnothing} & \{At least one s is not p.\}\\
\multicolumn{3}{l}{\textsc{~Proof.}}\\[4pt]
S3. & (ms)p'\neq\bm{\varnothing} & (P1), (P2)\\[4pt]
S4. & m(sp')\neq\bm{\varnothing} & (S3),~\begin{tabular}{@{}l@{}}\{$b(cd)=(bc)d$\}\\~$b\leftarrow m, c\leftarrow s, d\leftarrow p'$\end{tabular}\\[7pt]
S5. & \mathbf{sp'\neq\bm{\varnothing}} & (S4),~\begin{tabular}{@{}l@{}}\begin{tabular}{@{}c@{}}\{$bc\neq\bm{\varnothing} \metale c\neq\bm{\varnothing}$\}\\$b\leftarrow m, c\leftarrow sp'$.\end{tabular}\\Therefore: (C1).\end{tabular}
\end{tabular}
%
%\color{white}\qedsymbol
\end{syl}
%\hrule
\makebox[0.5\textwidth][c]{\textbf{*~*~*}}

\begin{syl}
\textbf{Darii-1:}\\
\scriptsize
\begin{tabular}{N @{\hspace{1.2\tabcolsep}} L @{\hspace{1.2\tabcolsep}} l}
P1. & sm\neq\bm{\varnothing} & \{At least one s is m.\}\\
P2. & mp=m & \{Every m is p.\}\\
\cline{1-2}
C1. & sp\neq\bm{\varnothing} & \{At least one s is p.\}\\
\multicolumn{3}{l}{\textsc{~Proof.}}\\[4pt]
S3. & s(mp)\neq\bm{\varnothing} & (P1), (P2)\\[4pt]
S4. & s(pm)\neq\bm{\varnothing} & (S3),~\begin{tabular}{@{}l@{}}\{$bc=cb$\}\\~$b\leftarrow m, c\leftarrow p$\end{tabular}\\[7pt]
S5. & (sp)m\neq\bm{\varnothing} & (S4),~\begin{tabular}{@{}l@{}}\{$b(cd)=(bc)d$\}\\~$b\leftarrow s, c\leftarrow p, d\leftarrow m$\end{tabular}\\[7pt]
S6. & \mathbf{sp\neq\bm{\varnothing}} & (S5),~\begin{tabular}{@{}l@{}}\begin{tabular}{@{}c@{}}\{$bc\neq\bm{\varnothing} \metale b\neq\bm{\varnothing}$\}\\$b\leftarrow sp, c\leftarrow m$.\end{tabular}\\Therefore: (C1).\end{tabular}
\end{tabular}
%
%\color{white}\qedsymbol
\end{syl}
%\hrule
\makebox[0.5\textwidth][c]{\textbf{*~*~*}}

\begin{syl}
\textbf{Ferio-1:}\\
\scriptsize
\begin{tabular}{N @{\hspace{1.2\tabcolsep}} L @{\hspace{1.2\tabcolsep}} l}
P1. & sm\neq\bm{\varnothing} & \{At least one s is m.\}\\
P2. & mp'=m & \{No m is p.\}\\
\cline{1-2}
C1. & sp'\neq\bm{\varnothing} & \{At least one s is not p.\}\\
\multicolumn{3}{l}{\textsc{~Proof.}}\\[4pt]
S3. & s(mp')\neq\bm{\varnothing} & (P1), (P2)\\[4pt]
S4. & s(p'm)\neq\bm{\varnothing} & (S3),~\begin{tabular}{@{}l@{}}\{$bc=cb$\}\\~$b\leftarrow m, c\leftarrow p'$\end{tabular}\\[7pt]
S5. & (sp')m\neq\bm{\varnothing} & (S4),~\begin{tabular}{@{}l@{}}\{$b(cd)=(bc)d$\}\\~$b\leftarrow s, c\leftarrow p', d\leftarrow m$\end{tabular}\\[7pt]
S6. & \mathbf{sp'\neq\bm{\varnothing}} & (S5),~\begin{tabular}{@{}l@{}}\begin{tabular}{@{}c@{}}\{$bc\neq\bm{\varnothing} \metale b\neq\bm{\varnothing}$\}\\$b\leftarrow sp', c\leftarrow m$.\end{tabular}\\Therefore: (C1).\end{tabular}
\end{tabular}
%
%\color{white}\qedsymbol
\end{syl}
%\hrule
\makebox[0.5\textwidth][c]{\textbf{*~*~*}}

\vspace{-5pt}\begin{syl}
\textbf{Baroko-2:}\\
\scriptsize
\begin{tabular}{N @{\hspace{1.2\tabcolsep}} L @{\hspace{1.2\tabcolsep}} l}
P1. & sm'\neq\bm{\varnothing} & \{At least one s is not m.\}\\
P2. & pm=p & \{Every p is m.\}\\
\cline{1-2}
C1. & sp'\neq\bm{\varnothing} & \{At least one s is not p.\}\\
\multicolumn{3}{l}{\textsc{~Proof.}}\\
S3. & m'p'=m' & (P2),~\begin{tabular}{@{}c@{}}\{$bc=b \metaeq c'b'=c'$\}\\$b\leftarrow p, c\leftarrow m$\end{tabular}\\[7pt]
S4. & s(m'p')\neq\bm{\varnothing} & (S3), (P1)\\[4pt]
S5. & s(p'm')\neq\bm{\varnothing} & (S4),~\begin{tabular}{@{}l@{}}\{$bc=cb$\}\\~$b\leftarrow m', c\leftarrow p'$\end{tabular}\\[7pt]
S6. & (sp')m'\neq\bm{\varnothing} & (S5),~\begin{tabular}{@{}l@{}}\{$b(cd)=(bc)d$\}\\~$b\leftarrow s, c\leftarrow p', d\leftarrow m'$\end{tabular}\\[7pt]
S7. & \mathbf{sp'\neq\bm{\varnothing}} & (S6),~\begin{tabular}{@{}l@{}}\begin{tabular}{@{}c@{}}\{$bc\neq\bm{\varnothing} \metale b\neq\bm{\varnothing}$\}\\$b\leftarrow s, c\leftarrow p'$.\end{tabular}\\Therefore: (C1).\end{tabular}
\end{tabular}
\end{syl}
\makebox[0.5\textwidth][c]{\textbf{*~*~*}}

\begin{syl}
\textbf{Dimatis-4:}\\
\scriptsize
\begin{tabular}{N @{\hspace{1.2\tabcolsep}} L @{\hspace{1.2\tabcolsep}} l}
P1. & ms=m & \{Every m is s.\}\\
P2. & pm\neq\bm{\varnothing} & \{At least one p is m.\}\\
\cline{1-2}
C1. & sp\neq\bm{\varnothing} & \{At least one s is p.\}\\
\multicolumn{3}{l}{\textsc{~Proof.}}\\[4pt]
P2a. & mp\neq\bm{\varnothing} & (P2),~\begin{tabular}{@{}l@{}}\{$bc=cb$\}\\~$b\leftarrow p, c\leftarrow m$\end{tabular}\\[7pt]
S3. & (ms)p\neq\bm{\varnothing} & (P2a), (P1)\\[4pt]
S4. & m(sp)\neq\bm{\varnothing} & (S3),~\begin{tabular}{@{}l@{}}\{$b(cd)=(bc)d$\}\\~$b\leftarrow m, c\leftarrow s, d\leftarrow p$\end{tabular}\\[7pt]
S5. & \mathbf{sp\neq\bm{\varnothing}} & (S4),~\begin{tabular}{@{}l@{}}\begin{tabular}{@{}c@{}}\{$bc\neq\bm{\varnothing} \metale c\neq\bm{\varnothing}$\}\\$b\leftarrow m, c\leftarrow sp$.\end{tabular}\\Therefore: (C1).\end{tabular}
\end{tabular}
\end{syl}
\makebox[0.5\textwidth][c]{\textbf{*~*~*}}

\begin{syl}
\textbf{Datisi-3:}\\
\scriptsize
\begin{tabular}{N @{\hspace{1.2\tabcolsep}} L @{\hspace{1.2\tabcolsep}} l}
P1. & ms\neq\bm{\varnothing} & \{At least one m is s.\}\\
P2. & mp=m & \{Every m is p.\}\\
\cline{1-2}
C1. & sp\neq\bm{\varnothing} & \{At least one s is p.\}\\
\multicolumn{3}{l}{\textsc{~Proof.}}\\[4pt]
S3. & (mp)s\neq\bm{\varnothing} & (P1), (P2)\\[4pt]
S4. & m(ps)\neq\bm{\varnothing} & (S4),~\begin{tabular}{@{}l@{}}\{$b(cd)=(bc)d$\}\\~$b\leftarrow m, c\leftarrow p, d\leftarrow s$\end{tabular}\\[7pt]
S5. & ps\neq\bm{\varnothing} & (S4),~\begin{tabular}{@{}c@{}}\{$bc\neq\bm{\varnothing} \metale c\neq\bm{\varnothing}$\}\\$b\leftarrow m, c\leftarrow ps$\end{tabular}\\[7pt]
S6. & \mathbf{sp\neq\bm{\varnothing}} & (S5),~\begin{tabular}{@{}l@{}}\begin{tabular}{@{}l@{}}\{$bc=cb$\}\\~$b\leftarrow p, c\leftarrow s$.\end{tabular}\\Therefore: (C1).\end{tabular}
\end{tabular}
\end{syl}
\makebox[0.5\textwidth][c]{\textbf{*~*~*}}

\begin{syl}
\textbf{Ferison-3:}\\
\scriptsize
\begin{tabular}{N @{\hspace{1.2\tabcolsep}} L @{\hspace{1.2\tabcolsep}} l}
P1. & ms\neq\bm{\varnothing} & \{At least one m is s.\}\\
P2. & mp'=m & \{No m is p.\}\\
\cline{1-2}
C1. & sp'\neq\bm{\varnothing} & \{At least one s is not p.\}\\
\multicolumn{3}{l}{\textsc{~Proof.}}\\[4pt]
S3. & (mp')s\neq\bm{\varnothing} & (P1), (P2)\\[4pt]
S4. & m(p's)\neq\bm{\varnothing} & (S3),~\begin{tabular}{@{}l@{}}\{$b(cd)=(bc)d$\}\\~$b\leftarrow m, c\leftarrow p', d\leftarrow s$\end{tabular}\\[7pt]
S5. & p's\neq\bm{\varnothing} & (S4),~\begin{tabular}{@{}c@{}}\{$bc\neq\bm{\varnothing} \metale c\neq\bm{\varnothing}$\}\\$b\leftarrow m, c\leftarrow p's$\end{tabular}\\[7pt]
S6. & \mathbf{sp'\neq\bm{\varnothing}} & (S5),~\begin{tabular}{@{}l@{}}~\begin{tabular}{@{}l@{}}\{$bc=cb$\}\\~$b\leftarrow p', c\leftarrow s$.\end{tabular}\\Therefore: (C1).\end{tabular}
\end{tabular}
\end{syl}
\makebox[0.5\textwidth][c]{\textbf{*~*~*}}

\columnbreak\begin{syl}
\textbf{Festino-2:}\\
\scriptsize
\begin{tabular}{N @{\hspace{1.2\tabcolsep}} L @{\hspace{1.2\tabcolsep}} l}
P1. & sm\neq\bm{\varnothing} & \{At least one s is m.\}\\
P2. & pm'=p & \{No p is m.\}\\
\cline{1-2}
C1. & sp'\neq\bm{\varnothing} & \{At least one s is not p.\}\\
\multicolumn{3}{l}{\textsc{~Proof.}}\\[4pt]
P2a. & mp'=m & (P2),~\begin{tabular}{@{}c@{}}\{$bc'=b \metaeq cb'=c$\}\\$b\leftarrow p, c\leftarrow m$\end{tabular}\\[7pt]
S3. & s(mp')\neq\bm{\varnothing} & (P1), (P2a)\\[4pt]
S4. & s(p'm)\neq\bm{\varnothing} & (S3),~\begin{tabular}{@{}l@{}}\{$bc=cb$\}\\~$b\leftarrow m, c\leftarrow p'$\end{tabular}\\[7pt]
S5. & (sp')m\neq\bm{\varnothing} & (S4),~\begin{tabular}{@{}l@{}}\{$b(cd)=(bc)d$\}\\~$b\leftarrow s, c\leftarrow p', d\leftarrow m$\end{tabular}\\[7pt]
S6. & \mathbf{sp'\neq\bm{\varnothing}} & (S5),~\begin{tabular}{@{}l@{}}\begin{tabular}{@{}c@{}}\{$bc\neq\bm{\varnothing} \metale b\neq\bm{\varnothing}$\}\\$b\leftarrow sp', c\leftarrow m$.\end{tabular}\\Therefore: (C1).\end{tabular}
\end{tabular}
\end{syl}
\makebox[0.5\textwidth][c]{\textbf{*~*~*}}

\begin{syl}
\textbf{Fresison-4:}\\
\scriptsize
\begin{tabular}{N @{\hspace{1.2\tabcolsep}} L @{\hspace{1.2\tabcolsep}} l}
P1. & ms\neq\bm{\varnothing} & \{At least one m is s.\}\\
P2. & pm'=p & \{No p is m.\}\\
\cline{1-2}
C1. & sp'\neq\bm{\varnothing} & \{At least one s is not p.\}\\
\multicolumn{3}{l}{\textsc{~Proof.}}\\[4pt]
P2a. & mp'=m & (P2),~\begin{tabular}{@{}c@{}}\{$bc'=b \metaeq cb'=c$\}\\$b\leftarrow p, c\leftarrow m$\end{tabular}\\[7pt]
S3. & (mp')s\neq\bm{\varnothing} & (P1), (P2a)\\[4pt]
S4. & m(p's)\neq\bm{\varnothing} & (S3),~\begin{tabular}{@{}l@{}}\{$b(cd)=(bc)d$\}\\~$b\leftarrow m, c\leftarrow p', d\leftarrow s$\end{tabular}\\[7pt]
S5. & p's\neq\bm{\varnothing} & (S4),~\begin{tabular}{@{}c@{}}\{$bc\neq\bm{\varnothing} \metale c\neq\bm{\varnothing}$\}\\$b\leftarrow m, c\leftarrow p's$\end{tabular}\\[7pt]
S6. & \mathbf{sp'\neq\bm{\varnothing}} & (S5),~\begin{tabular}{@{}l@{}}~\begin{tabular}{@{}l@{}}\{$bc=cb$\}\\~$b\leftarrow p', c\leftarrow s$.\end{tabular}\\Therefore: (C1).\end{tabular}
\end{tabular}
\end{syl}
\makebox[0.5\textwidth][c]{\textbf{*~*~*}}

\begin{syl}
\textbf{Fesapo-4:}\\
\scriptsize
\begin{tabular}{N @{\hspace{1.2\tabcolsep}} L @{\hspace{1.2\tabcolsep}} l}
P1. & ms=m & \{Every m is s.\}\\
P2. & pm'=p & \{No p is m.\}\\
P3. & m\neq\bm{\varnothing} & \{At least one m exists.\}\\
\cline{1-2}
C1. & sp'\neq\bm{\varnothing} & \{At least one s is not p.\}\\
\multicolumn{3}{l}{\textsc{~Proof.}}\\[4pt]
P2a. & mp'=m & (P2),~\begin{tabular}{@{}l@{}}\{$bc'=b \metaeq cb'=c$\}\\~$b\leftarrow p, c\leftarrow m$\end{tabular}\\[7pt]
S4. & (ms)p'=m & (P2a), (P1)\\[4pt]
S5. & m(sp')=m & (S4),~\begin{tabular}{@{}l@{}}\{$b(cd)=(bc)d$\}\\~$b\leftarrow m, c\leftarrow s, d\leftarrow p'$\end{tabular}\\[7pt]
S6. & m(sp')\neq\bm{\varnothing} & (S5), (P3)\\[4pt]
S7. & \mathbf{sp'\neq\bm{\varnothing}} & (S6),~\begin{tabular}{@{}l@{}}\begin{tabular}{@{}c@{}}\{$bc\neq\bm{\varnothing} \metale c\neq\bm{\varnothing}$\}\\$b\leftarrow m, c\leftarrow sp'$.\end{tabular}\\Therefore: (C1).\end{tabular}
\end{tabular}
\end{syl}
\makebox[0.5\textwidth][c]{\textbf{*~*~*}}

\columnbreak\vspace{-5pt}\begin{syl}
\textbf{Cesare-2:}\\
\scriptsize
\begin{tabular}{N @{\hspace{1.2\tabcolsep}} L @{\hspace{1.2\tabcolsep}} l}
P1. & sm=s & \{Every s is m.\}\\
P2. & pm'=p & \{No p is m.\}\\
\cline{1-2}
C1. & sp'=s & \{No s is p.\}\\
\multicolumn{3}{l}{\textsc{~Proof.}}\\
P2a. & mp'=m & (P2),~\begin{tabular}{@{}c@{}}\{$bc'=b \metaeq cb'=c$\}\\$b\leftarrow p, c\leftarrow m$\end{tabular}\\[7pt]
S3. & s(mp')=s & (P1), (P2a)\\[4pt]
S4. & (sm)p'=s & (S3),~\begin{tabular}{@{}l@{}}\{$b(cd)=(bc)d$\}\\~$b\leftarrow s, c\leftarrow m, d\leftarrow p'$\end{tabular}\\[7pt]
S5. & \mathbf{sp'=s} & \begin{tabular}{@{}l@{}}(S4), (P1).\\Therefore: (C1).\end{tabular}
\end{tabular}
\end{syl}
\vspace{-5pt}\makebox[0.5\textwidth][c]{\textbf{*~*~*}}

\vspace{-5pt}\begin{syl}
\textbf{Cesaro-2:}\\
\scriptsize
\begin{tabular}{N @{\hspace{1.2\tabcolsep}} L @{\hspace{1.2\tabcolsep}} l}
P1. & sm=s & \{Every s is m.\}\\
P2. & pm'=p & \{No p is m.\}\\
P3. & s\neq\bm{\varnothing} & \{At least one s exists.\}\\
\cline{1-2}
C1. & sp'=s & \{No s is p.\}\\
C2. & sp'\neq\bm{\varnothing} & \{At least one s is not p.\}\\
\multicolumn{3}{l}{\textsc{~Proof.}}\\
P2a. & mp'=m & (P2),~\begin{tabular}{@{}c@{}}\{$bc'=b \metaeq cb'=c$\}\\$b\leftarrow p, c\leftarrow m$\end{tabular}\\[7pt]
S4. & s(mp')=s & (P1), (P2a)\\[4pt]
S5. & (sm)p'=s & (S4),~\begin{tabular}{@{}l@{}}\{$b(cd)=(bc)d$\}\\~$b\leftarrow s, c\leftarrow m, d\leftarrow p'$\end{tabular}\\[7pt]
S6. & \mathbf{sp'=s} & \begin{tabular}{@{}l@{}}(S5), (P1).\\Therefore: (C1).\end{tabular}\\
S7. & \mathbf{sp'\neq\bm{\varnothing}} & \begin{tabular}{@{}l@{}}(S6), (P3).\\Therefore: (C2).\end{tabular}
\end{tabular}
\end{syl}
\vspace{-5pt}\makebox[0.5\textwidth][c]{\textbf{*~*~*}}

\vspace{-5pt}\begin{syl}
\textbf{Calemes-4:}\\
\scriptsize
\begin{tabular}{N @{\hspace{1.2\tabcolsep}} L @{\hspace{1.2\tabcolsep}} l}
P1. & ms'=m & \{No m is s.\}\\
P2. & pm=p & \{Every p is m.\}\\
\cline{1-2}
C1. & sp'=s & \{No s is p.\}\\
\multicolumn{3}{l}{\textsc{~Proof.}}\\
S3. & p(ms')=p & (P2), (P1)\\[4pt]
S4. & (pm)s'=p & (S3), ~\begin{tabular}{@{}l@{}}\{$b(cd)=(bc)d$\}\\~$b\leftarrow p, c\leftarrow m, d\leftarrow s'$\end{tabular}\\[7pt]
S5. & ps'=p & (S4), (P2)\\[4pt]
S6. & \mathbf{sp'=s} & (S5),~\begin{tabular}{@{}l@{}}\begin{tabular}{@{}c@{}}\{$bc'=b \metaeq cb'=c$\}\\~$b\leftarrow p, c\leftarrow s$.\end{tabular}\\Therefore: (C1).\end{tabular}
\end{tabular}
\end{syl}
\vspace{-5pt}\makebox[0.5\textwidth][c]{\textbf{*~*~*}}

\vspace{-5pt}\begin{syl}
\textbf{Calemos-4:}\\
\scriptsize
\begin{tabular}{N @{\hspace{1.2\tabcolsep}} L @{\hspace{1.2\tabcolsep}} l}
P1. & ms'=m & \{No m is s.\}\\
P2. & pm=p & \{Every p is m.\}\\
P3. & s\neq\bm{\varnothing} & \{At least one s exists.\}\\
\cline{1-2}
C1. & sp'=s & \{No s is p.\}\\
C2. & sp'\neq\bm{\varnothing} & \{At least one s is not p.\}\\
\multicolumn{3}{l}{\textsc{~Proof.}}\\
S4. & p(ms')=p & (P2), (P1)\\[4pt]
S5. & (pm)s'=p & (S4), ~\begin{tabular}{@{}l@{}}\{$b(cd)=(bc)d$\}\\~$b\leftarrow p, c\leftarrow m, d\leftarrow s'$\end{tabular}\\[7pt]
S6 & ps'=p & (S5), (P2)\\[4pt]
S7. & \mathbf{sp'=s} & (S6),~\begin{tabular}{@{}l@{}}\begin{tabular}{@{}c@{}}\{$bc'=b \metaeq cb'=c$\}\\~$b\leftarrow p, c\leftarrow s$.\end{tabular}\\Therefore: (C1).\end{tabular}\\[7pt]
S8. & \mathbf{sp'\neq\bm{\varnothing}} & \begin{tabular}{@{}l@{}}(S7), (P3).\\Therefore: (C2).\end{tabular}
\end{tabular}
\end{syl}

\begin{flushright}\qedsymbol\end{flushright}

\end{multicols}

Thus, we have shown that the LC axioms are sufficiently powerful for proving all 24 classic categorical syllogisms.

%\subsection{Syllogism-axiom matrix for LC and recap of the fundamental ingredients}
\subsection{Syllogism-axiom matrix for LC}
\label{subsec:LC_matrix}

Table~\ref{tab:LC_matrix} summarizes the axioms that have been adopted in the proof of each valid categorical syllogism in the preceding subsection.

\begin{table}[ht]
    \centering
    \begin{threeparttable}
    \footnotesize
    \begin{tabular}{|m{0.12\textwidth}|M{0.15\textwidth}|@{}M{0.15\textwidth}@{}|@{}M{0.15\textwidth}|@{~}M{0.15\textwidth}|M{0.15\textwidth}|}
        \hline
        Syllogism & (LC2)\linebreak associativity & (LC3)\linebreak predicative\linebreak inhabitation & (LC1)\linebreak commutativity & (LC5)\linebreak disjointness conversion & (LC4)\linebreak subsumption contraposition\\\hline
        Barbara-1 & \checkmark &  &  &  &  \\\hline 
        Barbari-1 & \checkmark &  &  &  &  \\\hline
        Celarent-1 & \checkmark &  &  &  &  \\\hline
        Celaront-1 & \checkmark &  &  &  &  \\\hline
        Camestres-2 & \checkmark &  &  &  & \checkmark \\\hline
        Camestros-2 & \checkmark &  &  &  & \checkmark \\\hline
        Bamalip-4 & \checkmark &  & \checkmark &  &  \\\hline
        Darapti-3 & \checkmark & \checkmark &  &  &  \\\hline
        Felapton-3 & \checkmark & \checkmark &  &  &  \\\hline
        Disamis-3 & \checkmark & \checkmark &  &  &  \\\hline
        Bokardo-3 & \checkmark & \checkmark &  &  &  \\\hline
        Darii-1* & \checkmark & \checkmark & \checkmark &  &  \\\hline
        Ferio-1* & \checkmark & \checkmark & \checkmark &  &  \\\hline
        Baroko-2* & \checkmark & \checkmark & \checkmark &  & \checkmark \\\Xhline{1pt}%\hline\hline

        Dimatis-4 & \checkmark & \checkmark & \checkmark &  &  \\\hline
        Datisi-3 & \checkmark & \checkmark & \checkmark &  &  \\\hline
        Ferison-3 & \checkmark & \checkmark & \checkmark &  &  \\\hline
        Festino-2* & \checkmark & \checkmark & \checkmark & \checkmark &  \\\hline
        Fresison-4 & \checkmark & \checkmark & \checkmark & \checkmark &  \\\hline
        Fesapo-4 & \checkmark & \checkmark &  & \checkmark &  \\\hline
        Cesare-2 & \checkmark &  &  & \checkmark &  \\\hline
        Cesaro-2 & \checkmark &  &  & \checkmark &  \\\hline
        Calemes-4 & \checkmark &  &  & \checkmark &  \\\hline
        Calemos-4 & \checkmark &  &  & \checkmark &  \\\hline
    \end{tabular}
    \begin{tablenotes}
        \footnotesize
        \item[*] The subject inhabitation (LC6) lemma, convenient for the proof of this mood-figure pair, requires the axioms of predicative inhabitation (LC3) and commutativity (LC1).
        \hrule
    \end{tablenotes}
    \end{threeparttable}
    \caption{Syllogism-axiom matrix for the Leibniz-Cayley system.}
    \label{tab:LC_matrix}
\end{table}

Notice that the proof of every mood requires LC2. In addition, the proof of every mood having a particular premise requires LC3.

LC5 is not needed for proving syllogisms in the ``basic'' set, and is only used where there is an E-conversion that maps certain syllogisms in the ``derived'' set into syllogisms in the ``basic'' set.

It is remarkable that the proof of 14 categorical syllogisms in LC requires only bounded (meet-)semilattice axioms (L1 to L3). Among them, all the 8 affirmative categorical syllogisms --which employ only ``$\termlogicA$'' or ``$\termlogicI$'' relations in their premises and conclusions-- require in their proof neither LC4 nor LC5; for them, the monadic complementation operation ``$'$'' is superfluous. (In contrast, categorical syllogisms which employ ``$\termlogicE$'' or ``$\termlogicO$'' already require complementation in the premise or conclusion, whether their proofs make use of LC4/LC5 or not.)
%LC4/LC5 is often required when we have P2 as $pm=p$ or $pm'=p$. This is due to the fact that, to replace m in P1 ($sR_1m$) by some expression involving p, thus yielding sR_3p, we need m (rather than p) isolated in P2. (In addition to these cases, see also Calemes-4 and Calemos-4.)

Table~\ref{tab:LC_matrix} would have been less parsimonious in the application of axioms had LC adopted subject inhabitation (LC6) as an axiom instead of predicative inhabitation (LC3): 5 categorical syllogisms --Darapti-3, Felapton-3, Disamis-3, Bokardo-3, Fesapo-4-- would require commutativity (LC1).

Many logical facts were \textbf{not} needed to prove all the 24 categorical syllogisms in LC, such as:

\begin{enumerate}[(a)]
    \item ``$\bm{\neq}$'' is the negation of ``$\bm{=}$'': this fact is not used in any proof. Formally, we could have replaced ``$\neq$'' by an arbitrary dyadic relation ``$R$'' which we know no property of -- the proofs would have been the same.
%At no point in the proofs we make use of the fact that $=$ and $\neq$ are mutually complementary dyadic relations. Obversion, conversion and contraposition are reflected in the representations and axioms, but contradiction is not: it is superfluous for the proofs.

    \item The representation of universal categorical assertions in terms of ``$\bm{\varnothing}$'':\\$b\termlogicA c \;\;\metale\;\;\; bc'\!=\!\bm{\varnothing};\qquad b\termlogicE c \;\;\metale\;\;\; bc\!=\!\bm{\varnothing}$.

    \item The symbolization and characterization of the properties of the universe class $\mathbf{I}$.

    \item Involution of complementation: $(b')'=b$.

    \item Disjointness of complements: $bb'=\bm{\varnothing}$\footnote{This is the algebraic form of the ``law of thought'' known as non-contradiction (Section~\ref{sec:algebraic_and_relational_representations}). The fact that a supposedly fundamental ``law of thought'' is superfluous for proving classic categorical syllogisms cannot escape our attention.}.

    \item Idempotence of combination/intersection: $bb=b$\footnote{This is remarkable. \citet[p.~49]{boole:1854} claims this is \textbf{the} fundamental law of thought. This is the special law that distinguishes his algebra of logic (subordinated to numerical algebra with 0 and 1 only) from numerical algebra over $\mathbb{N}$ or $\mathbb{Z}$. He even derives non-contradiction --widely held by many logicians up to the 19th century to be one of the fundamental laws of thought (see Section~\ref{sec:algebraic_and_relational_representations}), but also unnecessary in categorical syllogistic-- from it. Nevertheless, we have shown here that idempotence is a superfluous law for proving classic categorical syllogisms.}.

    \item The operation ``$\cup$'' (union of two classes) and its properties.
\end{enumerate}

Interestingly, the two operations adopted in LC, intersection and complementation, form together a functionally complete set of Boolean operations.

\section{McColl-Ladd (ML) system}
\label{sec:ML}
%Ladd (1889) eulogizes McColl - pp.~562-563: https://www.jstor.org/stable/1411857?seq=20
%Ladd (1916) laments Couturat's death - p.~721: https://www.jstor.org/stable/2012321?seq=7

\subsection{McColl-Ladd system representations}

%With respect to the linguistic classification adopted by Table~\ref{tab:extended_syllogistic_relations}, the
The McColl-Ladd system makes use only of symbols that represent particular or universal relations that are \emph{affirmative} for both the subject and the predicative (``$\bm{\subseteq}$'' and ``$\bm{\Cap}$'')\footnote{It would be fair to argue that this system is relational, instead of algebraic in a strict sense, since its object of study is a relational structure, not an algebraic structure with operations/functions only. In general, the signature of a mathematical structure can include special values and/or operations/functions and/or relations, and our position is that we consider their study algebraic in a wider sense. The study of the interplay between operations and relations is not uncommon in algebra, for instance, in lattice theory.}. Each fundamental categorical relation is represented in ML as follows:%https://en.wikipedia.org/wiki/Structure_(mathematical_logic) https://en.wikipedia.org/wiki/Signature_(logic)

\begin{tabular}{LlC}
\mathbf{A} & \makecell[l]{Every \textbf{b} is \textbf{c}.\\} & b\subseteq c\\\hline
\mathbf{E} & \makecell[l]{No \textbf{b} is \textbf{c}.\\(Every \textbf{b} is not-\textbf{c}.)} & b\subseteq c'\\\hline
\mathbf{I} & \makecell[l]{At least one \textbf{b} is \textbf{c}.\\} & b\Cap c\\\hline
\mathbf{O} & \makecell[l]{At least one \textbf{b} is not \textbf{c}.\\(At-least-one \textbf{b} is not-\textbf{c}.)} & b\Cap c'\\\hline
\mathbf{*} & \makecell[l]{At least one \textbf{b} exists.\\(At least one \textbf{b} is \textbf{b}.)} & b\Cap b
\end{tabular}

The symbolic representations of universal categorical assertions are by \citet[p.~181]{mccoll:1877} and reproduced by \citet[p.~24]{ladd:1883}, whereas the representations of the particular ones are by \citet[p.~26]{ladd:1883}. In addition, \citet[p.~29]{ladd:1883} employs ``$b\Cap \mathbf{I}$'' for ``*''; we employ instead the equivalent assertion ``$b\Cap b$'' for economy of concepts -- we are not strictly required to postulate a universe class, and the McColl-Ladd system as we present it is saved from an extra axiom ``$b\Cap \mathbf{I} \metale b\Cap b$''.

\subsection{McColl-Ladd system axioms}

In order to prove all the classic categorical syllogisms, we have selected the following axioms to form the McColl-Ladd system\footnote{\citeauthor{mccoll:1877}'s original system (\citeyear{mccoll:1877}) --which represented categorical assertions by means of the relations ``$\subseteq$'' and ``$\nsubseteq$''-- adopted the following laws: ML2 (p.~177, rule~11); ML3 (p.~181); ML4 (p.~180, rule~15); ``$b\nsubseteq c \metale c'\nsubseteq b'$'' (p.~180, rule~16) rather than ML1; Bokardo-3 --``$b\subseteq c,\ b\nsubseteq d \metale c\nsubseteq d$'' (p.~180, rule~17)-- rather than ML5.}:
%; and $(b')'=b$ --which he implicitly employs (pp.~178, rule 11; 180; 181; 182)-- rather than ML3

\begin{tabular}{lLl}
\textbf{(ML1)} & b\Cap c \metaeq c\Cap b & \{I-conversion\}\\
\textbf{(ML2)} & b\subseteq c \metaeq c'\subseteq b' & \{A-transposition\footnote{In the representation of categorical relations by means of single vowels, it corresponds to ``A-contraposition'' ($\mathbf{b}\termlogicA \mathbf{c} \metaeq \mathbf{c'}\termlogicA \mathbf{b'}$).}\}\\
\textbf{(ML3)} & b\subseteq c' \metaeq c\subseteq b' & \{E-transposition\footnote{In the representation of categorical relations by means of single vowels, it corresponds to ``E-conversion'' ($\mathbf{b}\termlogicE \mathbf{c} \metaeq \mathbf{c}\termlogicE \mathbf{b}$).\\\indent\citet[p.~399, point~200]{leibniz:1686a} also stated (in intensional/contravariant language) both ML3 and ML2, with the explicit \emph{algebraic} employment of the term negation (class complement) operation.}\}\\
\textbf{(ML4)} & b\subseteq c,\ c\subseteq d \;\metale\;\; b\subseteq d & \{Barbara-1\footnote{It symbolically represents transitivity of subclasshood.}\}\\
\textbf{(ML5)} & b\Cap c,\ \;c\subseteq d \;\metale\;\; b\Cap d & \{Darii-1\footnote{It symbolically represents covariance/monotonicity of conjointness: if two classes are conjoint, then one of them is conjoint with any superclass of the other.}\}\\
\end{tabular}
%Axioms: elimination, and swap [/ flip / double shift --a kind of rearrangement--] to prepare for elimination
%In mathematics, there is swap with relation change vs. without relation change (and compensate on terms/expressions):
%https://www.mathsisfun.com/algebra/inequality-solving.html#:~:text=swap%20the%20left%20and%20right
%https://en.wikipedia.org/wiki/Converse_(logic)

The following is a convenient lemma to shorten the proofs of some valid mood/figure pairs:

\begin{tabular}{lLl}
\textbf{(ML6)} & c\Cap b,\ \;c\subseteq d \;\metale\; d\Cap b & \quad\{Disamis-3\}\\
\end{tabular}

\begin{proof}

\begin{tabular}{lCl}
\textbf{S1.} & b\Cap d \metaeq d\Cap b & \begin{tabular}{@{}c}\{(ML1): $b\Cap c \metaeq c\Cap b$\}\\{$c\leftarrow d$}\end{tabular}\\
\textbf{S2.} & c\Cap b,\ \;c\subseteq d \;\metale\;\; b\Cap d & (ML5), \{(ML1): $b\Cap c \metaeq c\Cap b$\}\\
\textbf{S3.} & c\Cap b,\ \;c\subseteq d \;\metale\;\; d\Cap b & (S2), (S1)
\end{tabular}

\end{proof}

These ML axioms form the subset\footnote{A-transposition, taken here as an axiom, is a consequence of the axioms of E-transposition and involution of complementation from the mentioned sources.} of the axioms presented by \citet[p.~21, Figure~9]{moss:2007}(\citeyear[p.~31, Figure~3.4]{moss:2010})(\citeyear[p.~181, Figure~11.1]{moss:2011}) --and reused by \citet[p.~3]{hemann:2015}-- that is needed for proving all the classic categorical syllogisms. In addition, \citet[pp.~7-8]{reichenbach:1952} informally justifies why Barbara-1 and Darii-1 are the ``primitive'' categorical syllogisms which the other 22 ones are reducible to when the the four classic categorical relations are expressed in terms of A/$\subseteq$ and I/$\Cap$ by obversion.%See also:
%pp. 1,40,43
%Larry Moss. Basic Syllogistic Logics. In: ESSLLI 2010. http://www.indiana.edu/~iulg/moss/unit1.pdf
%pp. 22,13,10,19-20
%https://iulg.sitehost.iu.edu/moss/nasslli/logicslogics.pdf

The comma (``,'') is typically interpreted as the metalogical ``\emph{and}'' operator. The relational character of the ML system is enhanced if, in the ML4 and ML5 axioms, we reinterpret the ``,'' as the operator for composition of relations (from relation algebra) instead\footnote{\citet[pp.~331,355]{de_morgan:1860} is to be credited for noticing that the deduction of the conclusion in a syllogism can be seen as an application of composition of relations, the premises.}.
%See also:
%pp.~199,214
%https://archive.org/details/transactionsofca10camb/page/173/mode/1up
%"Deductive proof, or deduction, or proof a priori, is the *composition* of separate previous propositions, from which the same unavoidable necessity follows."
Intriguingly, the metalogical ``\emph{and}'' is the operator for a commutative operation, whereas composition of relations is not necessarily commutative. On the other hand, the fact that ML4 and ML5 are ``composition-friendly'', with the middle term occupying the position of a ``bridge'' between two ``endpoints'', is perhaps a reason why syllogism moods from the first figure were seen by Aristotle as ``perfect'' \citep[pp.~50-59]{patzig:1968}\citep[book~IV, chapter~17, §§~4 and 8, pp.~405--413]{locke:1700}\citep[p.~217]{de_morgan:1858}\citep{lorenzen:1957}\footnote{Once we recognize all the categorical relations in \citeauthor{de_morgan:1846}'s syllogistic, any categorical syllogism can be reduced into a first-figure syllogism by applying the conversion operation, e.g. $\textbf{p}\termlogicA\textbf{m} \metaeq \textbf{m}\termlogicAumlaut\textbf{p}$, as $\breve{\termlogicA}=\termlogicAumlaut$.}.

Notice that ML didn't adopts axioms of subalternation:

%$\textbf{b}\subseteq\textbf{c} \quad\metaeq\quad \textbf{b}\Cap\textbf{c}$.
$b\subseteq c \quad\metaeq\quad b\Cap c$

$b\subseteq c' \ \ \metaeq\quad b\Cap c'$.
%nor an axiom of existential import for universal categorical assertions:
%$\textbf{b}\subseteq\textbf{c} \quad\metaeq\quad \textbf{b}\Cap\textbf{b}$.

The lack of this axiom makes ML fully compatible to Boolean algebra, as we will prove in Section~\ref{sec:connection_between_systems}. However, due to this intentional omission, 9 of the 24 classic categorical syllogisms require not just two but three premises for them to be valid in this system, just like in the Euler system (Section~\ref{sec:euler}) and in LC (Section~\ref{sec:LC}).

ML can straightforwardly derive the subalternation laws of traditional Aristotelic logic from Darii-1 (ML5) if the subject term is inhabited:

$b\Cap b,\ b\subseteq c \quad\metaeq\quad b\Cap c$

$b\Cap b,\ b\subseteq c' \ \ \metaeq\quad b\Cap c'$.

As ML is a relational system, it is closer in spirit to the original Aristotelic syllogistic \citep{owen:1853a}(\citeyear{owen:1853b})\footnote{Though one important departure is that, in ML, intermediate proof steps which are the result of A-transpositions are not directly expressed as traditional Aristotelic relations $\{\termlogicA, \termlogicE, \termlogicI, \termlogicO\}$ with only positive subjects, namely, ``Every p is m. $\!\metale$ Every non-m is non-p'' in Camestres-2, Camestros-2 and Baroko-2, as we will see in Section~\ref{subsec:ML_proofs}.} than LC --an algebraic system-- is. The following categorical syllogism proofs --none of which has more than 3 steps-- reinforce this point.
%c belongs-to-all b.
%c does-not-belong-to-some b.
%c belongs-to-some b.
%c does-not-belong-to-all b.

%c is-said-of-all b.
%c is-not-said-of-some b.
%c is-said-of-some b.
%c is-not-said-of-some b.

%[https://usercontent.one/wp/denlillemann.no/wp-content/uploads/2021/06/ON_SYLLOGISMS_Prior_Analytics_A_TRANSLAT.pdf#page=19]

\subsection{Syllogism proofs in the McColl-Ladd system}
\label{subsec:ML_proofs}

Here are the proofs of categorical syllogisms in ML.

\begin{multicols}{2}
\begin{syl}
\textbf{Barbara-1:}\\
\scriptsize
\begin{tabular}{N @{\hspace{1.2\tabcolsep}} L @{\hspace{1.2\tabcolsep}} l}
P1. & s\subseteq m & \{Every s is m.\}\\
P2. & m\subseteq p & \{Every m is p.\}\\
\cline{1-2}
C1. & s\subseteq p & \{Every s is p.\}\\
\multicolumn{3}{l}{\textsc{~Proof.}}\\
S3. & \mathbf{s\subseteq p} & \begin{tabular}{@{}l@{}}~\enskip(P1), (P2),\\\begin{tabular}{@{}c@{}}\{$b\subseteq c, c\subseteq d \metale b\subseteq d$\}\\$b\leftarrow s, c\leftarrow m, d\leftarrow p$.\end{tabular}\\Therefore: (C1).\end{tabular}
\end{tabular}
%
%\color{white}\qedsymbol
\end{syl}
%\hrule width0.7\textwidth
\makebox[0.5\textwidth][c]{\textbf{*~*~*}}

\begin{syl}
\textbf{Barbari-1:}\\
\scriptsize
\begin{tabular}{N @{\hspace{1.2\tabcolsep}} L @{\hspace{1.2\tabcolsep}} l}
P1. & s\subseteq m & \{Every s is m.\}\\
P2. & m\subseteq p & \{Every m is p.\}\\
P3. & s\Cap s & \{At least one s exists.\}\\
\cline{1-2}
C1. & s\subseteq p & \{Every s is p.\}\\
C2. & s\Cap p & \{At least one s is p.\}\\
\multicolumn{3}{l}{\textsc{~Proof.}}\\
S4. & \mathbf{s\subseteq p} & \begin{tabular}{@{}l@{}}~\enskip(P1), (P2),\\\begin{tabular}{@{}c@{}}\{$b\subseteq c, c\subseteq d \metale b\subseteq d$\}\\$b\leftarrow s, c\leftarrow m, d\leftarrow p$.\end{tabular}\\Therefore: (C1).\end{tabular}\\[16pt]
S5. & \mathbf{s\Cap p} & \begin{tabular}{@{}l@{}}\enskip(P3), (S4),\\\begin{tabular}{@{}c@{}}\{$b\Cap c, c\subseteq d \metale b\Cap d$\}\\$b\leftarrow s, c\leftarrow s, d\leftarrow p$.\end{tabular}\\Therefore: (C2).\end{tabular}
\end{tabular}
%
%\color{white}\qedsymbol
\end{syl}
%\hrule width0.7\textwidth
\makebox[0.5\textwidth][c]{\textbf{*~*~*}}

\columnbreak\begin{syl}
\textbf{Celarent-1:}\\
\scriptsize
\begin{tabular}{N @{\hspace{1.2\tabcolsep}} L @{\hspace{1.2\tabcolsep}} l}
P1. & s\subseteq m & \{Every s is m.\}\\
P2. & m\subseteq p' & \{No m is p.\}\\
\cline{1-2}
C1. & s\subseteq p' & \{No s is p.\}\\
\multicolumn{3}{l}{\textsc{~Proof.}}\\
\multicolumn{3}{l}{(by \citet[p.~194]{lewis:1918}:)}\\
S3. & \mathbf{s\subseteq p'} & \begin{tabular}{@{}l@{}}~\enskip(P1), (P2),\\\begin{tabular}{@{}c@{}}\{$b\subseteq c, c\subseteq d \metale b\subseteq d$\}\\~$b\leftarrow s, c\leftarrow m, d\leftarrow p'$.\end{tabular}\\Therefore: (C1).\end{tabular}
\end{tabular}
%
%\color{white}\qedsymbol
\end{syl}
%\hrule width0.7\textwidth
\makebox[0.5\textwidth][c]{\textbf{*~*~*}}

\begin{syl}
\textbf{Celaront-1:}\\
\scriptsize
\begin{tabular}{N @{\hspace{1.2\tabcolsep}} L @{\hspace{1.2\tabcolsep}} l}
P1. & s\subseteq m & \{Every s is m.\}\\
P2. & m\subseteq p' & \{No m is p.\}\\
P3. & s\Cap s & \{At least one s exists.\}\\
\cline{1-2}
C1. & s\subseteq p' & \{No s is p.\}\\
C2. & s\Cap p' & \{At least one s is not p.\}\\
\multicolumn{3}{l}{\textsc{~Proof.}}\\
S4. & \mathbf{s\subseteq p'} & \begin{tabular}{@{}l@{}}~\enskip(P1), (P2),\\\begin{tabular}{@{}c@{}}\{$b\subseteq c, c\subseteq d \metale b\subseteq d$\}\\~$b\leftarrow s, c\leftarrow m, d\leftarrow p'$.\end{tabular}\\Therefore: (C1).\end{tabular}\\[16pt]
S5. & \mathbf{s\Cap p'} & \begin{tabular}{@{}l@{}}~\enskip(P3), (S4),\\\begin{tabular}{@{}c@{}}\{$b\Cap c, c\subseteq d \metale b\Cap d$\}\\~$b\leftarrow s, c\leftarrow s, d\leftarrow p'$.\end{tabular}\\Therefore: (C2).\end{tabular}
\end{tabular}
%
%\color{white}\qedsymbol
\end{syl}
%\hrule width0.7\textwidth
\makebox[0.5\textwidth][c]{\textbf{*~*~*}}

\columnbreak\begin{syl}
\textbf{Camestres-2:}\\%https://math.stackexchange.com/questions/1149926/prove-that-if-x-%E2%88%89-b-and-a-%E2%8A%86-b-then-x-%E2%88%89-a
\scriptsize
\begin{tabular}{N @{\hspace{1.2\tabcolsep}} L @{\hspace{1.2\tabcolsep}} l}
P1. & s\subseteq m' & \{No s is m.\}\\
P2. & p\subseteq m & \{Every p is m.\}\\
\cline{1-2}
C1. & s\subseteq p' & \{No s is p.\}\\
\multicolumn{3}{l}{\textsc{~Proof.}}\\
S3. & m'\subseteq p' & (P2),~\begin{tabular}{@{}c@{}}\{$b\subseteq c \metaeq c'\subseteq b'$\}\\$b\leftarrow p, c\leftarrow m$\end{tabular}\\[8pt]
S4. & \mathbf{s\subseteq p'} & \begin{tabular}{@{}l@{}}~\enskip(P1), (S3),\\\begin{tabular}{@{}l@{}}\{$b\subseteq c, c\subseteq d \metale b\subseteq d$\}\\~$b\leftarrow s, c\leftarrow m', d\leftarrow p'$.\end{tabular}\\Therefore: (C1).\end{tabular}
\end{tabular}
%
%\color{white}\qedsymbol
\end{syl}
%\hrule width0.7\textwidth
\makebox[0.5\textwidth][c]{\textbf{*~*~*}}

\begin{syl}
\textbf{Camestros-2:}\\
\scriptsize
\begin{tabular}{N @{\hspace{1.2\tabcolsep}} L @{\hspace{1.2\tabcolsep}} l}
P1. & s\subseteq m' & \{No s is m.\}\\
P2. & p\subseteq m & \{Every p is m.\}\\
P3. & s\Cap s & \{At least one s exists.\}\\
\cline{1-2}
C1. & s\subseteq p' & \{No s is p.\}\\
C2. & s\Cap p' & \{At least one s is not p.\}\\
\multicolumn{3}{l}{\textsc{~Proof.}}\\
S4. & m'\subseteq p' & (P2),~\begin{tabular}{@{}c@{}}\{$b\subseteq c \metaeq c'\subseteq b$'\}\\$b\leftarrow p, c\leftarrow m$\end{tabular}\\[8pt]
S5. & \mathbf{s\subseteq p'} & \begin{tabular}{@{}l@{}}~\enskip(P1), (S4),\\\begin{tabular}{@{}l@{}}\{$b\subseteq c, c\subseteq d \metale b\subseteq d$\}\\~$b\leftarrow s, c\leftarrow m', d\leftarrow p'$.\end{tabular}\\Therefore: (C1).\end{tabular}\\[16pt]
S6. & \mathbf{s\Cap p'} & \begin{tabular}{@{}l@{}}~\enskip(P3), (S5),\\\begin{tabular}{@{}c@{}}\{$b\Cap c, c\subseteq d \metale b\Cap d$\}\\~$b\leftarrow s, c\leftarrow s, d\leftarrow p'$.\end{tabular}\\Therefore: (C2).\end{tabular}
\end{tabular}
%
%\color{white}\qedsymbol
\end{syl}
%\hrule width0.7\textwidth
\makebox[0.5\textwidth][c]{\textbf{*~*~*}}

\begin{syl}
\textbf{Bamalip-4:}\\
\scriptsize
\begin{tabular}{N @{\hspace{1.2\tabcolsep}} L @{\hspace{1.2\tabcolsep}} l}
P1. & m\subseteq s & \{Every m is s.\}\\
P2. & p\subseteq m & \{Every p is m.\}\\
P3. & p\Cap p & \{At least one p exists.\}\\
\cline{1-2}
C1. & s\Cap p & \{At least one s is p.\}\\
\multicolumn{3}{l}{\textsc{~Proof.}}\\
S4. & p\subseteq s & \begin{tabular}{@{}l@{}}~\enskip(P2), (P1),\\\begin{tabular}{@{}c@{}}\{$b\subseteq c, c\subseteq d \metale b\subseteq d$\}\\$b\leftarrow p, c\leftarrow m, d\leftarrow s$\end{tabular}\end{tabular}\\[12pt]
S5. & \mathbf{s\Cap p} & \begin{tabular}{@{}l@{}}~\enskip(P3), (S4),\\\begin{tabular}{@{}c@{}}\{$c\Cap b, c\subseteq d \metale d\Cap b$\}\\$c\leftarrow p, b\leftarrow p, d\leftarrow s$.\end{tabular}\\Therefore: (C1).\end{tabular}
\end{tabular}
%
%\color{white}\qedsymbol
\end{syl}
%\hrule width0.7\textwidth
\makebox[0.5\textwidth][c]{\textbf{*~*~*}}

\begin{syl}
\textbf{Darapti-3:}\\
\scriptsize
\begin{tabular}{N @{\hspace{1.2\tabcolsep}} L @{\hspace{1.2\tabcolsep}} l}
P1. & m\subseteq s & \{Every m is s.\}\\
P2. & m\subseteq p & \{Every m is p.\}\\
P3. & m\Cap m & \{At least one m exists.\}\\
\cline{1-2}
C1. & s\Cap p & \{At least one s is p.\}\\
\multicolumn{3}{l}{\textsc{~Proof.}}\\
S4. & m\Cap p & \begin{tabular}{@{}l@{}}~\enskip(P3), (P2),\\\begin{tabular}{@{}l@{}}\{$b\Cap c, c\subseteq d \metale b\Cap d$\}\\~$b\leftarrow m, c\leftarrow m, d\leftarrow p$\end{tabular}\end{tabular}\\[12pt]
S5. & \mathbf{s\Cap p} & \begin{tabular}{@{}l@{}}~\enskip(S4), (P1),\\\begin{tabular}{@{}l@{}}\{$c\Cap b, c\subseteq d \metale d\Cap b$\}\\~$c\leftarrow m, b\leftarrow p, d\leftarrow s$.\end{tabular}\\Therefore: (C1).\end{tabular}
\end{tabular}
%
%\color{white}\qedsymbol
\end{syl}
%\hrule width0.7\textwidth
\makebox[0.5\textwidth][c]{\textbf{*~*~*}}

\begin{syl}
\textbf{Felapton-3:}\\
\scriptsize
\begin{tabular}{N @{\hspace{1.2\tabcolsep}} L @{\hspace{1.2\tabcolsep}} l}
P1. & m\subseteq s & \{Every m is s.\}\\
P2. & m\subseteq p' & \{No m is p.\}\\
P3. & m\Cap m & \{At least one m exists.\}\\
\cline{1-2}
C1. & s\Cap p' & \{At least one s is not p.\}\\
\multicolumn{3}{l}{\textsc{~Proof.}}\\
S4. & m\Cap p' & \begin{tabular}{@{}l@{}}~\enskip(P3), (P2),\\\begin{tabular}{@{}l@{}}\{$b\Cap c, c\subseteq d \metale b\Cap d$\}\\~$b\leftarrow m, c\leftarrow m, d\leftarrow p'$\end{tabular}\end{tabular}\\[12pt]
S5. & \mathbf{s\Cap p'} & \begin{tabular}{@{}l@{}}~\enskip(S4), (P1),\\\begin{tabular}{@{}l@{}}\{$c\Cap b, c\subseteq d \metale d\Cap b$\}\\~$c\leftarrow m, b\leftarrow p', d\leftarrow s$.\end{tabular}\\Therefore: (C1).\end{tabular}
\end{tabular}
%
%\color{white}\qedsymbol
\end{syl}
%\hrule width0.7\textwidth
\makebox[0.5\textwidth][c]{\textbf{*~*~*}}

\begin{syl}
\textbf{Disamis-3:}\\
\scriptsize
\begin{tabular}{N @{\hspace{1.2\tabcolsep}} L @{\hspace{1.2\tabcolsep}} l}
P1. & m\subseteq s & \{Every m is s.\}\\
P2. & m\Cap p & \{At least one m is p.\}\\
\cline{1-2}
C1. & s\Cap p & \{At least one s is p.\}\\
\multicolumn{3}{l}{\textsc{~Proof.}}\\
S3. & \mathbf{s\Cap p} & \begin{tabular}{@{}l@{}}~\enskip(P2), (P1),\\\begin{tabular}{@{}l@{}}\{$c\Cap b, c\subseteq d \metale d\Cap b$\}\\~$c\leftarrow m, b\leftarrow p, d\leftarrow s$.\end{tabular}\\Therefore: (C1).\end{tabular}
\end{tabular}
%
%\color{white}\qedsymbol
\end{syl}
%\hrule width0.7\textwidth
\makebox[0.5\textwidth][c]{\textbf{*~*~*}}

\begin{syl}
\textbf{Bokardo-3:}\\
\scriptsize
\begin{tabular}{N @{\hspace{1.2\tabcolsep}} L @{\hspace{1.2\tabcolsep}} l}
P1. & m\subseteq s & \{Every m is s.\}\\
P2. & m\Cap p' & \{At least one m is not p.\}\\
\cline{1-2}
P3. & s\Cap p' & \{At least one s is not p.\}\\
\multicolumn{3}{l}{\textsc{~Proof.}}\\
S3. & \mathbf{s\Cap p'} & \begin{tabular}{@{}l@{}}~\enskip(P2), (P1),\\\begin{tabular}{@{}l@{}}\{$c\Cap b, c\subseteq d \metale d\Cap b$\}\\~$c\leftarrow m, b\leftarrow p', d\leftarrow s$.\end{tabular}\\Therefore: (C1).\end{tabular}
\end{tabular}
%
%\color{white}\qedsymbol
\end{syl}
%\hrule width0.7\textwidth
\makebox[0.5\textwidth][c]{\textbf{*~*~*}}

\begin{syl}
\textbf{Darii-1:}\\
\scriptsize
\begin{tabular}{N @{\hspace{1.2\tabcolsep}} L @{\hspace{1.2\tabcolsep}} l}
P1. & s\Cap m & \{At least one s is m.\}\\
P2. & m\subseteq p & \{Every m is p.\}\\
\cline{1-2}
C1. & s\Cap p & \{At least one s is p.\}\\
\multicolumn{3}{l}{\textsc{~Proof.}}\\
S3. & \mathbf{s\Cap p} & \begin{tabular}{@{}l@{}}~\enskip(P1), (P2),\\\begin{tabular}{@{}l@{}}\{$b\Cap c, c\subseteq d \metale b\Cap d$\}\\~$b\leftarrow s, c\leftarrow m, d\leftarrow p$.\end{tabular}\\Therefore: (C1).\end{tabular}
\end{tabular}
%
%\color{white}\qedsymbol
\end{syl}
%\hrule width0.7\textwidth
\makebox[0.5\textwidth][c]{\textbf{*~*~*}}

\begin{syl}
\textbf{Ferio-1:}\\
\scriptsize
\begin{tabular}{N @{\hspace{1.2\tabcolsep}} L @{\hspace{1.2\tabcolsep}} l}
P1. & s\Cap m & \{At least one s is m.\}\\
P2. & m\subseteq p' & \{No m is p.\}\\
\cline{1-2}
C1. & s\Cap p' & \{At least one s is not p.\}\\
\multicolumn{3}{l}{\textsc{~Proof.}}\\
S3. & \mathbf{s\Cap p'} & \begin{tabular}{@{}l@{}}~\enskip(P1), (P2),\\\begin{tabular}{@{}l@{}}\{$b\Cap c, c\subseteq d \metale b\Cap d$\}\\~$b\leftarrow s, c\leftarrow m, d\leftarrow p'$.\end{tabular}\\Therefore: (C1).\end{tabular}
\end{tabular}
%
%\color{white}\qedsymbol
\end{syl}
%\hrule width0.7\textwidth
\makebox[0.5\textwidth][c]{\textbf{*~*~*}}

\columnbreak\begin{syl}
\textbf{Baroko-2:}\\
\scriptsize
\begin{tabular}{N @{\hspace{1.2\tabcolsep}} L @{\hspace{1.2\tabcolsep}} l}
P1. & s\Cap m' & \{At least one s is not m.\}\\
P2. & p\subseteq m & \{Every p is m.\}\\
\cline{1-2}
C1. & s\Cap p' & \{At least one s is not p.\}\\
\multicolumn{3}{l}{\textsc{~Proof.}}\\
S3. & m'\subseteq p' & (P2),~\begin{tabular}{@{}c@{}}\{b$\subseteq c \metaeq c'\subseteq b'$\}\\$b\leftarrow p, c\leftarrow m$\end{tabular}\\[8pt]
S4. & \mathbf{s\Cap p'} & \begin{tabular}{@{}l@{}}~\enskip(P1), (S3),\\\begin{tabular}{@{}l@{}}\{$b\Cap c, c\subseteq d \metale b\Cap d$\}\\~$b\leftarrow s, c\leftarrow m', d\leftarrow p'$.\end{tabular}\\Therefore: (C1).\end{tabular}
\end{tabular}
\end{syl}
\makebox[0.5\textwidth][c]{\textbf{*~*~*}}

\begin{syl}
\textbf{Dimatis-4:}\\
\scriptsize
\begin{tabular}{N @{\hspace{1.2\tabcolsep}} L @{\hspace{1.2\tabcolsep}} l}
P1. & m\subseteq s & \{Every m is s.\}\\
P2. & p\Cap m & \{At least one p is m.\}\\
\cline{1-2}
C1. & s\Cap p & \{At least one s is p.\}\\
\multicolumn{3}{l}{\textsc{~Proof.}}\\
S3. & p\Cap s & \begin{tabular}{@{}l@{}}~\enskip(P2), (P1),\\\begin{tabular}{@{}l@{}}\{$b\Cap c, c\subseteq d \metale b\Cap d$\}\\~$b\leftarrow p, c\leftarrow m, d\leftarrow s$\end{tabular}\end{tabular}\\[12pt]
S4. & \mathbf{s\Cap p} & \begin{tabular}{@{}l@{}}(S3),~\begin{tabular}{@{}c@{}}\{$b\Cap c \metaeq c\Cap b$\}\\$b\leftarrow p, c\leftarrow s.$\end{tabular}\\Therefore: (C1).\end{tabular}
%\\[16pt]\multicolumn{3}{l}{\textsc{~Alternative proof.}}\\
%P2a. & m\Cap p & (P2),~\begin{tabular}{@{}c@{}}\{$b\Cap c \metaeq c\Cap b$\}\\$b\leftarrow p, c\leftarrow m$\end{tabular}\\[8pt]
%S3. & \mathbf{s\Cap p} & \begin{tabular}{@{}l@{}}~\enskip(P2a), (P1),\\\begin{tabular}{@{}l@{}}\{$c\Cap b, c\subseteq d \metale d\Cap b$\}\\~$c\leftarrow m, b\leftarrow p, d\leftarrow s$.\end{tabular}\\Therefore: (C1).\end{tabular}
\end{tabular}
\end{syl}
\makebox[0.5\textwidth][c]{\textbf{*~*~*}}

\begin{syl}
\textbf{Datisi-3:}\\
\scriptsize
\begin{tabular}{N @{\hspace{1.2\tabcolsep}} L @{\hspace{1.2\tabcolsep}} l}
P1. & m\Cap s & \{At least one m is s.\}\\
P2. & m\subseteq p & \{Every m is p.\}\\
\cline{1-2}
C1. & s\Cap p & \{At least one s is p.\}\\
\multicolumn{3}{l}{\textsc{~Proof.}}\\
P1a. & s\Cap m & (P1),~\begin{tabular}{@{}c@{}}\{$b\Cap c \metaeq c\Cap b$\}\\$b\leftarrow m, c\leftarrow s$\end{tabular}\\[8pt]
S3. & \mathbf{s\Cap p} & \begin{tabular}{@{}l@{}}~\enskip(P1a), (P2),\\\begin{tabular}{@{}l@{}}\{$b\Cap c, c\subseteq d \metale b\Cap d$\}\\~$b\leftarrow s, c\leftarrow m, d\leftarrow p$.\end{tabular}\\Therefore: (C1).\end{tabular}
\end{tabular}
\end{syl}
\makebox[0.5\textwidth][c]{\textbf{*~*~*}}

\begin{syl}
\textbf{Ferison-3:}\\
\scriptsize
\begin{tabular}{N @{\hspace{1.2\tabcolsep}} L @{\hspace{1.2\tabcolsep}} l}
P1. & m\Cap s & \{At least one m is s.\}\\
P2. & m\subseteq p' & \{No m is p.\}\\
\cline{1-2}
C1. & s\Cap p' & \{At least one s is not p.\}\\
\multicolumn{3}{l}{\textsc{~Proof.}}\\
P1a. & s\Cap m & (P1),~\begin{tabular}{@{}c@{}}\{$b\Cap c \metaeq c\Cap b$\}\\$b\leftarrow m, c\leftarrow s$\end{tabular}\\[8pt]
S3. & \mathbf{s\Cap p'} & \begin{tabular}{@{}l@{}}~\enskip(P1a), (P2),\\\begin{tabular}{@{}l@{}}\{$b\Cap c, c\subseteq d \metale b\Cap d$\}\\~$b\leftarrow s, c\leftarrow m, d\leftarrow p'$.\end{tabular}\\Therefore: (C1).\end{tabular}
\end{tabular}
\end{syl}
\makebox[0.5\textwidth][c]{\textbf{*~*~*}}

\columnbreak\begin{syl}
\textbf{Festino-2:}\\
\scriptsize
\begin{tabular}{N @{\hspace{1.2\tabcolsep}} L @{\hspace{1.2\tabcolsep}} l}
P1. & s\Cap m & \{At least one s is m.\}\\
P2. & p\subseteq m' & \{No p is m.\}\\
\cline{1-2}
C1. & s\Cap p' & \{At least one s is not p.\}\\
\multicolumn{3}{l}{\textsc{~Proof.}}\\
P2a. & m\subseteq p' & (P2),~\begin{tabular}{@{}c@{}}\{$b\subseteq c' \metaeq c\subseteq b'$\}\\$b\leftarrow p, c\leftarrow m$\end{tabular}\\[12pt]
S3. & \mathbf{s\Cap p'} & \begin{tabular}{@{}l@{}}~\enskip(P1), (P2a),\\\begin{tabular}{@{}l@{}}\{$b\Cap c, c\subseteq d \metale b\Cap d$\}\\~$b\leftarrow s, c\leftarrow m, d\leftarrow p'$.\end{tabular}\\Therefore: (C1).\end{tabular}
\end{tabular}
\end{syl}
\makebox[0.5\textwidth][c]{\textbf{*~*~*}}

\begin{syl}
\textbf{Fresison-4:}\\
\scriptsize
\begin{tabular}{N @{\hspace{1.2\tabcolsep}} L @{\hspace{1.2\tabcolsep}} l}
P1. & m\Cap s & \{At least one m is s.\}\\
P2. & p\subseteq m' & \{No p is m.\}\\
\cline{1-2}
C1. & s\Cap p' & \{At least one s is not p.\}\\
\multicolumn{3}{l}{\textsc{~Proof.}}\\
P1a. & s\Cap m & (P1),~\begin{tabular}{@{}c@{}}\{$b\Cap c \metaeq c\Cap b$\}\\$b\leftarrow m, c\leftarrow s$\end{tabular}\\[8pt]
P2a. & m\subseteq p' & (P2),~\begin{tabular}{@{}c@{}}\{$b\subseteq c' \metaeq c\subseteq b'$\}\\$b\leftarrow p, c\leftarrow m$\end{tabular}\\[12pt]
S3. & \mathbf{s\Cap p'} & \begin{tabular}{@{}l@{}}~\enskip(P1a), (P2a),\\\begin{tabular}{@{}l@{}}\{$b\Cap c, c\subseteq d \metale b\Cap d$\}\\~$b\leftarrow s, c\leftarrow m, d\leftarrow p'$.\end{tabular}\\Therefore: (C1).\end{tabular}
\end{tabular}
\end{syl}%Among all the 2-premise categorical syllogisms, only Fresison-4 required 3 proof steps.
\makebox[0.5\textwidth][c]{\textbf{*~*~*}}

\begin{syl}
\textbf{Fesapo-4:}\\
\scriptsize
\begin{tabular}{N @{\hspace{1.2\tabcolsep}} L @{\hspace{1.2\tabcolsep}} l}
P1. & m\subseteq s & \{Every m is s.\}\\
P2. & p\subseteq m' & \{No p is m.\}\\
P3. & m\Cap m & \{At least one m exists.\}\\
\cline{1-2}
C1. & s\Cap p' & \{At least one s is not p.\}\\
\multicolumn{3}{l}{\textsc{~Proof.}}\\
P2a. & m\subseteq p' & (P2),~\begin{tabular}{@{}c@{}}\{$b\subseteq c' \metaeq c\subseteq b'$\}\\$b\leftarrow p, c\leftarrow m$\end{tabular}\\[12pt]
S4. & m\Cap p' & \begin{tabular}{@{}l@{}}~\enskip(P3), (P2a),\\\begin{tabular}{@{}l@{}}\{$b\Cap c, c\subseteq d \metale b\Cap d$\}\\~$b\leftarrow m, c\leftarrow m, d\leftarrow p'$\end{tabular}\end{tabular}\\[16pt]
S5. & \mathbf{s\Cap p'} & \begin{tabular}{@{}l@{}}~\enskip(S4), (P1),\\\begin{tabular}{@{}l@{}}\{$c\Cap b, c\subseteq d \metale d\Cap b$\}\\~$c\leftarrow m, b\leftarrow p', d\leftarrow s$.\end{tabular}\\Therefore: (C1).\end{tabular}
\end{tabular}
\end{syl}
\makebox[0.5\textwidth][c]{\textbf{*~*~*}}

\begin{syl}
\textbf{Cesare-2:}\\
\scriptsize
\begin{tabular}{N @{\hspace{1.2\tabcolsep}} L @{\hspace{1.2\tabcolsep}} l}
P1. & s\subseteq m & \{Every s is m.\}\\
P2. & p\subseteq m' & \{No p is m.\}\\
\cline{1-2}
C1. & s\subseteq p' & \{No s is p.\}\\
\multicolumn{3}{l}{\textsc{~Proof.}}\\
P2a. & m\subseteq p' & (P2),~\begin{tabular}{@{}c@{}}\{$b\subseteq c' \metaeq c\subseteq b'$\}\\$b\leftarrow p, c\leftarrow m$\end{tabular}\\[8pt]
S3. & \mathbf{s\subseteq p'} & \begin{tabular}{@{}l@{}}~\enskip(P1), (P2a),\\\begin{tabular}{@{}c@{}}\{$b\subseteq c, c\subseteq d \metale b\subseteq d$\}\\~$b\leftarrow s, c\leftarrow m, d\leftarrow p'$.\end{tabular}\\Therefore: (C1).\end{tabular}
\end{tabular}
\end{syl}
\makebox[0.5\textwidth][c]{\textbf{*~*~*}}

\columnbreak\begin{syl}
\textbf{Cesaro-2:}\\
\scriptsize
\begin{tabular}{N @{\hspace{1.2\tabcolsep}} L @{\hspace{1.2\tabcolsep}} l}
P1. & s\subseteq m & \{Every s is m.\}\\
P2. & p\subseteq m' & \{No p is m.\}\\
P3. & s\Cap s & \{At least one s exists.\}\\
\cline{1-2}
C1. & s\subseteq p' & \{No s is p.\}\\
C2. & s\Cap p' & \{At least one s is not p.\}\\
\multicolumn{3}{l}{\textsc{~Proof.}}\\
P2a. & m\subseteq p' & (P2),~\begin{tabular}{@{}c@{}}\{$b\subseteq c' \metaeq c\subseteq b'$\}\\$b\leftarrow p, c\leftarrow m$\end{tabular}\\[8pt]
S4. & \mathbf{s\subseteq p'} & \begin{tabular}{@{}l@{}}~\enskip(P1), (P2a),\\\begin{tabular}{@{}c@{}}\{$b\subseteq c, c\subseteq d \metale b\subseteq d$\}\\~$b\leftarrow s, c\leftarrow m, d\leftarrow p'$.\end{tabular}\\Therefore: (C1).\end{tabular}\\[16pt]
S5. & \mathbf{s\Cap p'} & \begin{tabular}{@{}l@{}}~\enskip(P3), (S4),\\\begin{tabular}{@{}c@{}}\{$b\Cap c, c\subseteq d \metale b\Cap d$\}\\~$b\leftarrow s, c\leftarrow s, d\leftarrow p'$.\end{tabular}\\Therefore: (C2).\end{tabular}
\end{tabular}
\end{syl}
\makebox[0.5\textwidth][c]{\textbf{*~*~*}}

\columnbreak\begin{syl}
\textbf{Calemes-4:}\\
\scriptsize
\begin{tabular}{N @{\hspace{1.2\tabcolsep}} L @{\hspace{1.2\tabcolsep}} l}
P1. & m\subseteq s' & \{No m is s.\}\\
P2. & p\subseteq m & \{Every p is m.\}\\
\cline{1-2}
C1. & s\subseteq p' & \{No s is p.\}\\
\multicolumn{3}{l}{\textsc{~Proof.}}\\
\multicolumn{3}{l}{(by \citet[p.~194]{lewis:1918}:)}\\
S3. & p\subseteq s' & \begin{tabular}{@{}l@{}}~\enskip(P2), (P1),\\\begin{tabular}{@{}l@{}}\{$b\subseteq c, c\subseteq d \metale b\subseteq d$\}\\~$b\leftarrow p, c\leftarrow m, d\leftarrow s'$\end{tabular}\end{tabular}\\[12pt]
S4. & \mathbf{s\subseteq p'} & \begin{tabular}{@{}l@{}}(S3), \begin{tabular}{@{}c@{}}\{$b\subseteq c' \metaeq c\subseteq b'$\}\\$b\leftarrow p, c\leftarrow s$.\end{tabular}\\Therefore: (C1).\end{tabular}
\end{tabular}
\end{syl}
\begin{center}\textbf{*~*~*}\end{center}

\begin{syl}
\textbf{Calemos-4:}\\
\scriptsize
\begin{tabular}{N @{\hspace{1.2\tabcolsep}} L @{\hspace{1.2\tabcolsep}} l}
P1. & m\subseteq s' & \{No m is s.\}\\
P2. & p\subseteq m & \{Every p is m.\}\\
P3. & s\Cap s & \{At least one s exists.\}\\
\cline{1-2}
C1. & s\subseteq p' & \{No s is p.\}\\
C2. & s\Cap p' & \{At least one s is not p.\}\\
\multicolumn{3}{l}{\textsc{~Proof.}}\\
S4. & p\subseteq s' & \begin{tabular}{@{}l@{}}~\enskip(P2), (P1),\\\begin{tabular}{@{}l@{}}\{$b\subseteq c, c\subseteq d \metale b\subseteq d$\}\\~$b\leftarrow p, c\leftarrow m, d\leftarrow s'$\end{tabular}\end{tabular}\\[12pt]
S5. & \mathbf{s\subseteq p'} & \begin{tabular}{@{}l@{}}(S4), \begin{tabular}{@{}c@{}}\{$b\subseteq c' \metaeq c\subseteq b'$\}\\$b\leftarrow p, c\leftarrow s$.\end{tabular}\\Therefore: (C1).\end{tabular}\\[12pt]
S6. & \mathbf{s\Cap p'} & \begin{tabular}{@{}l@{}}~\enskip(P3), (S5),\\\begin{tabular}{@{}c@{}}\{$b\Cap c, c\subseteq d \metale b\Cap d$\}\\~$b\leftarrow s, c\leftarrow s, d\leftarrow p'$.\end{tabular}\\Therefore: (C2).\end{tabular}
\end{tabular}
\end{syl}

\begin{flushright}\qedsymbol\end{flushright}

\end{multicols}

Thus, we have shown that the ML axioms are sufficiently powerful for proving all 24 classic categorical syllogisms.

%\subsection{Syllogism-axiom matrix for ML and recap of the fundamental ingredients}
\subsection{Syllogism-axiom matrix for ML}
\label{subsec:ML_matrix}

Table~\ref{tab:ML_matrix} summarizes the axioms that have been adopted in the proof of each valid categorical syllogism in the preceding subsection.

\begin{table}[ht]
    \centering
    \begin{threeparttable}
    \footnotesize
    \begin{tabular}{|m{0.12\textwidth}|M{0.15\textwidth}|M{0.15\textwidth}|M{0.15\textwidth}|@{}M{0.15\textwidth}@{}|M{0.15\textwidth}|}
        \hline
        Syllogism & (ML4)\linebreak Barbara-1 & (ML5)\linebreak Darii-1 & (ML1)\linebreak ~I-conversion & (ML3)\linebreak E-transposition & (ML2)\linebreak A-transposition\\\hline
        Barbara-1 & \checkmark &  &  &  &  \\\hline 
        Barbari-1 & \checkmark & \checkmark &  &  &  \\\hline
        Celarent-1 & \checkmark &  &  &  &  \\\hline
        Celaront-1 & \checkmark & \checkmark &  &  &  \\\hline
        Camestres-2 & \checkmark &  &  &  & \checkmark \\\hline
        Camestros-2 & \checkmark & \checkmark &  &  & \checkmark \\\hline
        Bamalip-4* & \checkmark & \checkmark & \checkmark &  &  \\\hline
        Darapti-3* &  & \checkmark & \checkmark &  &  \\\hline
        Felapton-3* &  & \checkmark & \checkmark &  &  \\\hline
        Disamis-3* &  & \checkmark & \checkmark &  &  \\\hline
        Bokardo-3* &  & \checkmark & \checkmark &  &  \\\hline
        Darii-1 &  & \checkmark &  &  &  \\\hline
        Ferio-1 &  & \checkmark &  &  &  \\\hline
        Baroko-2 &  & \checkmark &  &  & \checkmark \\\Xhline{1pt}
        Dimatis-4 &  & \checkmark & \checkmark &  &  \\\hline
        Datisi-3 &  & \checkmark & \checkmark &  &  \\\hline
        Ferison-3 &  & \checkmark & \checkmark &  &  \\\hline
        Festino-2 &  & \checkmark &  & \checkmark &   \\\hline
        Fresison-4 &  & \checkmark & \checkmark & \checkmark &  \\\hline
        Fesapo-4* &  & \checkmark & \checkmark & \checkmark &  \\\hline
        Cesare-2 & \checkmark &  &  & \checkmark &  \\\hline
        Cesaro-2 & \checkmark & \checkmark &  & \checkmark &  \\\hline
        Calemes-4 & \checkmark &  &  & \checkmark &  \\\hline
        Calemos-4 & \checkmark & \checkmark &  & \checkmark &  \\\hline
    \end{tabular}
    \begin{tablenotes}
        \footnotesize
        \item[*] The Disamis-3 (ML6) lemma, convenient for the proof of this mood-figure pair, requires the axioms Darii-1 (ML5) and I-conversion (ML1).
        \hrule
    \end{tablenotes}
    \end{threeparttable}
    \caption{Syllogism-axiom matrix for the McColl-Ladd system.}
    \label{tab:ML_matrix}
\end{table}

Notice that the proof of every mood requires ML4 (when there is a universal conclusion) or ML5 (when there is a particular conclusion). If the proof of a mood requires both ML4 and ML5, then the mood has more than two premises (not assuming existential import for universal assertions). Therefore, the proof of every mood having exactly two premises requires either ML4 or ML5. In other words, for our proofs of 2-premise moods, ML4 and ML5 are mutually exclusive and collectively exhaustive.

If ML1 is required in the proof of a mood, then ML5 is also required.

ML3 is not needed for proving syllogisms in the ``basic'' set, and is only used where there is an E-conversion that maps certain syllogisms in the ``derived'' set into syllogisms in the ``basic'' set.

All the 8 affirmative categorical syllogisms, which employ only ``$\termlogicA$'' or ``$\termlogicI$'' relations in their premises and conclusions, require in their proof neither ML2 nor ML3; for them, the monadic complementation operation ``$'$'' is superfluous. (In contrast, categorical syllogisms which employ ``$\termlogicE$'' or ``$\termlogicO$'' already require complementation in the premise or conclusion, whether their proofs make use of ML2/ML3 or not.)

Column ML3 has the same mood-figure pairs as column LC5 (Section~\ref{subsec:LC_matrix}), which corresponds to the same law expressed in a different form; likewise, column ML2 and column LC4 perfectly coincide.

%To recap, proving all 24 classic categorical syllogisms in ML requires discovering and assembling together \emph{\textbf{all}} the following philosophical and technical puzzle pieces:

%\begin{enumerate}
%    \item \textbf{Pure extensionality}, like in LC (see Section \ref{subsec:LC_matrix}).

%    \item \textbf{Lack of existential import} for universal categorical assertions\footnote{\citet{mccoll:1877} originally embraced Boethius' connexive thesis\footref{fn:boethius_connexive_thesis} in order to prove with only 2 premises all the classic categorical syllogisms which, in our ML system, require an additional existential premise.}.

%    \item \textbf{The dyadic relation ``$\bm{\subseteq}$''}, by \citet{mccoll:1877}.

%    \item \textbf{The dyadic relation ``$\bm{\Cap}$''}\footnote{McColl's original system (\citeyear[p.~180, def.~13]{mccoll:1877}) adopted ``$\nsubseteq$'' instead, and it worked just fine.}, by \citet{ladd:1883}.

%    \item \textbf{The adoption of (some of the) axioms stated by \citet{moss:2007}}, as an alternative to the axioms stated for the pure \citet{mccoll:1877} system.

%    \item \textbf{The replacement and inference rules} -- see Section~\ref{sec:inference_rules}.

%    \item \textbf{The step-by-step proofs} provided in Section~\ref{subsec:ML_proofs}, most of which are ours.
%\end{enumerate}

Many logical facts were \textbf{not} needed to prove all the 24 categorical syllogisms in ML, such as:

\begin{enumerate}[(a)]
    \item The postulation and symbolization of the empty class ($\bm{\varnothing}$) and its properties:\\$\bm{\varnothing}\subseteq\mathbf{b},\ \bm{\varnothing}\!\bm{\not}\!\Cap\,\mathbf{b}$ for any $\mathbf{b}$.

    \item The symbolization and characterization of the properties of the universe class $\mathbf{I}$.

    \item Reflexivity and antisymmetry of ``$\subseteq$''.

    \item How ``$\subseteq$'' is connected to ``$\Cap$'':\qquad$\mathbf{b}\subseteq\mathbf{c} \;\metaeq\; \mathbf{b}\!\bm{\not}\!\Cap\,\mathbf{c}';\qquad\mathbf{b}\Cap\mathbf{c} \;\metaeq\; \mathbf{b}\nsubseteq\mathbf{c}'$.

    \item The term inhabitation law (which is neither an axiom nor a theorem in ML):\\$\mathbf{b}\Cap\mathbf{c} \quad\metale\quad \mathbf{b}\Cap\mathbf{b}$.

    \item Laws involving the term combination (or class intersection) operation, such as:\\$\mathbf{b}\mathbf{c}\subseteq\mathbf{b}$;\qquad$\mathbf{b}\subseteq\mathbf{c} \;\metale\; \mathbf{b}\mathbf{d}\subseteq\mathbf{c}\mathbf{d}$\footnote{Stated by \citet[p.~178, Rule~12]{mccoll:1877}.};\qquad$\mathbf{b}\subseteq\mathbf{c},\ \mathbf{b}\subseteq\mathbf{d} \;\metale\;\; \mathbf{b}\subseteq\mathbf{c}\mathbf{d}$.

    \item Involution of complementation: $(b')'=b$.

    \item The 4 misnamed ``laws of thought'' (Section~\ref{sec:algebraic_and_relational_representations})\footnote{Aristotle himself recognized that the non-contradiction law is superfluous for proving the classic categorical syllogisms:

\vspace{-\topsep}\begin{quote}\footnotesize\guillemotleft The law that it is impossible to affirm and deny simultaneously the same predicate of the same subject is not expressly posited by any demonstration except when the conclusion also has to be expressed in that form [...].\guillemotright\citep[77a10]{mure:1926}\vspace{-\topsep}\end{quote}}.
\end{enumerate}

\section{Connection between the symbolic axiomatic systems}
\label{sec:connection_between_systems}
%https://en.wikipedia.org/wiki/Metatheorem
%https://mathworld.wolfram.com/Metatheorem.html 

\vspace{-10pt}An important characteristic of the symbolic representations selected for the fundamental relations in LC and ML is that they make obverse relations evident (the pairs A/E and I/O), but curiously they don't make contradictory relations evident (the pairs A/O and E/I). In contrast, for instance, in a pure Cayley system (with $=$/$\neq \bm{\varnothing}$), in a pure Ladd system (with $\Cap$/$\bm{\not}\!\Cap$), or in a pure McColl system (with $\subseteq$/$\nsubseteq$), contradictory pairs would also be made evident. Thus we have shown that the information about contradictory relations is superfluous to prove the 24 classic categorical syllogisms -- an information-poorer context is sufficient for that.

How can we compare the expressive power of the ML and LC systems? This is what we show in the next subsections.

\vspace{-15pt}\subsection{Deriving ML from LC}

Let's study the relation between the McColl-Ladd and Leibniz-Cayley algebraic axiomatic systems. Are these two axiomatic systems capable of proving exactly the same theorems?

These systems adopt very different representations, so we need definitions to bridge them. Let's adopt the following definitions\footnote{Intuitively justifiable by Venn and Euler diagrams.}:

\begin{tabular}{lL}
\textbf{(D1)} & b\subseteq c \:\metaeq\; bc=\:b\\
\textbf{(D2)} & b\,\Cap c \;\metaeq\; bc\neq\bm{\varnothing}
\end{tabular}

These definitions are enough to almost perfectly\footnote{When mapping the ML representation for ``At least one \textbf{b} exists.'' to the LC representation which we have chosen in Section~\ref{subsec:LC_representations} ($b\neq\bm{\varnothing}$), we have to make use of definition D2 and the idempotence law from Boolean algebra ($bb=b$). This is a minor detail; had we represented ``At least one \textbf{b} exists.'' in LC by $bb\neq\bm{\varnothing}$ rather than $b\neq\bm{\varnothing}$ --with the corresponding minor adaptation to the LC3 and LC6 laws--, the proofs in LC would have remained the same, and there would be no need of assuming the idempotence law, at the (small) cost of the loss of a more direct, more intuitive justification for the alternative LC representation ($bb\neq\bm{\varnothing}$).} derive the mapping from the McColl-Ladd system representation to the Leibniz-Cayley system representation:

\begin{tabular}{LlCC}
 & & \text{\textbf{ML}} & \text{\textbf{LC}}\\\hline
\mathbf{A} & \makecell[l]{Every \textbf{b} is \textbf{c}.\\} & b\subseteq c & bc=b\\\hline
\mathbf{E} & \makecell[l]{No \textbf{b} is \textbf{c}.\\(Every \textbf{b} is not-\textbf{c}.)} & b\subseteq c' & bc'\!=\!b\\\hline
\mathbf{I} & \makecell[l]{At least one \textbf{b} is \textbf{c}.\\} & b\Cap c & bc\neq\bm{\varnothing}\\\hline
\mathbf{O} & \makecell[l]{At least one \textbf{b} is not \textbf{c}.\\(At-least-one \textbf{b} is not-\textbf{c}.)} & b\Cap c' & bc'\!\neq\!\bm{\varnothing}\\\hline
\mathbf{*} & \makecell[l]{At least one \textbf{b} exists.\\(At least one \textbf{b} is \textbf{b}.)\\(\textbf{b} is-not empty.)} & b\Cap b & bb\neq\bm{\varnothing}
\end{tabular}\vspace{5pt}

Does LC$\metale$ ML? In other words, can we derive all ML axioms from LC axioms? Let's derive each individual ML axiom from LC.

\vspace{5pt}\begin{tabular}{lLl}
\textbf{ML1:} & b\Cap c \metaeq bc\neq\bm{\varnothing} \metaeq cb\neq\bm{\varnothing} \metaeq c\Cap b & \{(D2), (LC1)\}\\
\textbf{ML2:} & b\subseteq c \metaeq bc=b \metaeq c'b'=c' \metaeq c'\subseteq b' & \{(D1), (LC4)\}\\
\textbf{ML3:} & b\subseteq c' \metaeq bc'=b \metaeq cb'=c \metaeq c\subseteq b' & \{(D1), (LC5)\}\\
\textbf{ML4:} & \text{(D1), Barbara-1 proof in LC} & \{(D1), (LC2)\}\\
\textbf{ML5:} & \text{(D2), (D1), Darii-1 proof in LC} & \begin{tabular}{@{}l@{}}\{(D2), (D1),\\(LC1), (LC2), (LC3)\}\end{tabular}\\
\end{tabular}

Therefore, as long as we additionally assume D1 and D2 as bridge definitions enriching LC, \textbf{LC$\metale$ ML}.

What about the converse? Does ML$\metale$ LC?

If it is true, we should be able to prove every LC axiom from the ML axioms. If it is false, we should be able to find a consequence/theorem from LC which we cannot prove true or false (that is, which is undecidable) in ML.

It turns out that ``$b\Cap c \;\metale\; c\Cap c$'' cannot be proved from ML axioms alone\footnote{By the way, we would need ML4 and that extra assumption to prove covariance/monotonicity of inhabitation -- if a class is inhabited, then any of its superclasses is also inhabited:

\begin{tabular}{ll}
\textbf{S1.} $b\Cap b, b\subseteq c \metale b\Cap c$ & \begin{tabular}{@{}c@{}}\{$b\Cap c, c\subseteq d \metale b\Cap d$\}\\$c\leftarrow b, d\leftarrow c$\end{tabular}\\
\textbf{S2.} $b\Cap b, b\subseteq c \metale c\Cap c$ & (S1), \{$b\Cap c \metale c\Cap c$\}
%	c^b |= c^c
%	b<=c, b^b |= c^c
\end{tabular}

The corresponding theorem in LC ($b\neq\bm{\varnothing}, bc=b \metale c\neq\bm{\varnothing}$) can be proved by the application of substitution and LC3.}, whereas it is a straightforward consequence of LC3 and the correspondence of definitions of ``At least one \textbf{b} is \textbf{c}.'' and ``At least one \textbf{c} exists.'' in LC and ML:

\begin{tabular}{L}
bc\neq\bm{\varnothing} \quad\!\metale\quad c\neq\bm{\varnothing}\\
b\;\;\Cap\; c \quad\metale\quad c\,\Cap\; c
\end{tabular}

ML has no expressive apparatus to state sentences involving the intersection operation, such as ``$\mathbf{b}\subseteq\mathbf{c}\mathbf{d}$''. In contrast, in the more expressive LC system we can prove some theorems which are translatable to laws involving ``$\subseteq$'' and intersection, such as greatest lower bound:

$\mathbf{b}\subseteq\mathbf{c},~\mathbf{b}\subseteq\mathbf{d}\;\enskip\metale\enskip \enskip\,\mathbf{b}\subseteq\mathbf{c}\mathbf{d}$

(Here, $\mathbf{b}$ is a lower bound of both $\mathbf{c}$ and $\mathbf{d}$, and $\mathbf{c}\mathbf{d}$ is the \emph{greatest} lower bound of $\mathbf{c}$ and $\mathbf{d}$ taken together.)

Proof:\\
	1. $\mathbf{b}=\mathbf{b}\mathbf{c}$ \{$\mathbf{b}\subseteq\mathbf{c}$\}\\
	2. $\mathbf{b}=\mathbf{b}\mathbf{d}$ \{$\mathbf{b}\subseteq\mathbf{d}$\}\\
	---------------------------------------------------\\
	3. $\mathbf{b}=(\mathbf{b}\mathbf{c})\mathbf{d}$ \{(2), (1)\}\\
	4. $\mathbf{b}=\mathbf{b}(\mathbf{c}\mathbf{d})$ \{(3), LC2. Therefore: $\mathbf{b}\subseteq\mathbf{c}\mathbf{d}$\}

Therefore, \textbf{ML$\metanleq$ LC}.

By taking together the facts that LC$\metale$ ML and ML$\metanleq$ LC, we conclude that LC is strictly more ``powerful'' than ML, in the sense that we can prove more theorems in LC than in ML.

Note that Barbara-1 (subclasshood transitivity -- ML4) is a consequence of the associativity of intersection (one of the semilattice laws, as we will see in Section~\ref{subsec:LC_from_Boolean_algebra}). Associativity is more general than transitivity because the involved classes are not required to participate together in a subclasshood relation.

With the axioms we have chosen, every categorical syllogism proof in ML has shown to be shorter than (or at least as short as) the corresponding proof in LC.

\vspace{-15pt}\subsection{Deriving LC from Boolean algebra}
\label{subsec:LC_from_Boolean_algebra}

\vspace{-10pt}\subsubsection{Boolean lattice}
\label{subsubsec:Boolean_lattice}

There are various equivalent axiomatic systems --entry points-- for Boolean algebra. One of them is the Boolean lattice (BL) axiomatic system, which we will adopt in this paper\footnote{One could adopt instead the definition of Boolean ring \citep{stone:1935}, one of Huntington's axiomatic systems (\citeyear{huntington:1933, huntington:1904}), or one of the two alternative versions of \citeauthor{wolfram:2018}'s (\citeyear{wolfram:2018})\citep{mccune:2002} axiom, among various other equivalent axiomatic systems.}. ``Boolean lattice'' is usually defined as ``complemented distributive lattice'' \citep[p.~88]{birkhoff:1940}, with the signature $\langle S,\sqcap,\sqcup,',\bot,\top\rangle$. Does BL$\metale$ LC? To answer this, let's first enumerate the BL axioms, that is, the lattice, distributive lattice, and complemented lattice axioms\footnote{Boolean lattice, as a formal (abstract) algebraic structure, was \emph{axiomatically} defined by Ernst \citet[pp.~8-12]{schroeder:1877}, who called it simply ``\emph{Logikkalkul}'', with an axiomatic system very close to the one we adopt here (with the minor difference that Schröder was more parsimonious, since he proved the bounded lattice law ``$b\sqcup\bot=b$'' and the two absorption laws as theorems), long before Lattice Theory was systematized as a topic of study.
%Schröder's axioms summarized in a single page: https://books.google.com/books?id=74-g3vSAbnoC&pg=PA372
%[...] Schröder entwickelte 1877 das erste formale Axiomensystem einer booleschen Algebra in additiver Schreibweise. [...]
%[https://de.wikipedia.org/wiki/Boolesche_Algebra#Geschichte]
%Schröder's axioms, repeated by Peano
%https://books.google.com/books?id=DDZ-GQAACAAJ&pg=PA3
%https://de.wikipedia.org/wiki/Boolesche_Algebra#Definition
%Most of these laws (except distributivity of v over ^ and absorption with ^ over v) also appear in a scattered form in \citep[pp.~4-5,11,15-16]{grassmann:1872}.
%All of the Boolean lattice axioms were enumerated in a survey by \citet[pp.~33-36]{peirce:1880}, though as a scattered collection of laws which were not then recognized as axioms, and together with some theorems.
%The symbols $\cap$, \cup$, and $\sqcap$ (though not $\sqcup$) are employed by Leibniz ("De la methode de l'universalité", 1675?<https://archive.org/details/opusculesetfrag00coutgoog/page/97/mode/1up>) in a different context. Later: <$\sqcap$, 1674: https://archive.org/details/opusculesetfrag00coutgoog/page/170/mode/1up><$\sqcap$, 1678: https://archive.org/details/opusculesetfrag00coutgoog/page/148/mode/1up><$\sqcap$: https://archive.org/details/opusculesetfrag00coutgoog/page/324/mode/1up>

The axiomatic system as presented here is didactically clear (since it follows the typical progression of the study of algebraic structures in abstract algebra) but redundant. For instance, each idempotence law becomes a theorem when both absorption laws hold:\newline
$b\sqcap b= b\sqcap (b\sqcup (b\sqcap b)) = b$;\newline
$b\sqcup b= b\sqcup (b\sqcap (b\sqcup b)) = b$.\newline
For further examples of redundancy in this axiomatic system, see the Section ``§ 1. The First Set of Postulates'' by \citet{huntington:1904}.}:% (Table~\ref{tab:BL_axioms}).

%\vspace{-28pt}
%\begin{table}[bt]
\begin{tabular}{lLL}
\multicolumn{3}{l}{\textbf{Semilattice laws\footnotemark:}}\\
Idempotence\footnotemark & b\sqcap b=b & b\sqcup b=b\\%b<=b, b>=b
Commutativity & b\sqcap c=c\sqcap b & b\sqcup c=c\sqcup b\\
Associativity & \begin{tabular}{@{}L@{}}(b\sqcap c)\sqcap d=\\b\sqcap (c\sqcap d)\end{tabular} & \begin{tabular}{@{}L@{}}(b\sqcup c)\sqcup d=\\b\sqcup (c\sqcup d)\end{tabular}\\
\multicolumn{3}{l}{\textbf{Lattice laws:}}\\
Absorption\footnotemark & b\sqcap (b\sqcup c)=b & b\sqcup (b\sqcap c)=b\\%b <= bUc; b >= b^c
\multicolumn{3}{l}{\textbf{Distributive lattice laws:}}\\
Distributivity\footnotemark & \begin{tabular}{@{}L@{}}b\sqcap (c\sqcup d)=\\(b\sqcap c)\sqcup(b\sqcap d)\end{tabular} & \begin{tabular}{@{}L@{}}b\sqcup(c\sqcap d)=\\(b\sqcup c)\sqcap (b\sqcup d)\end{tabular}\\
\multicolumn{3}{l}{\textbf{Bounded lattice laws}}\\
Identity element & b\sqcap\top=b & b\sqcup\bot=b\\%b<=\top; b>=\bot
\multicolumn{3}{l}{\textbf{Complemented lattice laws}}\\
\begin{tabular}{@{}l@{}}Complementarity/\\Contradiction\footnotemark\end{tabular} & b\sqcap b'=\bot & b\sqcup b'=\top%bb'<=\bot; bUb'>=\top
\\\\
\end{tabular}
%\caption{Boolean lattice axioms.}
%\label{tab:BL_axioms}
%\end{table}

\addtocounter{footnote}{-4}\footnotetext{The (semi)lattice axioms were collected together into lattice theory --a branch of both abstract algebra and order theory-- by \citet[p.~795]{birkhoff:1938}.}\stepcounter{footnote}\footnotetext{Each idempotence law was algebraically enunciated by Leibniz, respectively in \citep{leibniz:1679c} and \citep[\emph{Axioma}~1 \& \emph{Scholium}]{leibniz:1686f} -- drafts published years before \citeauthor{boole:1847}'s (\citeyear{boole:1847}) pamphlet. Boole restated $b\sqcap b=b$ but did not tolerate $b\sqcup b=b$ because $\top\sqcup \top=\top$ would translate to $1+1=1$ in his syntax, which attempted to imitate as closely as possible certain ordinary operations and values from numerical algebra. Few years after \citeauthor{boole:1847}'s pamphlet was published, \citet[p.~26, point~69; pp.~82-83, points 191-193]{jevons:1864} restated $b\sqcup b=b$.
%[Leibniz]
%Axioma 1. Si idem secum ipso sumatur, nihil constituitur novum, seu A+A=A.
%Schol. Equidem in numeris 2+2 facit 4, seu bini nummi binis additi faciunt quatuor nummos, sed tunc bini additi sunt alii a prioribus; si iidem essent, nihil novi prodiret, et perinde esset ac si joco ex tribus ovis facere vellemus sex, numerando primum 3 ova, deinde uno sublato residua 2, ac deinde uno rursus sublato residuum.
%[https://archive.org/details/operaphilosophic00leibuoft/page/95/mode/1up https://archive.org/details/diephilosophisc01leibgoog/page/230/mode/1up]
}\stepcounter{footnote}\footnotetext{The earliest recognition we could find for an absorption law in Boolean algebra, $b\sqcup (b\sqcap c)=b$, is by \citet[p.~26, point~70]{jevons:1864}\citep[p.~454]{valencia:2004}\citep[p.~74]{lewis:1918}. Boole impeded himself from discovering it, since his partial ``union'' operation was valid only for disjoint classes, and it is not necessarily the case that $b$ and $b\sqcap c$ are mutually disjoint. For the other absorption law, $b\sqcap (b\sqcup c)=b$, the obstacle for its universal validity in \citeauthor{boole:1847}'s original algebra is that $b$ and $c$ (in the partial ``union'' inside parentheses) are not necessarily disjoint.
%Jevons:
%https://books.google.com/books?id=WVMOAAAAYAAJ&pg=PA26&dq=%22any+alternative+may%22
%Valencia:
%https://books.google.com/books?id=74-g3vSAbnoC&pg=PA454&dq=%22Laws+of+Logic%22+%22by+Jevons%22+absorption
%Proof of the absorption laws:
%bUbc=bIUbc=b(IUc)=b(cUI)=bI=b
%$b\sqcup(b\sqcap c)=(b\sqcap\mathbf{I})\sqcup(b\sqcap c)=b\sqcap(\mathbf{I}\sqcup c)=b\sqcap(c\sqcup\mathbf{I})=b\mathbf{I}=b$.
%b(bUc)=bbUbc=bUbc=b
%$b\sqcap(b\sqcup c)=(b\sqcap b)\sqcup(b\sqcap c)=b\sqcup(b\sqcap c)=b$.
%b(bUc)=(bU0)(bUc)=bU(0c)=bU(c0)=b
%$b\sqcap(b\sqcup c)=(b\sqcup\bot)\sqcap(b\sqcup c)=b\sqcup(\bot\sqcap c)=b\sqcup(c\sqcap\bot)=b$.
%Another author who says Boole didn't state any of the absorption laws: https://books.google.com/books?id=QnbQDwAAQBAJ&pg=PA153&dq=%22un+teorema+che+non%22

We didn't succeed in finding an excerpt of some draft where Leibniz makes this law explicit or at least makes use of it implicitly, but we may have inattentively passed over it in our reading. \citet[p.~709]{malink:2016} studied an important manuscript of Leibniz's on logic and couldn't find this law there either -- though the possibility that it might be present somewhere else remains.%p. 24 of the PDF file
}\stepcounter{footnote}\footnotetext{One of the distributive laws --$b\sqcap (c\sqcup d)=(b\sqcap c)\sqcup(b\sqcap d)$-- was stated by \citet[p.~17]{boole:1847}. The other one was stated by \citet[p.~251]{peirce:1867}, and wasn't anticipated by Boole likely because it is not universally valid for numerical algebra with ``$+$'' and ``$*$'', to which his algebra of logic is subordinated; it is not universally valid for non-trivial Boolean rings either.
%For rings, b+(c*d) = (b+c)*(b+d) iff b*(c+d)=0 <p. 9: https://www.cle.unicamp.br/eprints/index.php/CLE_e-Prints/article/view/880/740>. Example for Boolean rings: 1+(1*0)=/=(1+1)*(1+0): 1+0=/=0*1: 1=/=0 (in a non-trivial ring).

\citet[pp.~280,282,285-287,643]{schroeder:1890} discovered that not all lattices are distributive. By constructing an example of non-distributive lattice, we can prove the independence of the distributive laws from the other lattice axioms. The (modular, bounded) diamond lattice $M_5$ and the (non-modular, bounded) pentagon lattice $N_5$, both with 5 points, are examples of non-distributive lattices. %https://en.wikipedia.org/wiki/Distributive_lattice#Characteristic_properties https://math.stackexchange.com/questions/95392/what-elements-of-the-pentagon-lattice-n-5-do-not-satisfy-the-distributive-law
In addition, set partition lattices and noncrossing partition lattices are not distributive in general.%In the following set partition lattice, there are various $M_5$ sublattices, whose top is any green-colored element, and whose bottom is the original lattice's bottom: https://commons.wikimedia.org/wiki/File:Set_partitions_4;_Hasse;_circles.svg https://math.stackexchange.com/questions/4219642/can-you-provide-a-symmetric-presentation-of-this-partition-lattice https://doc.sagemath.org/html/en/reference/combinat/sage/combinat/set_partition.html#:~:text=graphics%20primitives
%Noncrossing set partition lattice: https://en.wikipedia.org/wiki/Noncrossing_partition https://www.researchgate.net/figure/The-noncrossing-partition-lattice-NC-4_fig1_328953253

\citeauthor{schroeder:1890}'s (\citeyear[pp.~643]{schroeder:1890}) diagrammatic example of non-distributive lattice contains two non-modular pentagon sublattices $N_5$: $\{AB, B, AB+AC, A(B+C), B+C\}$ and $\{AC, C, AB+AC, A(B+C), B+C\}$.}\stepcounter{footnote}\footnotetext{Lodovico Ignazio \citet[p.~48]{richeri:1761}, in an article where he attempts to construct a \emph{characteristica universalis} along lines similar to Leibniz's, adopted semiotically opposite symbols similar to ``\rotatebox[origin=c]{180}{\large\textomega}'' and ``{\large\textomega}'' for false/empty and true/universe, respectively --to which the modern lattice theory symbols ``$\bot$'' and ``$\top$'' are analogous--, in order to symbolically represent the complementarity laws -- disjointness and exhaustion \mbox{\citep[pp.~86-87]{ladd:1890}}\citep{peirce:1901}:

\begin{tabular}{@{}c@{\:}c@{\:}c@{\:}c@{}}
~ & \& & ~ & \rotatebox[origin=c]{180}{\large\textomega} \\
a & ~ & non a & ~\\
~ & vel & ~ & {\large\textomega}
\end{tabular}

What his symbolic representation of the complementarity laws had: literal placeholder for a term; a symbol for ``and''; symbols for false/contradiction and true/necessity.

What was missing: a symbol for ``or''; a symbol for ``non-''; a symbol for equality/identity.

With some notational adaptations to better exploit mirror reflection symmetry in order to show a kind of ``partial cancellation'' and ``combination'' of symbols, we obtain:

\begin{tabular}{@{}c@{\:}c@{\:}c@{\:}c@{\:}c@{}}
~ & $\bm{\cap}$ & ~ & ~ & \rotatebox[origin=c]{180}{\large\textomega} \\
%\underline{~{\tiny i}~} & ~ & $\overline{~\text{\tiny i}~}$ & = & ~\\
\underline{~{\tiny i}~} & ~ & \rotatebox[origin=c]{180}{\underline{~{\tiny i}~}} & = & ~\\
~ & $\bm{\cup}$ & ~ & ~ & {\large\textomega}
\end{tabular}

This judicious choice of semiotically opposite symbols for ``$\bot$'' and ``$\top$'' stressed a symmetry that later would suggest the duality principle of Boolean algebras \citep{ncatlab:2019} and the two \citeauthor{de_morgan:1847}'s laws -- features Leibniz \textbf{didn't} seem to have grasped the importance of \citep[p.~127]{levey:2011}\citep[p.~263-265]{lenzen:2018b}.

\begin{minipage}{\textwidth}
It is a pity that \citeauthor{richeri:1761} missed the opportunity to present the identity element laws with his layout and a nice symmetrical notation where a kind of ``cancellation'' of almost antagonistic symbols becomes apparent \citep[p.~218, 22$\cdot$05:IIa-IIb]{whitehead:1910}:

\begin{tabular}{@{}c@{\:}c@{\:}c@{\:}c@{\:}c@{}}
~ & $\bm{\cap}$ & {\large\textomega} & ~ & ~ \\
\underline{~{\tiny i}~} & ~ & ~ & = & \underline{~{\tiny i}~}\\
~ & $\bm{\cup}$ & \rotatebox[origin=c]{180}{\large\textomega} & ~ & ~
\end{tabular}
%a^a'=m {0}
%ava'=w {I}
%Laws from <https://books.google.com/books?id=OG0VDAAAQBAJ&pg=PA147>:
%avm=a
%a^w=a
\end{minipage}

The character ``T'' (from which the ``$\top$'' symbol surely comes from) was adopted to represent ``Totalität'' (``totality'', the universe of discourse) in logic by Robert \citet[p.~15]{grassmann:1872}.}From these axioms, we can derive the following theorems.

\begin{multicols}{2}

\begin{tabular}{@{\hspace{1.2\tabcolsep}} L @{\hspace{1.2\tabcolsep}} l}
%\begin{longtable}[l]{@{\hspace{1.2\tabcolsep}} L @{\hspace{1.2\tabcolsep}} l}
\multicolumn{2}{l}{~}\\
\multicolumn{2}{l}{\begin{tabular}{@{}l@{}}Dominating element for $\sqcap$ /\\Least element:\\$b\sqcap\bot=\bot$\end{tabular}}\\%Least element $\bot$ and greatest element $\top$: $\bottom \subseteq b \subseteq \top$
\multicolumn{2}{l}{\textsc{~Proof.}}\\[4pt]
b\sqcap\bot &  \\
= b\sqcap(b\sqcap b') & \{$b\sqcap b'=\bot$\}\\
= (b\sqcap b)\sqcap b' & \begin{tabular}{@{}l@{}}\{$(b\sqcap c)\sqcap d=$\\~~$b\sqcap(c\sqcap d)$\}\end{tabular}\\
= b\sqcap b' & \{$b\sqcap b=b$\}\\
=\bot & \{$b\sqcap b'=\bot$\}
%\multicolumn{2}{l}{\textsc{~Alternative proof.}}\\[4pt]
%b\sqcap\bot &  \\
%=(b\sqcap\bot)\sqcup\bot & \{$b\sqcup \bot=b$\}\\
%=(b\sqcap\bot)\sqcup (b\sqcap b') & \{$b\sqcap b'=\bot$\}\\
%=b\sqcap(\bot\sqcup b') & \begin{tabular}{@{}l@{}}\{$b\sqcap (c\sqcup d)=$\\$(b\sqcap c)\sqcup (b\sqcap d$\}\end{tabular}\\
%=b\sqcap (b'\sqcup\bot) & \{$b\sqcup c=c\sqcup b$\}\\
%=b\sqcap b' & \{$b\sqcup \bot=b$\}\\
%=\bot & \{$b\sqcap b'=\bot$\}
%\end{longtable}
\end{tabular}

\begin{tabular}{@{\hspace{1.2\tabcolsep}} L @{\hspace{1.2\tabcolsep}} l}
%\begin{longtable}[l]{@{\hspace{1.2\tabcolsep}} L @{\hspace{1.2\tabcolsep}} l}
\multicolumn{2}{l}{~}\\
\multicolumn{2}{l}{\begin{tabular}{@{}l@{}}Dominating element for $\sqcup$ /\\Greatest element:\\$b\sqcup\top=\top$\end{tabular}}\\
\multicolumn{2}{l}{\textsc{~Proof.}}\\[4pt]
b\sqcup\top &  \\
= b\sqcup(b\sqcup b') & \{$b\sqcup b'=\top$\}\\
= (b\sqcup b)\sqcup b' & \begin{tabular}{@{}l@{}}\{$(b\sqcup c)\sqcup d=$\\~~$b\sqcup(c\sqcup d)$\}\end{tabular}\\
= b\sqcup b' & \{$b\sqcup b=b$\}\\
=\top & \{$b\sqcup b'=\top$\}\\
%\multicolumn{2}{l}{\textsc{~Alternative proof.}}\\[4pt]
%b\sqcup\top &  \\
%=(b\sqcup\top)\sqcap\top & \{$b\sqcap \top=b$\}\\
%=(b\sqcup\top)\sqcap (b\sqcup b') & \{$b\sqcup b'=\top$\}\\
%=b\sqcup (\top\sqcap b') & \begin{tabular}{@{}l@{}}\{$b\sqcup (c\sqcap d)=$\\$(b\sqcup\! c)\sqcap (b\sqcup\! d)$\}\end{tabular}\\
%=b\sqcup (b'\sqcap \top) & \{$b\sqcap c=c\sqcap b$\}\\
%=b\sqcup b' & \{$b\sqcap \top=b$\}\\
%=\top & \{$b\sqcup b'=\top$\}
%\end{longtable}
\end{tabular}

\begin{tabular}{@{\hspace{1.2\tabcolsep}} L @{\hspace{1.2\tabcolsep}} l}
%\begin{longtable}[l]{@{\hspace{1.2\tabcolsep}} L @{\hspace{1.2\tabcolsep}} l}
\multicolumn{2}{l}{~}\\
\multicolumn{2}{l}{\begin{tabular}{@{}l@{}}Subsumption with $\sqcap$ and $\sqcup$:\\$b\sqcap c=b \enskip\metaeq\enskip b\sqcup c=c$\end{tabular}}\\%b<=c |=| c>=b
\multicolumn{2}{l}{\textsc{~Proof.}}\\[4pt]
\multicolumn{2}{L}{b\sqcap c=b \enskip\metale\enskip b\sqcup c=c}\\
b\sqcup c &  \\
=(b\sqcap c)\sqcup c & \{assumption: $b\sqcap c=b$\}\\
=c\sqcup (b\sqcap c) & \{$b\sqcup c=c\sqcup b$\}\\
=c\sqcup (c\sqcap b) & \{$b\sqcap c=c\sqcap b$\}\\
=c & \{$b\sqcup (b\sqcap c)=b$\}\\
\multicolumn{2}{l}{~}\\
\multicolumn{2}{L}{b\sqcup c=c \enskip\metale\enskip b\sqcap c=b}\\
b\sqcap c &  \\
=b\sqcap (b\sqcup c) & \{assumption: $b\sqcup c=c$\}\\
=b & \{$b\sqcap (b\sqcup c)=b$\}\\
\multicolumn{2}{l}{~}\\
\multicolumn{2}{l}{Therefore, $b\sqcap c=b \enskip\metaeq\enskip b\sqcup c=c$.}
%\end{longtable}
\end{tabular}

\begin{tabular}{@{\hspace{1.2\tabcolsep}} L @{\hspace{1.2\tabcolsep}} l}
%\begin{longtable}[l]{@{\hspace{1.2\tabcolsep}} L @{\hspace{1.2\tabcolsep}} l}
\multicolumn{2}{l}{~}\\
\multicolumn{2}{l}{\begin{tabular}{@{}l@{}}Subsumption and supersumption:\\$b\sqcap c=b,~b\sqcup c=b \enskip\metale\enskip b=c$\end{tabular}}\\%b<=c, b>=c |= b=c
\multicolumn{2}{l}{\textsc{~Proof.}}\\[4pt]
b &  \\
=b\sqcup c & \{assumption: $b\sqcup c=b$\}\\
=(b\sqcap c)\sqcup c & \{assumption: $b\sqcap c=b$\}\\
=(c\sqcap b)\sqcup c & \{$b\sqcap c=c\sqcap b$\}\\
=c\sqcup (c\sqcap b) & \{$b\sqcup c=c\sqcup b$\}\\
=c & \{$b\sqcup (b\sqcap c)=b$\}
%\end{longtable}
\end{tabular}

\end{multicols}

%\begin{tabular}{@{\hspace{1.2\tabcolsep}} L @{\hspace{1.2\tabcolsep}} l}
\begin{longtable}[l]{@{\hspace{1.2\tabcolsep}} L @{\hspace{1.2\tabcolsep}} l}
\multicolumn{2}{l}{~}\\
\multicolumn{2}{l}{\begin{tabular}{@{}l@{}}Involution:\\$(b')'=b$\end{tabular}}\\
\multicolumn{2}{l}{\textsc{~Proof.}}\\[4pt]
b\sqcup b'=\top &  \\
(b')'\sqcap (b\sqcup b')=(b')'\sqcap\top & \begin{tabular}{@{}c@{}}\{$b=c \enskip\metale\enskip f(b)=f(c)$\}\\$f(x)=(b')'\sqcap x$\end{tabular}\\
(b')'\sqcap (b\sqcup b')=(b')' & \{$b\sqcap\top=b$\}\\
((b')'\sqcap b)\sqcup ((b')'\sqcap b')=(b')' & \begin{tabular}{@{}l@{}}\{$b\sqcap (c\sqcup d)$=\\$~(b\sqcap c)\sqcup (b\sqcap d)$\}\end{tabular}\\
((b')'\sqcap b)\sqcup (b'\sqcap (b')')=(b')' & \{$b\sqcap c=c\sqcap b$\}\\
((b')'\sqcap b)\sqcup\bot=(b')' & \{$b\sqcap b'=\bot$\}\\
(b')'\sqcap b=(b')' & \{$b\sqcup\bot=b$\}\\
\multicolumn{2}{l}{~}\\
b\sqcap b'=\bot &  \\
(b')'\sqcup (b\sqcap b')=(b')'\sqcup\bot & \begin{tabular}{@{}c@{}}\{$b=c \enskip\metale\enskip f(b)=f(c)$\}\\$f(x)=(b')'\sqcup x$\end{tabular}\\
(b')'\sqcup (b\sqcap b')=(b')' & \{$b\sqcup\bot=b$\}\\
((b')'\sqcup b)\sqcap ((b')'\sqcup b')=(b')' & \begin{tabular}{@{}l@{}}\{$b\sqcup (c\sqcap d)=$\\~$(b\sqcup c)\sqcap (b\sqcup d)$\}\end{tabular}\\
((b')'\sqcup b)\sqcap (b'\sqcup (b')')=(b')' & \{$b\sqcup c=c\sqcup b$\}\\
((b')'\sqcup b)\sqcap\top=(b')' & \{$b\sqcup b'=\top$\}\\
(b')'\sqcup b=(b')' & \{$b\sqcap\top=b$\}\\
\multicolumn{2}{l}{~}\\
\multicolumn{2}{l}{Therefore,}\\
(b')'=b & \{$b\sqcap c=b,~b\sqcup c=b \metale\enskip b=c$\}
\end{longtable}
%\end{tabular}

%\begin{tabular}{@{\hspace{1.2\tabcolsep}} L @{\hspace{1.2\tabcolsep}} l}
\begin{longtable}[l]{@{\hspace{1.2\tabcolsep}} L @{\hspace{1.2\tabcolsep}} l}
\multicolumn{2}{l}{~}\\
\multicolumn{2}{l}{\begin{tabular}{@{}l@{}}Complementation of equivalent terms:\\$b=c \metaeq b'=c'$\end{tabular}}\\
\multicolumn{2}{l}{\textsc{~Proof} (by Leibniz\footref{fn:positive_and_negative_identity}).}\\[4pt]
b=c \metale b'=c' & \begin{tabular}{@{}c@{}}\{$b=c \enskip\metale\enskip f(b)=f(c)$\}\\$f(x)=x'$\end{tabular}\\
\multicolumn{2}{l}{~}\\
\multicolumn{2}{l}{Conversely,}\\
b'=c' \metale (b')'=(c')' & \begin{tabular}{@{}c@{}}\{$b=c \enskip\metale\enskip f(b)=f(c)$\}\\$b\leftarrow b', c\leftarrow c', f(x)=x'$\end{tabular}\\
b'=c' \metale b=c & \{$(b')'=b$\}\\
\multicolumn{2}{l}{~}\\
\multicolumn{2}{l}{Therefore,}\\
b=c \metaeq b'=c'
\end{longtable}
%\end{tabular}

\newpage
%\begin{tabular}{@{\hspace{1.2\tabcolsep}} L @{\hspace{1.2\tabcolsep}} l}
\begin{longtable}[l]{@{\hspace{1.2\tabcolsep}} L @{\hspace{1.2\tabcolsep}} l}
\multicolumn{2}{l}{~}\\
\multicolumn{2}{l}{\begin{tabular}{@{}l@{}}Unique complementation 1:\\$b\sqcap c=\bot,~b\sqcup c=\top \enskip\metale\enskip c=b'$\end{tabular}}\\
\multicolumn{2}{l}{\textsc{~Proof\footnote{Adapted from:\newline
\guillemotleft\url{https://proofwiki.org/wiki/Complement\_in\_Boolean\_Algebra\_is\_Unique}\guillemotright.\newline
See also:\newline
\guillemotleft\url{https://proofwiki.org/wiki/Complement\_in\_Distributive\_Lattice\_is\_Unique}\guillemotright\newline
\guillemotleft\url{https://math.stackexchange.com/questions/3239464/how-to-prove-the-uniqueness-of-complement-in-the-algebra-of-sets-without-using\#3239493}\guillemotright}.}}\\[4pt]
\multicolumn{2}{l}{\begin{tabular}{@{}l@{}}In a Boolean lattice, an element $b$ necessarily has a\\ complement, $b'$, by the complemented lattice laws.\end{tabular}}\\
%\multicolumn{2}{l}{In a Boolean lattice, $b$ necessarily has a complement, $b'$, by the complemented lattice laws.}\\
\multicolumn{2}{l}{Suppose $c$ is also a complement of $b$, that is:}\\
b\sqcap c=\bot & $b\sqcup c=\top$\\
\multicolumn{2}{l}{Then}\\
c &  \\
=c\sqcap\top & \{$b\sqcap\top=b$\}\\
=c\sqcap (b\sqcup b') & \{$b\sqcup b'=\top$\}\\
=(c\sqcap b)\sqcup (c\sqcap b') & \{$b\sqcap (c\sqcup d)=(b\sqcap c)\sqcup (b\sqcap d)$\}\\
=(b\sqcap c)\sqcup (c\sqcap b') & \{$b\sqcap c=c\sqcap b$\}\\
=\bot\sqcup (c\sqcap b') & \{assumption: $b\sqcap c=\bot$\}\\
=(b\sqcap b')\sqcup (c\sqcap b') & \{$b\sqcap b'=\bot$\}\\
=(b'\sqcap b)\sqcup (c\sqcap b') & \{$b\sqcap c=c\sqcap b$\}\\
=(b'\sqcap b)\sqcup (b'\sqcap c) & \{$b\sqcap c=c\sqcap b$\}\\
=b'\sqcap (b\sqcup c) & \{$b\sqcap (c\sqcup d)=(b\sqcap c)\sqcup (b\sqcap d)$\}\\
=b'\sqcap\top & \{assumption: $b\sqcup c=\top$\}\\
=b' & \{$b\sqcap\top=b$\}
\end{longtable}
%\end{tabular}

%\begin{tabular}{@{\hspace{1.2\tabcolsep}} L @{\hspace{1.2\tabcolsep}} l}
\begin{longtable}[l]{@{\hspace{1.2\tabcolsep}} L @{\hspace{1.2\tabcolsep}} l}
\multicolumn{2}{l}{~}\\
\multicolumn{2}{l}{\begin{tabular}{@{}l@{}}Unique complementation 2:\\$b\sqcap c'=\bot,~b\sqcup c'=\top \enskip\metale\enskip c=b$\end{tabular}}\\
\multicolumn{2}{l}{\textsc{~Proof.}}\\[4pt]
b\sqcap c'=\bot,~b\sqcup c'=\top \enskip\metale\enskip c'=b' & \begin{tabular}{@{}c@{}}\{$b\sqcap c=\bot,~b\sqcup c=\top \enskip\metale\enskip c=b'$\}\\$c\leftarrow c'$\end{tabular}\\
b\sqcap c'=\bot,~b\sqcup c'=\top \enskip\metale\enskip (c')'=(b')' & \{$b=c \metaeq b'=c'$\}\\
b\sqcap c'=\bot,~b\sqcup c'=\top \enskip\metale\enskip c=(b')' & \{$(b')'=b$\}\\
b\sqcap c'=\bot,~b\sqcup c'=\top \enskip\metale\enskip c=b & \{$(b')'=b$\}
\end{longtable}
%\end{tabular}

%\begin{tabular}{@{\hspace{1.2\tabcolsep}} L @{\hspace{1.2\tabcolsep}} l}
\begin{longtable}[l]{@{\hspace{1.2\tabcolsep}} L @{\hspace{1.2\tabcolsep}} l}
\multicolumn{2}{l}{~}\\
\multicolumn{2}{l}{\begin{tabular}{@{}l@{}}\citeauthor{de_morgan:1847}'s law\footnote{The two \citeauthor{de_morgan:1847}'s laws state or imply that ``$'$'' is a monoid homomorphism from ``$\sqcap$'' to ``$\sqcup$'' and vice versa~-- or, more precisely, from $\langle\mathbb{B}, \sqcap, \top\rangle$ to $\langle\mathbb{B}, \sqcup, \bot\rangle$ and vice versa, as it is an involution.\newline The \citeauthor{de_morgan:1847}'s (\citeyear[pp.~118,59]{de_morgan:1847}) laws were anticipated in prose by \citet{ockham:1323}.} for $\overline\sqcap$:\\$(b\sqcap c)'=b'\sqcup c'$\end{tabular}}\\
\multicolumn{2}{l}{\textsc{~Proof\footnote{Adapted from:\newline
\guillemotleft\url{https://www.geeksforgeeks.org/proof-of-de-morgans-laws-in-boolean-algebra/}\guillemotright.}.}}\\[4pt]
%See also:
%https://books.google.com/books?id=wEa9DwAAQBAJ&pg=PA361
%https://vixra.org/abs/2209.0054
%After the two De Morgan's laws are stated, we have the following corollaries:
%Identity element law for monoid homomorphism (``$'$'') from ``$\cup$'' to ``$\cap$'':
%\top' = \bot. Proof:
%(bUb')' = b'^(b')'
%\top' = b'^b
%\top' = b^b'
%\top' = \bot.
%Identity element law for monoid homomorphism (``$'$'') from ``$\cap$'' to ``$\cup$'':
%\bot' = \top. Proof:
%(b^b')' = b'U(b')'
%\bot' = b'Ub
%\bot' = bUb'
%\bot' = \top.
%As ``$'$'' is an involution:
%(b^c)' = b'Uc'
%(bUc)' = b'^c'
%b^c = ((b^c)')' = (b'Uc')' = (b')'^(c')' = b^c
%bUc = ((bUc)')' = (b'^c')' = (b')'U(c')' = bUc
%\text{According to ``unique complementation 2'':}\\
%b\sqcap c'=\bot,~b\sqcup c'~=\top\enskip \enskip\metale\enskip c=b &  \\
%d\sqcap\! f'=\bot,\;d\sqcup f'=\top\enskip \enskip\metale\enskip f\!=d &  \\
%\multicolumn{2}{l}{~}\\
f\leftarrow (b\sqcap c)',~d\leftarrow b'\sqcup c' &  \\
\multicolumn{2}{l}{~}\\
d\sqcap f' &  \\
=(b'\sqcup c')\sqcap ((b\sqcap c)')' &  \\
=(b'\sqcup c')\sqcap (b\sqcap c) & \{$(b')'=b$\}\\
=(b\sqcap c)\sqcap (b'\sqcup c') & \{$b\sqcap c=c\sqcap b$\}\\
=((b\sqcap c)\sqcap b')\sqcup ((b\sqcap c)\sqcap c') & \begin{tabular}{@{}l@{}}\{$b\sqcap (c\sqcup d)=$\\~$(b\sqcap c)\sqcup (b\sqcap d$\}\end{tabular}\\
=((c\sqcap b)\sqcap b')\sqcup ((b\sqcap c)\sqcap c') & \{$b\sqcap c=c\sqcap b$\}\\
=(c\sqcap (b\sqcap b'))\sqcup ((b\sqcap c)\sqcap c') & \begin{tabular}{@{}l@{}}\{$(b\sqcap c)\sqcap d=$\\~$b\sqcap (c\sqcap d)$\}\end{tabular}\\
=(c\sqcap\bot)\sqcup ((b\sqcap c)\sqcap c') & \{$b\sqcap b'=\bot$\}\\
=\bot\sqcup ((b\sqcap c)\sqcap c') & \{$b\sqcap \bot=\bot$\}\\
=((b\sqcap c)\sqcap c')\sqcup\bot & \{$b\sqcup c=c\sqcup b$\}\\
=(b\sqcap c)\sqcap c' & \{$b\sqcup\bot=b$\}\\
=b\sqcap (c\sqcap c') & \begin{tabular}{@{}l@{}}\{$(b\sqcap c)\sqcap d=$\\~$b\sqcap (c\sqcap d)$\}\end{tabular}\\
=b\sqcap\bot & \{$b\sqcap b'=\bot$\}\\
=\bot & \{$b\sqcap \bot=\bot$\}\\
\multicolumn{2}{l}{~}\\
d\sqcup f' &  \\
=(b'\sqcup c')\sqcup ((b\sqcap c)')' &  \\
=(b'\sqcup c')\sqcup (b\sqcap c) & \{$(b')'=b$\}\\
=((b'\sqcup c')\sqcup b)\sqcap ((b'\sqcup c')\sqcup c) & \begin{tabular}{@{}l@{}}\{$b\sqcup (c\sqcap d)=$\\~$(b\sqcup c)\sqcap (b\sqcup d)$\}\end{tabular}\\
=(b\sqcup (b'\sqcup c'))\sqcap ((b'\sqcup c')\sqcup c) & \{$b\sqcup c=c\sqcup b$\}\\
=((b\sqcup b')\sqcup c')\sqcap ((b'\sqcup c')\sqcup c) & \begin{tabular}{@{}l@{}}\{$(b\sqcup c)\sqcup d=$\\~$b\sqcup (c\sqcup d)$\}\end{tabular}\\
=(\top\sqcup c')\sqcap ((b'\sqcup c')\sqcup c) & \{$b\sqcup b'=\top$\}\\
=(c'\sqcup\top)\sqcap ((b'\sqcup c')\sqcup c) & \{$b\sqcup c=c\sqcup b$\}\\
=\top\sqcap ((b'\sqcup c')\sqcup c) & \{$b\sqcup\top=\top$\}\\
=((b'\sqcup c')\sqcup c)\sqcap\top & \{$b\sqcap c=c\sqcap b$\}\\
=(b'\sqcup c')\sqcup c & \{$b\sqcap\top=b$\}\\
=b'\sqcup (c'\sqcup c) & \begin{tabular}{@{}l@{}}\{$(b\sqcup c)\sqcup d=$\\~$b\sqcup (c\sqcup d)$\}\end{tabular}\\
=b'\sqcup (c\sqcup c') & \{$b\sqcup c=c\sqcup b$\}\\
=b'\sqcup\top & \{$b\sqcup b'=\top$\}\\
=\top & \{$b\sqcup\top=\top$\}\\
\multicolumn{2}{l}{~}\\
\multicolumn{2}{l}{Therefore,}\\
f=d & \begin{tabular}{@{}c@{}}\{$b\sqcap c'=\bot,~b\sqcup c'=\top \enskip\metale\enskip c=b$\}\\$b\leftarrow d,~c\leftarrow f$\end{tabular}\\
(b\sqcap c)'=b'\sqcup c' & $f\leftarrow (b\sqcap c)',~d\leftarrow b'\sqcup c'$\\
\end{longtable}
%\end{tabular}

%\begin{tabular}{@{\hspace{1.2\tabcolsep}} L @{\hspace{1.2\tabcolsep}} l}
\begin{longtable}[l]{@{\hspace{1.2\tabcolsep}} L @{\hspace{1.2\tabcolsep}} l}
\multicolumn{2}{l}{~}\\
\multicolumn{2}{l}{\begin{tabular}{@{}l@{}}\citeauthor{de_morgan:1847}'s law for $\overline\sqcup$:\\$(b\sqcup c)'=b'\sqcap c'$\end{tabular}}\\
\multicolumn{2}{l}{\textsc{~Proof.}}\\[4pt]

(b'\sqcap c')'=(b')'\sqcup (c')' & \begin{tabular}{@{}c@{}}\{\citeauthor{de_morgan:1847}'s law for $\overline\sqcap$\}\\$b\leftarrow b', c\leftarrow c'$\end{tabular}\\
(b'\sqcap c')'=b\sqcup c & \{$(b')'=b$\}\\
b\sqcup c=(b'\sqcap c')' & \{$b=c\metaeq c=b$\}\\
(b\sqcup c)'=((b'\sqcap c')')' & \{$b=c \metaeq b'=c'$\}\\
(b\sqcup c)'=b'\sqcap c' & \{$(b')'=b$\}\\
\end{longtable}
%\end{tabular}

%\end{multicols}

\subsubsection{Deriving LC from Boolean lattice axioms}
\label{subsubsec:LC_from_Boolean_lattice}

Does BL$\metale$ LC? Let's derive LC axioms from BL axioms and theorems.

The first thing to do is to map the abstract Boolean lattice signature $\langle S,\sqcap,\sqcup,',\bot,\top\rangle$ to the signature we need: $\langle \mathcal{P}(\mathbf{I}),\cap,\cup,',\bm{\varnothing},\mathbf{I}\rangle$, where ``$\mathcal{P}$'' is the ``powerclass-of'' function. After this mapping, we obtain the facts that follow.

LC1 (commutativity) and LC2 (associativity) are semilattice axioms.

LC3 (predicative inhabitation) can be proved as follows:

%\begin{tabular}{@{\hspace{1.2\tabcolsep}} L @{\hspace{1.2\tabcolsep}} l}
\begin{longtable}[c]{@{\hspace{1.2\tabcolsep}} L @{\hspace{1.2\tabcolsep}} l}
\qquad\qquad\qquad\!\mathbf{b}\bm{\varnothing}=\bm{\varnothing} & \{$b\sqcap\bot=\bot$\}\\
\enskip\,\mathbf{c}=\bm{\varnothing}\enskip \!\metale\enskip \mathbf{b}\mathbf{c}=\bm{\varnothing} & \{substitution of similars\}\\
\mathbf{b}\mathbf{c}\neq\bm{\varnothing}~\enskip\!\!\!\metale\enskip \enskip\, \mathbf{c}\neq\bm{\varnothing} & \{transposition (from propositional logic)\}%\footnote{\url{https://en.wikipedia.org/wiki/Transposition\_(logic)}}
\end{longtable}
%\end{tabular}

LC4 (subsumption contraposition) can be proved as follows:

%\begin{tabular}{@{\hspace{1.2\tabcolsep}} L @{\hspace{1.2\tabcolsep}} l}
\begin{longtable}[c]{@{\hspace{1.2\tabcolsep}} L @{\hspace{1.2\tabcolsep}} l}
\multicolumn{2}{l}{Suppose}\\
b\sqcap c=b. &  \\
\multicolumn{2}{l}{Then}\\
(b\sqcap c)'=b' & \{$b=c \metaeq b'=c'$\}\\
b'\sqcup c'=b' & \{$(b\sqcap c)'=b'\sqcup c'$\}\\
c'\sqcup b'=b' & \{$b\sqcup c=c\sqcup b$\}\\
c'\sqcap b'=c' & \{$b\sqcap c=b \enskip\metaeq\enskip b\sqcup c=c$\}\\
\multicolumn{2}{l}{Therefore, $\mathbf{b}\mathbf{c}=\mathbf{b}~\metaeq~\mathbf{c}'\mathbf{b}'=\mathbf{c}'$.}
\end{longtable}
%\end{tabular}

LC5 (disjointness conversion) can be proved as follows:

%\begin{tabular}{@{\hspace{1.2\tabcolsep}} L @{\hspace{1.2\tabcolsep}} l}
\begin{longtable}[c]{@{\hspace{1.2\tabcolsep}} L @{\hspace{1.2\tabcolsep}} l}
\multicolumn{2}{l}{Suppose}\\
b\sqcap c'=b. &  \\
\multicolumn{2}{l}{Then}\\
(b\sqcap c')'=b' & \{$b=c \metaeq b'=c'$\}\\
b'\sqcup (c')'=b' & \{$(b\sqcap c)'=b'\sqcup c'$\}\\
b'\sqcup c=b' & \{$(b')'=b$\}\\
c\sqcup b'=b' & \{$b\sqcup c=c\sqcup b$\}\\
c\sqcap b'=c & \{$b\sqcup c=b \;\metaeq\; b\sqcup c=c$\}\\
\multicolumn{2}{l}{Therefore, $\mathbf{b}\mathbf{c}'=\mathbf{b}~\metaeq~\mathbf{c}\mathbf{b}'=\mathbf{c}$.}
\end{longtable}
%\end{tabular}

Therefore, \textbf{BL$\metale$ LC}.

What about the converse? Does LC$\metale$ BL? No. The BL theorem $(b')'=b$ is neither an axiom nor a theorem that can be proved in LC. Its axioms are not sufficient to prove $bb=b$ either -- nor simple theorems like $(bc)b=cb=(cb)c$.

Therefore, \textbf{LC$\metanleq$ BL}.

By taking together the facts that BL$\metale$ LC and LC$\metanleq$ BL, we conclude that BL is strictly more ``powerful'' than LC, in the sense that we can prove more theorems in BL than in LC.

As we have shown, with the LC representation of an inhabited term it is straightforward to prove the predicative inhabitation law LC3 from Boolean lattice axioms. To our purposes here, this is an important advantage of LC over a pure Leibniz's system, which adopts Leibniz's representation of an inhabited term (Section~\ref{subsec:LC_representations}). No harm is done, since we can easily convert from such a pure Leibniz's system to LC and vice-versa through the following correspondence:

$\mathfrak{I}\langle s\rangle\ \quad\metaeq\quad s\neq\bm{\varnothing}$.

Therefore, Leibniz's sytem and LC have equivalent expressive power.

On a note that fits the importance of \citeauthor{cayley:1871}'s insight for the harmonization between Boolean algebra and the algebra of categorical syllogistic, \citet[p.~2]{green:1991} remarked that

\begin{quote}\guillemotleft[...] It was because of this difficulty of dealing with particular statements that a generally accepted solution of the elimination problem sufficient for a complete treatment of the syllogism came so late in the development of the algebra of logic.\guillemotright\footnote{When put in another context, this remark, in our view, would also be fitting to pay homage to \citeauthor{mccoll:1877}'s insight on the opposition ``$\subseteq$'' vs. ``$\nsubseteq$'', to \citeauthor{ladd:1883}'s insight on the complementary relations ``\!$\bm{\not}\!\Cap$'' vs. ``$\Cap$'', to \citeauthor{mitchell:1883}'s ``$\Cup$'' vs. ``\!$\bm{\not}\!\Cup$'', and to the opposition ``$\supseteq$'' vs. ``$\nsupseteq$'', as we saw in Section~\ref{sec:ML}.}\end{quote}

On the other hand, we should have in mind that Leibniz's system is conceptually more parsimonious than LC, as the former does not require postulating:

\begin{itemize}
    \item The complementary relation to ``$=$''. In no proof of a categorical syllogism in LC (Section~\ref{subsec:LC_proofs}) one needs to use the knowledge that ``$\neq$'' is the complement of ``$=$''; one could even have replaced it by a dyadic relation ``\!\emph{R}'' with unknown properties, and the proofs would have remained the same.

    \item The concept of empty class (``$\bm{\varnothing}$''). Nowhere in the proofs of categorical syllogisms in LC we make use of properties of ``$\bm{\varnothing}$'', such as ``$b\bm{\varnothing}=\bm{\varnothing}$'' -- many of these properties \citet[points~15-22,28-30,39]{leibniz:1686d} knew, by the way.
\end{itemize}

Indeed, in LC ``$\neq$'' and ``$\bm{\varnothing}$'' are only used together --namely in the representations of particular categorical assertions and in LC3--, and can be safely replaced by a monadic relation ``$\mathfrak{I}$'' (``\emph{est Ens}'') from Leibniz's original logic. \citeauthor{cayley:1871}'s contribution, valuable as it is, is not needed in order to prove the classic categorical syllogisms by an equational algebra of logic. His contribution is, above all, a bridge to what is external to the system: it enables an easy correspondence between ideas from Leibniz's pure system and Boolean algebra, allowing us to prove LC3 from BL.

\begin{figure}[ht]
    \centering
    \includegraphics[scale=0.6]{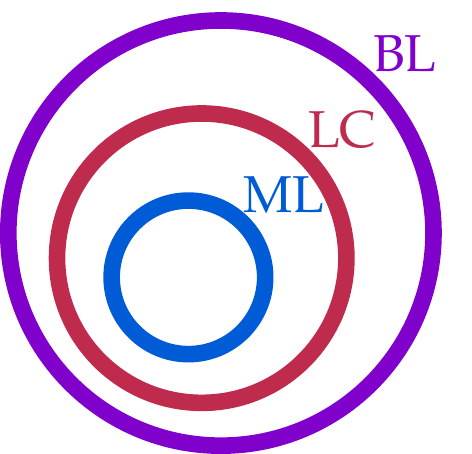}
    \caption{Euler diagram representing the relations between the Boolean lattice, Leibniz-Cayley, and McColl-Ladd axiomatic systems in terms of the set of theorems that can be proved in them.}
    \label{fig:theorem_sets}
\end{figure}

Figure~\ref{fig:theorem_sets} shows the connections we have proved between BL, LC and ML. Since the theorems of ML are a subset of the theorems of LC, the latter theorems are a subset of the theorems of BL, and BL axioms don't lead to (mutually) contradictory conclusions, it follows that ML and LC don't lead to (mutually) contradictory conclusions either.

BL (if interpreted as the Boolean algebra of terms, or of term logic) has the same expressive power as a fragment of \emph{monadic} first-order quantificational logic \citep{simons:2020}\citep[p.~7]{green:1991}\citep[pp.~25-30]{pratt-hartmann:2023}. Thus, we have proved how the hierarchy of expressiveness is constituted from some axiomatizations of classic Aristotelic categorical syllogistic (but without existential import for universal assertions) --a ``toy'' logic with a finite, small number of interesting theorems-- up to first-order logic.

\section{Future work}
\label{sec:future_work}

All proof techniques have their value to illuminating different aspects of categorical syllogistic, and we feel that the ``Euler system'' diagrammatic proof technique deserves as much respect as the algebraic proof techniques shown in this paper and the first-order logic proof technique shown in other catalogs. It has the didactic advantage of being easier to understand --almost intuitive-- and less intimidating for beginners in logic. Although the ``semiformal'' proofs we have reproduced here are well-known, appearing even in Wikimedia Commons \citep{piesk:2017}, what to the best of our knowledge is missing in the literature (and in this paper as well) is the description of all the axioms and inference rules that make the Euler system work. This would be useful for the full formalization of the proofs as a gapless sequence of steps, and would ensure a level of respectability of this diagrammatic proof technique similar to that of the fully formalized, gapless algebraic techniques.

\citet[pp.~54,16]{kraszewski:1956} noticed that, if we rewrite all the moods of the 24 traditional Aristotelic categorical syllogisms into the first figure, we see that only 6 fundamental categorical relations from \citeauthor{de_morgan:1846}'s syllogistic are used in the premises ($\termlogicA$, $\termlogicAumlaut$, $\termlogicE$, $\termlogicI$, $\termlogicO$, $\termlogicOumlaut$ for the premises in the form $\mathbf{s}\termlogicassertion\mathbf{m}$ and $\mathbf{m}\termlogicassertion\mathbf{p}$, with in some cases the addition of a premise $\mathbf{s}\termlogicI\mathbf{s}$, $\mathbf{m}\termlogicI\mathbf{m}$ or $\mathbf{p}\termlogicI\mathbf{p}$) and only 4 are used in the conclusion ($\termlogicA$, $\termlogicE$, $\termlogicI$, $\termlogicO$ in the form $\mathbf{s}\termlogicassertion\mathbf{p}$). Missing in the enumeration are valid categorical syllogisms with premises in the remaining 2 fundamental categorical relations ($\termlogicEumlaut$, $\termlogicIumlaut$) and the conclusion in the 4 fundamental categorical relations with umlaut. In a future work, they should be enumerated, and we should check whether the axioms presented in this paper --perhaps with the replacement of one of the axioms in each symbolic system by the axiom of involution of complementation: $(\mathbf{b}')'=\mathbf{b}$-- are sufficient to prove all the valid categorical syllogisms in \citeauthor{de_morgan:1846}'s syllogistic.

With our proof methods, whenever we are able to construct a proof, we do know a conclusion necessarily follows from the premises (if the proof is correct). However, when we are unable to construct a proof, our proof methods do not provide us tools to know whether there exists a proof (and we only lacked the skill to construct it) or whether the conclusion does not necessarily follow from the premises (in which case no proof of the necessity of that conclusion is possible). For instance:\\P1: Every \textbf{s} is \textbf{m}.\\P2: Every \textbf{m} is \textbf{p}.\\C1: Therefore: At least one \textbf{s} is \textbf{p}.\\The conclusion is not incompatible with the premises, though it is not necessary either. It might be the case that ``No \textbf{s} is \textbf{p}.'' instead -- namely, when $\mathbf{s}=\bm{\varnothing}$ / $\mathbf{s}\!\!\bm{\not}\!\Cap\:\mathbf{s}$. This limitation of our proof methods implies that they should also be complemented by techniques for rejection proofs of invalid categorical syllogisms, for completeness\footnote{The logician David Makinson, in private correspondence with the main author, claimed that ``it can be done semantically by considering models of up to 8 elements''.}. It would be even better if the rejection proofs helped us to discriminate invalid conclusions which are incompatible with the premises (and thus the complement of the former relations would be necessary conclusions) from invalid conclusions which are compatible with the premises but not necessary (and thus the complement of the former relations would neither be incompatible nor necessary conclusions as well). For instance, with the premises of Barbara-1 --$\mathbf{s}\termlogicA\mathbf{m},~\mathbf{m}\termlogicA\mathbf{p}$--, any conclusion $\mathbf{s}\termlogicassertion\mathbf{p}$ where ``$\termlogicassertion$'' is not ``$\termlogicA$'' is invalid; ``$\termlogicO$'' (the complementary relation to ``$\termlogicA$'') would be an incompatible conclusion, and remaining relations would be compatible but non-necessary conclusions -- we wish those facts to be proved in an style similar to our LC and ML systems. A system is said to be refutationally complete if each formula of its language is either a theorem or a rejected formula \citep[p.~10]{kulicki:2020}\citep[pp.~578--582]{wybraniec:2018}. Since this logic --categorical syllogistic-- is finite, given that we can enumerate all possible combinations of the 8 basic relations in the strict format ``premise 1 - premise 2 - candidate conclusion'', what we wish is defining decidable, refutationally complete systems extending LC and ML.

\section{Conclusion}

We have described two symbolic axiomatic systems --Leibniz-Cayley (Section~\ref{sec:LC}) and McColl-Ladd (Section~\ref{sec:ML})-- which are sufficiently powerful to prove all the 24 classic categorical syllogisms in the term logic tradition founded by Aristotle, as made visual by the diagrammatic Euler system (Section~\ref{sec:euler}), thus totalling three systems presented by this paper (one diagrammatic and two symbolic ones). Our main novel result is summarized in Figure~\ref{fig:theorem_sets}.

We have also unveiled new proofs of known theorems -- the 24 classic categorical syllogisms. All proofs are short and don't fly over lay people's heads. We claim that, unlike first-order logic proofs, the diagrammatic, algebraic and relational techniques we have adopted from the literature on the topic are understandable to mathematically curious students who are finishing the middle school and that, by providing them all the axioms and three or four proofs of sample theorems, they should be able to prove all remaining theorems\footnote{The categorical syllogisms Bamalip-4, Disamis-3, Camestres-2 and Cesare-2 have representative proofs that cover together all axioms proposed in this paper for LC. (For ML, Camestres-2, Dimatis-4 and Cesare-2.) As an exercise, a teacher could show her students the axioms and those proofs and ask them to prove the remaining classic categorical syllogisms.}. This is because our algebraic proofs can piggyback on their familiarity to elementary numerical algebra, which has many analogous properties to the algebra of classic logic (such as substitutability of equals, commutativity and associativity of certain operations), and they are short and don't impose much cognitive burden, thus reducing the likelihood that the learners will get lost. Moreover, we hypothesize that the Euler diagram technique is immediately ``intuitive'' even to ordinary students not skilled enough to reproduce the proofs, who would at a minimum be able to visually track the proof steps and grasp them. Thus, these diagrammatic proofs are particularly effective to mathematically-averse Philosophy students who are required to deal with Aristotelic logic in their course curriculum.

The conventional proofs in monadic first-order quantificational logic à la Frege and Peirce \citep{tennant:2014, metamath:2021, argyraki:2019} are a great way (or a gateway) to introduce categorical syllogistic students into first-order logic. Our proofs dispel the misconceptions that (\emph{a}) the mathematical treatment of categorical syllogisms requires methods from first-order logic, and that (\emph{b}) first-order logic makes Leibnizian/Boolean methods ``obsolete''.

\section*{Acknowledgements}

This research project received no funding other than the authors' ordinary salaries by their respective institutions, to which they are thankful. This paper wouldn't have been possible hadn't \citet{couturat:1903} worked so diligently as a book editor to bring Leibniz's spectacular drafts on Logic to the knowledge of the wider public. Previous versions of most Euler diagrams in this paper were made to order by Isis Azevedo Sylvestre by closely following the main author's sketches. Some paragraphs and footnotes of a draft of this paper were improved after recommendations by David Clement Makinson and the anonymous reviewer \#2 from the ``\emph{History and Philosophy of Logic}'' journal. Guillermo Badia kindly endorsed the main author to the \emph{math.LO} category of arXiv, which fulfilled a formal requirement for paper submission.

\section*{Disclosure statement}

Division of work: AGR wrote this survey and proved the results; EMD was AGR's formal research advisor and authorized the submission of the paper. This arrangement is typical and expected in the Brazilian academic culture, which may differ in particular points from established cultural practices in some other places. Without the advisor's kind invitation for AGR to join the graduate program to advance pharmaceutical drug traceability technology, this paper on Logic would never have been written; AGR is very thankful to EMD for the opportunity to join his research group.

The authors have no competing financial interests to report.

\bibliography{syllogistic}

\begin{thebibliography}{163}
\newcommand{\enquote}[1]{``#1''}
\providecommand{\natexlab}[1]{#1}
\providecommand{\url}[1]{\normalfont{#1}}
\providecommand{\urlprefix}{}

\bibitem[Anellis(2007)]{anellis:2007}
Anellis, Irving~H. 2007. ``{T}he Rise of Modern Logic: From {L}eibniz to
  {F}rege. Review-Essay of `{H}andbook of the History of Logic', Volume 3.''
  \emph{Review of Modern Logic} 10 (3/4): 13--90.
  \url{https://projecteuclid.org/journals/review-of-modern-logic/volume-10/issue-3-4/The-Rise-of-Modern-Logic--From-Leibniz-to-Frege/rml/1312203772.full}
  \url{https://projecteuclid.org/download/pdf_1/euclid.rml/1312203772}.

\bibitem[{Aristotle of Stagira}(ca. 350 BCE{\natexlab{a}})]{owen:1853a}
{Aristotle of Stagira}. ca. 350 BCE{\natexlab{a}}. ``On {I}nterpretation,
  chapter 7.'' In \emph{Organon, or logical treatises, of {A}ristotle},  edited
  by Octavius~Freire Owen, year 1853 ed., Vol.~1. London:UK: Henry G. Bohn.
  \url{https://en.wikisource.org/wiki/Organon_(Owen)/On_Interpretation}.

\bibitem[{Aristotle of Stagira}(ca. 350 BCE{\natexlab{b}})]{mure:1926}
{Aristotle of Stagira}. ca. 350 BCE{\natexlab{b}}. ``Posterior {A}nalytics.''
  In \emph{The Works of {A}ristotle},  edited by Geoffrey Reginald~Gilchrist
  Mure, year 1926 ed., 221--325. Oxford:UK: Clarendon Press.
  \url{https://www.logicmuseum.com/authors/aristotle/posterioranalytics/posterioranalytics.htm#:~:text=77a10}
  \url{https://archive.org/details/AristotleOrganon/page/n243/mode/1up}.

\bibitem[{Aristotle of Stagira}(ca. 350 BCE{\natexlab{c}})]{owen:1853b}
{Aristotle of Stagira}. ca. 350 BCE{\natexlab{c}}. ``Prior {A}nalytics, book 1,
  chapter 2.'' In \emph{Organon, or logical treatises, of {A}ristotle},  edited
  by Octavius~Freire Owen, year 1853 ed., Vol.~1. London:UK: Henry G. Bohn.
  \url{https://en.wikisource.org/wiki/Organon_(Owen)/Prior_Analytics/Book_1#Chapter_2}.

\bibitem[Bar-Am(2008)]{bar-am:2008}
Bar-Am, Nimrod. 2008. \emph{Extensionalism: The Revolution in Logic}. Springer.
  \url{https://books.google.com/books?id=dWp2HUQl-0UC}
  \url{https://doi.org/10.1007/978-1-4020-8168-2}.

\bibitem[Bennett(1994)]{bennett:1994}
Bennett, Brandon. 1994. ``Spatial Reasoning with Propositional Logics.'' In
  \emph{Principles of Knowledge Representation and Reasoning: Proceedings of
  the {F}ourth {I}nternational {C}onference ({KR} '94)}, 51--62.
  \url{https://doi.org/10.1016/B978-1-4832-1452-8.50102-0}
  \url{https://citeseerx.ist.psu.edu/doc_view/pid/4c45519c2db0dac5ceaa76e1b53b1ca3c0bfce00}.

\bibitem[Bennett(2015)]{bennett:2015}
Bennett, Deborah. 2015. ``Origins of the {V}enn Diagram.'' In \emph{Research in
  {H}istory and {P}hilosophy of {M}athematics}, 105--119. Springer.
  \url{https://logic-teaching.github.io/pred/texts/Bennett%202015%20-%20Origins%20of%20the%20Venn%20Diagram.pdf}.

\bibitem[Birkhoff(1938)]{birkhoff:1938}
Birkhoff, Garrett. 1938. ``Lattices and their applications.'' \emph{Bulletin of
  the {A}merican {M}athematical {S}ociety} 44 (12): 793--800.
  \url{https://projecteuclid.org/journals/bulletin-of-the-american-mathematical-society/volume-44/issue-12.P1/Lattices-and-their-applications/bams/1183500966.full}.

\bibitem[Birkhoff(1940)]{birkhoff:1940}
Birkhoff, Garrett. 1940. \emph{Lattice Theory}. American Mathematical Society.
  \url{https://books.google.com/books?id=RhxMAQAAMAAJ&q=%22complemented+distributive%22}.

\bibitem[Boole(1847)]{boole:1847}
Boole, George. 1847. \emph{The Mathematical Analysis of Logic, being an essay
  towards a calculus of deductive reasoning}. London:UK: Macmillan, Barclay, \&
  Macmillan. \url{https://archive.org/details/mathematicalanal00booluoft}
  \url{https://books.google.com/books?id=lD3DAgAAQBAJ&pg=PA45}.

\bibitem[Boole(1854)]{boole:1854}
Boole, George. 1854. \emph{An Investigation of the Laws of Thought, on which
  are founded the mathematical theories of Logic and Probabilities}. London:UK:
  Walton and Maberly.
  \url{https://archive.org/details/investigationofl00boolrich/page/49/mode/1up}.

\bibitem[Boolos(1985)]{boolos:1985}
Boolos, George. 1985. ``Nominalist Platonism.'' \emph{The Philosophical Review}
  94 (3): 327--344. \url{https://www.jstor.org/stable/2185003}.

\bibitem[Braem(1475)]{braem:1475}
Braem, Konrad, ed. 1475. \emph{Analytica {P}riora}. Löwen:Belgium.
  \url{https://books.google.com/books?id=dnUUIZ8YoBQC&pg=PT8&dq=%22fi+a+oni+b%22}.

\bibitem[Brentano(1874)]{brentano:1874}
Brentano, Franz Clemens Honoratus Hermann~Josef. 1874. \emph{Psychologie vom
  empirischen {S}tandpunkte}. Leipzig:Germany: Duncker \& Humblot.
  \url{https://archive.org/details/psychologievome02brengoog/page/n301/mode/1up}
  \url{https://books.google.com/books?id=8BUalavx0QQC&pg=PA283}
  \url{https://books.google.com/books?id=HEeSpAp7mtQC&pg=PA283}.

\bibitem[Brown(2012)]{brown:2012}
Brown, Richard~C. 2012. \emph{The Tangled Origins of the {L}eibnizian Calculus:
  A Case Study of a Mathematical Revolution}. World Scientific.
  \url{https://books.google.com/books?id=s4M7qp02cGMC&pg=PA165}.

\bibitem[Béziau(2012)]{beziau:2012}
Béziau, Jean-Yves. 2012. ``The Power of the Hexagon.'' \emph{{L}ogica
  {U}niversalis} 6: 1--43. \url{https://doi.org/10.1007/s11787-012-0046-9}.

\bibitem[Carroll(1886)]{carroll:1886}
Carroll, Lewis. 1886. \emph{The Game of Logic}. London:UK: Macmillan and Co.
  \url{https://archive.org/details/gameoflogic00carrrich}
  \url{https://en.wikisource.org/wiki/The_Game_of_Logic}.

\bibitem[Cayley(1871)]{cayley:1871}
Cayley, Arthur. 1871. ``Note on the calculus of logic.'' \emph{Quarterly
  Journal of Pure and Applied Mathematics} XI: 282--283.
  \url{https://archive.org/details/collmathpapers08caylrich/page/n128/mode/1up}
  \url{https://books.google.com/books?id=4gWd3fZi4zIC&pg=PA65}
  \url{https://doi.org/10.1017/CBO9780511703744.015}
  \url{https://geographiclib.sourceforge.io/geodesic-papers/cayley-V8.pdf#page=123}.

\bibitem[Church(1932)]{church:1932}
Church, Alonzo. 1932. ``A Set of Postulates for the Foundation of Logic.''
  \emph{Annals of Mathematics} 33 (2): 346--366.
  \url{https://www.jstor.org/stable/1968337?seq=10}
  \url{https://doi.org/10.2307/1968337}.

\bibitem[Cohn and Gotts(1994)]{cohn:1994}
Cohn, Anthony~G., and Nicholas~Mark Gotts. 1994. ``A Theory of Spatial Regions
  with Indeterminate Boundaries.'' In \emph{``{T}opological Foundations of
  Cognitive Science'' Workshop at the {F}irst {I}nternational {S}ummer
  {I}nstitute in {C}ognitive {S}cience ({FISI-CS})},  edited by Carola
  Eschenbach, Christopher Habel, and Barry Smith, Buffalo, NY:USA, July,
  131--150.
  \url{https://citeseerx.ist.psu.edu/doc_view/pid/b8cc6727ffd2b77a29130eb4d128901d115f7960}
  \url{https://www.researchgate.net/publication/2622707_A_Theory_of_Spatial_Regions_with_Indeterminate_Boundaries}
  \url{https://philarchive.org/archive/ESCTFO}
  \url{https://philarchive.org/archive/SMITFO-2}.

\bibitem[Copi, Cohen, and Flage(2016)]{copi:2016}
Copi, Irving, Carl Cohen, and Daniel Flage. 2016. \emph{Essentials of Logic},
  2nd ed., Chap. 3.3 {S}ymbolism and {V}enn Diagrams for Categorical
  Propositions. {T}aylor \& {F}rancis.
  \url{https://books.google.com/books?id=6eAqDwAAQBAJ&pg=PA106}.

\bibitem[Couturat(1901)]{couturat:1901}
Couturat, Louis. 1901. \emph{La logique de {Leibniz}: d'après des documents
  inédits}. Paris:France: Félix Alcan.
  \url{https://archive.org/details/lalogiquedeleib00coutgoog}.

\bibitem[Couturat(1903)]{couturat:1903}
Couturat, Louis, ed. 1903. \emph{Opuscules et fragments inédits de {L}eibniz:
  Extraits des manuscrits de la Bibliothèque royale de {H}anovre}.
  Paris:France: Félix Alcan.
  \url{https://archive.org/details/opusculesetfrag00coutgoog}.

\bibitem[{D}e {M}organ(1846)]{de_morgan:1846}
{D}e {M}organ, Augustus. 1846. ``On the Syllogism, {I}: On the Structure of the
  Syllogism, and on the Application of the Theory of Probabilities to Questions
  of Argument and Authority.'' In \emph{{T}ransactions of the {C}ambridge
  {P}hilosophical {S}ociety}, Vol.~8, November, 379--408.
  \url{https://archive.org/details/transactionsofca08camb/page/379/mode/1up}
  \url{https://books.google.com/books?id=t02wDwAAQBAJ&pg=PA1}
  \url{https://books.google.com/books?id=2vsIAAAAIAAJ&pg=PA379}.

\bibitem[{D}e {M}organ(1847)]{de_morgan:1847}
{D}e {M}organ, Augustus. 1847. \emph{Formal logic: or, The Calculus of
  Inference, Necessary and Probable}. London:UK: Taylor and Walton.
  \url{https://archive.org/details/formallogicorca01morggoog/page/n83/mode/2up}
  \url{https://books.google.com/books?id=HscAAAAAMAAJ&pg=PA60}
  \url{https://books.google.com/books?id=LOFAVGS-8oAC&pg=PA60}.

\bibitem[{D}e {M}organ(1858)]{de_morgan:1858}
{D}e {M}organ, Augustus. 1858. ``On the Syllogism, No. {III}, and on Logic in
  general.'' In \emph{{T}ransactions of the {C}ambridge {P}hilosophical
  {S}ociety}, Vol.~10, February, 173--230.
  \url{https://archive.org/details/transactionsofca10camb/page/173/mode/1up}
  \url{https://books.google.com/books?id=2vIIAAAAIAAJ&pg=PA217&dq=%22The+natural+character+of+the+first+figure%22}.

\bibitem[{D}e {M}organ(1860)]{de_morgan:1860}
{D}e {M}organ, Augustus. 1860. ``On the Syllogism, No. {IV}, and on the Logic
  of Relations.'' In \emph{{T}ransactions of the {C}ambridge {P}hilosophical
  {S}ociety}, Vol.~10, April, 331--358.
  \url{https://archive.org/details/transactionsofca10camb/page/331/mode/1up}
  \url{https://books.google.com/books?id=2vIIAAAAIAAJ&pg=PA331}.

\bibitem[de~Nemore(ca. 1225)]{de_nemore:1225}
de~Nemore, Jordanus. ca. 1225. \emph{De Numeris Datis}.
  \url{https://books.google.com/books?id=AUR0c5rR_yoC&pg=PA58}.

\bibitem[Dekker(2015)]{dekker:2015}
Dekker, Paul J.~E. 2015. ``Not Only {B}arbara.'' \emph{Journal of Logic,
  Language and Information} 24: 95--129.
  \url{https://doi.org/10.1007/s10849-015-9215-6#:~:text=its%20dual}
  \url{https://www.researchgate.net/publication/273526737_Not_Only_Barbara}.

\bibitem[Erdmann(1840)]{erdmann:1840}
Erdmann, Johann~Eduard, ed. 1840. \emph{God. {G}uil. {L}eibnitii Opera
  philosophica quae exstant {L}atina {G}allica {G}ermanica omnia}.
  Berlin:Prussia: Sumtibus G. Eichleri.
  \url{https://archive.org/details/godguilleibniti00erdmgoog}
  \url{https://books.google.com/books?id=SWOSd6gCCPYC}.

\bibitem[{Euclid of Alexandria}(ca. 300 BCE)]{euclid:300bce}
{Euclid of Alexandria}. ca. 300 BCE. \emph{Elements (Stoikheîa) of Geometry,
  Book 1}. \url{http://aleph0.clarku.edu/~djoyce/elements/bookI/bookI.html#cns}
  \url{http://aleph0.clarku.edu/~djoyce/elements/bookI/cn.html}.

\bibitem[Euler(1770)]{euler:1770}
Euler, Leonhard. 1770. \emph{Lettres a une princesse d'{A}llemagne sur divers
  sujets de physique \& de philosophie}. Vol.~2. Chez Steidel et Compagnie.
  \url{https://books.google.com/books?id=CIn1qBshhOQC&pg=PA99}
  \url{https://www.e-rara.ch/zut/content/zoom/2380253}
  \url{https://books.google.com/books?id=gxsAAAAAQAAJ&pg=PA95}.

\bibitem[Flage(2002)]{flage:2002}
Flage, Daniel. 2002. ``Boolean {E}uler diagrams.'' \emph{{APA} Newsletter on
  Teaching {P}hilosophy} 1 (2): 185--190.
  \url{https://cdn.ymaws.com/www.apaonline.org/resource/collection/808CBF9D-D8E6-44A7-AE13-41A70645A525/v01n2Teaching.pdf#page=11}.

\bibitem[Forrest(2020)]{forrest:2018}
Forrest, Peter. 2020. ``The Identity of Indiscernibles.'' In \emph{The
  {S}tanford Encyclopedia of Philosophy},  edited by Edward~N. Zalta.
  Metaphysics Research Lab, Stanford University.
  \url{https://plato.stanford.edu/archives/win2020/entries/identity-indiscernible/}.

\bibitem[Gerhardt(1890)]{gerhardt:1890}
Gerhardt, Carl~Immanuel, ed. 1890. \emph{Die philosophischen {S}chriften von
  {G}ottfried {W}ilhelm {L}eibniz}. Vol.~7. Berlin:Germany: Weidmann.
  \url{https://archive.org/details/diephilosophisc01leibgoog}.

\bibitem[Grassmann(1872)]{grassmann:1872}
Grassmann, Robert. 1872. \emph{{D}ie {F}ormenlehre oder {M}athematik}. Vol. 2:
  {D}ie {B}egriffslehre oder {L}ogik. Stettin:German Empire: Robert Grassmann.
  \url{https://books.google.com/books?id=UyVLAAAAYAAJ&pg=RA1-PA15&dq=%22die+Totalit%C3%A4t%22+%22ist+T%22}
  \url{https://books.google.com/books?id=YSY3AAAAYAAJ&pg=PA20&dq=40}
  \url{https://digitale-sammlungen.de/de/view/bsb11122944}.

\bibitem[Grattan-Guinness(2000)]{grattan-guinness:2000}
Grattan-Guinness, Ivor~Owen. 2000. \emph{The Search for Mathematical Roots,
  1870-1940: Logics, Set Theories and the Foundations of Mathematics from
  {C}antor Through {R}ussell to {G}ödel}. Princeton:USA: Princeton University
  Press.
  \url{https://books.google.com/books?id=2Y-_YTazbYUC&pg=PA158&dq=%22non-identity%22}
  \url{https://books.google.com/books?id=LcCx9MROpbUC&pg=PA158&dq=%22non-identity%22&redir_esc=y}
  \url{https://www.jstor.org/stable/j.ctt7rp8j.6}.

\bibitem[Green(1991)]{green:1991}
Green, Judy. 1991. ``The problem of elimination in the algebra of logic.'' In
  \emph{Perspectives on the history of mathematical logic},  edited by Thomas
  Drucker, 1--9. Boston:USA: Birkhäuser.
  \url{https://doi.org/10.1007/978-0-8176-4769-8_1}
  \url{https://books.google.com/books?id=iYHuBwAAQBAJ&pg=PA1}.

\bibitem[Hailperin(1986)]{hailperin:1986}
Hailperin, Theodore. 1986. \emph{Boole's Logic and Probability: A Critical
  Exposition From the Standpoint of Contemporary Algebra, Logic, and
  Probability Theory}. 2nd ed. Amsterdam:Netherlands: Elsevier.
  \url{https://books.google.com/books?id=TS8hSCfmBQgC&pg=PA139}
  \url{http://ndl.ethernet.edu.et/bitstream/123456789/77335/1/61.pdf}.

\bibitem[Halsted(1883)]{halsted:1883}
Halsted, George~Bruce. 1883. ``The modern logic.'' \emph{The Journal of
  Speculative Philosophy} 17 (2): 210--213.
  \url{https://www.jstor.org/stable/25667964}.

\bibitem[Hamilton(1843)]{hamilton:1843}
Hamilton, William~Rowan. 1843. ``On a new Species of Imaginary Quantities,
  connected with a theory of Quaternions.'' \emph{Proceedings of the {R}oyal
  {I}rish {A}cademy} 2: 424--434.
  \url{https://www.jstor.org/stable/20520177?seq=7}
  \url{https://www.maths.tcd.ie/pub/HistMath/People/Hamilton/Quatern1/Quatern1.html}.

\bibitem[Hammer(1995)]{hammer:1995}
Hammer, Eric. 1995. ``{P}eirce on Logical Diagrams.'' \emph{{T}ransactions of
  the {C}harles {S}. {P}eirce {S}ociety} 31 (4): 807--827.
  \url{https://www.jstor.org/stable/40320573}.

\bibitem[Hartshorne and Weiss(1960)]{cp:1960}
Hartshorne, Charles, and Paul Weiss, eds. 1960. \emph{{C}ollected {P}apers of
  {C}harles {S}anders {P}eirce}. Vol. 3-4. Cambridge,MA:USA: {H}arvard
  {U}niversity {P}ress.
  \url{https://books.google.com/books?id=3JJgOkGmnjEC&pg=RA1-PA308}
  \url{https://books.google.com/books?id=x2YNAQAAMAAJ&q=%22these+three+forms%22}.

\bibitem[Hemann, Swords, and Moss(2015)]{hemann:2015}
Hemann, Jason, Cameron Swords, and Lawrence~S. Moss. 2015. ``Two Advances in
  the Implementations of Extended Syllogistic Logics.'' In \emph{Joint
  Proceedings of the 2nd Workshop on Natural Language Processing and Automated
  Reasoning, and the 2nd International Workshop on Learning and Nonmonotonic
  Reasoning (Proceedings of the Joint {LNMR/NLPAR} Workshop) at {LPNMR} 2015},
  edited by Marcello Balduccini, Alessandra Mileo, Ekaterina Ovchinnikova,
  Alessandra Russo, and Peter Schüller, 1--14.
  \url{http://cswords.com/paper/syl.logic.pdf#page=3}
  \url{https://peterschueller.com//pub/2015/nlpar2015-proceedings.pdf#page=7}.

\bibitem[Hilbert(1922)]{hilbert:1922}
Hilbert, David. 1922. ``Neubegründung der {M}athematik. Erste {M}itteilung.''
  In \emph{Abhandlungen aus dem {M}athematischen {S}eminar der {U}niversität
  {H}amburg}, Vol.~1, Hamburg:Germany, 157--177. Springer.
  \url{https://doi.org/10.1007/BF02940589}.

\bibitem[Huntington(1904)]{huntington:1904}
Huntington, Edward~Vermilye. 1904. ``Sets of Independent Postulates for the
  Algebra of Logic.'' \emph{Transactions of the {A}merican {M}athematical
  {S}ociety} 5 (3): 288--309. \url{https://www.jstor.org/stable/1986459}
  \url{https://doi.org/10.2307/1986459}.

\bibitem[Huntington(1933)]{huntington:1933}
Huntington, Edward~Vermilye. 1933. ``New sets of independent postulates for the
  algebra of logic, with special reference to {W}hitehead and {R}ussell's
  {P}rincipia mathematica.'' \emph{Transactions of the {A}merican
  {M}athematical {S}ociety} 35 (1): 274--304.
  \url{https://doi.org/10.1090/S0002-9947-1933-1501684-X}
  \url{https://www.jstor.org/stable/1989325}.

\bibitem[{Icard, III}(2014)]{icard:2014}
{Icard, III}, Thomas~{F}. 2014. ``Higher-order syllogistics.'' In
  \emph{Proceedings of the 19th International Conference on Formal Grammar
  ({LNCS} 8612)},  edited by Glyn Morrill, Reinhard Muskens, Rainer Osswald,
  and Frank Richter, 1--14. Springer.
  \url{https://books.google.com/books?id=8uIkBAAAQBAJ&pg=PA11}.

\bibitem[Janssen-Lauret(2023)]{janssen-lauret:2023}
Janssen-Lauret, Frederique. 2023. ``Constance {J}ones, {C}hristine
  {L}add-{F}ranklin, and {V}ictoria {W}elby: Grandmothers of Analytic
  Philosophy in {M}ind, 1888-1911.'' In \emph{The Early Years of {M}ind: Making
  Contemporary {P}hilosophy and {P}sychology (forthcoming)},  edited by
  Lukas~M. Verburgt, Oxford:UK, 515--528. Oxford University Press.
  \url{https://research.manchester.ac.uk/en/publications/constance-jones-christine-ladd-franklin-and-victoria-welby-grandm}
  \url{https://pure.manchester.ac.uk/ws/portalfiles/portal/260437127/Jones_Ladd_Franklin_Welby.pdf}.

\bibitem[Jevons(1864)]{jevons:1864}
Jevons, William~Stanley. 1864. \emph{Pure Logic, or the Logic of Quality apart
  from Quantity: with Remarks on {B}oole's System, and on the Relation of Logic
  and Mathematics}. London:UK: Edward Stanford.
  \url{https://archive.org/details/purelogicorlogi00jevogoog}
  \url{https://books.google.com/books?id=WVMOAAAAYAAJ}.

\bibitem[Jevons(1869)]{jevons:1869}
Jevons, William~Stanley. 1869. \emph{The Substitution of Similars: the True
  Principle of Reasoning, Derived from a Modification of {A}ristotle's Dictum}.
  London:UK: McMillan and Co.
  \url{https://archive.org/details/substitutionofsi00jevorich}
  \url{https://books.google.com/books?id=TA5VAAAAMAAJ}.

\bibitem[Klement(2010)]{klement:2010}
Klement, Kevin~C. 2010. ``The functions of {R}ussell's no class theory.''
  \emph{The Review of Symbolic Logic} 3 (4): 633--664.
  \url{https://people.umass.edu/klement/noclass-published.pdf}
  \url{https://doi.org/10.1017/S1755020310000225}.

\bibitem[Koszkało(2017)]{koszkalo:2017}
Koszkało, Martyna. 2017. ``Scholastic sources of {G}ottfried {W}ilhelm
  {L}eibniz's treatise {D}isputatio Metaphysica de Principio Individui.''
  \emph{Roczniki Filozoficzne} 65 (2): 23--55.
  \url{http://dx.doi.org/10.18290/rf.2017.65.2-2}.

\bibitem[Koutsoukou-Argyraki(2019)]{argyraki:2019}
Koutsoukou-Argyraki, Angeliki. 2019. ``Aristotle's Assertoric Syllogistic.''
  \emph{Archive of Formal Proofs -- {I}sabelle}
  \url{https://isa-afp.org/entries/Aristotles_Assertoric_Syllogistic.html}.

\bibitem[Kraszewski(1956)]{kraszewski:1956}
Kraszewski, Zdzisław. 1956. ``Logika stosunków zakresowych ({R}achunek zdań
  zakresowych).'' \emph{Studia {L}ogica} 4: 63--116.
  \url{https://www.jstor.org/stable/20013548?seq=54}
  \url{https://www.jstor.org/stable/20013548?seq=16}.

\bibitem[Kulicki(2012)]{kulicki:2012}
Kulicki, Piotr. 2012. ``An Axiomatisation of a Pure Calculus of Names.''
  \emph{Studia Logica} 100 (5): 921--946.
  \url{https://doi.org/10.1007/s11225-012-9441-8}
  \url{https://www.jstor.org/stable/23324800}
  \url{https://www.researchgate.net/publication/257665696_An_Axiomatisation_of_a_Pure_Calculus_of_Names}.

\bibitem[Kulicki(2020)]{kulicki:2020}
Kulicki, Piotr. 2020. ``Aristotle's Syllogistic as a Deductive System.''
  \emph{Axioms} 9 (2). \url{https://doi.org/10.3390/axioms9020056}
  \url{http://hdl.handle.net/20.500.12153/1138}
  \url{https://www.researchgate.net/publication/341495044_Aristotle's_Syllogistic_as_a_Deductive_System}.

\bibitem[Ladd(1883)]{ladd:1883}
Ladd, Christine. 1883. ``On the algebra of logic.'' In \emph{Studies in {L}ogic
  by members of the {J}ohns {H}opkins {U}niversity},  edited by Charles~Sanders
  Peirce, 17--71. Boston:USA: Little, Brown, and Company.
  \url{https://archive.org/details/bub_gb_V7oIAAAAQAAJ/page/n25/mode/1up}
  \url{https://archive.org/details/studiesinlogic00gilmgoog/page/n34/mode/1up}
  \url{https://archive.org/details/studiesinlogic00peiruoft/page/17/mode/1up}.

\bibitem[{Ladd Franklin}(1890)]{ladd:1890}
{Ladd Franklin}, Christine. 1890. ``Some Proposed Reforms in Common Logic.''
  \emph{Mind} 15 (57): 75--88.
  \url{https://archive.org/details/Mind2/page/n104/mode/1up}
  \url{https://doi.org/10.1093/mind/os-XV.57.75}
  \url{https://www.jstor.org/stable/2247362?seq=12}.

\bibitem[Land(1876)]{land:1876}
Land, Jan Pieter~Nicolaas. 1876. ``Brentano's Logical Innovations.''
  \emph{Mind} 1 (2): 289--292.
  \url{http://www.logicmuseum.com/opposition/brentanoinnovations.htm}
  \url{https://www.jstor.org/stable/2246513}.

\bibitem[Lawvere(1991)]{lawvere:1991}
Lawvere, Francis~William. 1991. ``Intrinsic co-{H}eyting boundaries and the
  {L}eibniz rule in certain toposes.'' In \emph{Category Theory},  edited by
  Aurelio Carboni, Maria~Cristina Pedicchio, and Guiseppe Rosolini, 279--281.
  Springer.
  \url{https://github.com/mattearnshaw/lawvere/blob/master/pdfs/1991-intrinsic-co-heyting-boundaries-and-the-leibniz-rule-in-certain-toposes.pdf}.

\bibitem[Leibniz(1679{\natexlab{a}})]{leibniz:1679b}
Leibniz, Gottfried~Wilhelm. 1679{\natexlab{a}}. ``{E}lementa {C}alculi.'' In
  \emph{Opuscules et fragments inédits de {L}eibniz: Extraits des manuscrits
  de la Bibliothèque royale de {H}anovre},  edited by Louis Couturat, 1903rd
  ed., 49--57. Paris:France: Félix Alcan.
  \url{https://archive.org/details/opusculesetfrag00coutgoog/page/n79/mode/1up}
  \url{https://doi.org/10.1007/978-94-010-1426-7_27}.

\bibitem[Leibniz(1679{\natexlab{b}})]{leibniz:1679a}
Leibniz, Gottfried~Wilhelm. 1679{\natexlab{b}}. ``{E}lementa {C}haracteristicae
  universalis.'' In \emph{Opuscules et fragments inédits de {L}eibniz:
  Extraits des manuscrits de la Bibliothèque royale de {H}anovre},  edited by
  Louis Couturat, 1903rd ed., 42--49. Paris:France: Félix Alcan.
  \url{https://archive.org/details/opusculesetfrag00coutgoog/page/n72/mode/1up}.

\bibitem[Leibniz(1682)]{leibniz:1682}
Leibniz, Gottfried~Wilhelm. 1682. ``{D}e formis syllogismorum {M}athematice
  definiendis.'' In \emph{Opuscules et fragments inédits de {L}eibniz:
  Extraits des manuscrits de la Bibliothèque royale de {H}anovre},  edited by
  Louis Couturat, 1903rd ed., 410--416. Paris:France: Félix Alcan.
  \url{https://archive.org/details/opusculesetfrag00coutgoog/page/410/mode/1up}.

\bibitem[Leibniz(1690{\natexlab{a}})]{leibniz:1690c}
Leibniz, Gottfried~Wilhelm. 1690{\natexlab{a}}. ``{F}undamenta {C}alculi
  {L}ogici.'' In \emph{Opuscules et fragments inédits de {L}eibniz: Extraits
  des manuscrits de la Bibliothèque royale de {H}anovre},  edited by Louis
  Couturat, 1903rd ed., 421--423. Paris:France: Félix Alcan.
  \url{https://archive.org/details/opusculesetfrag00coutgoog/page/421/mode/1up}.

\bibitem[Leibniz(1690{\natexlab{b}})]{leibniz:1690b}
Leibniz, Gottfried~Wilhelm. 1690{\natexlab{b}}. ``{P}rimaria {C}alculi {L}ogici
  fundamenta.'' In \emph{Opuscules et fragments inédits de {L}eibniz: Extraits
  des manuscrits de la Bibliothèque royale de {H}anovre},  edited by Louis
  Couturat, 1903rd ed., 232--237. Paris:France: Félix Alcan.
  \url{https://archive.org/details/opusculesetfrag00coutgoog/page/232/mode/1up}
  \url{https://books.google.com/books?id=WXFLDwAAQBAJ&pg=PT249&dq=%22The+first+draft%22}.

\bibitem[Leibniz(ca. 1679)]{leibniz:1679c}
Leibniz, Gottfried~Wilhelm. ca. 1679. ``Addenda ad {S}pecimen calculi
  universalis.'' In \emph{God. {G}uil. {L}eibnitii Opera philosophica quae
  exstant {L}atina {G}allica {G}ermanica omnia},  edited by Johann~Eduard
  Erdmann, 1840th ed., 98--99. Berlin:Prussia: Sumtibus G. Eichleri.
  \url{https://archive.org/details/godguilleibniti00erdmgoog/page/98/mode/1up}
  \url{https://books.google.com/books?id=SWOSd6gCCPYC&pg=PA98&dq=%22Propositiones+per+se+verae%22}
  \url{https://archive.org/details/diephilosophisc01leibgoog/page/224/mode/1up}
  \url{https://doi.org/10.1007/978-94-010-1426-7_27}.

\bibitem[Leibniz(ca. 1686{\natexlab{a}})]{leibniz:1686e}
Leibniz, Gottfried~Wilhelm. ca. 1686{\natexlab{a}}. ``{D}e {F}ormae {L}ogicae
  comprobatione per linearum ductus.'' In \emph{Opuscules et fragments inédits
  de {L}eibniz: Extraits des manuscrits de la Bibliothèque royale de
  {H}anovre},  edited by Louis Couturat, 1903rd ed., 292--321. Paris:France:
  Félix Alcan.
  \url{https://archive.org/details/opusculesetfrag00coutgoog/page/n322/mode/1up}.

\bibitem[Leibniz(ca. 1686{\natexlab{b}})]{leibniz:1686b}
Leibniz, Gottfried~Wilhelm. ca. 1686{\natexlab{b}}. ``Essai de calcul logique,
  {P}hil., {VII}, {B}, {II}, 27.'' In \emph{Opuscules et fragments inédits de
  {L}eibniz: Extraits des manuscrits de la Bibliothèque royale de {H}anovre},
  edited by Louis Couturat, 1903rd ed., 250--251. Paris:France: Félix Alcan.
  \url{https://archive.org/details/opusculesetfrag00coutgoog/page/n280/mode/1up}.

\bibitem[Leibniz(ca. 1686{\natexlab{c}})]{leibniz:1686c}
Leibniz, Gottfried~Wilhelm. ca. 1686{\natexlab{c}}. ``Essais de calcul logique,
  {P}hil., {VII}, {B}, {II}, 62.'' In \emph{Opuscules et fragments inédits de
  {L}eibniz: Extraits des manuscrits de la Bibliothèque royale de {H}anovre},
  edited by Louis Couturat, 1903rd ed., 259--261. Paris:France: Félix Alcan.
  \url{https://archive.org/details/opusculesetfrag00coutgoog/page/n289/mode/1up}.

\bibitem[Leibniz(ca. 1686{\natexlab{d}})]{leibniz:1686d}
Leibniz, Gottfried~Wilhelm. ca. 1686{\natexlab{d}}. ``Essais de calcul logique,
  {P}hil., {VII}, {B}, {II}, 64-65.'' In \emph{Opuscules et fragments inédits
  de {L}eibniz: Extraits des manuscrits de la Bibliothèque royale de
  {H}anovre},  edited by Louis Couturat, 1903rd ed., 264--270. Paris:France:
  Félix Alcan.
  \url{https://archive.org/details/opusculesetfrag00coutgoog/page/n294/mode/1up}.

\bibitem[Leibniz(ca. 1686{\natexlab{e}})]{leibniz:1686a}
Leibniz, Gottfried~Wilhelm. ca. 1686{\natexlab{e}}. ``Generales Inquisitiones
  de Analysi Notionum et Veritatum.'' In \emph{Opuscules et fragments inédits
  de {L}eibniz: Extraits des manuscrits de la Bibliothèque royale de
  {H}anovre},  edited by Louis Couturat, 1903rd ed., 356--399. Paris:France:
  Félix Alcan.
  \url{https://archive.org/details/opusculesetfrag00coutgoog/page/356/mode/1up}.

\bibitem[Leibniz(ca. 1686{\natexlab{f}})]{leibniz:1686f}
Leibniz, Gottfried~Wilhelm. ca. 1686{\natexlab{f}}. ``Non inelegans specimen
  demonstrandi in abstractis.'' In \emph{God. {G}uil. {L}eibnitii Opera
  philosophica quae exstant {L}atina {G}allica {G}ermanica omnia},  edited by
  Johann~Eduard Erdmann, 1840th ed., 94--97. Berlin:Prussia: Sumtibus G.
  Eichleri.
  \url{https://archive.org/details/godguilleibniti00erdmgoog/page/94/mode/1up}
  \url{https://books.google.com/books?id=SWOSd6gCCPYC&pg=PA94}.

\bibitem[Leibniz(ca. 1687)]{leibniz:1687}
Leibniz, Gottfried~Wilhelm. ca. 1687. ``{XX}: {S}pecimen calculi coincidentium
  et inexistentium.'' In \emph{Die philosophischen {S}chriften von {G}ottfried
  {W}ilhelm {L}eibniz},  edited by Carl~Immanuel Gerhardt, 1890th ed., Vol.~7,
  237--247. Berlin:Germany: Weidmann.
  \url{https://archive.org/details/diephilosophisc01leibgoog/page/237/mode/1up}
  \url{https://commons.wikimedia.org/w/index.php?title=File%3ALeibniz_-_Die_philosophischen_Schriften_hg._Gerhardt_Band_7.djvu&page=251}
  \url{https://books.google.com/books?id=LwyKDwAAQBAJ&pg=PA124&dq=%22seu+transpositio+hic+nihil+mutat.%22}
  \url{https://books.google.com/books?id=KcryCQAAQBAJ&pg=PA833&dq=%22seu+transpositio+hic+nihil+mutat.%22}.

\bibitem[Leibniz(ca. 1690)]{leibniz:1690a}
Leibniz, Gottfried~Wilhelm. ca. 1690. ``{P}rincipia {C}alculi rationalis.'' In
  \emph{Opuscules et fragments inédits de {L}eibniz: Extraits des manuscrits
  de la Bibliothèque royale de {H}anovre},  edited by Louis Couturat, 1903rd
  ed., 229--231. Paris:France: Félix Alcan.
  \url{https://archive.org/details/opusculesetfrag00coutgoog/page/n259/mode/1up}.

\bibitem[Leibniz(ca. 1691{\natexlab{a}})]{leibniz:1691a}
Leibniz, Gottfried~Wilhelm. ca. 1691{\natexlab{a}}. ``Difficultates quaedam
  logicae.'' In \emph{God. {G}uil. {L}eibnitii Opera philosophica quae exstant
  {L}atina {G}allica {G}ermanica omnia},  edited by Johann~Eduard Erdmann,
  1840th ed., 101--104. Berlin:Prussia: Sumtibus G. Eichleri.
  \url{https://archive.org/details/godguilleibniti00erdmgoog/page/101/mode/1up}
  \url{https://books.google.com/books?id=SWOSd6gCCPYC&pg=PA101}.

\bibitem[Leibniz(ca. 1691{\natexlab{b}})]{leibniz:1691b}
Leibniz, Gottfried~Wilhelm. ca. 1691{\natexlab{b}}. ``Difficultates quaedam
  logicae.'' In \emph{Die philosophischen {S}chriften von {G}ottfried {W}ilhelm
  {L}eibniz},  edited by Carl~Immanuel Gerhardt, 1890th ed., Vol.~7, 211--214.
  Berlin:Germany: Weidmann.
  \url{https://archive.org/details/diephilosophisc01leibgoog/page/211/mode/1up}
  \url{https://commons.wikimedia.org/w/index.php?title=File%3ALeibniz_-_Die_philosophischen_Schriften_hg._Gerhardt_Band_7.djvu&page=225}.

\bibitem[Lemanski(2017)]{lemanski:2017}
Lemanski, Jens. 2017. ``Periods in the Use of {E}uler-Type Diagrams.''
  \emph{Acta {B}altica {H}istoriae et {P}hilosophiae {S}cientiarum} 5 (1):
  50--69. \url{http://www.bahps.org/03_Lemanski-2017-1-03.pdf}
  \url{https://www.academia.edu/34347207/Periods_in_the_Use_of_Euler_type_Diagrams}.

\bibitem[Lemanski(2018)]{lemanski:2018}
Lemanski, Jens. 2018. ``Logic Diagrams in the {W}eigel and {W}eise Circles.''
  \emph{History and {P}hilosophy of {L}ogic} 39 (1): 3--28.
  \url{https://www.researchgate.net/publication/318709501_Logic_Diagrams_in_the_Weigel_and_Weise_Circles}.

\bibitem[Lenzen(1986)]{lenzen:1986}
Lenzen, Wolfgang. 1986. ``\glq {N}on est\grq\ non est \glq est non\grq: {Z}u
  {L}eibnizens {T}heorie der {N}egation.'' \emph{Studia {L}eibnitiana} 18 (1):
  1--37. \url{https://www.jstor.org/stable/40694037}.

\bibitem[Lenzen(1987)]{lenzen:1987}
Lenzen, Wolfgang. 1987. ``Leibniz's Calculus of Strict Implication.'' In
  \emph{Initiatives in Logic},  edited by Jan T.~J. Srzednicki, 1--35.
  Dordrecht:Netherlands: Martinus Nijhoff Publishers.
  \url{https://books.google.com/books?id=DHmhBQAAQBAJ&pg=PT17&redir_esc=y}
  \url{https://doi.org/10.1007/978-94-009-3673-7_1}.

\bibitem[Lenzen(2004{\natexlab{a}})]{lenzen:2004a}
Lenzen, Wolfgang. 2004{\natexlab{a}}. ``Leibniz's logic.'' In \emph{Handbook of
  the History of Logic, volume 3 -- {T}he Rise of Modern Logic: from {L}eibniz
  to {F}rege},  edited by Dov~M. Gabbay and John Woods, 1--84.
  Amsterdam:Netherlands: Elsevier B.V.
  \url{https://books.google.com/books?id=74-g3vSAbnoC&pg=PA77&dq=%22Propositions+are+either%22}
  \url{https://www.philosophie.uni-osnabrueck.de/fileadmin/Allgemeine_Uploads/Publikationen/Lenzen/Lenzen_Leibniz_Logic.pdf#page=81}.

\bibitem[Lenzen(2004{\natexlab{b}})]{lenzen:2004b}
Lenzen, Wolfgang. 2004{\natexlab{b}}. ``Logical Criteria for
  Individual(concept)s.'' In \emph{Individuals, Minds and Bodies: Themes from
  {L}eibniz},  edited by Massimiliano Carrara, Antonio~Maria Nunziante, and
  Gabriele Tomasi, 87--107. Stuttgart:Germany: Franz Steiner Verlag.
  \url{https://books.google.com/books?id=lxtSYBtShVsC&pg=PA94&dq=%22est+ens%22+empty}
  \url{https://www.philosophie.uni-osnabrueck.de/fileadmin/Allgemeine_Uploads/Publikationen/Lenzen/Logical_criteria_for_individuals.pdf#page=6}.

\bibitem[Lenzen(2014)]{lenzen:2014a}
Lenzen, Wolfgang. 2014. ``Leibniz: Logic.'' In \emph{{I}nternet Encyclopedia of
  {P}hilosophy}, Section: ``The Plus-Minus-Calculus''.
  \url{https://iep.utm.edu/leib-log/#SH3c}.

\bibitem[Lenzen(2018{\natexlab{a}})]{lenzen:2018a}
Lenzen, Wolfgang. 2018{\natexlab{a}}. ``Leibniz and the Calculus
  Raciotinator.'' In \emph{Technology and Mathematics: Philosophical and
  Historical Investigations},  edited by Sven~Ove Hansson, 47--78. Springer.
  \url{https://books.google.com/books?id=nM10DwAAQBAJ&pg=PA67}.

\bibitem[Lenzen(2018{\natexlab{b}})]{lenzen:2018b}
Lenzen, Wolfgang. 2018{\natexlab{b}}. ``Two days in the life of a genius.'' In
  \emph{From Arithmetic to Metaphysics: A Path through Philosophical Logic},
  edited by Ciro de~Florio and Alessandro Giordani, 207--240. Berlin:Germany:
  Walter de Gruyter GmbH.
  \url{https://books.google.com/books?id=WXFLDwAAQBAJ&pg=PT263}
  \url{https://doi.org/10.1515/9783110529494-015}.

\bibitem[Levey(2011)]{levey:2011}
Levey, Samuel. 2011. ``Chapter 7: Logical theory in {L}eibniz.'' In \emph{The
  {C}ontinuum {C}ompanion to {L}eibniz},  edited by Brandon~C. Look, 110--135.
  London:UK: Continuum.
  \url{https://books.google.com/books?id=PTmmjB3nDyoC&pg=PA118}.

\bibitem[Lewis(1918)]{lewis:1918}
Lewis, Clarence~Irving. 1918. \emph{A survey of symbolic logic}. Berkeley:USA:
  {U}niversity of {C}alifornia {P}ress.
  \url{https://archive.org/details/cu31924028923451/page/n8/mode/1up}
  \url{https://archive.org/details/surveyofsymbolic00lewiiala/page/n6/mode/1up}
  \url{https://archive.org/details/asurveyofsymboli00lewiuoft/page/n8/mode/1up}.

\bibitem[Llull(1993)]{llull:1993}
Llull, Ramon. 1993. ``Raimundi {L}ulli opera latina, vol. {XIX}: 86-91.
  {P}arisiis, {B}arcinonae et in civitate {M}aioricensi annis {MCCXCIX-MCCC}
  composita.'' In \emph{Corpus {C}hristianorum: {C}ontinuatio {M}ediaevalis
  {CXI}},  edited by Fernando~Domínguez Reboiras. Turnhout:Belgium: Brepols
  Editores Pontificii.
  \url{https://books.google.com/books?id=pytKAAAAYAAJ&dq=%22Si+extra+intellectum+nullum+non+ens+est+ens%22}.

\bibitem[Locke(1700)]{locke:1700}
Locke, John. 1700. \emph{An Essay concerning Humane Understanding}. 4th ed.
  London:England: Black Swan and Ship.
  \url{https://archive.org/details/essayconcernin00lockuoft/page/405/mode/1up}.

\bibitem[Lorenzen(1957)]{lorenzen:1957}
Lorenzen, Paul. 1957. ``Über die {S}yllogismen als
  {R}elationenmultiplikationen.'' \emph{Archiv für mathematische {L}ogik und
  {G}rundlagenforschung} 3: 112--116.

\bibitem[Luciano(2012)]{luciano:2012}
Luciano, Erika. 2012. ``Peano and his school between {L}eibniz and {C}outurat:
  The influence in mathematics and in international language.'' In \emph{New
  Essays on {L}eibniz Reception: In Science and Philosophy of Science
  1800-2000},  edited by Ralph Krömer and Yannick Chin-Drian, 41--64. Springer
  Basel. \url{https://doi.org/10.1007/978-3-0346-0504-5_4}
  \url{https://books.google.com/books?id=n4LI9ZFHwCwC&pg=PA49}.

\bibitem[Maddux(1991)]{maddux:1991}
Maddux, Roger~D. 1991. ``The Origin of Relation Algebras in the Development and
  Axiomatization of the Calculus of Relations.'' \emph{Studia {L}ogica} 50
  (3-4): 421--455. \url{http://www.jstor.org/stable/20015596?seq=14}
  \url{https://www.researchgate.net/publication/2837299_The_Origin_Of_Relation_Algebras_In_The_Development_And_Axiomatization_Of_The_Calculus_Of_Relations}
  \url{https://web.archive.org/web/20090313163522/http://orion.math.iastate.edu/maddux/papers/Maddux1991.pdf#page=15}.

\bibitem[Makinson(2022)]{makinson:2022}
Makinson, David. 2022. ``Boole's indefinite symbols re-examined.'' \emph{The
  {A}ustralasian {J}ournal of {L}ogic} 19 (5): 167--181.
  \url{https://ojs.victoria.ac.nz/ajl/article/view/8011}
  \url{https://doi.org/10.26686/ajl.v19i5.8011}.

\bibitem[Malink and Vasudevan(2016)]{malink:2016}
Malink, Marko, and Anubav Vasudevan. 2016. ``The logic of {L}eibniz's
  {G}enerales {I}nquisitiones de {A}nalysi {N}otionum et {V}eritatum.''
  \emph{The Review of Symbolic Logic} 9 (4): 686--751.
  \url{https://doi.org/10.1017/S1755020316000137}
  \url{https://as.nyu.edu/content/dam/nyu-as/philosophy/documents/faculty-documents/malink/Malink_Logic-of-Leibniz.pdf}.

\bibitem[Malink and Vasudevan(2019)]{malink:2019}
Malink, Marko, and Anubav Vasudevan. 2019. ``Leibniz on the Logic of Conceptual
  Containment and Coincidence.'' In \emph{Leibniz and the Structure of
  Sciences},  edited by Vincenzo de~Risi, 1--46. Cham:Switzerland: Springer
  Nature Switzerland.
  \url{https://books.google.com/books?id=Q5PHDwAAQBAJ&pg=PA1}
  \url{https://as.nyu.edu/content/dam/nyu-as/faculty/documents/Leibniz%20on%20the%20Logic%20of%20Conceptual%20Containment%20and%20Coincidence.pdf}.

\bibitem[Marciszewski(1984)]{marciszewski:1984}
Marciszewski, Witold. 1984. ``The Principle of Comprehension as a Present-Day
  Contribution to {M}athesis {U}niversalis.'' \emph{Philosophia Naturalis} 21
  (2-4): 523--537.
  \url{https://calculemus.org/pub-libr/as-ref/MUpresent.pdf#page=4}.

\bibitem[Marciszewski and Murawski(1995)]{marciszewski:1995}
Marciszewski, Witold, and Roman Murawski. 1995. \emph{Mechanization of
  Reasoning in a Historical Perspective}. Amsterdam:Netherlands: Rodopi.
  \url{https://books.google.com/books?id=U1S_GUgdPO0C&pg=PA140&dq=%22The+introduced%22+%22It+seems+that%22}.

\bibitem[McCall(1967)]{mccall:1967}
McCall, Storrs. 1967. ``Connexive Implication and the Syllogism.'' \emph{Mind}
  76 (303): 346--356. \url{https://www.jstor.org/stable/2251844?seq=4}.

\bibitem[McColl(1877)]{mccoll:1877}
McColl, Hugh. 1877. ``The Calculus of Equivalent Statements (Second Paper).''
  \emph{{P}roceedings of the {L}ondon {M}athematical {S}ociety} s1--9 (1):
  177--186. \url{https://doi.org/10.1112/plms/s1-9.1.177}
  \url{https://zenodo.org/record/2401556}
  \url{https://academic.oup.com/plms/article-abstract/s1-9/1/177/1547621}.

\bibitem[McCune et~al.(2002)]{mccune:2002}
McCune, William, Robert Veroff, Branden Fitelson, Kenneth Harris, Andrew Feist,
  and Larry Wos. 2002. ``Short Single Axioms for {B}oolean Algebra.''
  \emph{Journal of Automated Reasoning} 29 (1): 1--16.
  \url{https://www.cs.unm.edu/~mccune/papers/basax/v12.pdf}
  \url{https://www.mcs.anl.gov/papers/P848.pdf}.

\bibitem[Menne(1957)]{menne:1957}
Menne, Albert. 1957. ``Implikation und {S}yllogistik.'' \emph{Zeitschrift für
  philosophische {F}orschung} 11 (3): 375--386.
  \url{https://www.jstor.org/stable/20480930}.

\bibitem[Menne(1962)]{menne:1962}
Menne, Albert. 1962. ``Some Results of Investigation of the Syllogism and Their
  Philosophical Consequences.'' In \emph{Logico-Philosophical Studies},  edited
  by Albert Menne, 55--63. Dordrecht:Netherlands: Springer Netherlands.
  \url{https://doi.org/10.1007/978-94-010-3649-8_4}
  \url{https://books.google.com/books?id=P_nwCAAAQBAJ&pg=PA61}.

\bibitem[Metamath(2021)]{metamath:2021}
Metamath. 2021. ``1.7.1 {A}ristotelian logic: {A}ssertic syllogisms.''
  \url{http://us.metamath.org/mpegif/mmtheorems26.html#mm2562s}
  \url{http://us.metamath.org/mpegif/barbara.html}.

\bibitem[Mitchell(1883)]{mitchell:1883}
Mitchell, Oscar~Howard. 1883. ``On a new algebra of logic.'' In \emph{Studies
  in logic by members of the {J}ohns {H}opkins {U}niversity},  edited by
  Charles~Sanders Peirce, 72--106. Boston:USA: Little, Brown, and Company.
  \url{https://archive.org/details/bub_gb_V7oIAAAAQAAJ/page/n80/mode/1up}
  \url{https://archive.org/details/studiesinlogic00gilmgoog/page/n91/mode/1up}
  \url{https://archive.org/details/studiesinlogic00peiruoft/page/72/mode/1up}.

\bibitem[Moktefi(2019)]{moktefi:2019}
Moktefi, Amirouche. 2019. ``The Social Shaping of Modern Logic.'' In
  \emph{Natural Arguments: A Tribute to {J}ohn {W}oods},  edited by Dov Gabbay,
  Lorenzo Magnani, Woosuk Park, and Athi-Veikko Pietarinen, 503--520. College
  Publications.
  \url{https://www.researchgate.net/publication/332112727_The_Social_Shaping_of_Modern_Logic}
  \url{https://www.academia.edu/38720002/The_Social_Shaping_of_Modern_Logic}.

\bibitem[Moss(2007)]{moss:2007}
Moss, Lawrence~S. 2007. ``Natural Logic Notes: Course Notes for {ESSLLI}
  2007.'' In \emph{{ESSLLI} 2007},
  \url{https://www.scss.tcd.ie/conferences/esslli2007/content/CD_Contents/content/id56/nlesslli.pdf#page=23}
  \url{https://iulg.sitehost.iu.edu/moss/nl/nlesslli.pdf#page=21}.

\bibitem[Moss(2010)]{moss:2010}
Moss, Lawrence~S. 2010. ``Logics for Natural Language Inference (expanded
  version of lecture notes from a course at {ESSLLI} 2010).'' In \emph{{ESSLLI}
  2010},
  \url{https://web.archive.org/web/20170705072341/http://www.indiana.edu/~iulg/moss/notes.pdf}.

\bibitem[Moss(2011)]{moss:2011}
Moss, Lawrence~S. 2011. ``Syllogistic Logic with Complements.'' In \emph{Games,
  Norms and Reasons: Logic at the Crossroads},  edited by J.~van Benthem,
  A.~Gupta, and E.~Pacuit, 179--197. Springer.
  \url{https://books.google.com/books?id=mzIqFJby6UoC&pg=PA181&dq=%22Syllogistic+Logic+with+Complement%22}
  \url{https://iulg.sitehost.iu.edu/moss/comp2.pdf#page=3}
  \url{https://citeseerx.ist.psu.edu/doc_view/pid/a3507d081ed298d823fdfe843ef0ed83c4f23abc}.

\bibitem[Mounyer and Fabri(1646)]{mounyer:1646}
Mounyer, Pierre, and Honoré Fabri. 1646. \emph{Philosophiae tomvs primvs ...:
  {L}ogicam Analyticam ...} Lyon:France: Ioannis Champion.
  \url{https://books.google.com/books?id=Vodj9a2dd34C&pg=PA254}
  \url{https://archive.org/details/bub_gb_AM8yj7c_qf0C/page/n320/mode/1up}.

\bibitem[nLab authors(2016)]{ncatlab:2016}
nLab authors. 2016. ``co-{H}eyting boundary.''
  \url{http://ncatlab.org/nlab/show/co-Heyting+boundary}.

\bibitem[nLab authors(2019)]{ncatlab:2019}
nLab authors. 2019. ``Boolean algebra.''
  \url{http://ncatlab.org/nlab/show/Boolean+algebra#principle_of_duality}.

\bibitem[Novak(1980)]{novak:1980}
Novak, Joseph~A. 1980. ``Some recent work on the assertoric syllogistic.''
  \emph{Notre Dame Journal of Formal Logic} 21 (2): 229--242.
  \url{https://projecteuclid.org/journals/notre-dame-journal-of-formal-logic/volume-21/issue-2/Some-recent-work-on-the-assertoric-syllogistic/10.1305/ndjfl/1093883042.full}.

\bibitem[Ongley and Carey(2013)]{ongley:2013}
Ongley, John, and Rosalind Carey. 2013. \emph{Russell: A Guide for the
  Perplexed}, Chap. 1: Introduction, sections 4--6, 7--14. New York:USA:
  Bloomsbury.
  \url{https://books.google.com/books?id=6oTHF95Y1IwC&pg=PA7&dq=%22Logical+analysis:+The+theory+of+descriptions%22}
  \url{https://users.drew.edu/~jlenz/brs-br-analytic-phil.html#:~:text=Logical%20analysis%3A%20The%20theory%20of%20descriptions}.

\bibitem[Pagliani(1998)]{pagliani:1998}
Pagliani, Piero. 1998. ``Intrinsic Co-{H}eyting Boundaries and Information
  Incompleteness in Rough Set Analysis.'' In \emph{Rough Sets and Current
  Trends in Computing: First International Conference}, June, 123--130.
  Springer.
  \url{https://www.researchgate.net/publication/28763376_Intrinsic_Co-Heyting_Boundaries_and_Information_Incompleteness_in_Rough_Set_Analysis}.

\bibitem[Parsons(2021)]{parsons:2021}
Parsons, Terence. 2021. ``The Traditional Square of Opposition.'' In \emph{The
  {S}tanford Encyclopedia of Philosophy},  edited by Edward~N. Zalta, {F}all
  2021 ed. Metaphysics Research Lab, Stanford University.
  \url{https://plato.stanford.edu/archives/fall2021/entries/square/}.

\bibitem[Patzig(1968)]{patzig:1968}
Patzig, Günther. 1968. \emph{Aristotle's Theory of the Syllogism: A
  Logico-Philological Study of {B}ook {A} of the {P}rior {A}nalytics}. 2nd ed.
  Dordrecht:Netherlands: Springer.
  \url{https://books.google.com/books?id=8g3wCAAAQBAJ&pg=PA12&dq=%22in+his+systematic+presentation+of+the+theory+of+the+syllogism+Aristotle%22+%22he+uses%22}.

\bibitem[Peckhaus(2018)]{peckhaus:2018}
Peckhaus, Volker. 2018. ``Leibniz's Influence on 19th Century Logic.'' In
  \emph{The {S}tanford Encyclopedia of {P}hilosophy},  edited by Edward~N.
  Zalta, {W}inter 2018 ed. Metaphysics Research Lab, Stanford University.
  \url{https://plato.stanford.edu/archives/win2018/entries/leibniz-logic-influence/}.

\bibitem[Peirce(1867)]{peirce:1867}
Peirce, Charles~Sanders. 1867. ``On an Improvement in {B}oole's Calculus of
  Logic.'' \emph{Proceedings of the {A}merican {A}cademy of {A}rts and
  {S}ciences} 7: 250--261.
  \url{https://archive.org/details/proceedingsofam07amer/page/250/mode/1up}
  \url{https://books.google.com/books?id=nG8UAAAAYAAJ&pg=PA250}.

\bibitem[Peirce(1870)]{peirce:1870}
Peirce, Charles~Sanders. 1870. ``Description of a Notation for the Logic of
  Relatives, Resulting from an Amplification of the Conceptions of {B}oole's
  Calculus of Logic.'' \emph{Memoirs of the American Academy of Arts and
  Sciences 9 (1870)} 9: 317--378.
  \url{https://books.google.com/books?id=fFnWmf5oLaoC&pg=PA57&dq=%22Particular+propositions%22}.

\bibitem[Peirce(1883)]{peirce:1883}
Peirce, Charles~Sanders, ed. 1883. \emph{Studies in logic by members of the
  {J}ohns {H}opkins {U}niversity}. Boston:USA: Little, Brown, and Company.
  \url{https://archive.org/details/bub_gb_V7oIAAAAQAAJ}
  \url{https://archive.org/details/studiesinlogic00gilmgoog/page/n8/mode/1up}
  \url{https://archive.org/details/studiesinlogic00peiruoft/page/n6/mode/1up}.

\bibitem[Peirce and Ladd-Franklin(1901)]{peirce:1901}
Peirce, Charles~Sanders, and Christine Ladd-Franklin. 1901. ``Laws of
  Thought.'' In \emph{Dictionary of Philosophy and Psychology},  edited by
  James~Mark Baldwin, Vol.~1, 641--644. New York:USA: The Macmillan Company.
  \url{https://archive.org/details/dictionaryphilo01baldgoog/page/641/mode/1up}
  \url{https://www.jfsowa.com/peirce/baldwin.htm#Laws%20of%20Thought}.

\bibitem[Piesk(2017)]{piesk:2017}
Piesk, Tilman. 2017. \emph{Syllogism charts with all {E}uler diagrams}.
  {W}ikimedia {C}ommons.
  \url{https://commons.wikimedia.org/wiki/Template:Syllogism\_charts\_with\_all\_Euler\_diagrams}.

\bibitem[Pietarinen(2016)]{pietarinen:2016}
Pietarinen, Ahti-Veikko. 2016. ``{E}xtensions of {E}uler Diagrams in {P}eirce's
  Four Manuscripts on Logical Graphs.'' In \emph{{D}iagrammatic
  {R}epresentation and {I}nference, 9th {I}nternational {C}onference,
  {D}iagrams 2016, {LNAI} 9781}, 139--154.
  \url{https://www.researchgate.net/publication/305621831\_Extensions\_of\_Euler\_Diagrams\_in\_Peirce's\_Four\_Manuscripts\_on\_Logical\_Graphs}
  \url{https://www.academia.edu/22213024/Extensions\_of\_Euler\_Diagrams\_in\_Peirces\_Four\_Manuscripts\_on\_Logical\_Graphs}.

\bibitem[Pietarinen(2021)]{pietarinen:2021}
Pietarinen, Ahti-Veikko. 2021. \emph{The Logical Tracts}, Chap. Selection 31:
  The Logical Tracts No. 2 Part II. On Logical Graphs [Euler's Diagrams].
  {W}alter de {G}ruyter.
  \url{https://books.google.com/books?id=2JY9EAAAQBAJ&pg=PA84}.

\bibitem[Pietarinen and Chevalier(2015)]{pietarinen:2015}
Pietarinen, Ahti-Veikko, and Jean-Marie Chevalier. 2015. ``The {J}ohns
  {H}opkins {M}etaphysical {C}lub and Its Impact on the Development of the
  Philosophy and Methodology of Sciences in the Late 19th-Century {U}nited
  {S}tates.'' \emph{The {C}ommens Working Papers: Preprints, Research Reports
  \& Scientific Communications} 1--35.
  \url{https://www.researchgate.net/publication/288181937_The_Johns_Hopkins_Metaphysical_Club_and_Its_Impact_on_the_Development_of_the_Philosophy_and_Methodology_of_Sciences_in_the_Late_19th-Century_United_States}
  \url{https://web.archive.org/web/20211127113437/http://www.commens.org/papers/paper/pietarinen-ahti-veikko-chevalier-jean-marie-2014-johns-hopkins-metaphysical-club-and}
  \url{https://web.archive.org/web/20220419174857/http://www.commens.org/sites/default/files/working_papers/metaphysical-club-pietarinen-chevalier.pdf}.

\bibitem[Pratt-Hartmann(2023)]{pratt-hartmann:2023}
Pratt-Hartmann, Ian. 2023. \emph{Fragments of First-Order Logic}. Oxford:UK:
  Oxford University Press.
  \url{https://books.google.com/books?id=O9mzEAAAQBAJ&pg=PA25}.

\bibitem[Quine(1950)]{quine:1950}
Quine, Willard Van~Orman. 1950. \emph{Methods of Logic}. 1st ed. {H}enry {H}olt
  and {C}ompany.
  \url{https://books.google.com/books?id=D2MYAAAAIAAJ&pg=PA70&dq=%22For+nonemptiness+another+symbol+is+used%22}.

\bibitem[Quine(1982)]{quine:1982}
Quine, Willard Van~Orman. 1982. \emph{Methods of Logic}. 4th ed. {H}arvard
  {U}niversity {P}ress.
  \url{https://books.google.com/books?id=liHivlUYWcUC&pg=PA98&dq=%22For+nonemptiness+another+symbol+is+used%22}.

\bibitem[Randell, Cui, and Cohn(1992)]{randell:1992}
Randell, David~A., Zhan Cui, and Anthony~G. Cohn. 1992. ``A Spatial Logic based
  on Regions and Connection.'' In \emph{3rd {I}nternational {C}onference on
  {K}nowledge {R}epresentation and {R}easoning ({KR}'92)}, Morgan Kaufmann.
  \url{https://www.researchgate.net/publication/221393453_A_Spatial_Logic_based_on_Regions_and_Connection}.

\bibitem[Reichenbach(1952)]{reichenbach:1952}
Reichenbach, Hans. 1952. ``The syllogism revised.'' \emph{Philosophy of
  Science} 19 (1): 1--16. \url{https://www.jstor.org/stable/185095?seq=7}.

\bibitem[Rescher(1954)]{rescher:1954}
Rescher, Nicholas. 1954. ``Leibniz's interpretation of his logical calculi.''
  \emph{The Journal of Symbolic Logic} 19 (1): 1--13.
  \url{https://www.jstor.org/stable/2267644?seq=11}.

\bibitem[Richeri(1761)]{richeri:1761}
Richeri, Lodovico~Ignazio. 1761. ``Algebrae philosophicae in usum artis
  inveniendi specimen primum.'' In \emph{Mélanges de Philosophie et de
  Mathématique de la {S}ociété {R}oyale de {T}urin por les Années
  1760-1761}, Vol. Tomus 2, section 3, Torino:Regno di Sardegna, 46--63.
  Imprimerie Royale.
  \url{https://www.biodiversitylibrary.org/page/7650150#page/626/mode/1up}
  \url{https://archive.org/details/mlangesdephilo02soci/page/46/mode/1up}
  \url{http://digitale.beic.it/BEIC:RD01:39bei_hostART-AAIpc-cpt-1_001574}
  \url{https://books.google.com/books?id=XQRlAAAAcAAJ&pg=RA1-PA46}.

\bibitem[Schröder(1877)]{schroeder:1877}
Schröder, Ernst. 1877. \emph{Der {O}perationskreis des {L}ogikkalkuls}.
  Leipzig:Germany: B. G. Teubner.
  \url{https://archive.org/details/bub_gb_Q5YXAAAAIAAJ/page/n16/mode/1up}
  \url{https://archive.org/details/deroperationskr00schrgoog/page/n19/mode/1up}.

\bibitem[Schröder(1890)]{schroeder:1890}
Schröder, Ernst. 1890. \emph{Vorlesungen über die {A}lgebra der {L}ogik
  ({E}xakte {L}ogik)}. Vol.~1. Leipzig:Germany: B. G. Teubner.
  \url{https://books.google.com/books?id=UzNLAAAAMAAJ&pg=PA280}
  \url{https://www.deutschestextarchiv.de/book/view/schroeder_logik01_1890?p=300}
  \url{https://archive.org/details/voalgebraderlogi01schrrich/page/n299/mode/2up}
  \url{https://gdz.sub.uni-goettingen.de/id/PPN717192873?tify={%22pages%22:[298],%22view%22:%22info%22}}.

\bibitem[Schröder(1895)]{schroeder:1895}
Schröder, Ernst. 1895. \emph{Vorlesungen über die {A}lgebra der {L}ogik
  ({E}xakte {L}ogik)}. Vol. 3: {A}lgebra und {L}ogik der {R}elative.
  Leipzig:Germany: B. G. Teubner.
  \url{https://www.deutschestextarchiv.de/book/view/schroeder_logik03_1895/?p=256}
  \url{https://archive.org/details/voalgebraderlogi03schrrich/page/n255/mode/2up}
  \url{https://gdz.sub.uni-goettingen.de/id/PPN717195317?tify={%22pages%22:[256],%22view%22:%22info%22}}.

\bibitem[Shin, Lemon, and Mumma(2018)]{shin:2018}
Shin, Sun-Joo, Oliver Lemon, and John Mumma. 2018. ``Diagrams.'' In \emph{The
  {S}tanford Encyclopedia of Philosophy},  edited by Edward~N. Zalta.
  Metaphysics Research Lab, Stanford University.
  \url{https://plato.stanford.edu/archives/win2018/entries/diagrams/}.

\bibitem[Simons(2020)]{simons:2020}
Simons, Peter. 2020. ``Term Logic.'' \emph{Axioms} 9 (1).
  \url{https://doi.org/10.3390/axioms9010018}.

\bibitem[Sotirov(1999)]{sotirov:1999}
Sotirov, Vladimir. 1999. ``Various Syllogistics from the Algebraic Point of
  View.'' In \emph{Proceedings of the 2nd {P}anhellenic {L}ogic {S}ymposium},
  Delphi:Greece, July, 197--200.
  \url{http://www.math.bas.bg/~vlsot/algebra.pdf#page=3}.

\bibitem[Stone(1935)]{stone:1935}
Stone, Marshall~Harvey. 1935. ``Subsumption of the Theory of {B}oolean Algebras
  under the Theory of Rings.'' \emph{Proceedings of the National Academy of
  Sciences} 21 (2): 103--105. \url{https://doi.org/10.1073/pnas.21.2.103}.

\bibitem[Sturm(1661)]{sturm:1661}
Sturm, Johann~Christoph. 1661. \emph{Universalia {E}uclidea ... {N}ovi
  {S}yllogizandi {M}odi}. Den Haag:Netherlands: Adriaan Vlacq.
  \url{https://books.google.com/books?id=Q1H3blJBSQAC&pg=PA86}.

\bibitem[Tamaki(1974)]{tamaki:1974}
Tamaki, Itsuo. 1974. ``Syllogistic and Calculus of Classes.'' \emph{Logique et
  Analyse} 17 (65/66): 191--196. \url{https://www.jstor.org/stable/44083658}.

\bibitem[Tarski(1933)]{tarski:1933}
Tarski, Alfred. 1933. ``The Concept of Truth in Formalized Languages.'' In
  \emph{Logic, Semantics, Metamathematics: Papers from 1923 to 1938},
  Warsaw:Poland, 152--278. Towarzystwa Naukowego Warszawskiego.
  \url{https://books.google.com/books?id=2uhra9PEFZsC&pg=PA167&dq=%22In+contrast+to+natural+languages%22+%22we+must+always+distinguish+clearly%22+object+%22meta+language%22}
  \url{http://www.thatmarcusfamily.org/philosophy/Course_Websites/Readings/Tarski%20-%20The%20Concept%20of%20Truth%20in%20Formalized%20Languages.pdf}.

\bibitem[Tarski(1936)]{tarski:1936}
Tarski, Alfred. 1936. ``Grundlegung der wissenschaftlichen {S}emantik ({T}he
  establishment of scientific semantics).'' In \emph{Actes du Congrès
  International de Philosophie Scientifique; Logic, Semantics, Metamathematics:
  Papers from 1923 to 1938}, Paris:France, 401--408. Hermann \& Cie.
  \url{https://books.google.com/books?id=2uhra9PEFZsC&pg=PA402&dq=Le%C5%9Bniewski}
  \url{https://books.google.com/books?id=2uhra9PEFZsC&pg=PA407&dq=metalogical+metalanguage}.

\bibitem[Tennant(2014)]{tennant:2014}
Tennant, Neil. 2014. ``Aristotle's Syllogistic and Core Logic.'' \emph{History
  and {P}hilosophy of {L}ogic} 35 (2): 120--147.
  \url{https://doi.org/10.1080/01445340.2013.867144}
  \url{https://philpapers.org/archive/tenasa.pdf#page=9}
  \url{http://u.osu.edu/tennant.9/files/2014/07/tennant_hpl2014-pqsi5u.pdf#page=10}.

\bibitem[Turner(1983)]{turner:1983}
Turner, G. 1983. ``Review -- {J}ordanus de {N}emore: {D}e numeris datis.''
  \emph{The Antiquaries Journal} 63 (1): 183--184.
  \url{https://doi.org/10.1017/S0003581500014797}.

\bibitem[Vacca(1899)]{vacca:1899}
Vacca, Giovanni. 1899. ``Sui manoscritti inediti di {L}eibniz.'' In
  \emph{Bollettino di bibliografia e storia delle scienze matematiche},  edited
  by Gino Loria, Torino:Italia, October/November/December, 113--116. Carlo
  Clausen.
  \url{http://digitale.bnc.roma.sbn.it/tecadigitale/giornale/TO00179204/1899/unico/00000121}
  \url{http://digitale.bnc.roma.sbn.it/tecadigitale/visore/#/main/viewer?idMetadato=12114666}.

\bibitem[Valencia(2004)]{valencia:2004}
Valencia, Victor~Sánchez. 2004. ``The algebra of logic.'' In \emph{{H}andbook
  of the History of Logic, Volume 3, {T}he Rise of Modern Logic: From {L}eibniz
  to {F}rege},  edited by Dov~M. Gabbay and John Woods, 389--544.
  Amsterdam:Netherlands: Elsevier.
  \url{https://books.google.com/books?id=74-g3vSAbnoC&pg=PA473}.

\bibitem[Venn(1880)]{venn:1880}
Venn, John. 1880. ``I. On the Diagrammatic and Mechanical Representation of
  Propositions and Reasonings.'' \emph{The {L}ondon, {E}dinburgh, and {D}ublin
  Philosophical Magazine and Journal of Science, fifth series} 9 (59): 1--18.
  \url{https://doi.org/10.1080/14786448008626877}
  \url{https://zenodo.org/record/1431145/files/article.pdf}
  \url{https://www.cis.upenn.edu/~bhusnur4/cit592_fall2014/venn%20diagrams.pdf#page=3}.

\bibitem[Venn(1881)]{venn:1881}
Venn, John. 1881. \emph{Symbolic Logic}. London:UK: Macmillan and Co.
  \url{https://archive.org/details/symboliclogic00venniala/page/167/mode/1up}.

\bibitem[Venn(1883)]{venn:1883}
Venn, John. 1883. ``Reviewed work: {S}tudies in Logic by members of the {J}ohns
  {H}opkins {U}niversity.'' \emph{Mind} 8 (32): 594--603.
  \url{https://www.jstor.org/stable/2246489}.

\bibitem[von Plato(2021)]{von_plato:2021}
von Plato, Jan. 2021. ``Logic as Calculus and Logic as Language: Too Suggestive
  to be Truthful?'' \emph{{P}hilosophia {S}cientiae} 25 (1): 35--47.
  \url{https://www.cairn.info/revue-philosophia-scientiae-2021-1-page-35.htm}
  \url{https://www.cairn.info/load_pdf.php?ID_ARTICLE=SCIE_251_0035}.

\bibitem[von Segner(1740)]{segner:1740}
von Segner, Johann~Andreas. 1740. \emph{Specimen Logicae Vniversaliter
  Demonstratae}. Universität Jena.
  \url{https://books.google.com/books?id=GBvZgeWHKBYC&pg=PA71}
  \url{https://books.google.com/books?id=YFNcvkwwOmIC&pg=PA71}.

\bibitem[Waragai and Oyamada(2007)]{waragai:2007}
Waragai, Toshiharu, and Keiichi Oyamada. 2007. ``A System of Ontology Based on
  Identity and Partial Ordering as an Adequate Logical Apparatus for Describing
  Taxonomical Structures of Concepts.'' \emph{Annals of the Japan Association
  for Philosophy of Science} 15 (2): 123--149.
  \url{https://doi.org/10.4288/jafpos1956.15.123}
  \url{https://www.researchgate.net/publication/266707810_A_System_of_Ontology_Based_on_Identity_and_Partial_Ordering_as_an_Adequate_Logical_Apparatus_for_Describing_Taxonomical_Structures_of_Concepts}.

\bibitem[Wege et~al.(2020)]{wege:2020}
Wege, Theresa~Elise, Sophie Batchelor, Matthew Inglis, Honali Mistry, and Dirk
  Schlimm. 2020. ``Iconicity in Mathematical Notation: Commutativity and
  Symmetry.'' \emph{Journal of Numerical Cognition} 6 (3): 378--392.
  \url{https://doi.org/10.5964/jnc.v6i3.314}
  \url{https://d-nb.info/1229754148/34}.

\bibitem[Whitehead(1898)]{whitehead:1898}
Whitehead, Alfred~North. 1898. \emph{A Treatise on Universal Algebra, with
  Applications}. London:UK: Cambridge University Press.
  \url{https://archive.org/details/atreatiseonuniv00goog/page/102/mode/2up}.

\bibitem[Whitehead and Russell(1910)]{whitehead:1910}
Whitehead, Alfred~North, and Bertrand Russell. 1910. \emph{Principia
  {M}athematica, volume {I}}, 1st ed., Chap. 22 Calculus of Classes. Cambridge
  University Press.
  \url{https://commons.wikimedia.org/w/index.php?title=File%3ARussell%2C_Whitehead_-_Principia_Mathematica%2C_vol._I%2C_1910.djvu&page=240}
  \url{https://books.google.com/books?id=B_5ubiPgOmgC&q=%22There+is+an+element%22}.

\bibitem[{William of Ockham}(ca. 1323)]{ockham:1323}
{William of Ockham}. ca. 1323. \emph{Summa Logicae}, Vol.~2, Chap. 32 and 33.
  Oxford:England.
  \url{http://www.logicmuseum.com/wiki/Authors/Ockham/Summa_Logicae/Book_II/Chapter_32#:~:text=the%20contradictory%20opposite%20of%20a%20copulative}
  \url{http://www.logicmuseum.com/wiki/Authors/Ockham/Summa_Logicae/Book_II/Chapter_33#:~:text=the%20contradictory%20opposite%20of%20a%20disjunctive}
  \url{https://books.google.com/books?id=Z-ZOAAAAcAAJ&pg=PA35&dq=oppofita+oppo+fita}
  \url{https://books.google.com/books?id=lC08AAAAcAAJ&pg=RA2-PA43&dq=oppofita}.

\bibitem[Wolfram(2018)]{wolfram:2018}
Wolfram, Stephen. 2018. ``Logic, Explainability and the Future of
  Understanding.''
  \url{https://writings.stephenwolfram.com/2018/11/logic-explainability-and-the-future-of-understanding/}.

\bibitem[Wundt(1880)]{wundt:1880}
Wundt, Wilhelm~Maximilian. 1880. \emph{Logik: Eine {U}ntersuchung der
  {P}rincipien der {E}rkenntniss und der {M}ethoden wissenschaftlicher
  {F}orschung}. Vol.~1. Stuttgart:Germany: Ferdinand Enke.
  \url{https://books.google.com/books?id=cL05AQAAMAAJ&pg=PA244&dq=Kreuzung}.

\bibitem[Wybraniec-Skardowska(2018)]{wybraniec:2018}
Wybraniec-Skardowska, Urszula. 2018. ``{R}ejection in {Ł}ukasiewicz's and
  {S}łupecki's Sense.'' In \emph{{T}he {L}vov-{W}arsaw School. {P}ast and
  Present},  edited by Ángel Garrido and Urszula Wybraniec-Skardowska,
  575--597. Cham:Switzerland: Birkhäuser.
  \url{https://philpapers.org/rec/URSRIU}
  \url{https://books.google.com/books?id=ueZfDwAAQBAJ&pg=PA578}
  \url{https://philarchive.org/rec/WYBRIU}
  \url{https://philpapers.org/rec/WYBRIU-2}
  \url{https://web.archive.org/web/20211020011251/https://www.logic.ifil.uz.zgora.pl/refutation/files/wybraniec-skardowska-rejection.pdf}.

\bibitem[Łukasiewicz(1957)]{lukaziewicz:1957}
Łukasiewicz, Jan. 1957. \emph{Aristotle's Syllogistic from the Standpoint of
  Modern Formal Logic}. 2nd ed. Oxford:UK: Oxford University Press.

\bibitem[Łukasiewicz and Tarski(1930)]{lukaziewicz:1930}
Łukasiewicz, Jan, and Alfred Tarski. 1930. ``Untersuchungen über den
  {A}ussagenkalkül.'' In \emph{Sprawozdania z {P}osiedzeń {T}owarzystwa
  {N}aukowego {W}arszawskiego}, Warsaw:Poland, 30--50. Towarzystwa Naukowego
  Warszawskiego.
  \url{https://web.archive.org/web/20140821042636/http://chc60.fgcu.edu/images/articles/lukasiewicztarski.pdf}
  \url{https://rcin.org.pl/impan/dlibra/publication/edition/50601}
  \url{https://books.google.com/books?id=AOPWAAAAMAAJ&q=metalogic}
  \url{https://books.google.com/books?id=2uhra9PEFZsC&pg=PA38&dq=%22better+metalogic%22}
  \url{https://books.google.com/books?id=2uhra9PEFZsC&q=%22better+metalogic%22}.

\end{thebibliography}
\bibliographystyle{tfcad}
\end{document}